%% file: main.tex
\documentclass[
  aip
  12pt,
  superscriptaddress,
  amsfonts,
  amssymb,
  amsmath,
  letterpaper,
  nofootinbib,
  nobibnotes,
  noshowpacs
  noshowkeyws
  titlepage
  eqsecnum
  floatfix
]{revtex4-2}

\usepackage{graphicx}
\graphicspath{ {./IMAGES/}{../IMAGES/}{../IMAGES/Beamer/}{../IMAGES/Feynmann/} }

\usepackage{epsfig}
\usepackage{subfig}
\usepackage{sidecap}
\usepackage{keyval}
\usepackage{wrapfig}

\usepackage{dcolumn}
\usepackage{bm}

\usepackage{textcomp}

\usepackage{gensymb}

\usepackage{hyperref}

\usepackage{cleveref}
\crefname{figure}{Fig.\@}{Figs.\@}
\crefname{equation}{Eq.\@}{Eqs.\@}
\crefname{section}{Sec.\@}{Secs.\@}
\crefname{table}{Table}{Tables}

\usepackage{color}

\usepackage{lineno}

\usepackage{siunitx}

\makeatletter

\renewcommand*{\p@subsection}{}

\renewcommand*{\p@subsubsection}{}
\makeatother

\begin{document}


\title{Two-Photon EXchange - TPEX}

\author{R.~Alarcon}
\affiliation{Arizona State University, Tempe, AZ, USA}
\author{R.~Beck}
\affiliation{Friedrich Wilhelms Universit\"at, Bonn, Germany}
\author{J.C.~Bernauer}
\affiliation{Stony Brook University, Stony Brook, NY, USA}
\affiliation{Riken BNL Research Center, Upton, NY, USA}
\author{M.~Broering}
\affiliation{Massachusetts Institute of Technology, Cambridge, MA, USA}
\author{E.~Cline}
\affiliation{Stony Brook University, Stony Brook, NY, USA}
\author{B.~Dongwi}
\affiliation{Hampton University, Hampton, VA, USA}
\author{I.~Fernando}
\affiliation{Hampton University, Hampton, VA, USA}
\author{M.~Finger}
\affiliation{Charles University, Prague, Czech Republic}
\author{M.~Finger~Jr.\@}
\affiliation{Charles University, Prague, Czech Republic}
\author{I.~Fri{\v s}{\v c}i\'c}
\affiliation{Massachusetts Institute of Technology, Cambridge, MA, USA}
\author{T.~Gautam}
\affiliation{Hampton University, Hampton, VA, USA}
\author{D.K.~Hasell}
\affiliation{Massachusetts Institute of Technology, Cambridge, MA, USA}
\email[Corresponding author: ]{hasell@mit.edu}
\author{O.~Hen}
\affiliation{Massachusetts Institute of Technology, Cambridge, MA, USA}
\author{J.~Holmes}
\affiliation{Arizona State University, Tempe, AZ, USA}
\author{T.~Horn}
\affiliation{Catholic University of America, Washington, DC, USA}
\author{E.~Ihloff}
\affiliation{Massachusetts Institute of Technology, Cambridge, MA, USA}
\author{R.~Johnston}
\affiliation{Massachusetts Institute of Technology, Cambridge, MA, USA}
\author{J.~Kelsey}
\affiliation{Massachusetts Institute of Technology, Cambridge, MA, USA}
\author{M.~Kohl}
\affiliation{Hampton University, Hampton, VA, USA}
\author{T.~Kutz}
\affiliation{The George Washington University, Washington, DC, USA}
\author{I.~Lavrukhin}
\affiliation{University of Michigan, Ann Arbor, MI, USA}
\author{S.~Lee}
\affiliation{Massachusetts Institute of Technology, Cambridge, MA, USA}
\author{W.~Lorenzon}
\affiliation{University of Michigan, Ann Arbor, MI, USA}
\author{F.~Maas}
\affiliation{Johannes Gutenberg Universit\"at, Mainz, Germany}
\author{H.~Merkel}
\affiliation{Johannes Gutenberg Universit\"at, Mainz, Germany}
\author{R.G.~Milner}
\affiliation{Massachusetts Institute of Technology, Cambridge, MA, USA}
\author{P.~Moran}
\affiliation{Massachusetts Institute of Technology, Cambridge, MA, USA}
\author{J.~Nazeer}
\affiliation{Hampton University, Hampton, VA, USA}
\author{T.~Patel}
\affiliation{Hampton University, Hampton, VA, USA}
\author{M.~Rathnayake}
\affiliation{Hampton University, Hampton, VA, USA}
\author{R.~Raymond}
\affiliation{University of Michigan, Ann Arbor, MI, USA}
\author{R.P.~Redwine}
\affiliation{Massachusetts Institute of Technology, Cambridge, MA, USA}
\author{A.~Schmidt}
\affiliation{The George Washington University, Washington, DC, USA}
\author{U.~Schneekloth}
\affiliation{Deutsches Elektronen-Synchrotron, Hamburg, Germany}
\author{D.~Sokhan}
\affiliation{University of Glasgow, Glasgow, Scotland}
\author{M.~Suresh}
\affiliation{Hampton University, Hampton, VA, USA}
\author{C.~Vidal}
\affiliation{Massachusetts Institute of Technology, Cambridge, MA, USA}
\collaboration{The TPEX Collaboration}
\noaffiliation{}

\date{\today}

\begin{abstract}
    We propose a new measurement of the ratio of positron-proton to
    electron-proton, elastic scattering at DESY to determine the
    contributions beyond single-photon exchange, which are essential
    to the QED description of the most fundamental process in hadronic
    physics.  A 20~cm long liquid hydrogen target together with the
    extracted beam from the DESY synchrotron would yield an average
    luminosity of
    $2.12\times10^{35}$~cm$^{-2}\cdot$s$^{-1}\cdot$sr$^{-1}$
    ($\sim200$ times the luminosity achieved by OLYMPUS). A
    commissioning run at 2~GeV followed by measurements at 3~GeV would
    provide new data up to $Q^2=4.6$~(GeV/$c$)$^2$ (twice the range of
    current measurements).  Lead tungstate calorimeters would be used
    to detect the scattered leptons at polar angles of $30\degree$,
    $50\degree$, $70\degree$, $90\degree$, and $110\degree$.  The
    measurements could be scheduled to not interfere with the
    operation of PETRA.  We present rate estimates and simulations for
    the planned measurements including background considerations.
    Initial measurements at the DESY test beam facility using prototype lead
    tungstate calorimeters in 2019, 2021, and 2022 were made to check the Monte
    Carlo simulations and the performance of the calorimeters.  These
    tests also investigated different readout schemes (triggered and
    streaming). Various upgrades are possible to shorten the running
    time and to make higher beam energies and thus greater $Q^2$
    ranges accessible.
\end{abstract}

\maketitle

\tableofcontents

\clearpage


\section{Introduction}
\label{sec:intro}

Elastic lepton-proton scattering is a fundamental process that should
be well described by QED.  Understanding this interaction is important
to the scientific programs at FAIR, Jefferson Lab, and the future
electron-ion collider (EIC) planned for Brookhaven.  It is described
theoretically in the Standard Model by a perturbative expansion in
$\alpha=\frac{1}{137}$ with radiative corrections. For more than half
a century it has been assumed that the leading single-photon exchange
term adequately describes the scattering process and that higher-order
terms are negligible.  However, recent experiments at Jefferson Lab
have been widely interpreted as evidence that higher order terms are
significant in elastic electron-proton scattering and must be included
to correctly extract the proton elastic form factors.  Recent
experiments, including the OLYMPUS experiment at DESY, show little
evidence for significant contributions beyond single photon exchange
up to $Q^2\approx2.3$~(GeV/c)$^2$.  It is essential that the QED
expansion be studied experimentally at higher $Q^2$ comparing the
positron and electron scattering cross section to determine the
contribution of higher order terms not normally included in radiative
corrections.

\begin{figure}[!ht]
  \centering
  \includegraphics[width=0.85\textwidth, viewport=0 0 512 308, clip ]
    {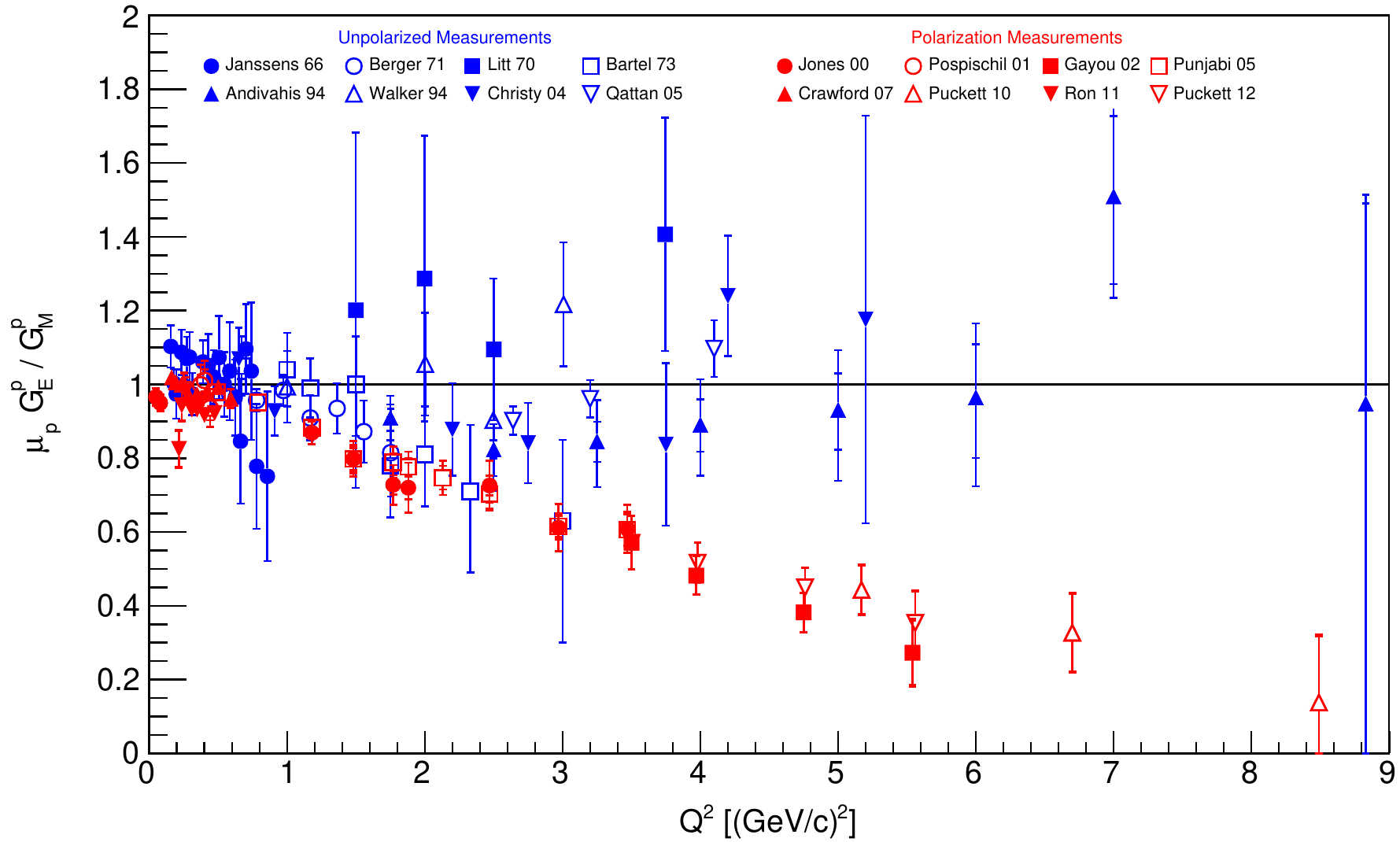}
  \caption{Proton form factor ratio measured using
    unpolarized~\citep{Janssens:1965kd, Berger:1971kr, Litt:1969my,
      Bartel:1973rf, Andivahis:1994rq, Walker:1993vj, Christy:2004rc,
      Qattan:2004ht} (blue) and polarized~\citep{Jones:1999rz,
      Pospischil:2001pp, Gayou:2001qt, Punjabi:2005wq,
      Crawford:2007dl, Puckett:2010ac, Ron:2011rd, Puckett:2011xg}
    (red) techniques.}
    \label{fig:ratio}
\end{figure}
The proton form factors, $G_{E}^{p}$ and $G_{M}^{p}$, have
historically been envisaged as very similar and are often modeled by
the same dipole form factor.  Measurements over the past 50~years
using the unpolarized Rosenbluth separation technique yielded a ratio,
$\mu^{p}\,G_{E}^{p}/G_{M}^{p}$, close to unity over a broad range in
$Q^{2}$ shown by the blue data points in~\cref{fig:ratio}.  This
supported the idea that $G_{E}^{p}$ and $G_{M}^{p}$ are
similar. However, recent measurements using polarization techniques
revealed a completely different picture with the ratio decreasing
rapidly with increasing $Q^{2}$ as shown by the red data points
in~\cref{fig:ratio}.

The most commonly proposed explanation for this discrepancy is
``hard'' two-photon exchange contributions beyond the standard
radiative corrections to one-photon exchange. 
\begin{figure}[!ht]
  \centering
  \includegraphics[width=0.4\textwidth, viewport=32 356 563 508, clip ]
    {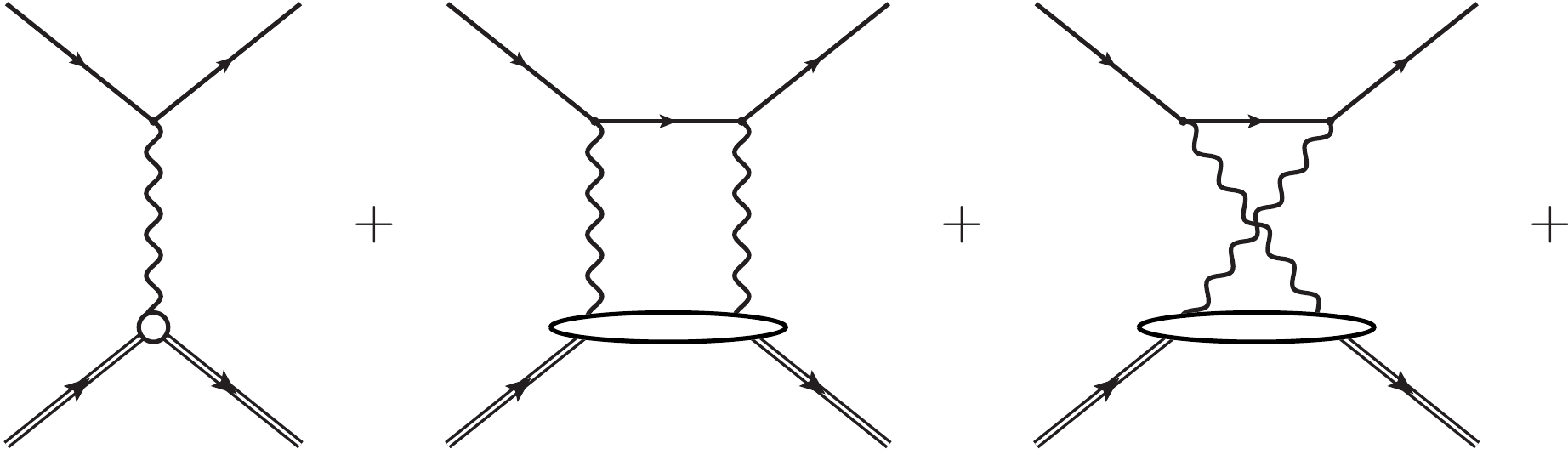}
  \caption{Feynman diagrams for one- and two-photon exchange. Further
    diagrams for bremsstrahlung, vertex, self-energy, and vacuum
    polarization radiative corrections are not shown but must also be
    included in calculations.}
  \label{fig:epLine1}
\end{figure}
Two-photon exchange, TPE, (see~\cref{fig:epLine1}) is generally included as 
part of the radiative corrections when analyzing electron-proton
scattering. However, it is usually only included in the ``soft'' limit
where one of the two photons, in the diagrams with two photons, is
assumed to carry negligible momentum and the intermediate hadronic
state remains a proton. To include ``hard'' two-photon exchange, a
model for the off-shell, intermediate hadronic state must also be
included, making the calculations difficult and model dependent.

In the Born or single photon exchange approximation the elastic
scattering cross section for leptons from protons is given by the
reduced Rosenbluth cross section,
\begin{equation}
    \frac{d\sigma_{e^\pm p}}{d\Omega}={\frac{d\sigma}{d\Omega}}_{Mott}
    \frac{\tau {G_M^p}^2 +\epsilon {G_E^p}^2}{\epsilon(1+\tau)},
\end{equation}
where: $\tau=\frac{Q^2}{4 M_p^2}$ and $\epsilon=(1 +2(1+\tau)
\tan^2{\frac{\theta_l}{2}} )^{-1}$.

To measure the ``hard'' two-photon contribution, one can measure the
ratio $R_{2\gamma}=\sigma_{e^{+}p}/\sigma_{e^{-}p}$ at different
values of $Q^{2}$ and $\epsilon$. Note, the interference terms between
one- and two-photon exchange change sign between positron and electron
scattering and this cross section ratio provides a measure of the
two-photon exchange contribution.

\begin{figure}[!ht]
  \centering
  \includegraphics[width=0.7\textwidth, viewport=0 34 272 217, clip ]
    {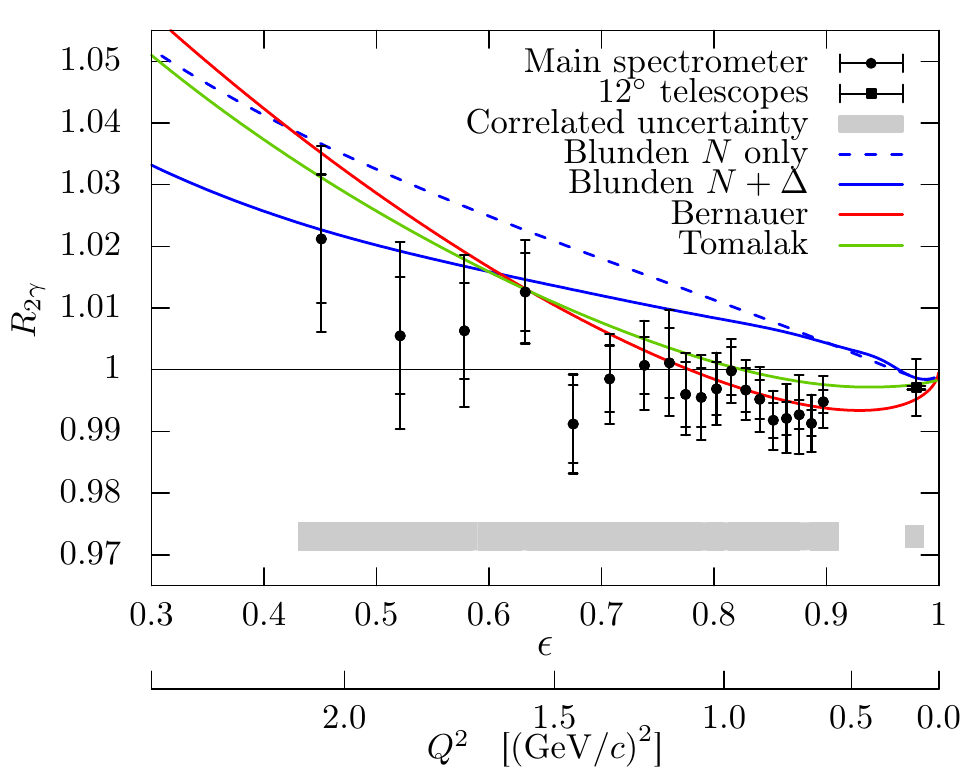}
  \caption{OLYMPUS results for $R_{2\gamma}$ as a function of
    $\epsilon$.  Inner error bars are statistical while the outer
    error bars include uncorrelated systematic uncertainties added in
    quadrature.  The gray band is correlated systematic uncertainty.}
  \label{fig:OLYMPUS}
\end{figure}

The results from the OLYMPUS experiment~\citep{Henderson:2016dea} are
shown in~\cref{fig:OLYMPUS} together with various calculations.
The deviation of the results from unity are small, on the order of
1\%, and are below unity at large $\epsilon$ and rising with
decreasing $\epsilon$.  The dispersive calculations of
Blunden~\citep{Blunden:2017nby} are systematically above the OLYMPUS
results in this energy regime.  The results below unity cannot be
explained by current QED calculations. The phenomenological prediction
from Bernauer~\citep{Bernauer:2013tpr} and the subtractive dispersion
calculation from Tomalak~\citep{Tomalak:2014sva} are in better
agreement with the OLYMPUS results but appear to rise too quickly as
$\epsilon$ decreases.  There is some indication that TPE increases
with decreasing $\epsilon$ or increasing $Q^{2}$, suggesting that a
significant ``hard'' two-photon contribution might be present at lower
$\epsilon$ or higher $Q^{2}$.

Two other experiments, VEPP-3~\citep{Rachek:2014fam} and
CLAS~\citep{Adikaram:2014ykv}, also measured the ``hard'' two-photon
exchange contribution to electron-proton elastic scattering.
\begin{figure}[!ht]
  \centering
  \begin{subfloat}[]{\label{fig:diffblunden}
    \includegraphics[width=0.48\textwidth, viewport=0 0 512 354, clip]
  {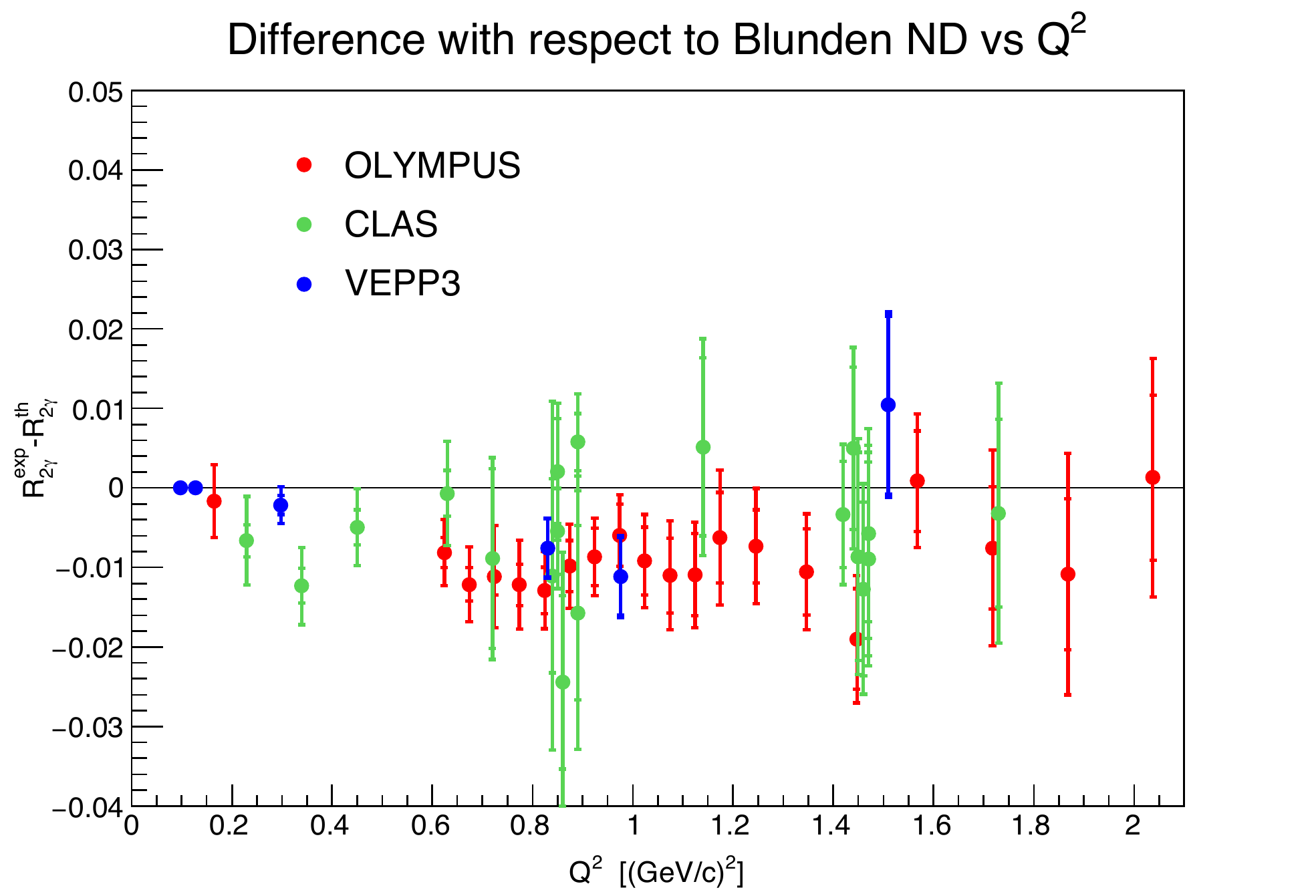}}
  \end{subfloat}
  \hfill
  \begin{subfloat}[]{\label{fig:diffbernauer}
    \includegraphics[width=0.48\textwidth, viewport=0 0 512 354, clip]
  {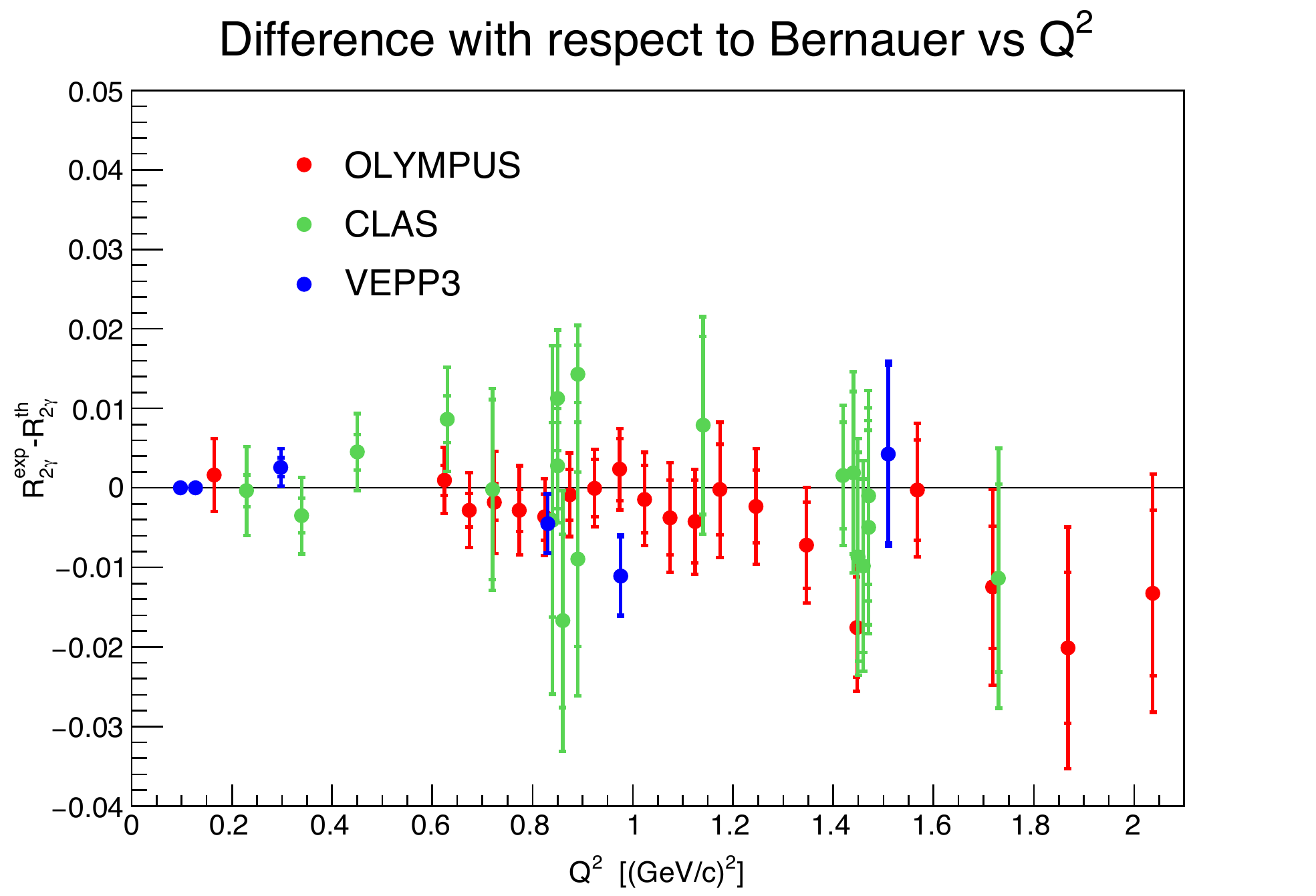}}
  \end{subfloat}
  \caption{Difference between the results from the three recent
    experiments and (a) Blunden's N+$\Delta$ calculation and (b)
    Bernauer's prediction.}.
\end{figure}
It is difficult to compare the results from the three experiments
directly since the measurements are at different points in the
$(\epsilon, Q^2)$ plane.  To partially account for this, we can
compare all the two-photon exchange results by taking the difference
with respect to a selected calculation evaluated at the correct
$(\epsilon, Q^2)$ for each data point.  This is shown
in~\cref{fig:diffblunden} for Blunden's calculation and in
~\cref{fig:diffbernauer} for Bernauer's phenomenological prediction,
plotted versus $Q^2$.  In these views, the results from the three
experiments are shown to be in reasonable agreement supporting the
previous conclusions.
 
The results from the three TPE experiments are all below
$Q^2=2.3$~(GeV/c)$^2$.  In this regime the discrepancy in the form
factor ratios is not obvious, so the small ``hard'' TPE contribution
measured is consistent with the measured form factor ratios.  The
suggested slope with $\epsilon$ indicates TPE may be important at
smaller $\epsilon$ or higher $Q^2$.  But, since this slope appears to
deviate from Bernauer's phenomenological prediction, which fits the
observed discrepancy, it may also suggest that ``hard'' TPE, while
contributing, may not explain all of the observed form factor
discrepancy.

\begin{figure}[!ht]
  \centering
  \includegraphics[width=0.6\textwidth] {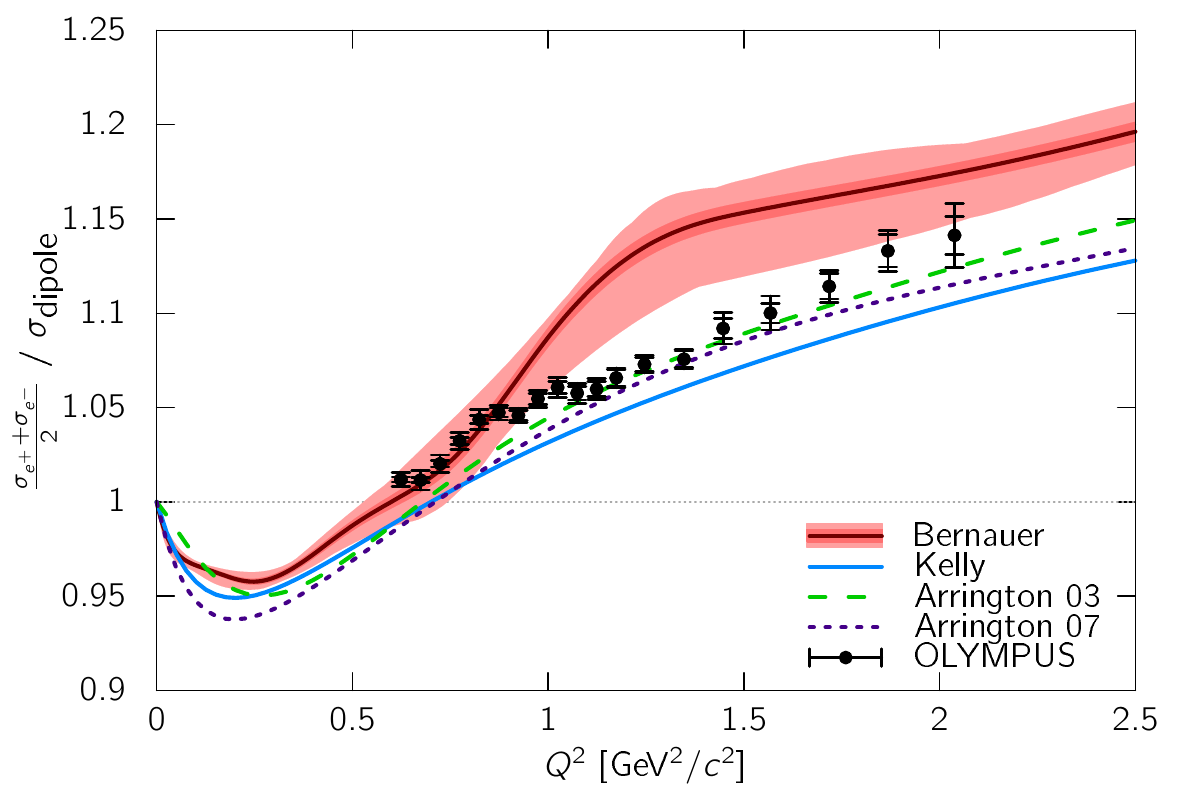}
  \caption{The charge-averaged yield for elastic lepton-proton
    scattering from the OLYMPUS experiment~\citep{Bernauer:2020vue}.}
  \label{fig:yield}
\end{figure}

Recently, the OLYMPUS data has also been analyzed to determine the
charge-averaged yield for elastic lepton-proton
scattering~\citep{Bernauer:2020vue}.  The result is shown in
\cref{fig:yield}.
This measurement is insensitive to any charge-odd radiative
corrections including ``hard'' two-photon exchange and thus provides a
better measure of the proton form factors.  The data shown covers an
important range of $Q^2$ where the $G_M$ form factor changes slope.
The calculations by Kelly~\citep{Kelly:2004hm} and
Arrington~\citep{Arrington:2003ck,Arrington:2007ux} appear to be in
better agreement with the data, but Bernauer's global
fit~\citep{Bernauer:2013tpr} should be redone to incorporate all the
OLYMPUS data.

\begin{figure}[htbp]
  \centering
  \includegraphics[width=0.7\textwidth] {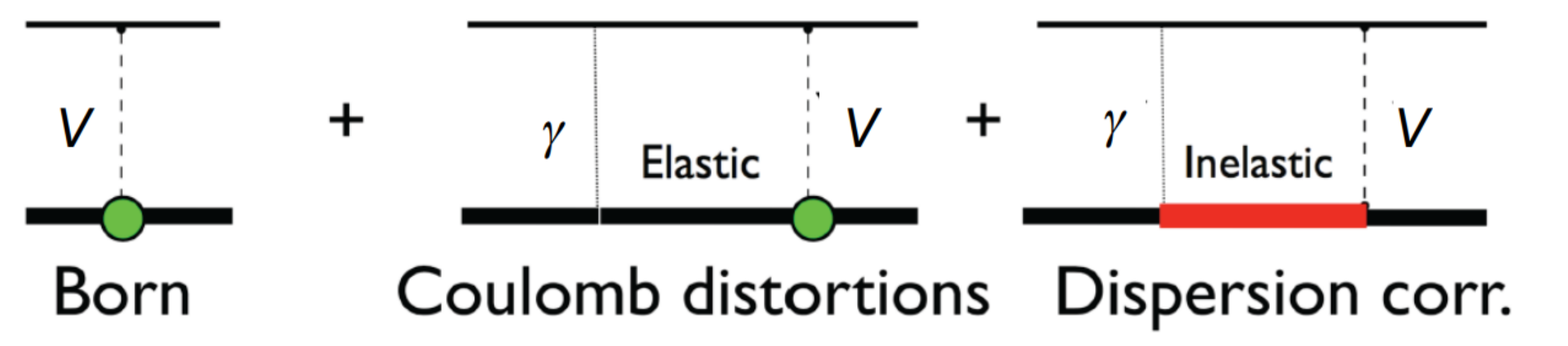}
  \caption{More general electroweak box diagram that is important in
    many fundamental nuclear physics processes.}
  \label{fig:ewbox}
\end{figure}

The two-photon exchange diagram in the QED expansion for electron
scattering is an example of the more generic electroweak photon-boson
diagram (see \cref{fig:ewbox} where $V = Z^0, W^\pm, {\rm or
}\ \gamma$) which enters into a number of fundamental processes in
subatomic physics.  The $\gamma-Z$ box is a significant contribution
to the asymmetry in parity-violating electron scattering and the
$\gamma-W^\pm$ box is an important radiative correction in
$\beta-$decay which must be implemented to extract $V_{ud}$ of the
Standard Model from $0^+ \rightarrow 0^+$ super-allowed nuclear
$\beta$-decays.  A workshop~\citep{EWBox:2017aa} was held at the
Amherst Center for Fundamental Interactions in September 2017,
attended by physicists from these different subfields, to discuss the
Electroweak Box. A white paper is in preparation.

The proton form factors are fundamental to hadronic physics.
Understanding the QED expansion, the role of two-photon exchange, and
the scale of radiative corrections at higher $Q^2$ will be crucial in
future studies at FAIR, JLab, EIC, and elsewhere.  The charge-averaged
yield eliminates all charge-odd radiative corrections including the
leading terms of two-photon exchange, which cannot be calculated with
current theories.  Measuring the ratio of positron-proton to
electron-proton scattering is sensitive to the charge-odd radiative
corrections and insensitive to the charge-even radiative corrections.
Together they help to study radiative corrections and unravel the
proton form factors.  TPEX, like OLYMPUS, will provide both these
measurements at higher $Q^2$.
    
The discrepancy in the form factor ratio has not been resolved and the
role played by two-photon exchange continues to be widely discussed
within the nuclear physics community~\citep{EWBox:2017aa,
  Afanasev:2017gsk, Blau:2017du, NSTAR:2017aa, JPos:2017aa}. Further
measurements and theoretical work on the role of two-photon exchange
on the proton form factors are clearly needed.  However, measurements
at higher $Q^2$ and smaller $\epsilon$, where the discrepancy is clear
and TPE are expected to be larger, are difficult as the cross sections
decrease rapidly.  In addition, there are not many laboratories
capable of providing both electron and positron beams with sufficient
intensity.

The best, and for the foreseeable future only, opportunity is at
DESY. This proposal outlines an experiment that could measure
$R_{2\gamma}$ at $Q^{2}$ up to 4.6~(GeV/c)$^{2}$ or higher, and
$\epsilon$ below 0.1 where the form factor discrepancy is clear
(see~\cref{fig:ratio}). Such an experiment would overlap with the
existing OLYMPUS data as a cross-check and would map out the
two-photon exchange contribution over a broad range in $Q^{2}$ and
$\epsilon$ to provide data to constrain theoretical calculations.

The following sections describe the proposed site for the TPEX
experiment at DESY, the experimental configuration with its liquid
hydrogen target, lead tungstate calorimeters, GEM detectors,
luminosity monitor, beamdump/Faraday cup, electronics and data
acquisition, and possible improvements.  Sources of background are
considered together with solutions and Monte Carlo simulations are
presented.  The appendices give more background
plots, properties of hydrogen and lead tungstate, some useful numbers
for this proposal, and references.


\section{DESY}
\label{sec:DESY}

One of the primary requirements for measuring $R_{2\gamma}$ is high
intensity positron and electron beams at energies of several GeV
available for nuclear and particle physics applications. DESY is
effectively the only high energy physics laboratory currently capable
of such intense positron beams. The DESY~II synchrotron can provide
extracted beams of up to 30~nA of positrons and up to 60~nA of
electrons at energies between 0.5 and 6.3~GeV with a bunch frequency
of 12.5~Hz.

A proposal, currently under consideration at DESY for an extension to
the present test beam facility~\citep{Diener:2018qap}, to include an
extracted lepton beam from DESY~II provides a unique opportunity to
investigate two-photon exchange. The extracted beam would only be
available when DESY-II is not needed for the operation of PETRA~III.
For our purposes the electron and positron beams would be used
directly at 2.0~GeV and 3.0~GeV with an option for higher energies in
the future.

The current operation of PETRA~III uses only electrons. That would
restrict the availability of positrons to times when PETRA~III is not
operating due to scheduled maintenance or shutdown periods. Hopefully
this is not an insurmountable problem and we believe our experiment
can be successfully carried out in the shutdown periods.
Commissioning can be done with just electrons if necessary.  If the
storage ring PETRA~III is running in ``top up'' mode (fills every
$\sim30$~s) we would not be able to run parasitically.  For ``non-top
up'' mode (fills every $\sim240$~s) it might be possible to have the
extracted beam for TPEX between fills for PETRA~III.

If the modification to the test beam facility in Hall~2 provides a
new, extracted beam area; this would allow a left/right symmetric
detector arrangement that is much preferred for this proposal to
reduce systematics and to increase count rate.

This proposal requires a significant effort from DESY:
\begin{enumerate}
  \item The positron production target has been removed.  This would
    need to be reinstalled.
  \item A new, extracted beam area would have to be assembled.  Two options are possible:
  \begin{itemize}
    \item[A -]Hall~2
    \begin{itemize}
      \item[-]The floor space is currently
    occupied by another group that would have to be relocated.
      \item[-] The ``kicker'' would have to be moved from its current
    location on DESY~II to one suitable for providing beams to the new
    area.
      \item[-]The shielding wall around DESY~II would have to be disassembled and reassembled with a beamline incorporated to the new area.
    \end{itemize}
    \item[B -]R-Weg
      \begin{itemize}
        \item[-]The transfer line previously used for DORIS would have to be re-established for a new experimental area.
        \item[-]A new area, possibly a specially designed experimental area would have to be developed.
      \end{itemize}
   \end{itemize}
   \item For both options beamline elements (quadrupoles, steering magnets, vacuum pumps, valves, collimators, beam dump, {\it etc.}) would have to
    be found or produced and then installed.
  \item The new extracted beam area would need shielding walls,
    infrastructure services like power and water, an access maze with
    interlocks, and a new counting hut.
  \item Everything would need to be surveyed and aligned and then
    tests performed to satisfy all safety requirements.
\end{enumerate}

In addition to enabling the TPEX experiment, an extracted beam
facility at DESY would allow other experiments, detector development,
and material studies to be performed.  Another interesting experiment
would be Deeply Virtual Compton Scattering, DVCS.  This could also use
the TPEX liquid hydrogen target and lead tungstate calorimeters but with
a different configuration to allow the scattered
lepton and recoil proton to be detected in coincidence.  Other nuclear physics
measurements could also benefit from comparing electron and positron
scattering.


\section{Proposed Experiment}
\label{sec:exper}

\begin{figure}[htbp]
  \centering
  \includegraphics[width=0.7\textwidth] {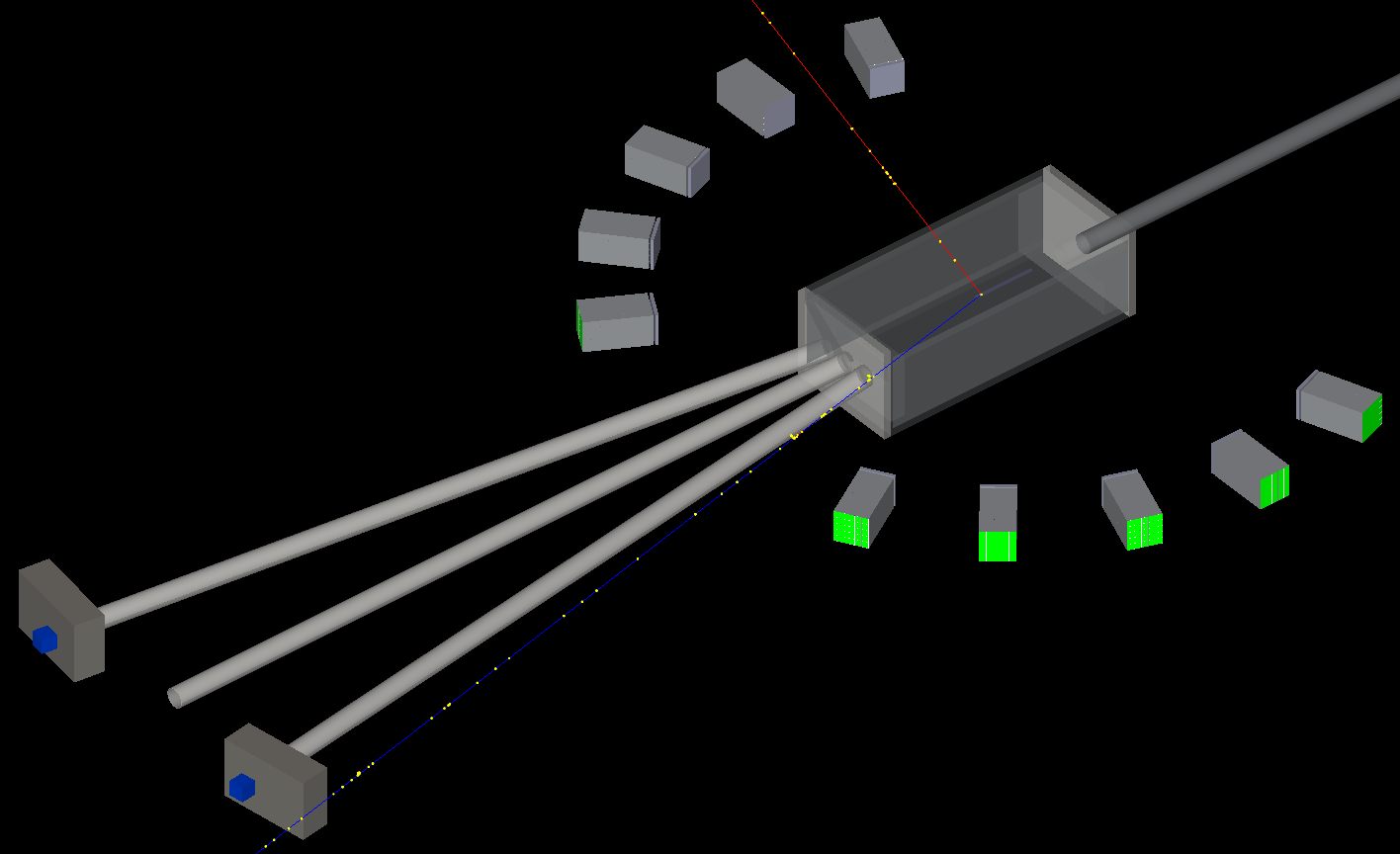}
  \caption{Geant4 simulation of a proposed TPEX target, scattering
    chamber, and detector configuration including the luminosity
    monitors and beamlines.  The lepton beam would enter through the
    beamline in the upper-right, traverse the target cell, and scatter
    into the detectors or continue straight to the beamdump.}
  \label{fig:TPEX}
\end{figure}
The proposed experimental configuration has ten $5\times5$~arrays of
lead tungstate crystals at polar angles of $30\degree$, $50\degree$,
$70\degree$, $90\degree$, and $110\degree$ left and right of the beam
axis with the front face of the calorimeter modules at a radius of 1~m
from the target. Other configurations are possible and can be
optimized with Monte Carlo studies. A simple schematic for this
arrangement is shown in~\cref{fig:TPEX}.  The electron or positron
beam enters the scattering chamber along the beamline (upper-right)
and passes through the 20~cm long liquid hydrogen target before
exiting the scattering chamber into another section of beam line
leading to the beamdump.  At $\pm8\deg$ there are 3~m long beampipes
connecting the scattering chamber to the lead collimators before the
Cherenkov detectors used to monitor the luminosity.  These beamlines
are under vacuum and are used to reduce the multiple scattering for
the relatively low energy (30--50~MeV) M{\o}ller and Bhabha scattered
leptons.

Using just the central $3\times3$~array of the $5\times5$ array of crystals
to define the acceptance
yields a solid angle of 3.6~msr at each angle. With a 20~cm long
liquid hydrogen target the acceptance covers $\pm5.7\degree$ in polar
and azimuthal angle thus data is averaged over a small range in
$Q^{2}$ and $\epsilon$.

We propose to commission the experiment using 2~GeV electrons.  We do
this to debug the electronics, detectors, and data acquisition system
taking advantage of the relatively high cross section at 2~GeV.  We
would require about 2 weeks of beam time for this commissioning after
the experiment was installed and surveyed.  We would also like a brief
run (few days) with positrons to verify that the beam alignment and
performance do not change with positron running.  The commissioning
run (including a few days with positrons) would also allow a
crosscheck of the OLYMPUS data at $30\degree$, $50\degree$, and
$70\degree$ and give a modest extension in $Q^2$ up to
2.7~(GeV/$c$)$^2$.

\Cref{tb:2GeV} shows $Q^{2}$, $\epsilon$, differential cross section,
and event rate expected for one day of running for the proposed
left/right symmetric configuration with 2~GeV lepton beams averaging
40~nA on a 20~cm liquid hydrogen target and using just the central
$3\times3$ array of crystals to calculate the acceptance area.
\begin{table}[htbp]
  \centering
  \begin{tabular*}{0.6\textwidth}{@{\extracolsep{\fill}}ccccc}
    $\theta$&$Q^{2}$&$\epsilon$&$d\sigma/d\Omega$&Events/day\\
            &(GeV/c)$^{2}$&&fb\\
    \hline
    $30\degree$&0.834&0.849&$2.41\times10^{7}$&$3.16\times10^{6}$\\
    $50\degree$&1.62&0.611&$7.66\times10^{5}$&$1.01\times10^{5}$\\
    $70\degree$&2.19&0.386&$1.00\times10^{5}$&$1.32\times10^{4}$\\
    $90\degree$&2.55&0.224&$2.81\times10^{4}$&$3.70\times10^{3}$\\
    $110\degree$&2.78&0.120&$1.22\times10^{4}$&$1.61\times10^{3}$\\
  \end{tabular*}
  \caption{Kinematics, cross section, and events expected in one day
    for an incident lepton beam of 2~GeV and 40~nA averaged current on
    a 20~cm liquid hydrogen target.}
  \label{tb:2GeV}
\end{table}

The TPEX experiment proper would run at 3.0~GeV and would require
approximately 6~weeks (2 weeks with electrons and 4 weeks with
positrons in total) to collect the required statistics.  \Cref{tb:3GeV} shows
$Q^{2}$, $\epsilon$, differential cross section, and event rate
expected for one day of running for the proposed configuration with
3~GeV lepton beams. This would extend the measurements to
$Q^2=4.57$~(GeV/$c$)$^2$ where the form factor ratio discrepancy is
large. The 6~weeks could be divided into two three-week periods if
that was more convenient.  To minimize systematic we would like to switch
between positron and electron running as frequently as possible ({\it{e.g.}} 
1~day positron, 1~day electron, and 1~day positron repeating).
\begin{table}[htbp]
  \centering
  \begin{tabular*}{0.6\textwidth}{@{\extracolsep{\fill}}ccccc}
    $\theta$&$Q^{2}$&$\epsilon$&$d\sigma/d\Omega$&Events/day\\
            &(GeV/c)$^{2}$&&fb\\
    \hline
    $30\degree$&1.69&0.825&$2.41\times10^{6}$&$3.16\times10^{5}$\\
    $50\degree$&3.00&0.554&$6.51\times10^{4}$&$8.55\times10^{3}$\\
    $70\degree$&3.82&0.329&$8.94\times10^{3}$&$1.17\times10^{3}$\\
    $90\degree$&4.29&0.184&$2.65\times10^{3}$&$3.48\times10^{2}$\\
    $110\degree$&4.57&0.096&$1.20\times10^{3}$&$1.58\times10^{2}$\\
  \end{tabular*}
  \caption{Kinematics, cross section, and events expected in one day
    for an incident lepton beam of 3~GeV and 40~nA averaged current on
    a 20~cm liquid hydrogen target and 3.6~msr acceptance and a
    left/right symmetric detector configuration.}
  \label{tb:3GeV}
\end{table}

\begin{figure}[htbp]
  \centering \includegraphics[width=0.9\textwidth] {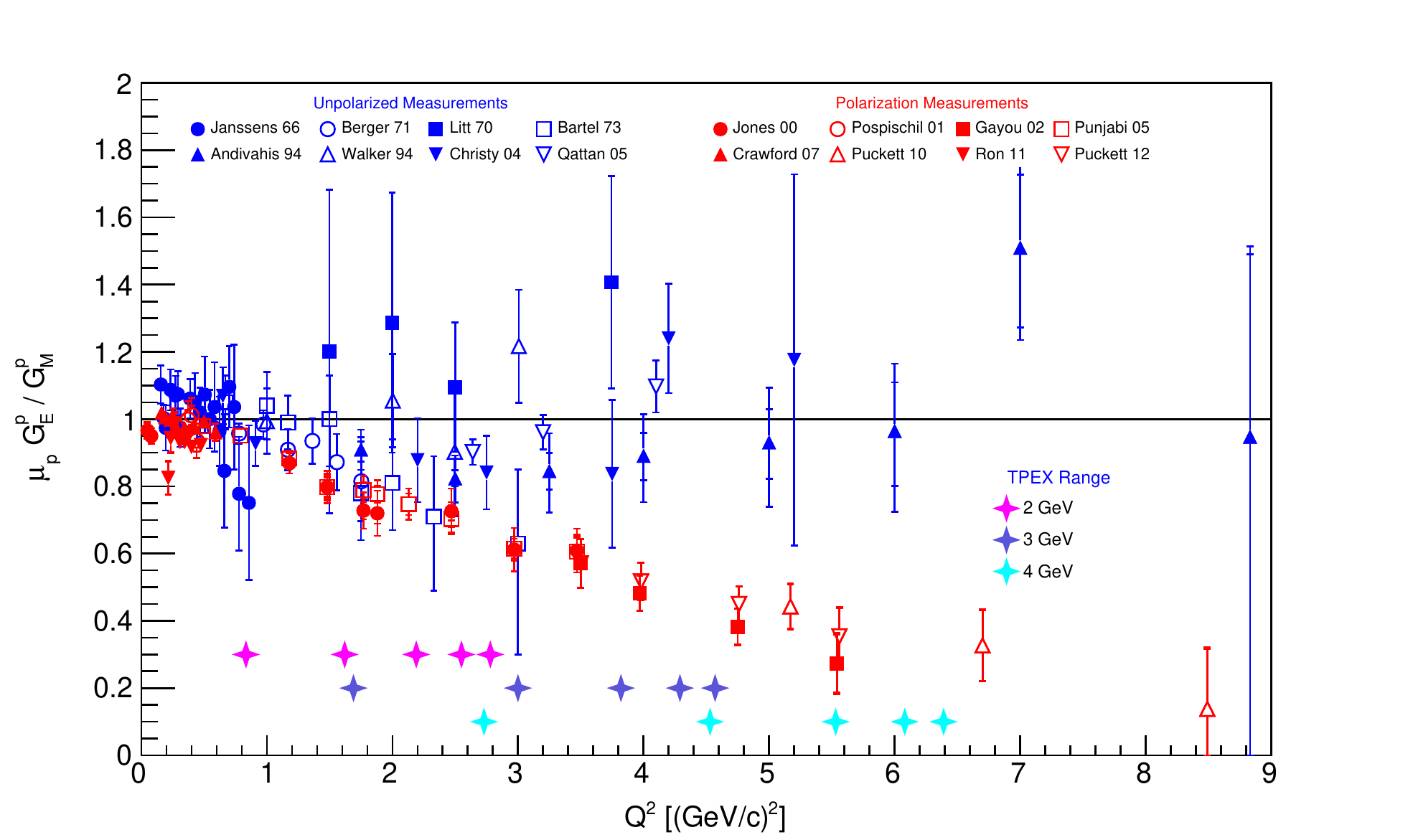}
  \caption{Proton form factor ratio as before but also showing the
    $Q^2$ range accessible with the proposed TPEX configuration at 2
    and 3~GeV.  The 4~GeV range would be possible with additional
    crystals.}
  \label{fig:Q2TPEX}
\end{figure}

The $Q^2$ range that the proposed TPEX experiment would be capable of
reaching is shown in \cref{fig:Q2TPEX} for the 2 and 3~GeV runs of
this proposal. The reach with TPEX can be seen in relation to the
discrepancy in the form factor ratio.  With additional crystals at
back angles the 4~GeV runs would also be possible in a reasonable time
frame.

\begin{figure}[htbp]
  \centering \includegraphics[width=0.9\textwidth] {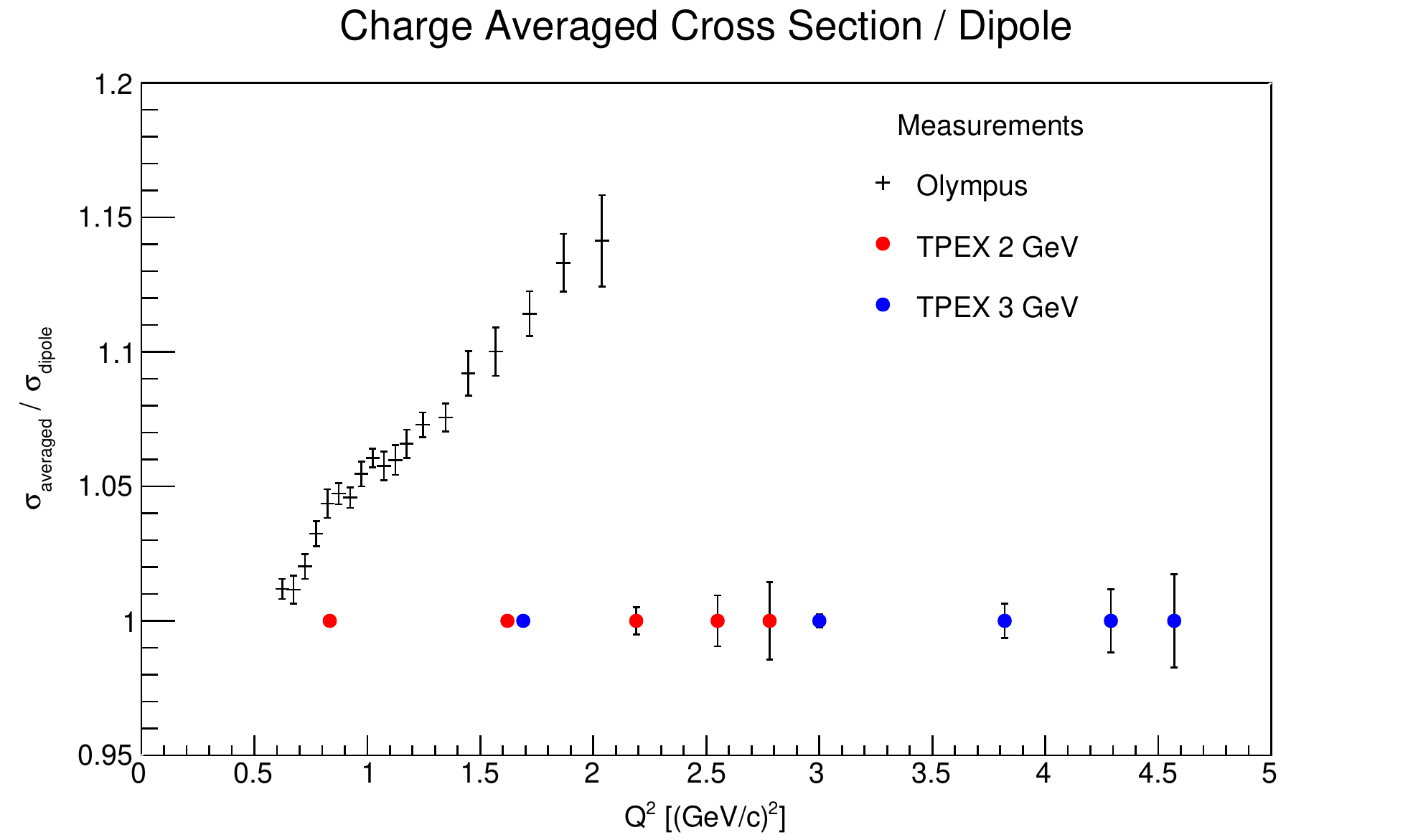}
  \caption{Charge-averaged cross section divided by the dipole cross
    section from OLYMPUS and expected uncertainties and coverage from
    TPEX at 2 and 3~GeV.}
  \label{fig:Qaverage}
\end{figure}

The TPEX experiment at DESY would also measure the charge-averaged
cross section just like the recent result from
OLYMPUS~\cref{fig:yield}.  As mentioned above this cross section is
insensitive to charge-odd radiative corrections including ``hard''
two-photon exchange terms.  Thus, it provides a more robust measure of
the proton form factors.  The expected charge-averaged cross section
uncertainties (assuming dipole cross section) are shown in
\cref{fig:Qaverage} for TPEX assuming 6 days of running at 2~GeV and 6
weeks of running at 3~GeV with only 50\% data collection efficiency.
The recent OLYMPUS results are also shown.


\section{Liquid Hydrogen Target and Scattering Chamber}
\label{sec:LH2}



\input cryotarget_WL


\section{Lead Tungstate Calorimeters}
\label{sec:cal}

For the proposed experiment we are leveraging the R\&D
experience~\citep{Zhu:1996tt,Zhu:2004dg} from the CMS experiment and
subsequent applications by the Bonn and Mainz groups at
CEBAF~\citep{Neyret:1999tr} and for PANDA~\citep{Albrecht:2015zma}. We
would start with ten $5\times5$~arrays of lead tungstate (PbWO$_{4}$)
crystals for a total of 250~crystals, some of which may be available
from Mainz. Other configurations are possible and will be investigated
with more detailed Monte Carlo simulations. 

Some properties for lead tungstate are provided in
\cref{sec:PbWOprop}.  We plan to use crystals
$2\times2\times20$~cm$^{3}$. The density is 8.3~g$\cdot$cm$^{-3}$ so
each crystal weighs around 664~g, or 16.6~kg for a $5\times5$ array of
crystals.  Lead tungstate has a radiation length $X_0=0.8904$~cm, so
these crystals are approximately 22.5~$X_{0}$ for good longitudinal
electromagnetic shower confinement.  The Moli\`{e}re radius is
1.959~cm, so using just the central $3\times3$~array of crystals for
acceptance, the outer ring of crystals contains the transverse shower
adequately. The nuclear interaction length for lead tungstate is
$\lambda_I=20.28$~cm, so the crystals are roughly
$0.986\ \lambda_I$. Again, other configurations are possible and
further studies and simulations are in progress. The energy resolution
obtained with lead tungstate for the lepton energy range of interest
is approximately 2\%. The Mainz Panda readout design uses Avalanche
Photo-Diodes (APD).  We are also considering SiPM and PMT readout
schemes.

Lead tungstate resolution varies with temperature.  To achieve the
best energy resolution, the crystal arrays should be maintained at a
constant temperature.  The best energy resolution has been obtained at
$-25\degree$~C. This requires refrigeration and complicates what
would otherwise be very simple and compact calorimeter modules.
Results from the test runs at the DESY test beam facility on prototype
lead tungstate calorimeters will be used to determine whether or not
such cooling is required or if adequate resolution can be obtained
with more modest cooling to have a stable temperature.

An alternative to lead tungstate is being investigated by T.~Horn at
Catholic University of America.  She is developing high-density,
ceramic glass crystals.  These would be approximately 15\% less dense
than lead tungstate, so a larger crystal might be required.  But the
ceramic glass is much easier to produce and would be 5--10 times
cheaper.  In addition, the ceramic glass is not as sensitive to
temperature, which would simplify the design.  We will be testing both
lead tungstate and ceramic glass in the future.

Clearly, the lead tungstate crystals will be a crucial part of the
TPEX experiment.  It will therefore be very important to test and
maintain the quality of the crystals whether they are produced by the
firm Crytur in the Czech Republic or obtained from existing supplies
in Europe or America.  The collaboration has colleagues from Charles
University in Prague, Czech Republic who have volunteered to take
responsibility for testing the crystals, verifying the quality and
maintaining a database for tracking the crystals from delivery to the
final calorimeter modules.


\section{GEM Detectors}
\label{sec:GEMs}

It is proposed to stack two Gas Electron Multiplier (GEM) detector
layers with two-dimensional readout in front of each calorimeter.
Thin absorbers will be placed between the target and the GEMs to stop
low-energy M{\o}ller or Bhabha leptons.  The GEMs provide spatial
information of the traversing charged particle at the 100 micrometer
precision level. The hits on two GEM elements are used to form a track
segment providing directional information between the impact point on
the calorimeter and the event origin in the target.  This serves to
suppress charged-particle backgrounds from regions other than the
target. Also, the GEMs are insensitive to neutral particles, hence
they provide a veto against photons and neutrons. Using the
calorimeter hit as a third tracking point will allow a measure of the
efficiency of each.

An active area of slightly more than 20x20 cm$^2$ is required to fully
cover the area of the calorimeter entrance. A total of 20 elements is
required to instrument ten calorimeter arms. The standard readout
strip pitch of 0.4 mm results in 500 channels per axis, or 1,000
channels per GEM element. The full experiment would have 20,000
channels. Since the occupancy will be at the few percent level at
most, zero suppression will reduce the amount of recorded data
substantially.

The Hampton group has developed GEM detectors for OLYMPUS, MUSE and
DarkLight. Recently, the group has established the novel scheme (NS2)
of assembling GEM detectors without gluing, while stretching GEM foils
mechanically within a double frame structure, for the first time for
nuclear physics applications. The scheme makes the assembly fast and
low risk, such that even a larger number of GEM elements can be
produced fairly easily.

\section{Luminosity and Beam Alignment Monitor}
\label{sec:Lumi}

The relative luminosity between the electron and positron running
modes is the crucial normalization for the proposed measurement. The
luminosity could be monitored by a pair of small-angle detectors
positioned downstream on either side of the beamline. This approach
was also used in the OLYMPUS experiment \cite{Benito:2016cmp}, and
based on the lessons learned from that experiment, could be improved
substantially. Given the running conditions of the proposed
measurement, we favor a pair of quartz Cherenkov counters positioned
$8\degree$ from the beamline to monitor the rates of M{\o}ller and
Bhabha scattering from atomic electrons in the target.

\begin{figure}
    \centering
    \includegraphics{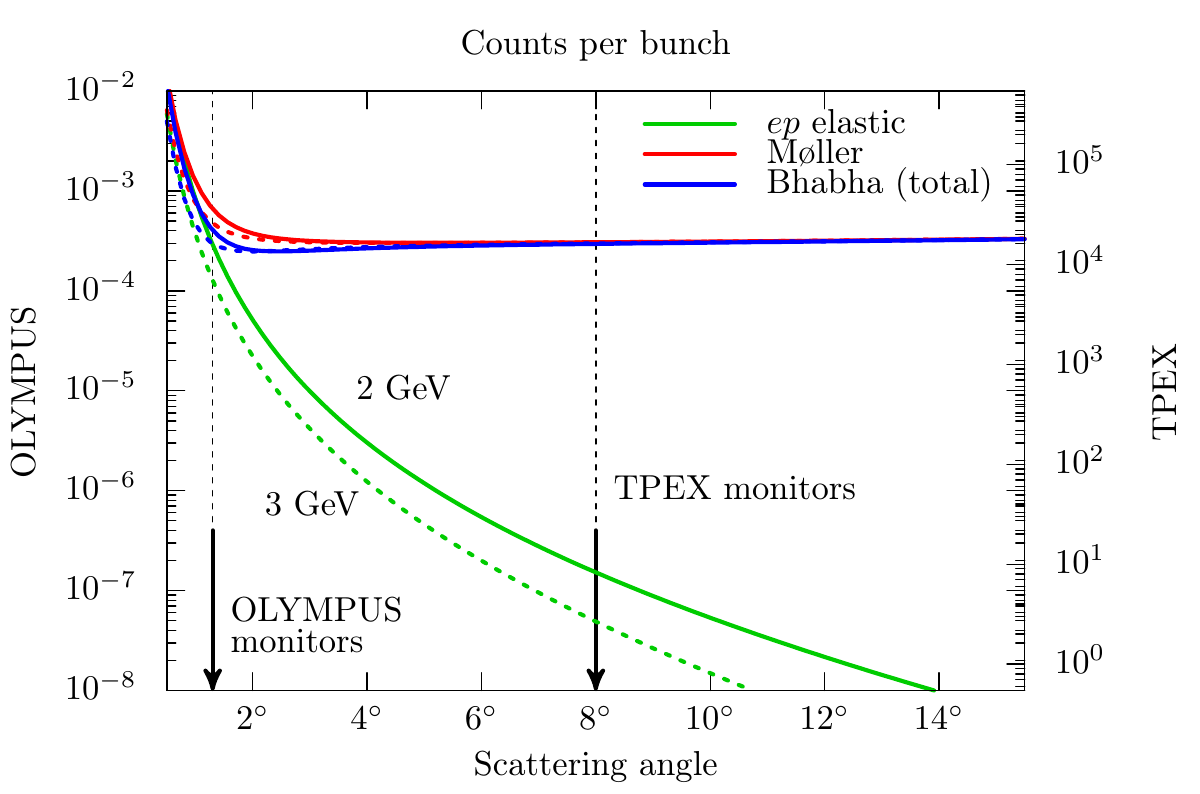}
    \caption{Whereas the forward monitors in OLYMPUS had an event per
      bunch rate well below 1, the TPEX monitors will see $10^4$
      M{\o}ller or Bhabha events per bunch crossing.}
    \label{fig:mb_rates}
\end{figure}

In OLYMPUS, the most accurate determination of the relative luminosity
was obtained from the rates of multi-interaction events---in which a
M{\o}ller or Bhabha event occurred in the same bunch as a forward
elastic $e^\pm p$ event~\cite{Schmidt:2017jby}.  This method had an
overall uncertainty of 0.36\% and looked promising for future
measurements. Unfortunately it is not feasible for the proposed
measurement because of the higher rate per bunch crossing, as seen in
Fig.~\ref{fig:mb_rates}. The multi-interaction event method requires
that the event per bunch rate to be much less than unity. However, the
monitors for the proposed measurement will see approximately $10^4$
M{\o}ller or Bhabha events per bunch crossing.

Instead, the proposed monitor can work by integrating the signal from
all particles produced during each bunch. A monitor placed at
$8\degree$ has a number of advantages relative to the $1.3\degree$
placement of the OLYMPUS luminosity monitors. First, at $8\degree$,
the M{\o}ller and Bhabha cross sections are only a few percent
different, whereas for the OLYMPUS monitors, which covered the
symmetric angle ($90\degree$ in the center-of-mass frame), the two
cross sections differed by over 50\%, with significant angular
dependence. Second, the M{\o}ller/Bhabha rate completely dwarfs the
$e^\pm p$ elastic scattering rate, meaning that it is really only
sensitive to QED processes. No form factors or any other hadronic
corrections\footnote{other than the radiative correction from vacuum
polarization} are needed to calculate the M{\o}ller and Bhabha cross
sections. Third, the sensitivity to alignment scales as
$1/\sin\theta$, meaning the monitor will be much more robust to small
misalignments, which were a significant problem for the OLYMPUS
luminosity monitor.

\begin{figure}
    \centering
    \includegraphics[width=0.9\textwidth]{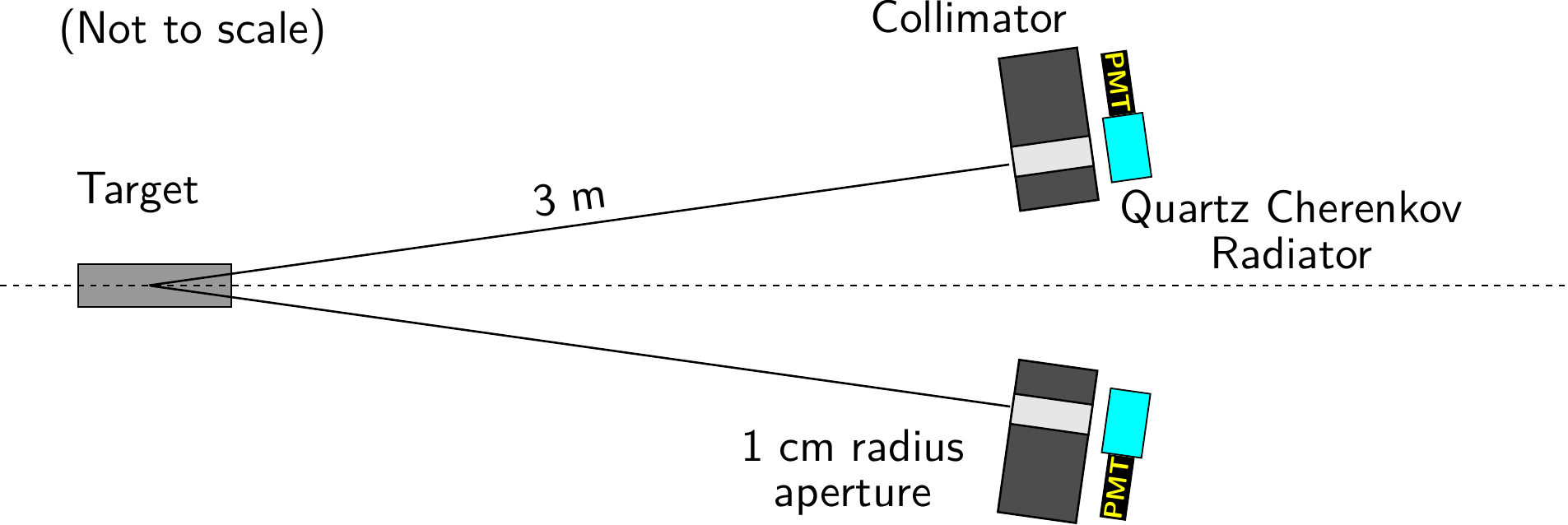}
    \caption{Schematic for the proposed luminosity monitor, consisting
      of two quartz Cherenkov detectors with an acceptance defined by
      1~cm radius apertures in high-Z collimators}
    \label{fig:lumi_monitor_design}
\end{figure}

To test the feasibility of the proposed luminosity monitor, we have
developed a preliminary design, and run a Monte Carlo simulation to
test the sensitivity to misalignments and beam position shifts, which
were the dominant systematic errors for the OLYMPUS luminosity
determination~\cite{Schmidt:2017jby}. A schematic of the design is
shown in Fig.~\ref{fig:lumi_monitor_design}. The monitor consists of
two quartz Cherenkov detectors, which act as independent
monitors. Cherenkov detectors were chosen because they are widely used
for monitoring in high-rate applications, such as in parity-violating
electron scattering~\cite{Abrahamyan:2012gp,Allison:2014tpu}. The two
detectors operate independently and can cross-check each other,
helping to reduce systematic errors from beam alignment. In this
design, the monitors are placed 3~m away from the center of the
target, along the $8\degree$ scattering angle. The acceptance is
defined by a collimator with a circular aperture with a radius of
1~cm.

\begin{figure}[!ht]
  \centering
  \begin{subfloat}[]{\label{fig:biasleft}
    \includegraphics[width=0.48\textwidth]
                    {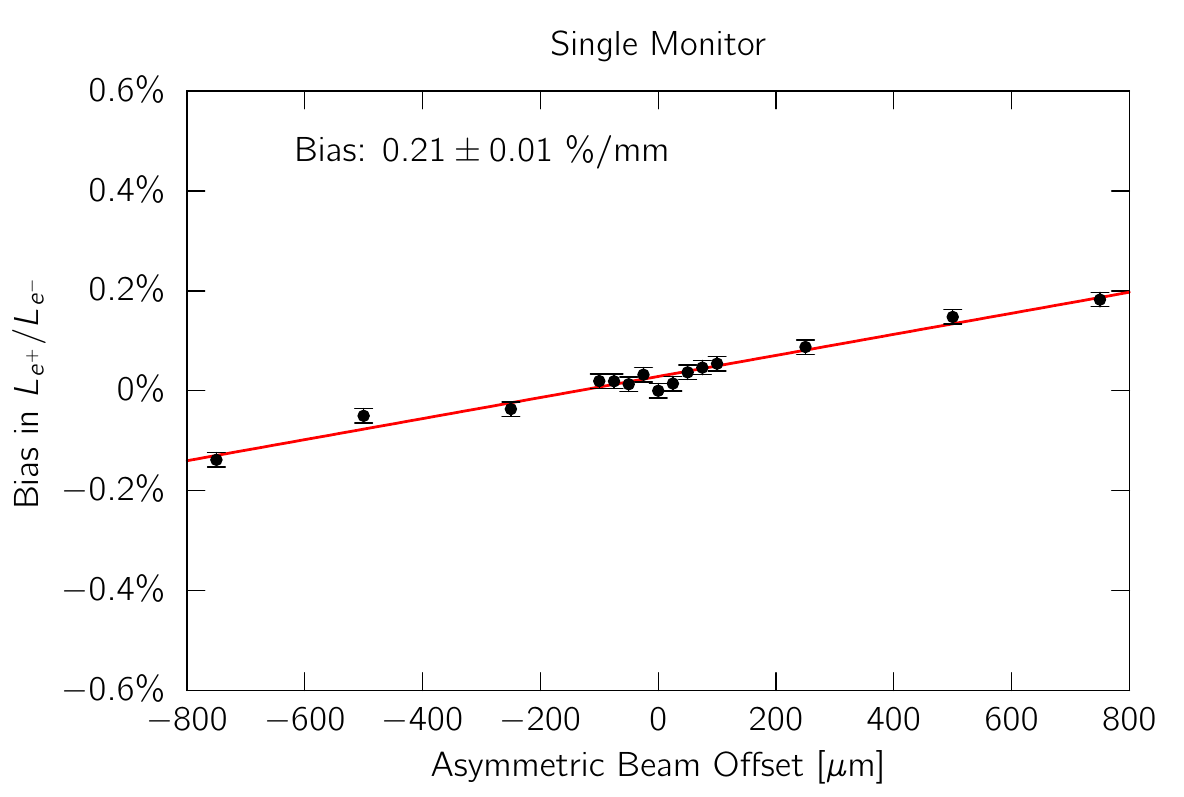}}
  \end{subfloat}
  \hfill
  \begin{subfloat}[]{\label{fig:biasright}
    \includegraphics[width=0.48\textwidth]
                    {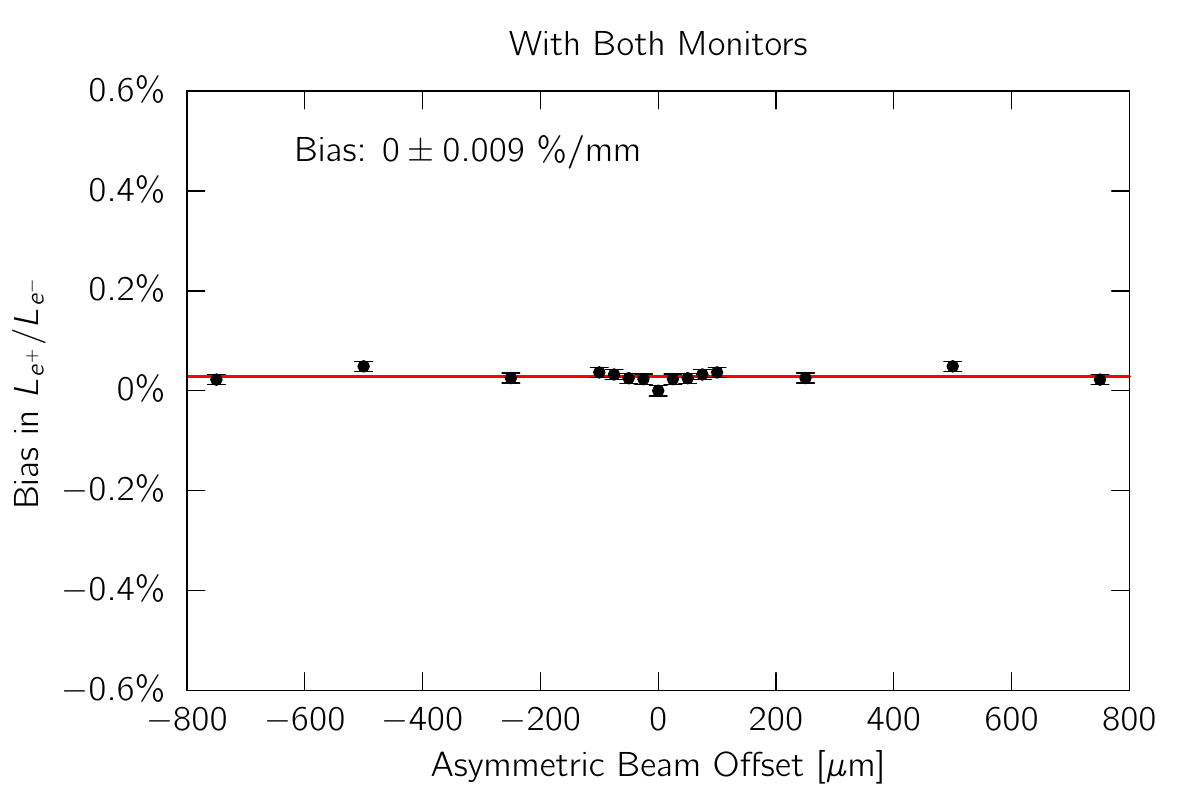}}
  \end{subfloat}
    \caption{The potential bias from a charge-asymmetric beam
      misalignment is a mere 0.2\%/mm in a single monitor (a), and
      this is completely eliminated by using a pair of monitors
      (b).}
    \label{fig:lumi_offset}
\end{figure}

Fig.~\ref{fig:lumi_offset} shows the effect on the luminosity
determination from beam misalignment. The most pernicious misalignment
would be one that is asymmetric between electron and positron modes,
and so this was the focus of this study. When using a single monitor,
an asymmetric misalignment would cause a mere $0.21\pm 0.1$~\%/mm bias
in the determination of the relative luminosity. When using the
combination of both a left and right monitor, this bias is completely
eliminated to the uncertainty of this simulation. For comparison, the
OLYMPUS luminosity monitor was sensitive to asymmetric misalignments
at the level of 5.7~\%/mm, and it was estimated that the beam position
monitors could control the asymmetric misalignment to within
20~$\mu$m~\cite{Schmidt:2017jby}. Such control will probably not be
possible in the proposed measurement due to the much smaller beam
current. However the simulation demonstrates that such control is not
needed. This is largely due to the flatness of the M{\o}ller and
Bhabha cross sections at $8\degree$ (see Fig.~\ref{fig:mb_rates}). One
downside of this robustness is that the proposed monitors are not very
sensitive as beam alignment monitors.

\begin{figure}
    \centering
    \includegraphics[width=0.6\textwidth]{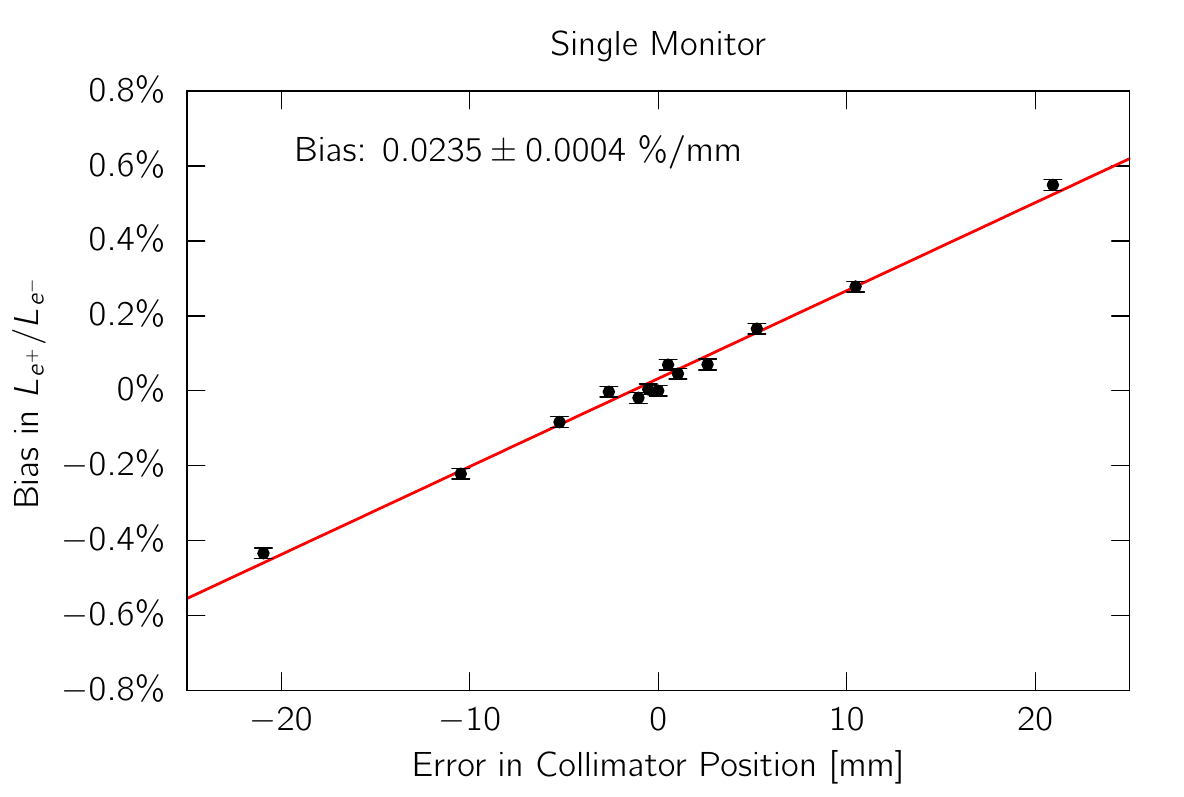}
    \caption{Errors in the positioning of the collimator aperture also
      have a minimal impact on the relative luminosity determination.}
    \label{fig:lumi_misalign}
\end{figure}

Fig.~\ref{fig:lumi_misalign} shows the simulation results for the
effect on the luminosity determination if one of the collimator
apertures were to be positioned in a different place than
expected. The maximum effect occurs when the collimator is shifted to
a larger or smaller scattering angle, and so this was the focus of the
study. For a single monitor, the effect is a mere 0.02~\%/mm. In
practice, both monitors can have positioning errors, and these effects
could end up adding or partially canceling. Regardless, because of the
positioning at $8^\circ$, the bias is minimal. For comparison, the
effect on OLYMPUS was approximately 0.13~\%/mm, with a survey accuracy
of approximately 0.5~mm. Based on the experience gained from OLYMPUS,
we can make improvements in the collimator positioning. One obvious
improvement is to include integral survey marks on the collimator
itself, since the aperture defines the monitor acceptance.

The simulation shows that two of the major systematic limitations of
the OLYMPUS luminosity monitor will be minimal for the proposed
design. The third major systematic, stemming from the residual
magnetic field along the beamline, will be irrelevant. The proposed
design does have other systematic limitations. The biggest concern
will be the amplitude stability of the photomultiplier. With about
$10^4$ particles passing through the aperture every bunch crossing,
there is no way to calibrate the light yield from the data itself. To
guard against gain drifts, an external calibration source will be
vital. A pulsed light source, {\it e.g.} a UV laser, coupled to a
fiber-optic distribution system can be used to monitor the gain of
both photomultipliers throughout the experiment, while an independent
photodiode can be used to cross check that the laser intensity itself
does not drift. Several of us have experience with laser calibration
systems used in previous
experiments~\cite{Denniston:2020gmc,Hasell:2009zza}. Sub-percent level
accuracy should be achievable, though this will almost certainly be
the limiting systematic effect.


\section{Beamdump / Faraday Cup}
\label{sec:beamdump}

A new extracted beam facility from DESY~II will need a beamdump.
M.~Schmitz (DESY) and C.~Tschalar (MIT) have looked into the
requirements and proposed very similar configurations with an aluminum
core and a copper shell.  M.~Schmitz's design was an aluminum
cylinder 10~cm in diameter and 50~cm long embedded in a copper shell
22~cm in diameter and 65~cm long and had water cooling. C.~Tschalar's
design was larger with 20~cm diameter and 50~cm long aluminum in a
32~cm diameter and 75~cm long copper shell but was air cooled.  Both
recommended that the beamdump be surrounded by neutron absorbing
material like cement blocks or borated polyethylene.

\begin{figure}[h]
    \centering
    \includegraphics[width=0.8\textwidth]{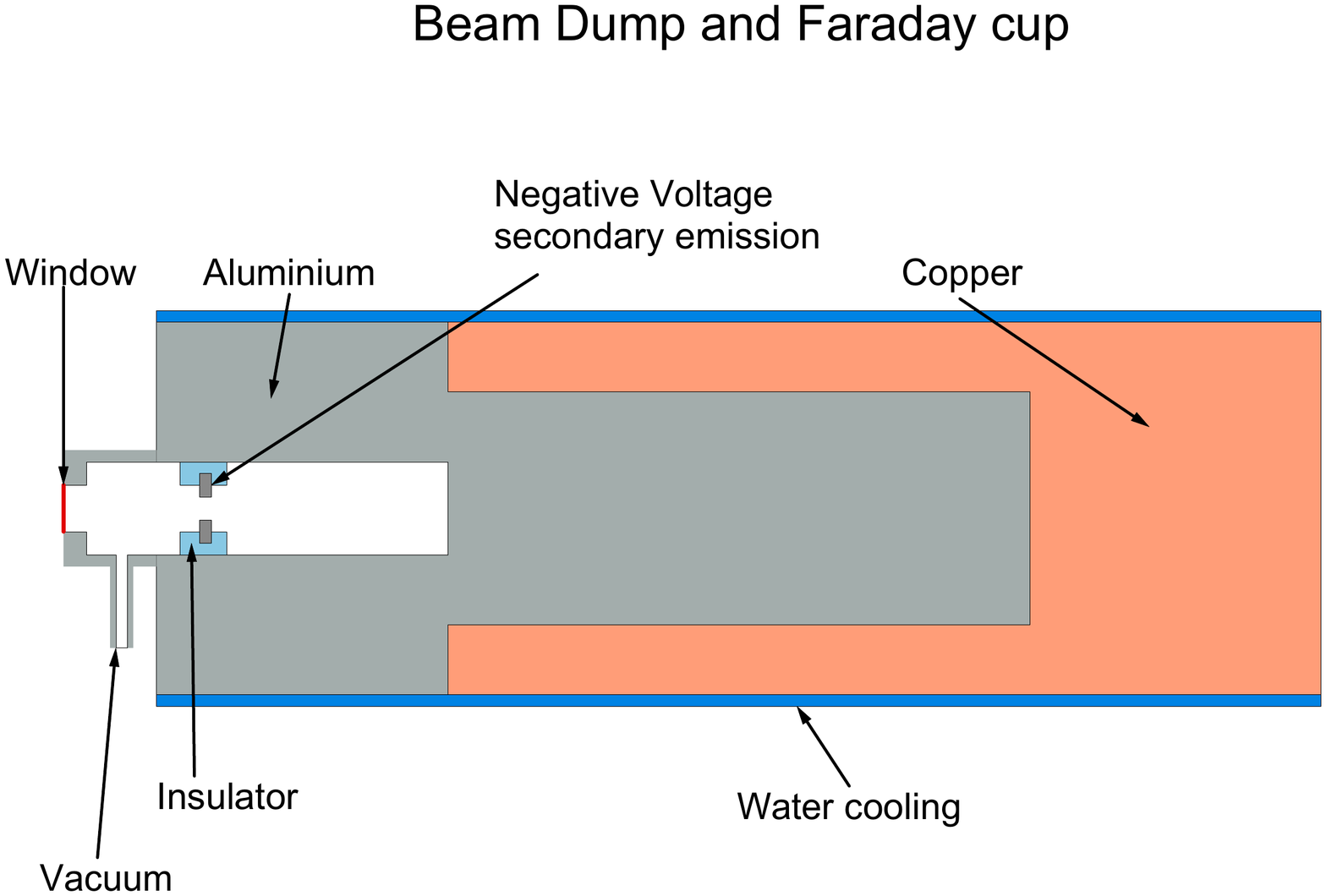}
    \caption{Schematic of a possible beamdump / Faraday cup for TPEX}
    \label{fig:beamdump}
\end{figure}

Assuming a maximum current of 100~nA and a beam energy of 7~GeV the
maximum power to be handled is 700~W.  To contain the showering you
want order of 5~Moli\'{e}re radii laterally and order of 25~radiation
lengths longitudinally.  To be conservative we have selected
C.~Tschalar's numbers as a starting point \cref{fig:beamdump}.
To augment the luminosity measurement proposed above we thought it
might be useful to modify the beamdump to function as a Faraday cup as
well to integrate the charge that passes through the target.  Then,
assuming the length of the target cell and density of liquid hydrogen
are known we can get a quick measure of the luminosity.  As shown in
the figure an insulated ring held at negative voltage of a few hundred
volts is needed to suppress secondary emission from back scattering
out of the Faraday cup.  The beamdump / Faraday cup is under vacuum
but this need only be roughing vacuum pressure.


\section{Electronics and Readout System}
\label{sec:Elect}

The requirements for data acquisition are comparatively modest. Less
than 300 channels have to be read out, including the luminosity
monitor. With a readout at a low and fixed frequency, no busy logic is
required.

\subsection {Trigger}
The beam has a very low and fixed bunch frequency of
\SI{12.5}{\hertz}, allowing us to trigger on the bunch clock instead
of a trigger detector. The only complication is that for proper gate
alignment, the beam bunch signal has to be shifted by up to 80 ms,
with stability on the 10 ns level. This can be achieved with an FPGA,
which can also generate the required gates. A V2495 module from CAEN
is available in the collaboration and would be adequate.

\subsection {Front end electronics}
For the calorimeter and luminosity monitors, all proposed readout
devices require the acquisition of a time-integrated current pulse
with a QDC. However, they differ in the required front end
electronics:
\begin{itemize}
    \item PMTs require only a base for the high-voltage
      distribution. Active bases can minimize power losses and improve
      stability. Such bases are available commercially, or can be
      manufactured by the collaboration. Circuit designs are readily
      available and can be adapted to fit the required form factor. No
      further signal conditioning (except attenuation in the case of
      the luminosity monitors) is required for interfacing with
      standard QDCs.
    \item APDs and SiPMs require driver and preamplifier circuits. In
      the case of APDs, we would copy the tested design for the PANDA
      detector \cite{pandatdr} from Mainz. For SiPMs, the MUSE
      collaboration produced an amplifier design which could be
      adapted \cite{musesipm}.
\end{itemize}

\subsection{Baseline DAQ hardware and software}
In addition to the V2495 for trigger generation, the baseline
configuration for 250+2 detectors would require eight 32-channel QDC
modules like CAEN's V792 or Mesytec's MQDC-32. This can be housed in a
single VME crate and read out via a single board computer (SBC) as the
VME controller.

The data rates are low, with about \SI{6}{kB/s} for the readout of 252
channels at \SI{12.5}{\hertz}. This makes it possible to store all
experiment data to a single server outside of the experimental area
via standard network file systems. This rate requires less than 4 GB
per week.

The SBU group already developed the DAQ software for the test beams,
which will be basis for the DAQ solution of the actual experiment.

For the GEMs, multiple readout solutions are possible:
\begin{itemize}
\item APV based readout, based on the MPD-4 readout boards already
  used for SBS@JLAB and MUSE \cite{MUSE_TDR}. While APVs are out of
  production, these collaborations have a significant number of APV
  readout cards and MPDs available, and experience in operating these
  components.
\item SAMPA based readout. This is currently in development at JLab
  for TDIS and other projects. Compared to GEMs, the signal quality is
  better and the wave form can be sampled as well. There has been a
  lot of progress on the testing of the chips, which will be in
  production for the foreseeable future. However, a switch would
  require the procurement of new hardware.
\item VMM based readout. The VMM chip is considered to be the
  successor to the APV chip in the Scalable Readout System (SRS). This
  new development is cost-effective and scalable, and has been adopted
  and recommended by RD-51 in the framework of the SRS readout scheme.
  The Mainz MAGIX collaboration recently decided to start using VMM
  for their GEM readout at MESA.  The VMM offers readout with time and
  pulse shape digitization directly on the front-end card. Two VMM
  chips are housed on one front-end card to process 128 readout
  channels.
\end{itemize}
The data rate from the GEMs is considerably higher, but still
manageable, particularly with zero suppression. With 1,000 channels
per detector and 20 detectors, the estimated rate is between
\SI{500}{kB/s} (1 sample per event and channel) to \si{5}{MB/s} (10
samples per event and channel), resulting in about 1.5~TB per week of
beam time. With zero suppression this could be reduced to a level of
300 GB per week.

\subsection {Possible improvements}
We are evaluating multiple improvements over this baseline design:
\begin{itemize}
    \item Higher trigger rate: Instead of triggering just on the beam
      clock, the FPGA can generate additional gates before and after
      each trigger, spaced so that all conversion and data
      transmission can happen before the real gate opens. This triples
      the trigger and thus data rate---easily handled by the proposed
      system---but would allow for baseline and background monitoring.
    \item Instead of QDCs, which only give information about the
      integrated charge, the signal wave forms could be digitized with
      high speed ADCs. This would allow even better baseline control,
      but would increase the bandwidth considerably. For example,
      sampling the signals for \SI{1}{\micro s} with
      \SI{250}{\mega\hertz} at 14 bit would result in about 6~kB/s per
      channel, less than \SI{1.6}{MB/s} in total. These data rates are
      still readily managed by the system outlined above, and about 1
      TB of storage per week. Commercial solutions for these
      digitizers exist, but are about factor four more costly than
      QDCs. Cost-effective alternatives are the 12-ch WaveBoard 2.0
      designed by INFN Roma/Genova or commercial boards using the DRS4
      chip~\cite{RittDRS4}, like CAEN's V1742. The DRS4 chip realizes
      an analog buffer to allow for the cost effective and high-speed
      (multi-GSamples/s) sampling of events. The trade-off is
      considerable dead-time for the conversion; however this is
      completely hidden in the proposed experiment by the comparably
      low trigger rate. The digitization of the waveform would provide
      additional insights into the detected particle and it's timing,
      allowing us to improve background rejection.
\end{itemize}
The decision on these improvements will be based on our experience
with these options in test beams planned for the near future.


\section{Upgrades / Improvements to the Proposed Experiment}
\label{sec:improve}

While the configuration proposed so far is possible and would allow
the two-photon exchange contribution to be investigated in a region
where the observed form factor discrepancy is clear; a number of
upgrades are possible.
\begin{enumerate}
\item The current configuration assumes that 250 lead tungstate
  crystals can be obtained.  Clearly if less or more crystals are
  possible the configuration would change.  Adding more crystals to
  the back angle calorimeters would increase the acceptance in a
  region of low count rate. With an additional $5\times5$~array above
  and below the current modules the solid angle would be increased
  from $3.6$~msr to $15.6$~msr an increase of 4.3.
\item If the showering in the $5\times5$~arrays of PbWO$_{4}$ is well
  understood; it may be possible to accept a larger area of the
  calorimeter, say an effective area of 6.4~msr rather than the
  3.6~msr using just the central $3\times3$ array.  This would
  increase the acceptance by $1.78$.  Placing a tracking detector,
  {\it e.g.}~GEM, immediately before the calorimeter would help to
  define the acceptance.  This needs to be investigated with test beam
  studies with a $5\times5$ calorimeter. 
\item Move the back angle calorimeters closer to the target. Going to
  a radius of 0.5~m would increase the count rate by a factor of four,
  though increasing the angular range subtended and thus reduce the
  $Q^2$ resolution.  Addition of GEM tracking may help this but needs
  further Monte Carlo simulation and study.
\end{enumerate}

These options could increase the count rate significantly, making even
higher beam energies accessible. \Cref{tb:HE} shows the kinematic
reach and differential cross section for just the back angle,
$110\degree$, for various lepton beam energies. A measurement at
lepton beam energy of 4~GeV would extend the two-photon exchange
measurements to 6.39~(GeV/$c$)$^2$, and only requires an improvement
by a factor of five to be comparable to the proposed measurement rate
at 3~GeV.
\begin{table}[htbp]
  \centering
  \begin{tabular*}{0.6\textwidth}{@{\extracolsep{\fill}}ccccc}
    $E_{beam}$&$\theta$&$Q^{2}$&$\epsilon$&$d\sigma/d\Omega$\\
    GeV&&(GeV/c)$^{2}$&&fb\\
    \hline
    2.0&$110\degree$&2.78&0.120&$1.22\times10^{4}$\\
    3.0&$110\degree$&4.57&0.096&$1.20\times10^{3}$\\
    4.0&$110\degree$&6.39&0.080&$2.23\times10^{2}$\\
    5.0&$110\degree$&8.23&0.068&$5.92\times10^{1}$\\
    6.0&$110\degree$&10.1&0.060&$2.00\times10^{1}$\\
  \end{tabular*}
  \caption{Kinematics and cross section for measurements at
    $110\degree$ for lepton beam energies possible with DESY-II}
  \label{tb:HE}
\end{table}


\section{Background Considerations}
\label{sec:bkgd}


\subsection{Protons from $e^{\pm}p$ elastic scattering}

As proposed, the experiment does not measure the scattered lepton and
proton in coincidence. While this would have some benefits it would
also require detectors at far forward angles where the event rates
from elastic lepton scattering, M{\o}ller and Bhabha scattering, and
pion production would be problematic.  Nevertheless, protons from
elastic lepton-proton scattering will strike the proposed detectors
and will be a source of background for the measurement.

\begin{figure}[!ht]
\centering
\begin{subfloat}[]{\label{fig:epangles}
  \includegraphics[width=0.48\textwidth, viewport=48 44 704 495, clip]
    {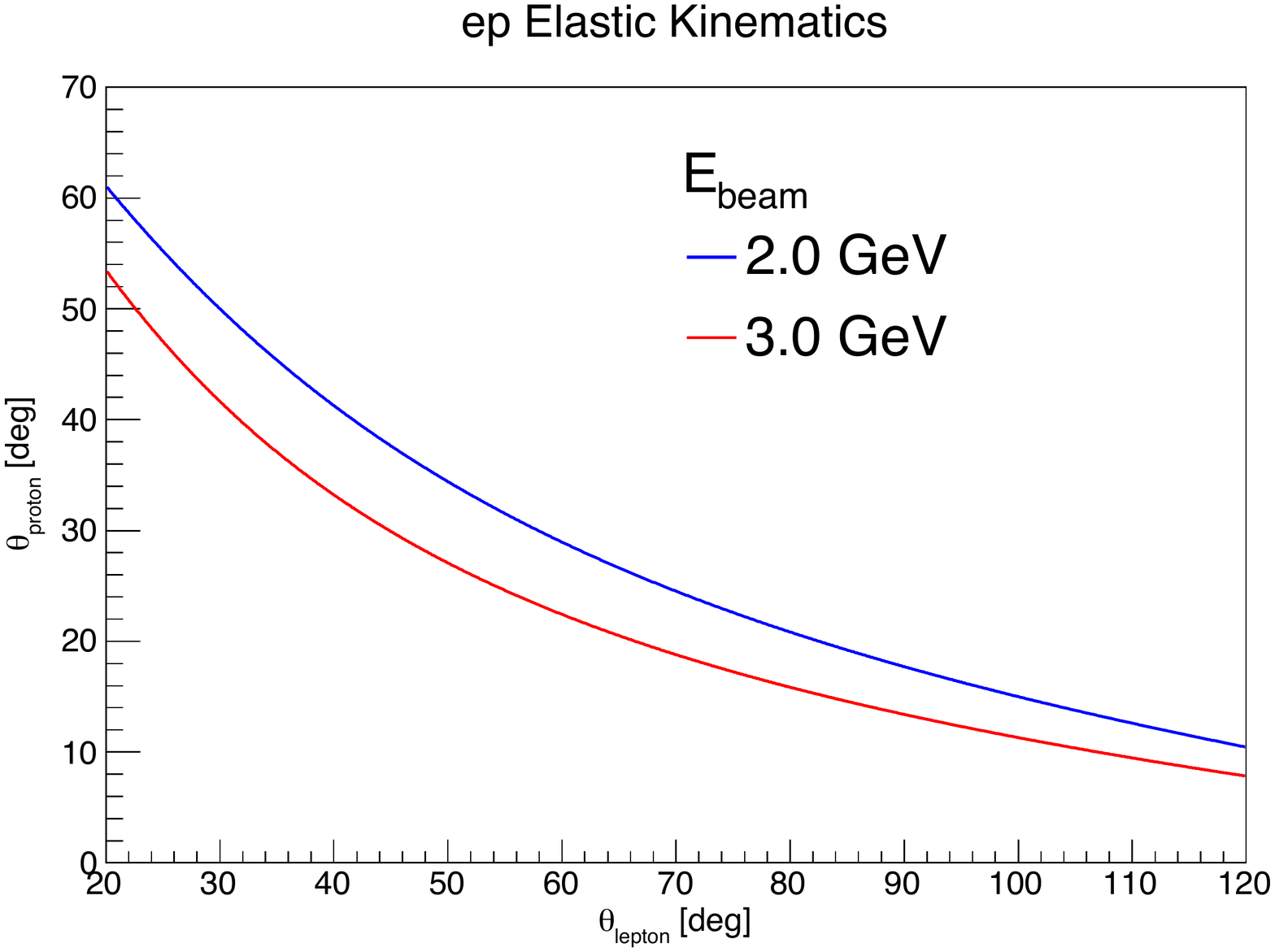}}
\end{subfloat}
\hfill
\begin{subfloat}[]{\label{fig:epmoment}
  \includegraphics[width=0.48\textwidth, viewport=48 44 704 495, clip]
    {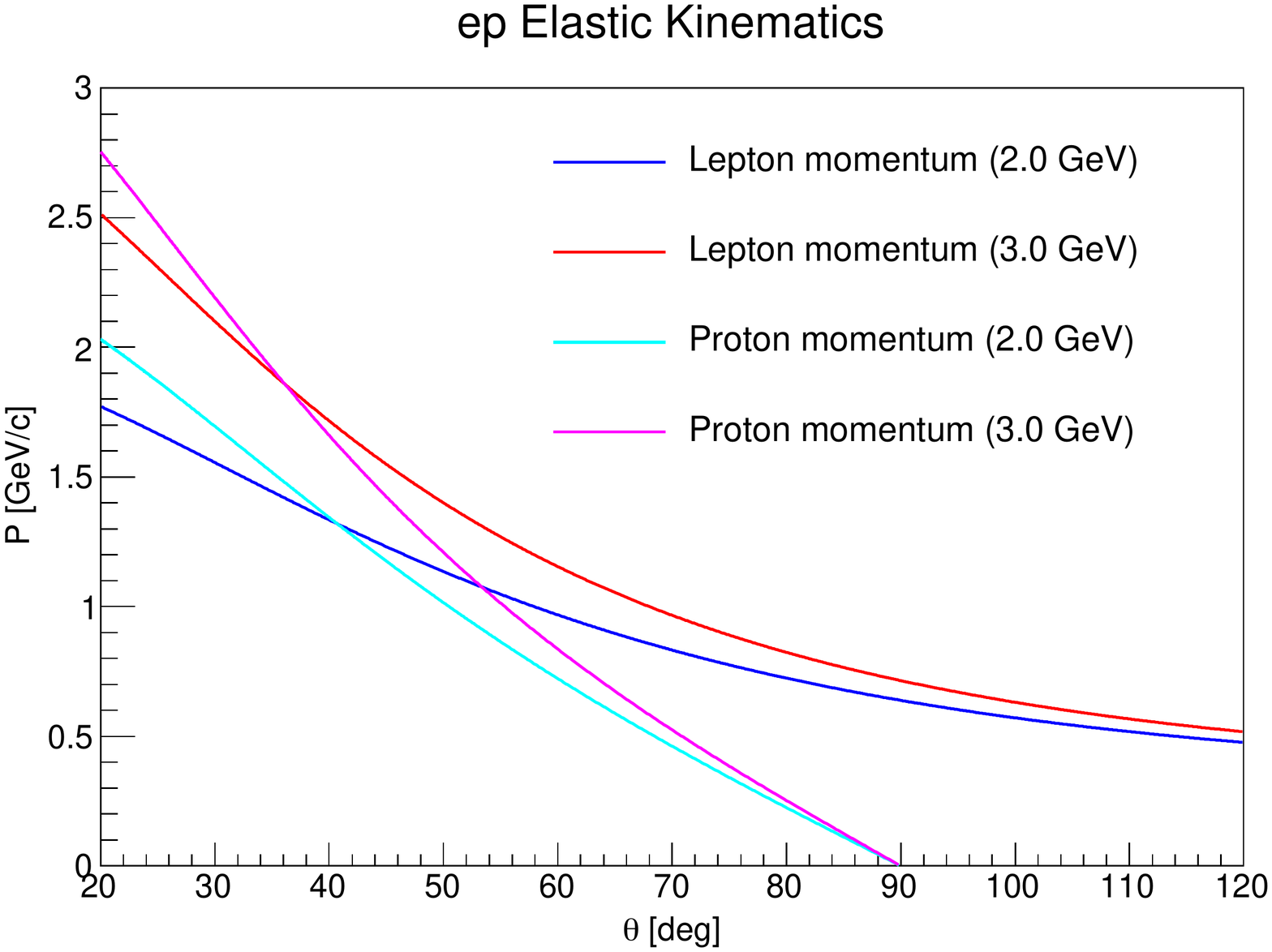}}
\end{subfloat}
\caption{Kinematics for $ep$ elastic scattering. (a) Proton scattering
  angle as a function of the lepton scattering angle. (b) Lepton and
  proton momenta as a function of their respective scatting angles.}
\label{fig:Kine}
\end{figure}

\Cref{fig:epangles} shows the relation between the proton polar
scattering angle and the lepton scattering angle.  \Cref{fig:epmoment}
shows the momentum for the lepton and proton as a function of their
scattering angle. The proton momentum is greater than that of the
lepton at forward angles but drops more rapidly as its scattering
angle increases. Protons will also be detected in the calorimeters and
will have to be identified and corrected for on an event by event
basis.  The calorimeter modules at over 22~$X_0$ will adequately
contain the electromagnetic showers and detect most of the lepton
energy, but the proton will not deposit its full energy as the
calorimeter is only around one nuclear interaction length in depth.
Monte Carlo studies (presented below) indicate that the proton will
deposit at most 300 to 400~MeV in the calorimeters at $30\degree$ and
$50\degree$ and significantly less at larger angles, particularly if
an absorber shield is placed in front of the calorimeters.  This will
allow the lepton signal to be clearly resolved from the proton signal.
\begin{figure}[!ht]
\centering
\begin{subfloat}[]{\label{fig:epsigma}
  \includegraphics[width=0.48\textwidth, viewport=48 44 704 495, clip]
    {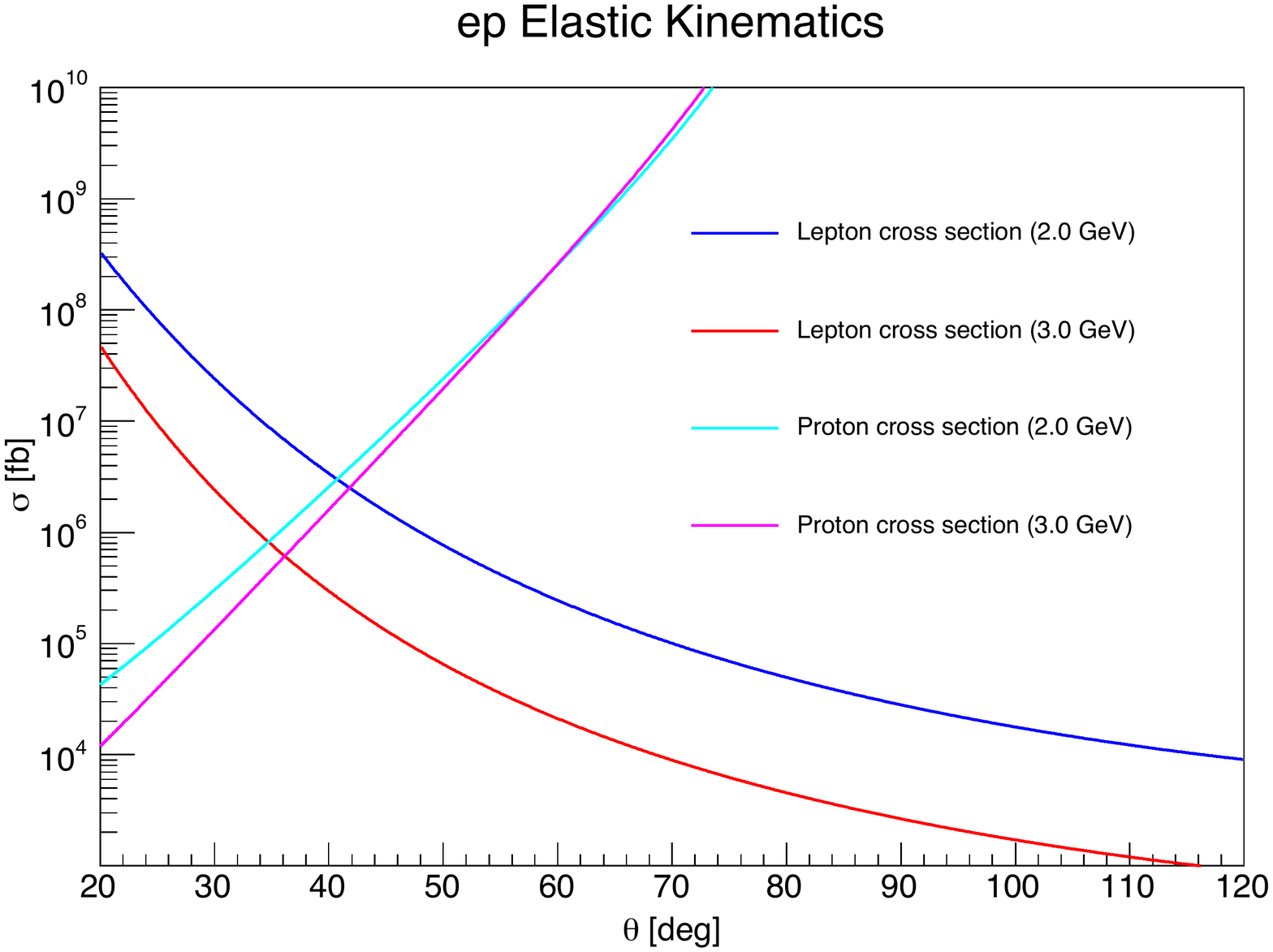}}
\end{subfloat}
\hfill
\begin{subfloat}[]{\label{fig:epbeta}
  \includegraphics[width=0.48\textwidth, viewport=48 44 704 495, clip]
    {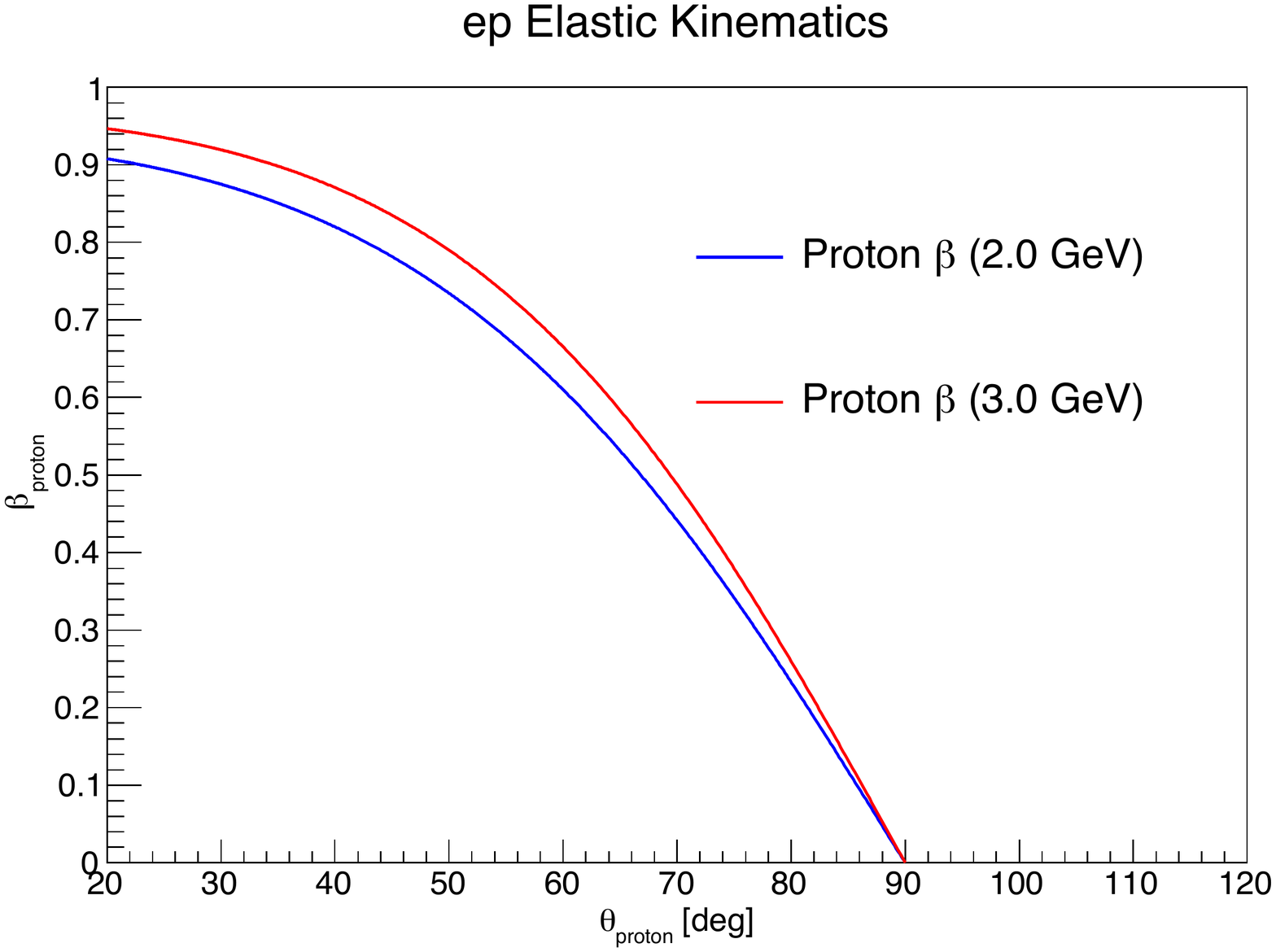}}
\end{subfloat}
\caption{Kinematics for $ep$ elastic scattering. (a) Cross sections
  for lepton and proton as a function of their polar angle. (b) Beta
  for the proton as a function of its polar angle.}
\end{figure}

At $30\degree$ the proton rate will be an occasional nuisance as it is
significantly less than the lepton rate as shown in
\cref{fig:epsigma}.  However, at $50\degree$ the proton rate will be
10--100 times that for the lepton.  This is still not a problem, as
the rate is manageable (approximately once every three beam spills)
plus the deposited proton energy will be more than 700~MeV lower than
the lepton's.  But, at $70\degree$ the rate for protons is $10^4-10^5$
greater than that for leptons resulting in multiple protons from every
beam spill. However, as discussed in the Monte Carlo section, with a
suitable absorber the protons can be stopped before the calorimeter
without significantly affecting the lepton signal.  It may also be
possible to eliminate this if the calorimeter timing and readout is
sufficiently fast.  The $\beta$ value for the proton is shown in
\cref{fig:epbeta} and is around 0.5 at $70\degree$.  That means the
protons will arrive around 3~ns after the lepton possibly allowing a
timing window to exclude them.  At $90\degree$ the proton rate is even
higher but the energy is much lower and can be handled by an absorber
and/or timing.


\subsection{M{\o}ller and Bhabha scattering}

The cross sections for M{\o}ller and Bhabha scattering into the
detector angles being considered in this proposal are given in
\cref{tb:moller}.  For an average lepton current of 40~nA incident on
a 20~cm liquid hydrogen target the luminosity is
$2.11\times10^{-4}$~fb$^{-1}\cdot$ s$^{-1}$.  If we consider the face
of the entire $5\times5$ array of each calorimeter module this
corresponds to 10~msr so the luminosity factor becomes
$2.11\times10^{-6}$.  Multiplying this factor through the cross
sections in \cref{tb:moller} yields rates ranging from around 4 to
$2\times10^{9}$ events per second.
\begin{table}[htbp!]
  \centering
  \begin{tabular*}{0.6\textwidth}{@{\extracolsep{\fill}}cccc}
    $\theta$&M{\o}ller&Bhabha $e^+$&Bhabha $e^-$\\
    &fb&fb&fb\\
    2.0~GeV\\
    \hline
     $30\degree$& $1.223\times10^{14}$& $2.863\times10^{8}$&
     $1.219\times10^{14}$\\
     $50\degree$& $2.991\times10^{14}$& $3.866\times10^{7}$&
     $2.989\times10^{14}$\\
     $70\degree$& $1.986\times10^{15}$& $9.089\times10^{6}$&
     $1.985\times10^{15}$\\
     $90\degree$& diverges&       0& diverges\\
    $110\degree$&        0&       0&        0\\
    \\
    3.0~GeV\\
    \hline
     $30\degree$& $1.223\times10^{14}$& $1.274\times10^{8}$&
     $1.220\times10^{14}$\\
     $50\degree$& $2.991\times10^{14}$& $1.719\times10^{7}$&
     $2.989\times10^{14}$\\
     $70\degree$& $1.985\times10^{15}$& $4.041\times10^{6}$&
     $1.985\times10^{15}$\\
     $90\degree$& diverges&       0& diverges\\
    $110\degree$&        0&       0&        0\\
  \end{tabular*}
  \caption{Cross section for M{\o}ller and Bhabha scattering as a
    function of the polar scattering angle.}
  \label{tb:moller}
\end{table}

The energies of these M{\o}ller and Bhabha scattered leptons are low,
less than 4~MeV at $30\degree$ and even lower at the larger angles.
For the most part they are still relativistic so a timing cut is not
possible except at $90\degree$ and possibly at $70\degree$ if the
calorimeter electronics are fast. To sweep these leptons away would
require a magnetic field around 400~G.  However, Monte Carlo studies
show that a simple 10~mm aluminum absorber before the calorimeter
modules will stop these particles from producing any signal in the
calorimeter without degrading the response to the higher energy
leptons of interest.  Since a 10~mm aluminum plate over the front face
of the calorimeter array would work well as part of the cooling
system; the high rate of M{\o}ller and Bhabha scattered leptons is not
a problem.


\subsection{Pion Production}

Another source of background comes from pion production.  There are
four reactions to consider:
\begin{eqnarray}
   e^- + p &\rightarrow& e^- + p + \pi^0\\
   e^- + p &\rightarrow& e^- + n + \pi^+\\
   e^+ + p &\rightarrow& e^+ + p + \pi^0\\
   e^+ + p &\rightarrow& e^+ + n + \pi^+
\end{eqnarray}

A Monte Carlo pion event generator was used to simulate these four
reactions at 2 and 3~GeV.  The calorimeter modules in the proposed
configuration will be struck by the leptons (electrons or positrons),
baryons (protons or neutrons), and pions ($\pi^0$ or $\pi^+$) from the
various pion production reactions.  In the case of $\pi^0$ production
the most likely decay to two photons must also be considered.

\begin{figure}[!ht]
\centering
\begin{subfloat}[]{\label{fig:Pi0_e_2_30a}
  \includegraphics[width=0.47\textwidth, viewport=10 5 525 390, clip]
    {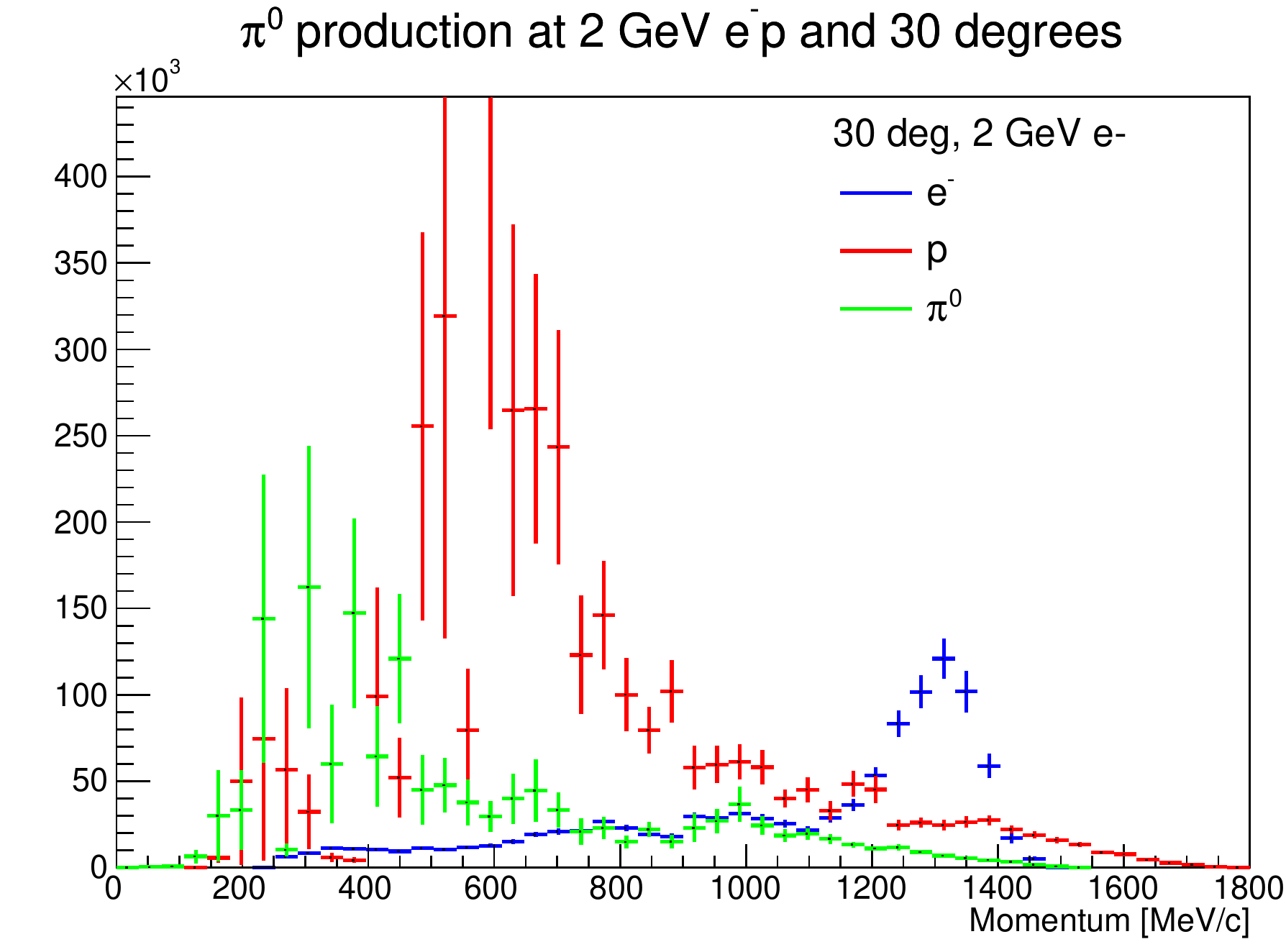}}
\end{subfloat}
\hfill
\begin{subfloat}[]{\label{fig:Pi+_e_2_30a}
  \includegraphics[width=0.47\textwidth, viewport=10 5 525 390, clip]
    {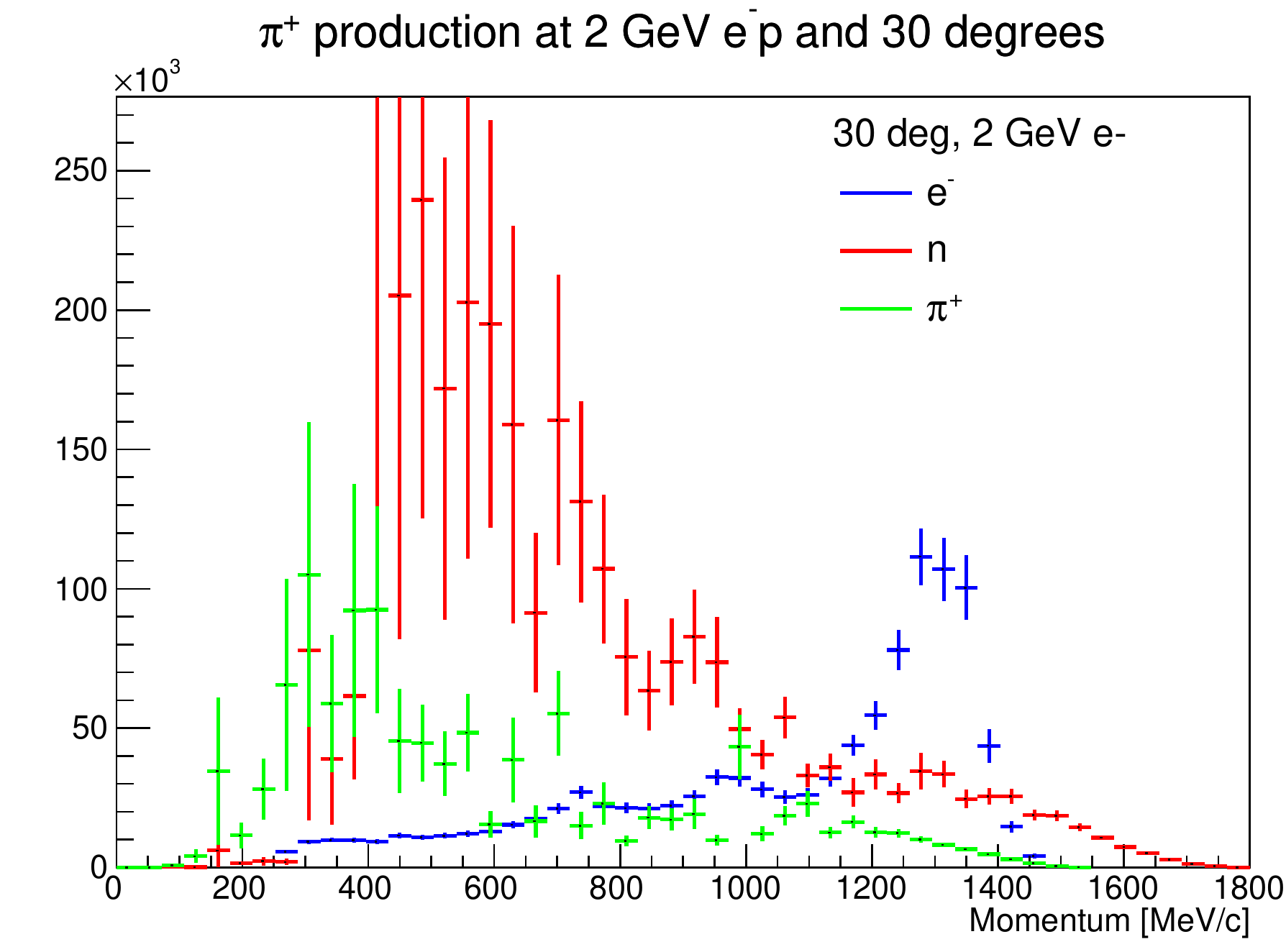}}
\end{subfloat}
\\
\begin{subfloat}[]{\label{fig:Pi0_e_3_30a}
  \includegraphics[width=0.47\textwidth, viewport=10 5 525 390, clip]
    {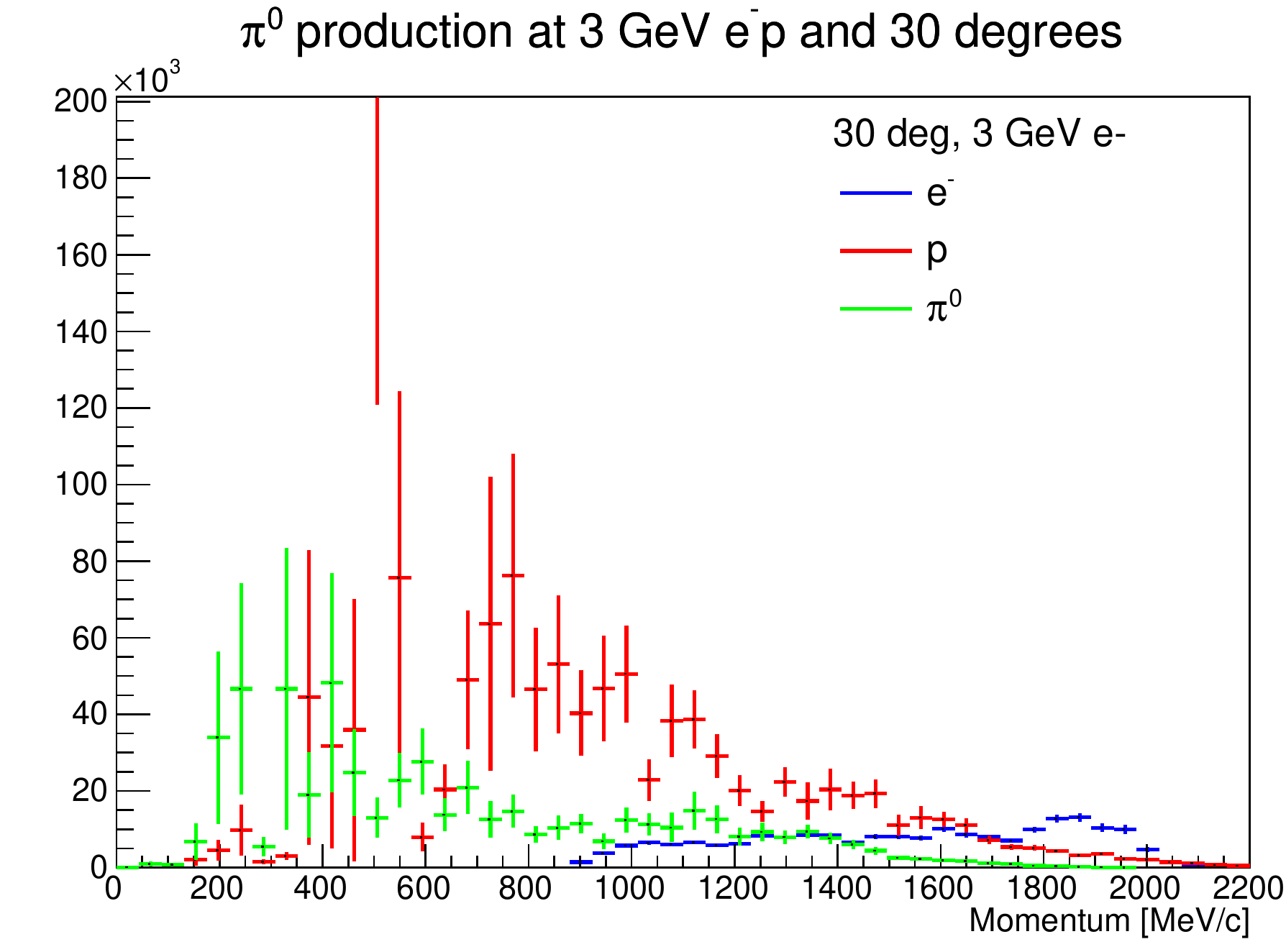}}
\end{subfloat}
\hfill
\begin{subfloat}[]{\label{fig:Pi+_e_3_30a}
  \includegraphics[width=0.47\textwidth, viewport=10 5 525 390, clip]
    {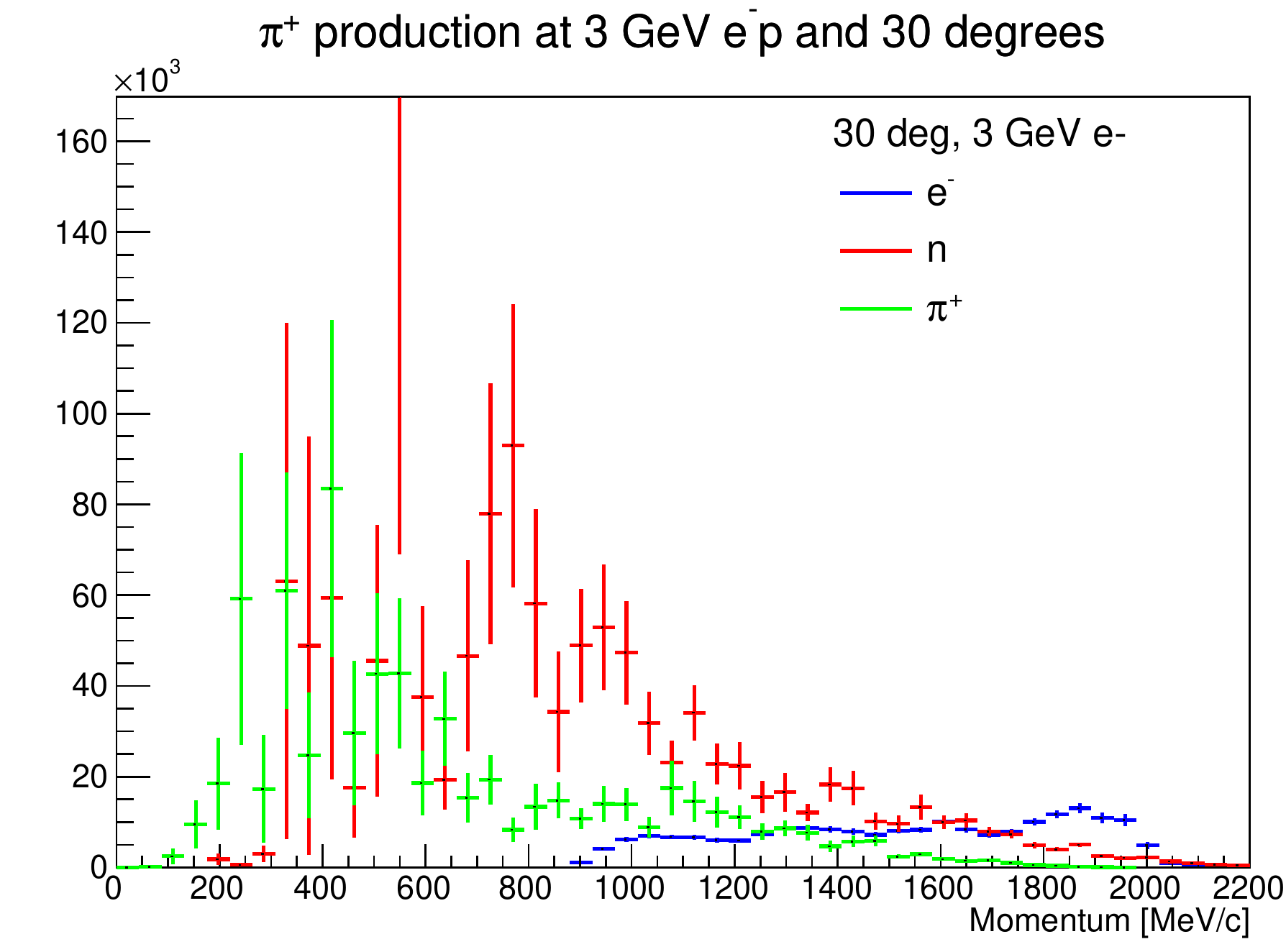}}
\end{subfloat}
\caption{\label{fig:Pi_e_2+3}Number of particles directed towards the
  $5\times5$ calorimeter array situated at $30\degree$ from the
  reactions (a) $e^-+p\rightarrow e^-+p+\pi^0$ and (b)
  $e^-+p\rightarrow e^-+p+\pi^+$ at 2~GeV and (c) $e^-+p\rightarrow
  e^-+p+\pi^0$ and (d) $e^-+p\rightarrow e^-+p+\pi^+$ at 3~GeV during
  one day of running at the nominal luminosity.}
\end{figure}

The event rate per day and the momentum distribution of the leptons,
baryons, and pions incident on the $5\times5$ calorimeter array at
$30\degree$ for the reactions $e^- + p \rightarrow e^- + p + \pi^0$
and $\rightarrow e^- + n + \pi^+$ for an incident electron beam energy
of 2 and 3~GeV are given in \cref{fig:Pi_e_2+3}.  ({\it N.B.}~No
accounting for $\pi^0$ decay or the energy actually deposited in the
calorimeters has been made.  A more complete Monte Carlo simulation is
in progress and further plots for pion production are provided in the
Appendix.)

The total event rate for electrons from pion production at 2~GeV
striking the $5\times5$ face of the calorimeter array at $30\degree$
is $2.07\times10^6$ per day.  This is comparable to the
$7.92\times10^5$ events per day expected in the central $3\times3$
array for the elastic scattering events we wish to detect.  However,
the elastic events are peaked around a momentum of 1.56~GeV/c while
the lepton momentum from pion production has a small peak around
1.35~GeV/c and a long tail to much lower momenta. With PbWO$_4$'s
excellent energy resolution this difference should be easily
resolved. The rates at this angle are such that we can expect one of
these pion events every beam spill. This background must be detected
and corrected on an event by event basis. The deposited energies of
the baryons and pions are significantly lower but will also contribute
to the background and will need to be handled in the analysis.

\begin{table}[!htb]
  \centering
  \begin{tabular}{m{2.5cm}m{2.5cm}m{2.5cm}m{2.5cm}m{2.5cm}}
    $\theta$&$e^-+p+\pi^0$&$e^-+n+\pi^+$&$e^++p+\pi^0$&$e^++n+\pi^+$\\
    \\
    2.0~GeV\\
    \hline
     $30\degree$&$2.08\times10^{6}$&$2.06\times10^{6}$&$2.06\times10^{6}$&$2.08\times10^{6}$\\ 
     $50\degree$&$2.64\times10^{5}$&$2.54\times10^{5}$&$2.60\times10^{5}$&$2.56\times10^{5}$\\ 
     $70\degree$&$7.22\times10^{4}$&$7.14\times10^{4}$&$7.28\times10^{4}$&$7.16\times10^{4}$\\ 
     $90\degree$&$2.86\times10^{4}$&$2.90\times10^{4}$&$2.84\times10^{4}$&$2.90\times10^{4}$\\ 
     $110\degree$&$1.40\times10^{4}$&$1.45\times10^{4}$&$1.41\times10^{4}$&$1.38\times10^{4}$\\ 
    \\
    3.0~GeV\\
    \hline
     $30\degree$&$4.02\times10^{5}$&$4.08\times10^{5}$&$4.08\times10^{5}$&$4.06\times10^{5}$\\ 
     $50\degree$&$6.60\times10^{4}$&$6.60\times10^{4}$&$6.64\times10^{4}$&$6.62\times10^{4}$\\ 
     $70\degree$&$6.58\times10^{3}$&$6.54\times10^{3}$&$6.64\times10^{3}$&$6.72\times10^{3}$\\ 
     $90\degree$&$2.24\times10^{3}$&$2.26\times10^{3}$&$2.20\times10^{3}$&$2.26\times10^{3}$\\ 
     $110\degree$&$9.22\times10^{2}$&$9.10\times10^{2}$&$9.50\times10^{2}$&$9.32\times10^{2}$\\ 
  \end{tabular}
  \caption{Event rates per day for leptons from pion production
    striking the $5\times5$ calorimeter detector arrays.}
  \label{tb:leptonrates}
\end{table}

\Cref{fig:Pi_e_2+3} shows the number of particles detected per day at
the calorimeter positioned at $30\degree$ for each particle produced
in pion production from $e^-p$ at 2 and 3~GeV.  The plots for $e^+p$
are similar and plots for all detector angles are given in the
appendix. At higher angles the lepton rates fall quickly and are
broadly distributed in momentum.  The rates for protons and neutrons
also fall quickly plus they deposit little energy in the
calorimeters.  Pion rates remain fairly uniform with angle but are at
a low momentum and as Monte Carlo studies indicate the deposited
energy is even less.  The $\pi^0$ decay to two photons however will be
a rather uniform, low energy background.  Further Monte Carlo studies
are needed to verify that the background events can be cleanly
resolved from the lepton signals that need to be measured.

\begin{table}[htbp!]
  \centering
  \begin{tabular}{m{2.5cm}m{2.5cm}m{2.5cm}m{2.5cm}m{2.5cm}}
    $\theta$&$e^-+p+\pi^0$&$e^-+n+\pi^+$&$e^++p+\pi^0$&$e^++n+\pi^+$\\
    \\
    2.0~GeV\\
    \hline
     $30\degree$&$6.96\times10^{6}$&$6.04\times10^{6}$&$5.08\times10^{6}$&$6.64\times10^{6}$\\ 
     $50\degree$&$7.14\times10^{6}$&$7.70\times10^{6}$&$7.16\times10^{6}$&$7.06\times10^{6}$\\ 
     $70\degree$&$5.14\times10^{5}$&$9.74\times10^{5}$&$5.30\times10^{5}$&$8.86\times10^{5}$\\ 
     $90\degree$&$0$&$0$&$0$&$0$\\ 
     $110\degree$&$0$&$0$&$0$&$0$\\ 
    \\
    3.0~GeV\\
    \hline
     $30\degree$&$2.42\times10^{6}$&$2.46\times10^{6}$&$2.62\times10^{6}$&$3.26\times10^{6}$\\ 
     $50\degree$&$3.76\times10^{6}$&$3.56\times10^{6}$&$3.38\times10^{6}$&$3.86\times10^{6}$\\ 
     $70\degree$&$7.02\times10^{4}$&$6.56\times10^{4}$&$7.26\times10^{4}$&$9.30\times10^{4}$\\ 
     $90\degree$&$0$&$0$&$0$&$0$\\ 
     $110\degree$&$0$&$0$&$0$&$0$\\ 
  \end{tabular}
  \caption{Event rates per day for baryons from pion production
    striking the $5\times5$ calorimeter detector arrays.}
  \label{tb:baryonrates}
\end{table}

\begin{table}[htbp!]
  \centering
  \begin{tabular}{m{2.5cm}m{2.5cm}m{2.5cm}m{2.5cm}m{2.5cm}}
    $\theta$&$e^-+p+\pi^0$&$e^-+n+\pi^+$&$e^++p+\pi^0$&$e^++n+\pi^+$\\
    \\
    2.0~GeV\\
    \hline
     $30\degree$&$2.78\times10^{6}$&$2.18\times10^{6}$&$2.02\times10^{6}$&$2.54\times10^{6}$\\ 
     $50\degree$&$4.50\times10^{6}$&$4.04\times10^{6}$&$4.36\times10^{6}$&$5.18\times10^{6}$\\ 
     $70\degree$&$3.36\times10^{6}$&$4.44\times10^{6}$&$4.10\times10^{6}$&$3.94\times10^{6}$\\ 
     $90\degree$&$2.90\times10^{6}$&$2.52\times10^{6}$&$2.50\times10^{6}$&$2.86\times10^{6}$\\ 
     $110\degree$&$1.68\times10^{6}$&$1.81\times10^{6}$&$1.21\times10^{6}$&$1.95\times10^{6}$\\ 
    \\
    3.0~GeV\\
    \hline
     $30\degree$&$1.04\times10^{6}$&$1.34\times10^{6}$&$1.35\times10^{6}$&$1.48\times10^{6}$\\ 
     $50\degree$&$2.10\times10^{6}$&$1.96\times10^{6}$&$1.46\times10^{6}$&$1.60\times10^{6}$\\ 
     $70\degree$&$1.53\times10^{5}$&$1.39\times10^{6}$&$1.81\times10^{6}$&$1.34\times10^{6}$\\ 
     $90\degree$&$8.16\times10^{5}$&$7.58\times10^{5}$&$6.10\times10^{5}$&$6.68\times10^{5}$\\ 
     $110\degree$&$4.20\times10^{5}$&$3.18\times10^{5}$&$3.22\times10^{5}$&$2.80\times10^{5}$\\ 
  \end{tabular}
  \caption{Event rates per day for pions from pion production striking
    the $5\times5$ calorimeter detector arrays.}
  \label{tb:pionrates}
\end{table}

\Cref{tb:leptonrates} gives the daily event rates for the leptons from
pion production striking the $5\times5$ array of each calorimeter.
These rates should be compared with those in \cref{tb:2GeV} and
\cref{tb:3GeV} that give the rates for the events of interest striking
the central $3\times3$ array.  The lepton rate from pion production is
generally higher than the elastic scattered events of interest.
However, the lepton events of interest, arising from elastic $ep$
scattering, are peaked at significantly higher energies while the
lepton energies from pion production are lower in energy and
distributed over a broad range.

\Cref{tb:baryonrates} and \cref{tb:pionrates} give the corresponding
daily rates for the baryons and pions striking the $5\times5$ array of
each calorimeter.  While these rates by themselves are comparable to
the elastic $ep$ events of interest the energy actually deposited in
the calorimeter will be significantly less and should be readily
distinguished from the elastic lepton signal.  Of course more detailed
Monte Carlo simulations are necessary and these are in progress.


\section{Monte Carlo Simulations}
\label{sec:MC}

In order to study the energy deposited in the $5\times5$ calorimeter
arrays proposed in this document a simple GEANT4 Monte
Carlo~\citep{ALLISON2016186} simulation was developed. Beams of
electrons, protons, and pions ($\pi^+$ and $\pi^0$) were directed
through 1~m of air at the center of a $5\times5$ calorimeter array at
normal incidence. Initial momenta of 100~MeV/c to 2500~MeV/c in
100~MeV/c steps were studied.

Four combinations of absorbers (none, 10~mm Al, 10~mm Al + 10~mm Pb,
and 10~mm Al + 20~mm Pb) were placed at the front face of the
calorimeter array to study the effect this would have.  The 10~mm
aluminum plate would naturally form part of the cooling system needed
to obtain a stable energy resolution from the PbWO$_4$ crystals.  The
various thicknesses of lead were introduced to study how this could be
used to reduce background from protons and pions and the effect this
would have on the lepton signal.

\begin{figure}[!ht]
\begin{subfloat}[][]{
  \includegraphics[width=0.48\textwidth,viewport=0 0 570 380, clip]
  {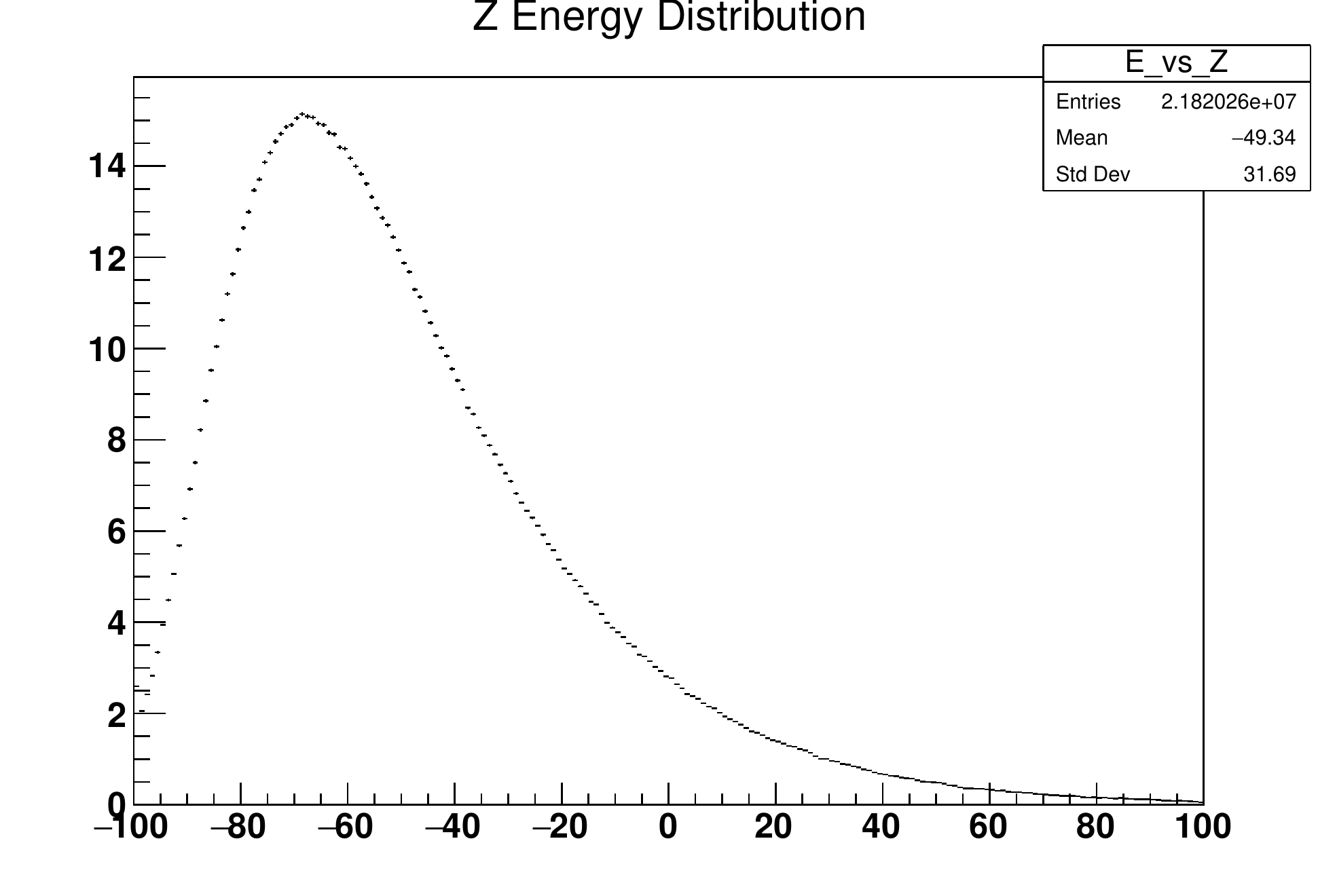}}
\end{subfloat}
\hfill
\begin{subfloat}[][]{
  \includegraphics[width=0.48\textwidth,viewport=0 0 570 380, clip]
  {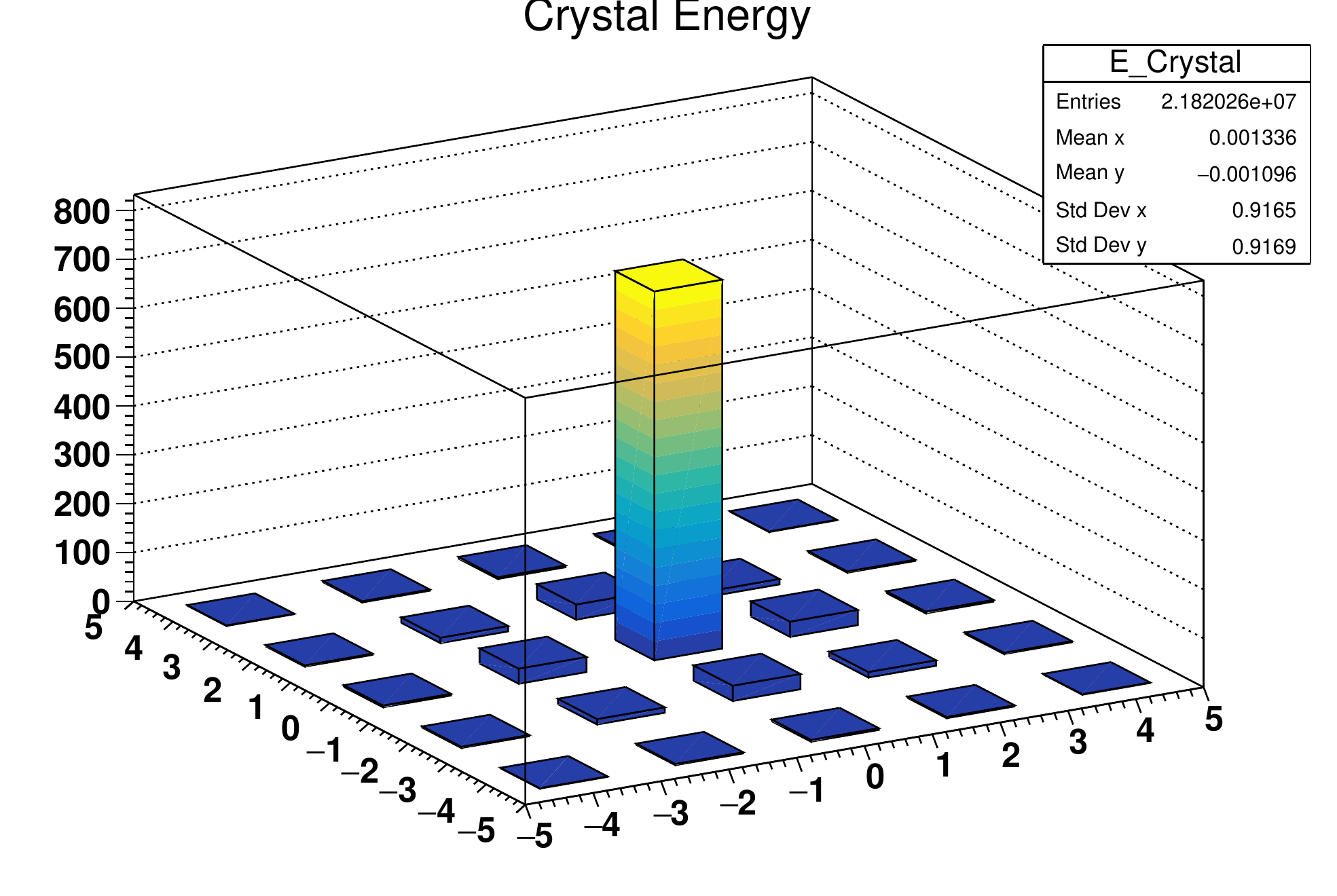}}
\end{subfloat}
\vskip0.3cm
\begin{subfloat}[][]{
  \includegraphics[width=0.48\textwidth,viewport=0 0 570 380, clip]
  {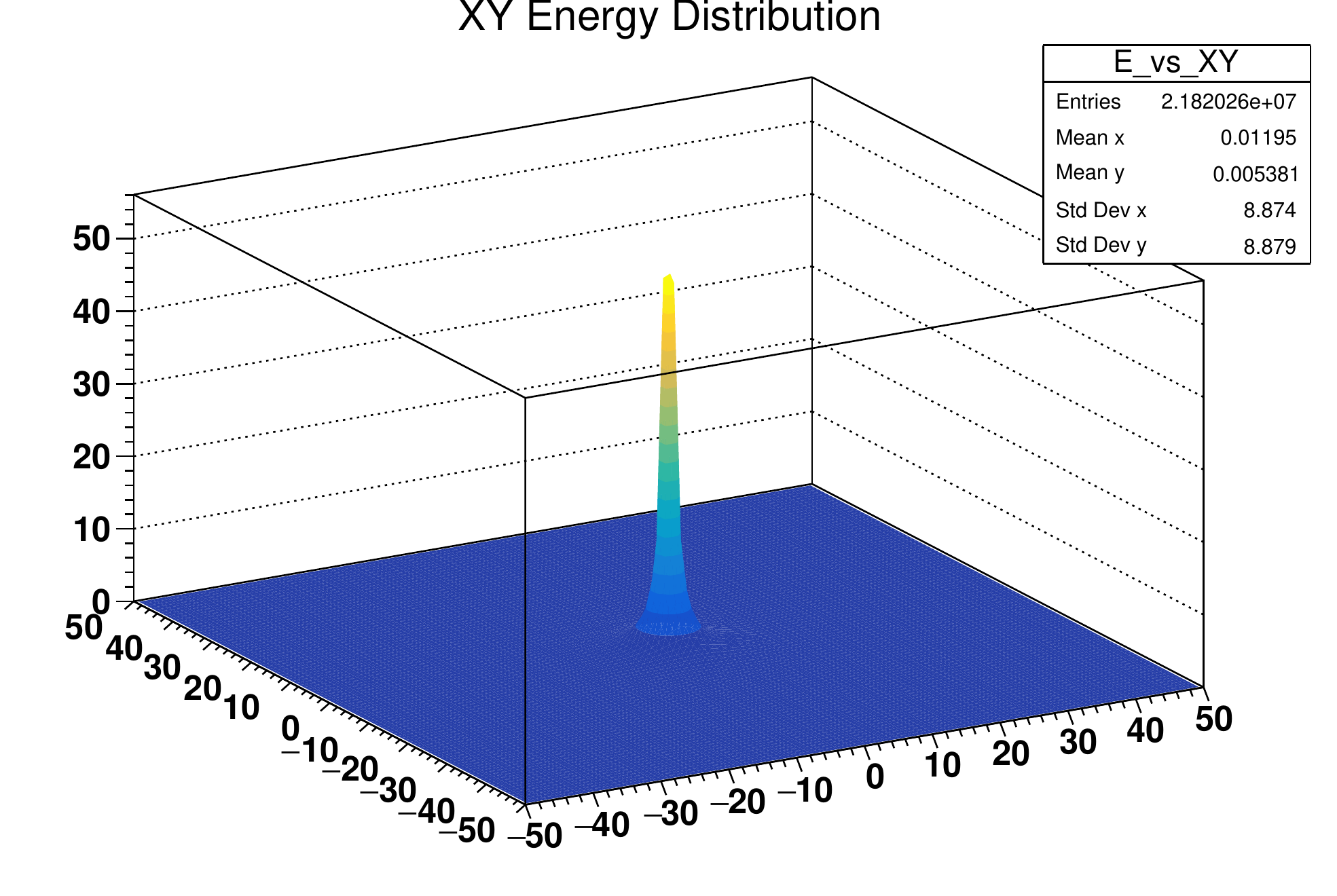}}
\end{subfloat}
\hfill
\parbox[b][0.35\textwidth]{0.48\textwidth} {\caption{Monte Carlo
    studies of electron showering in a $5\times5$ PbWO$_4$
    calorimeter.  Incident electron momentum was 1000~MeV and a 10~mm
    aluminum absorber was placed before the crystals. (a) Longitudinal
    energy distribution. (b) Total energy detected by each crystal. (c)
    Transverse energy distribution in the calorimeter.
\label{fig:E1000-Al10}}}
\end{figure}

\Cref{fig:E1000-Al10} illustrates the Monte Carlo studies performed.
With the simulations, details of the longitudinal and transverse
energy distributions can be studied though in an actual experiment
only the energy deposited in the individual crystals are
available. However, these studies show that the electron shower is
effectively contained longitudinally and the transverse distribution
is narrow. Further studies will investigate the reconstruction of
position and angle from the energy deposited in the crystals alone and
also unfolding events with multiple incident particles.

Results for various incident particles and absorbers are presented as
a function of the particle momentum and absorber thicknesses in the
following sections.

\clearpage


\subsection{Electrons and Positrons}

\begin{figure}[!ht]
\begin{subfloat}[][]{
  \includegraphics[width=0.47\textwidth,viewport=10 5 515 355, clip]
  {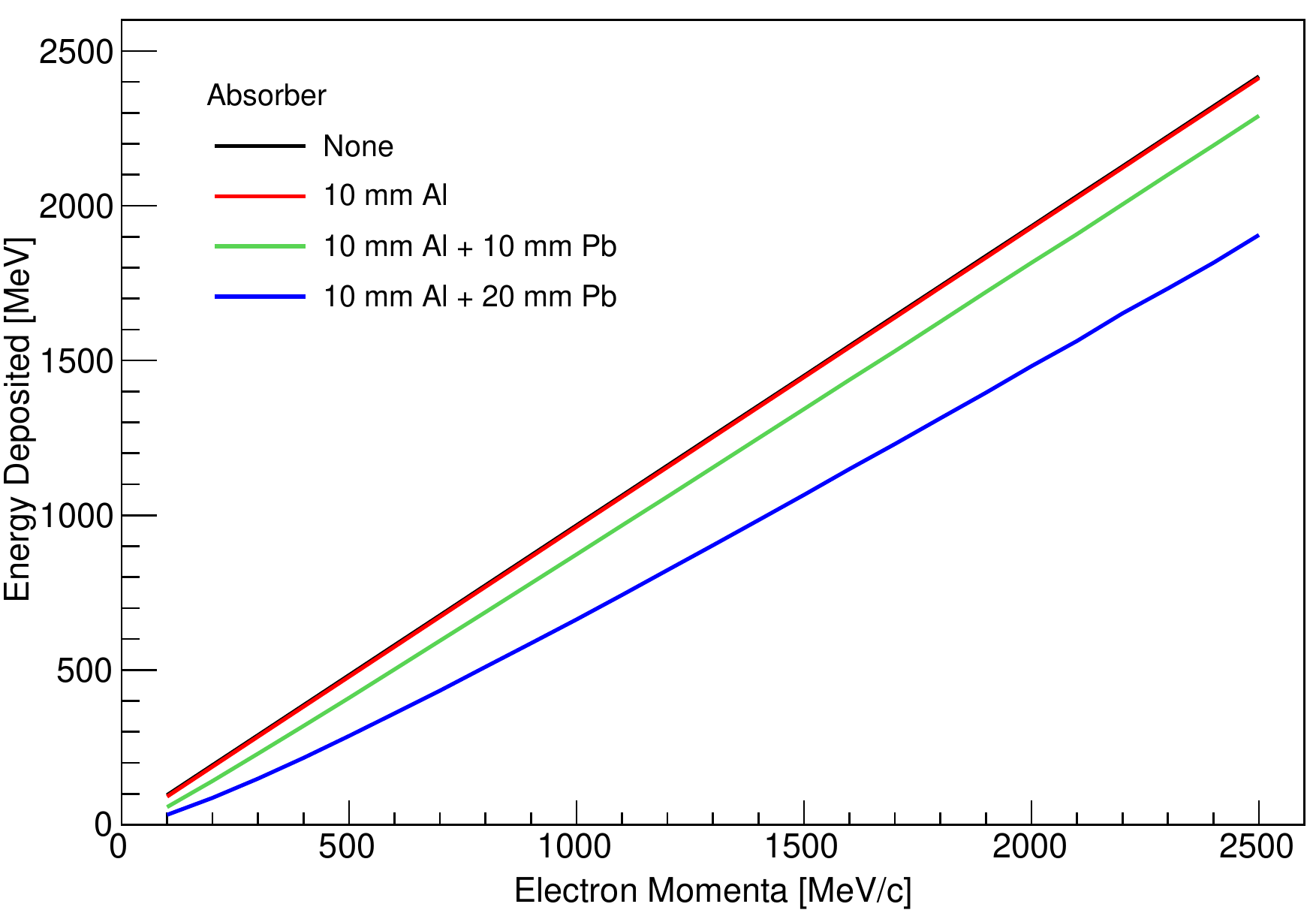}}
\end{subfloat}
\hfill
\begin{subfloat}[][]{
  \includegraphics[width=0.47\textwidth,viewport=10 5 515 355, clip]
  {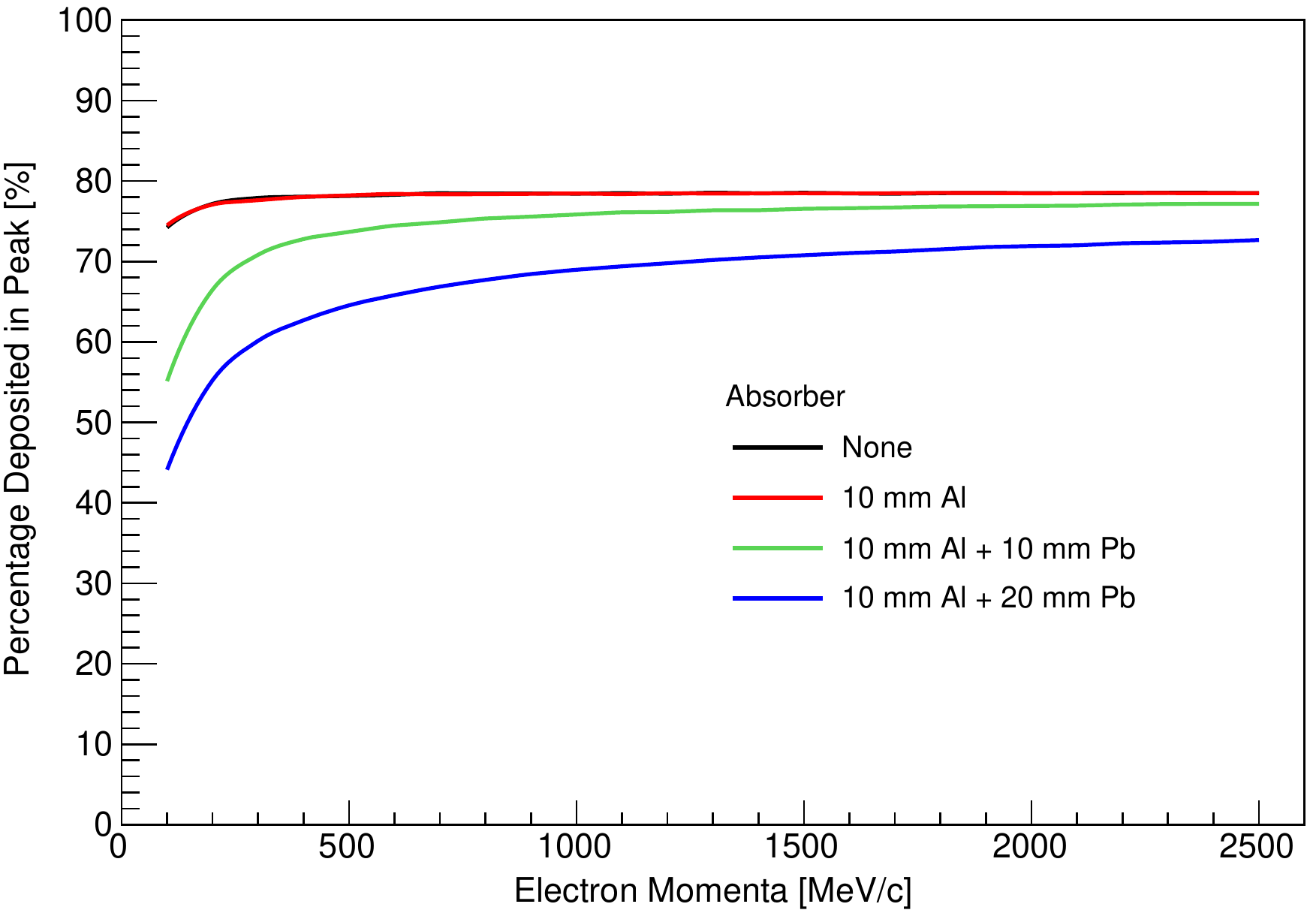}}
\end{subfloat}
\skip 0.3cm
\begin{subfloat}[][]{
  \includegraphics[width=0.47\textwidth,viewport=10 5 515 355, clip]
  {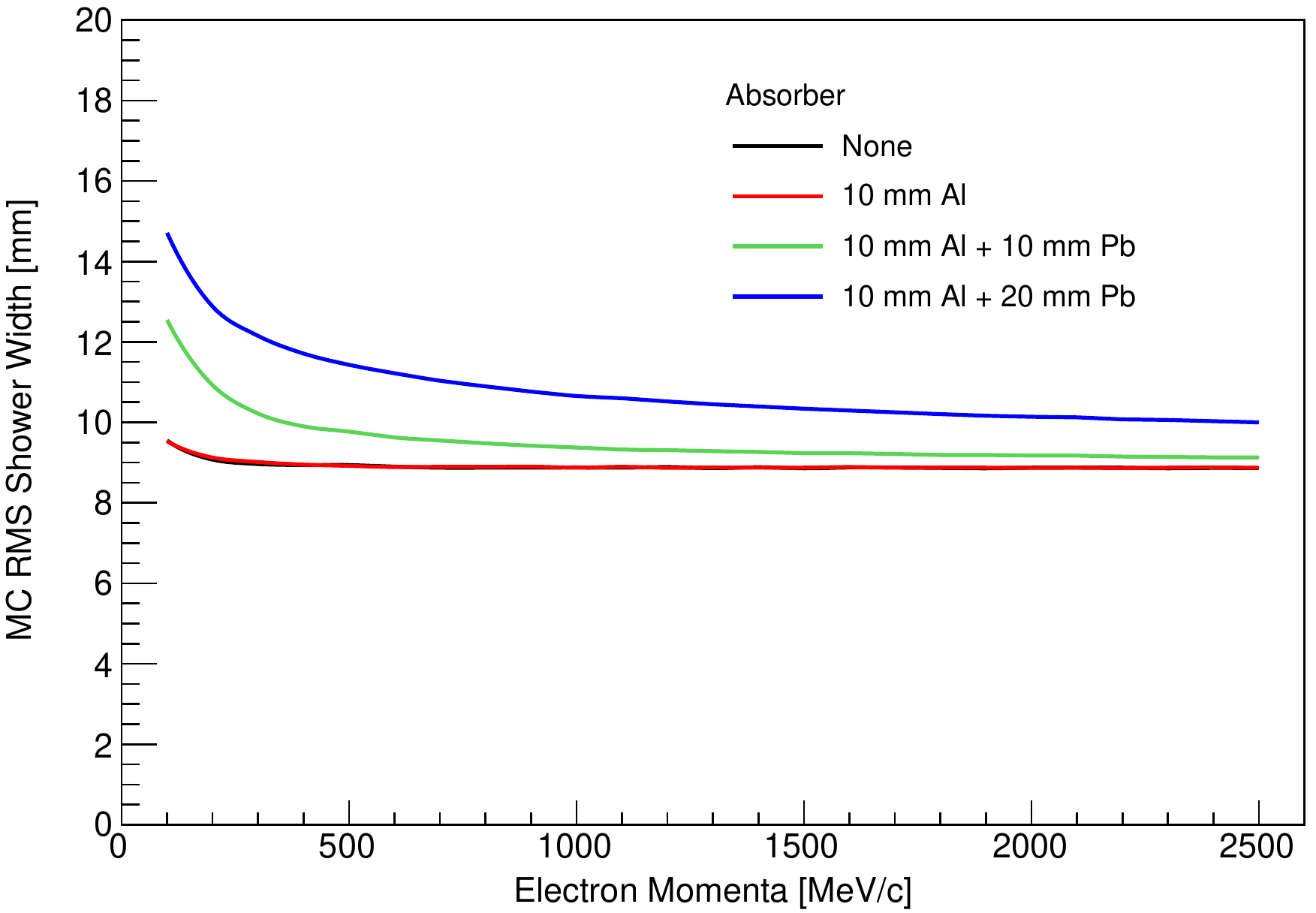}}
\end{subfloat}
\hfill
\parbox[b][0.35\textwidth]{0.47\textwidth} {\caption{Electron
    showering in a $5\times5$ PbWO$_4$ calorimeter array as a function
    of incident electron momentum with different absorbers. (a)
    Sum of energies in all 25 crystals, (b) Percentage of energy in
    the central crystal, and (c) RMS width of transverse shower
    development.
\label{fig:Electron}}}
\end{figure}

As shown in~\cref{fig:Electron} a lepton incident on the central
crystal of the calorimeter array deposits almost all its energy in the
calorimeter.  Most of that energy ($\sim80\%$) is in the central
crystal. The shower width is also quite narrow ($\sim10$~mm). Note
that the 10~mm aluminum absorber has almost no effect on the lepton
shower.  The lead absorber increases the transverse width of the
shower significantly at low momenta and to a lesser degree at higher
momenta resulting in some losses in total energy and the percentage
deposited in the central crystal.

Not surprisingly positrons have a virtually identical behavior and are
therefore not plotted separately here.

\clearpage


\subsection{Protons}

\begin{figure}[!ht]
\begin{subfloat}[][]{
  \includegraphics[width=0.47\textwidth,viewport=10 5 515 355, clip]
  {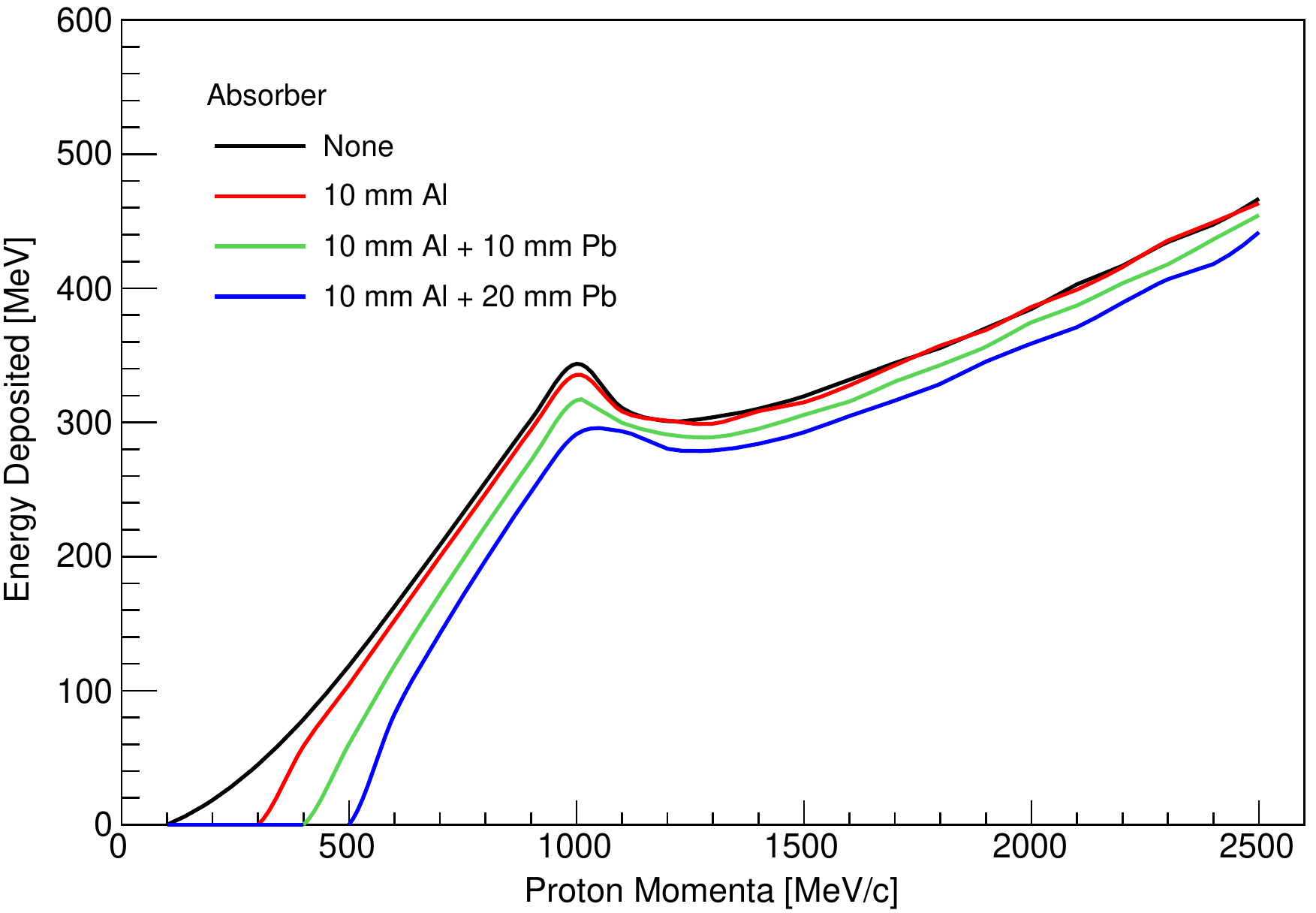}}
\end{subfloat}
\hfill
\begin{subfloat}[][]{
  \includegraphics[width=0.47\textwidth,viewport=10 5 515 355, clip]
  {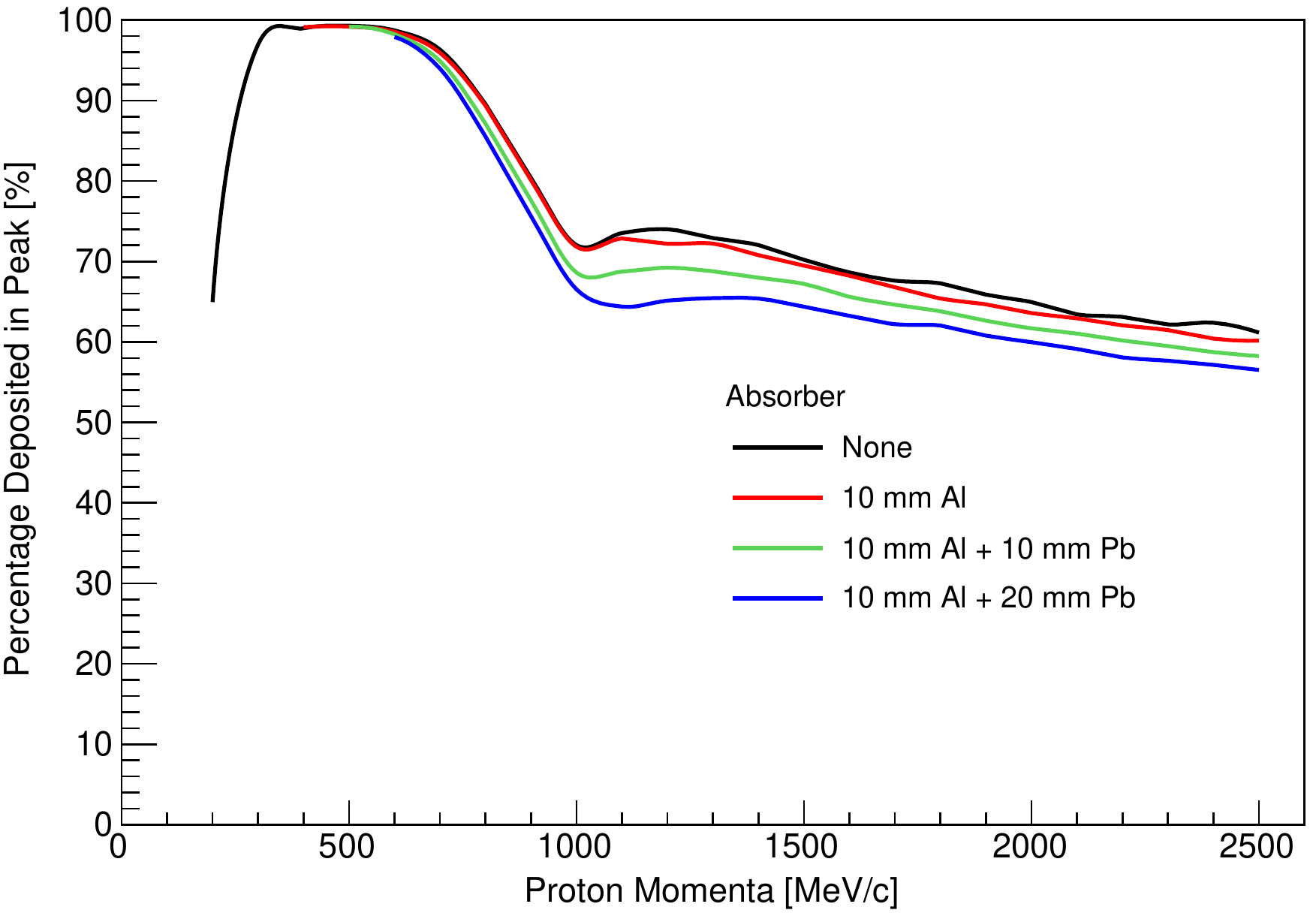}}
\end{subfloat}
\vskip 0.3cm
\begin{subfloat}[][]{
  \includegraphics[width=0.47\textwidth,viewport=10 5 515 355, clip]
  {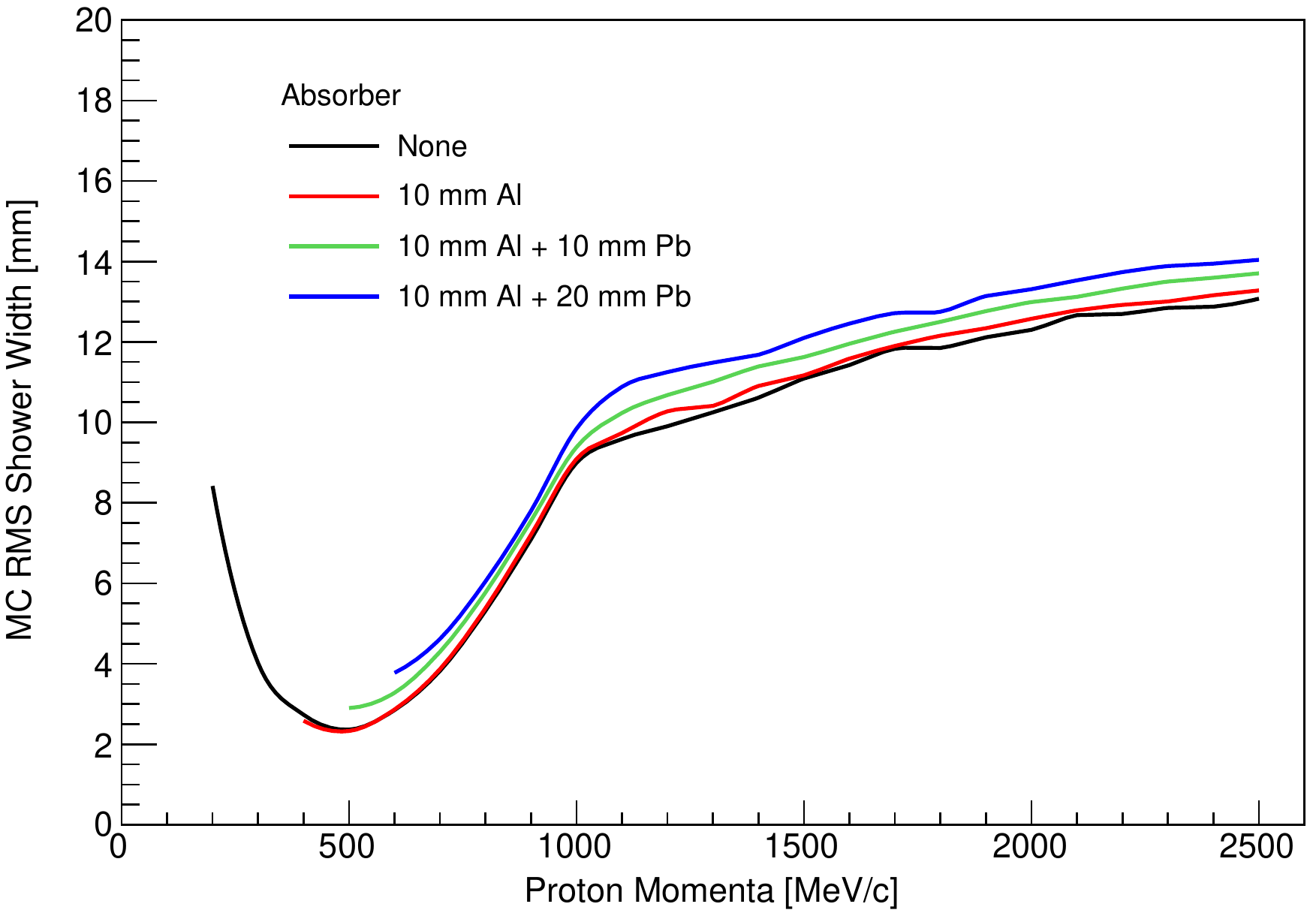}}
\end{subfloat}
\hfill
\parbox[b][0.35\textwidth]{0.47\textwidth} {\caption{Proton showering
    in a $5\times5$ PbWO$_4$ calorimeter array as a function of
    incident proton momentum and different absorbers. (a) Sum of
    energies deposited in all 25 crystals, (b) Percentage of energy in
    the central crystal, and (c) RMS width of transverse shower
    development.
\label{fig:Proton}}}
\end{figure}

\Cref{fig:Proton} shows the results for proton incident on the
calorimeter array. The total energy deposited in the calorimeter is
significantly less than the incident energies and for the most part is
a third for momenta below 1000~MeV/c and between 300 and 400~MeV/c for
higher momenta.  This is consistent with the calorimeter being only
one nuclear interaction length in depth so the proton has a tendency
to pass straight through depositing only a fraction of its energy.
Most of the energy deposited ($\sim70\%$) is in the struck
crystal. The absorbers have little effect except at low incident
momenta where the proton can be completely absorbed.  This may be
useful in stopping the large number of lower energy protons produced
at backward angles as well as low energy protons from pion production.

\clearpage


\subsection{Neutrons}

\begin{figure}[!ht]
\begin{subfloat}[][]{
  \includegraphics[width=0.47\textwidth,viewport=10 5 515 355, clip]
  {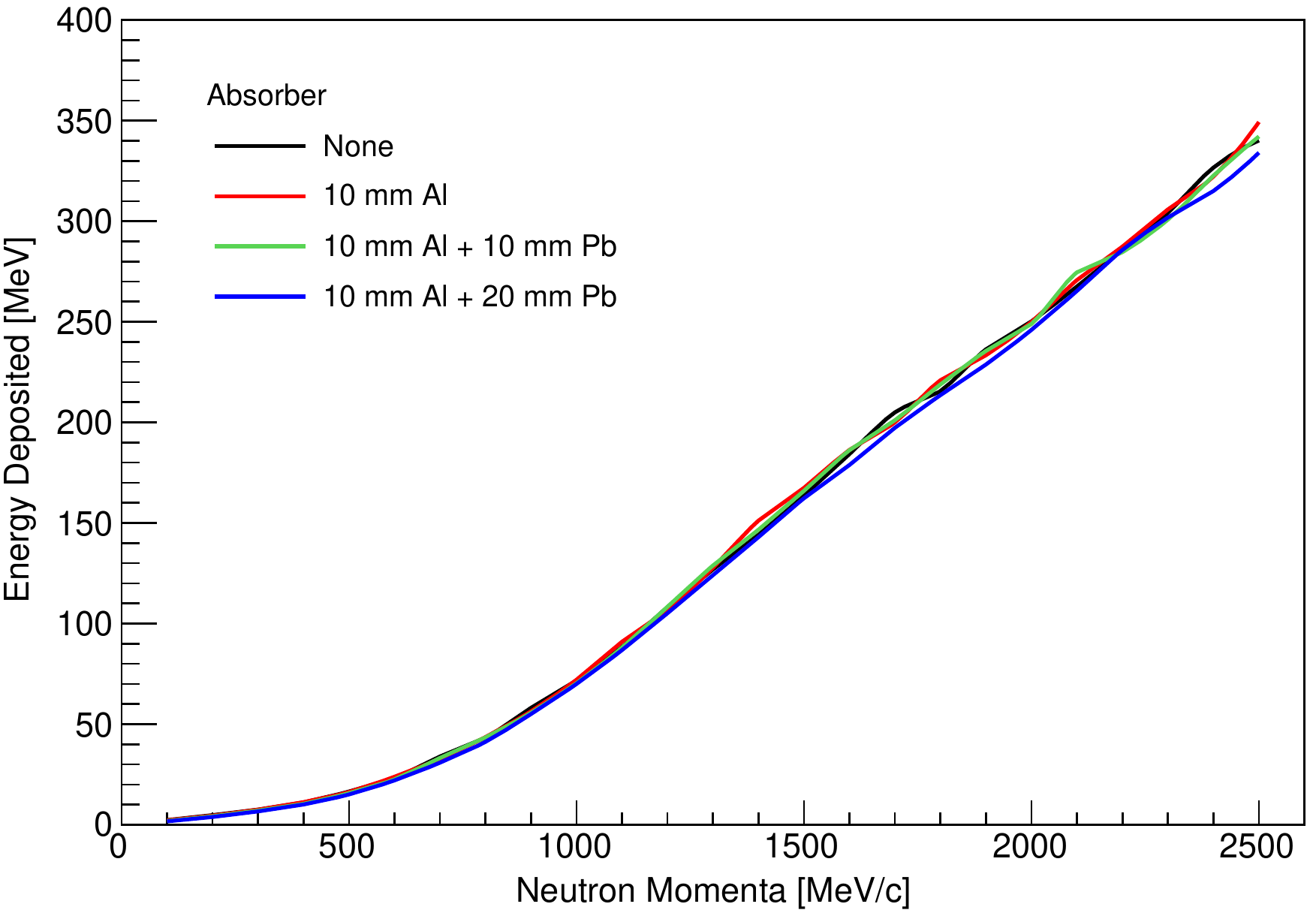}}
\end{subfloat}
\hfill
\begin{subfloat}[][]{
  \includegraphics[width=0.47\textwidth,viewport=10 5 515 355, clip]
  {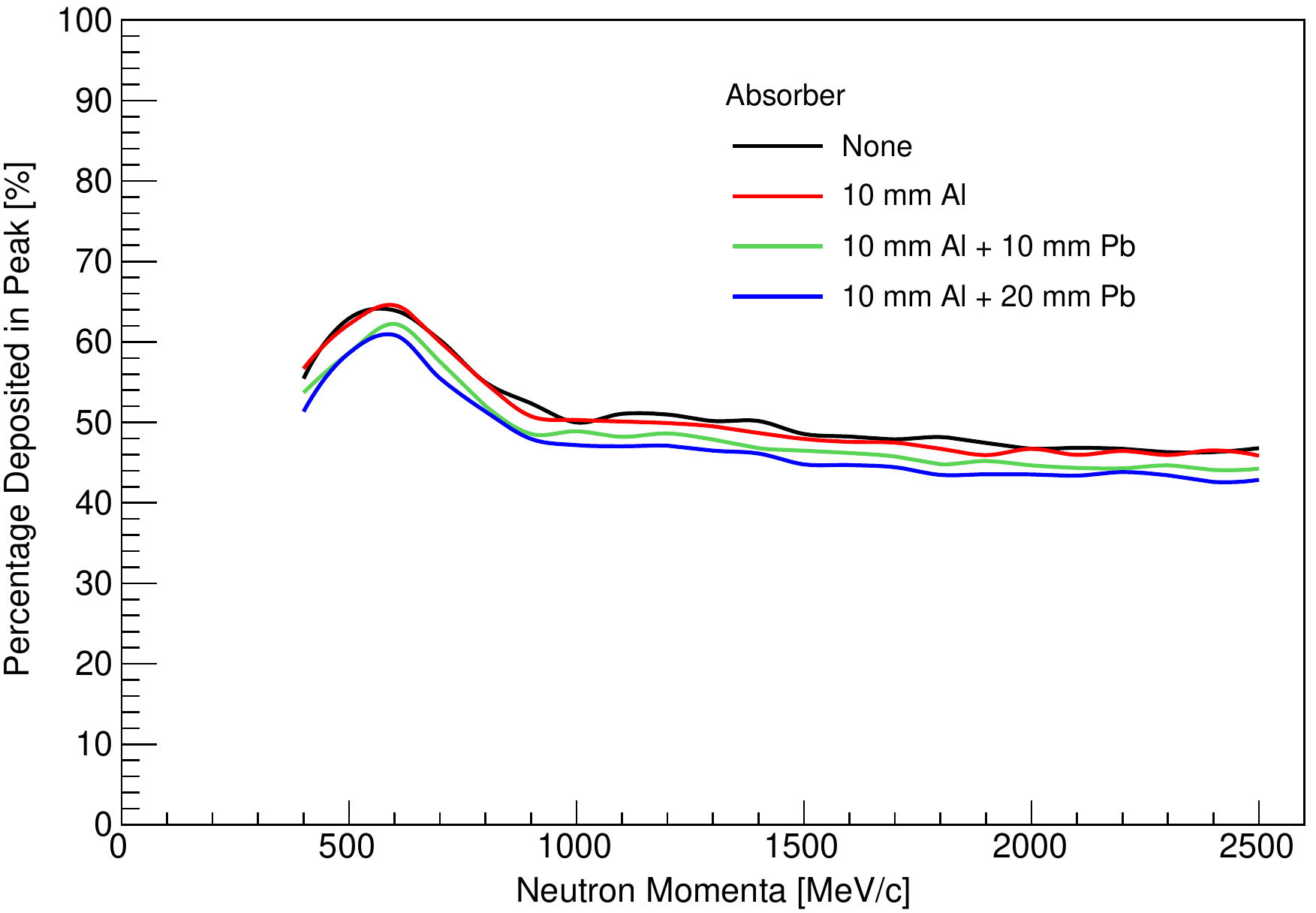}}
\end{subfloat}
\vskip 0.3cm
\begin{subfloat}[][]{
  \includegraphics[width=0.47\textwidth,viewport=10 5 515 355, clip]
  {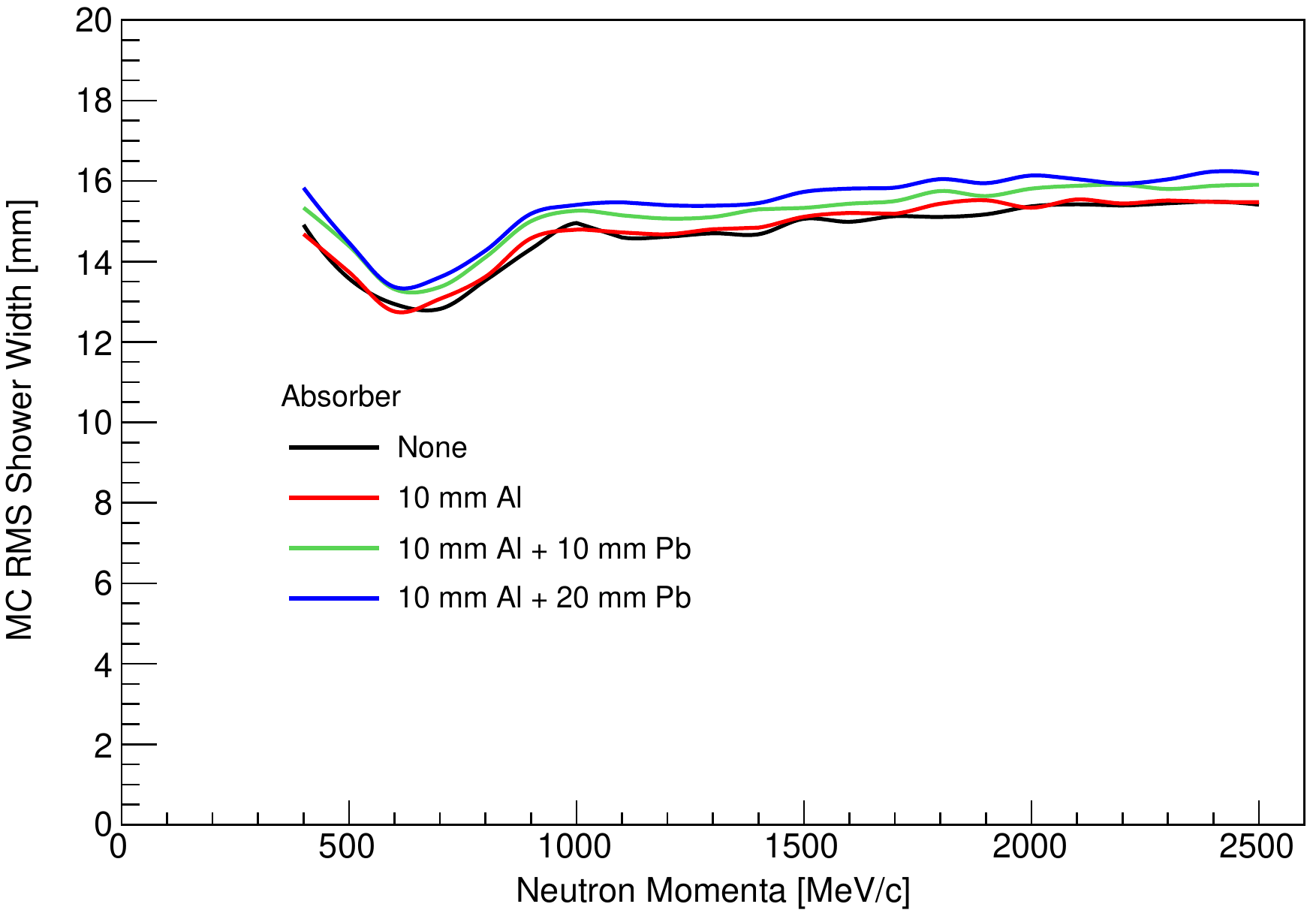}}
\end{subfloat}
\hfill
\parbox[b][0.35\textwidth]{0.47\textwidth} {\caption{Neutron showering
    in a $5\times5$ PbWO$_4$ calorimeter array as a function of
    incident neutron momentum and different absorbers. (a) Sum of
    energies deposited in all 25 crystals, (b) Percentage of energy in
    the central crystal, and (c) RMS width of transverse shower
    development.
\label{fig:Neutron}}}
\end{figure}

\Cref{fig:Neutron} shows the results for neutrons incident on the
calorimeter array. The total energy deposited in the calorimeter just
5\%--15\% of the incident energies.  About 50\% of the energy
deposited is in the struck crystal.  The absorbers have little effect.

\clearpage


\subsection{$\pi^+$}

\begin{figure}[!ht]
\begin{subfloat}[][]{
  \includegraphics[width=0.47\textwidth,viewport=10 5 515 355, clip]
  {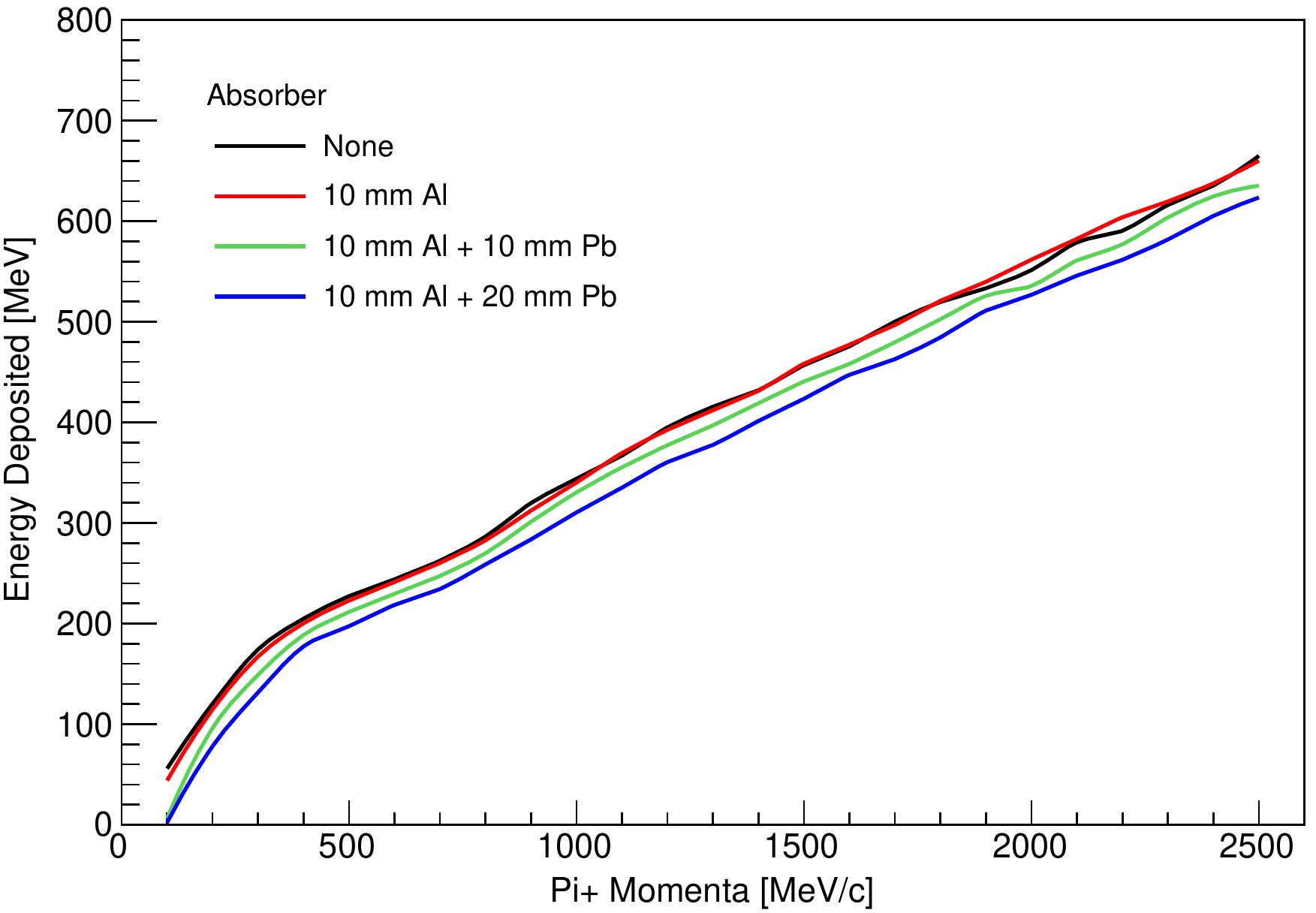}}
\end{subfloat}
\hfill
\begin{subfloat}[][]{
  \includegraphics[width=0.47\textwidth,viewport=10 5 515 355, clip]
  {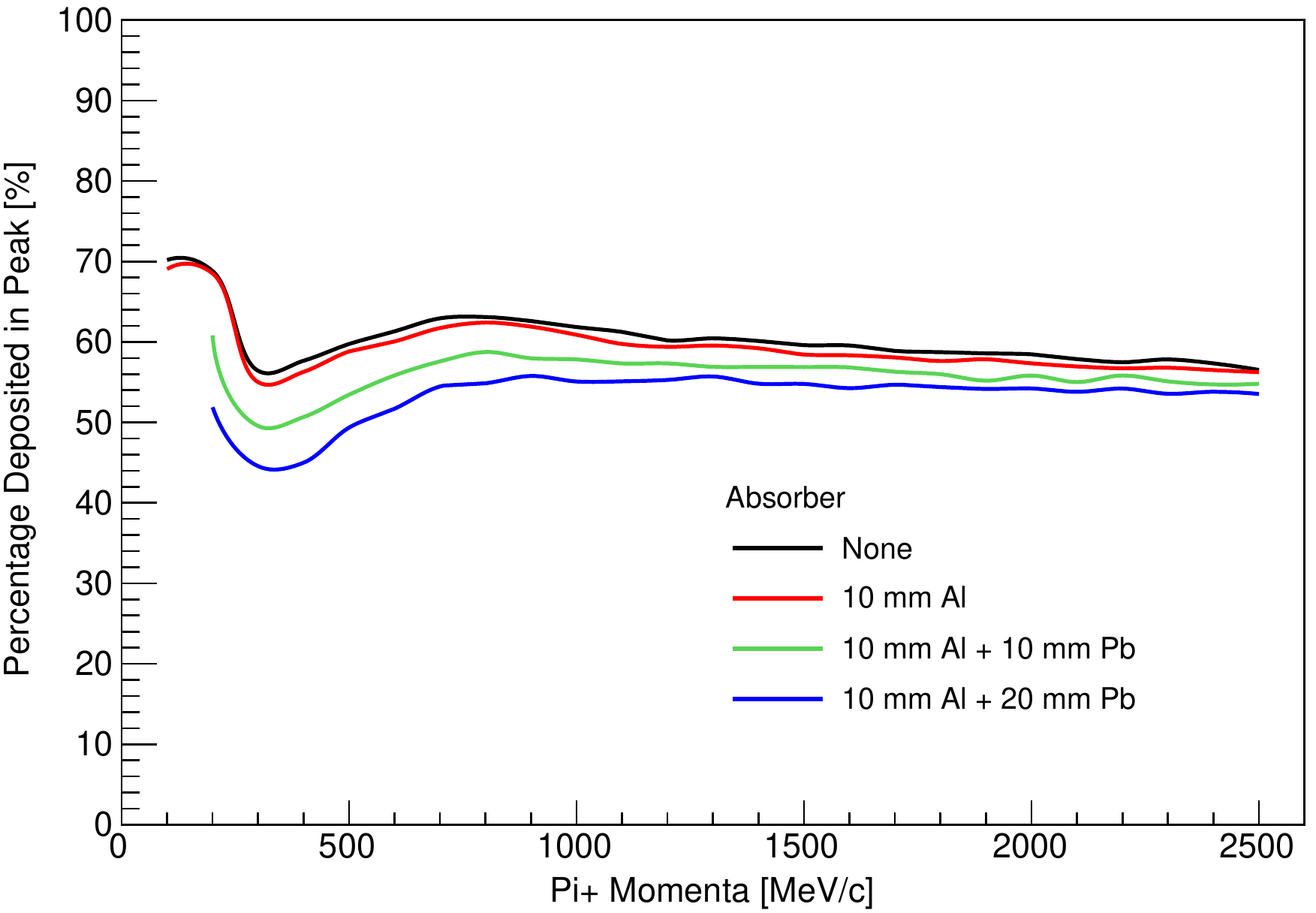}}
\end{subfloat}
\vskip 0.3cm
\begin{subfloat}[][]{
  \includegraphics[width=0.47\textwidth,viewport=10 5 515 355, clip]
  {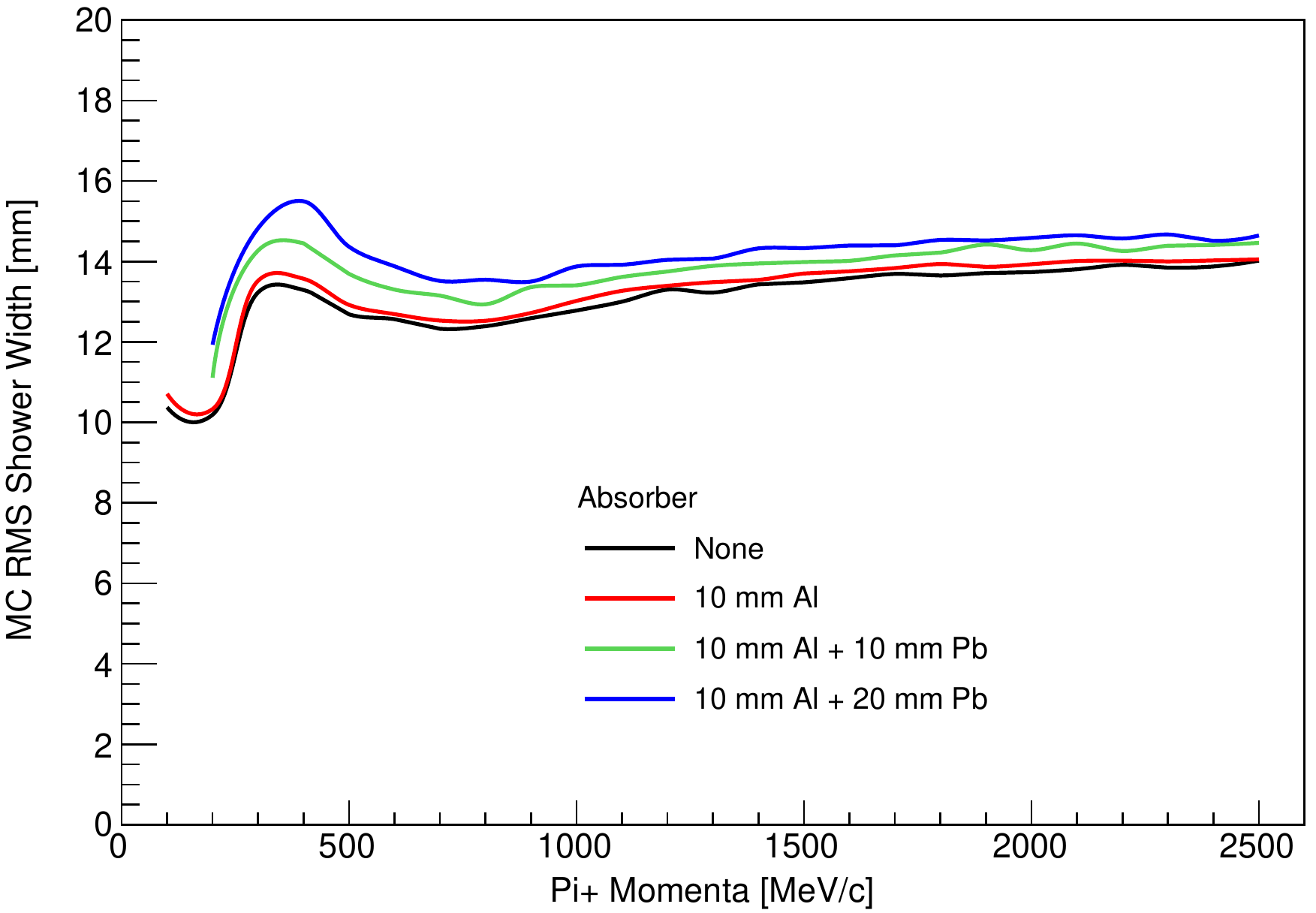}}
\end{subfloat}
\hfill
\parbox[b][0.35\textwidth]{0.47\textwidth} {\caption{$\pi^+$ showering
    in a $5\times5$ PbWO$_4$ calorimeter array as a function of
    incident pion momentum with different absorbers. (a) Sum of
    energies deposited in all 25 crystals, (b) Percentage of energy in
    the central crystal, and (c) RMS width of transverse shower
    development.
\label{fig:Pi+}}}
\end{figure}

\Cref{fig:Pi+} shows the calorimeter response to incident $\pi^+$
mesons.  The total energy deposited in the calorimeter array varies
from around 50\% of the incident momenta below 500~MeV to 25\% at
higher momenta.  The various absorber thicknesses have a small effect.
50\%--60\% is deposited in the central crystal and the RMS width for
the transverse shower development is fairly constant around
14~mm. This reduced signal from $\pi+$ will aid in distinguishing them
from the leptons of interest.  Response with $\pi^-$ mesons is
similar.

\clearpage


\subsection{$\pi^0$}

\begin{figure}[!ht]
\begin{subfloat}[][]{
  \includegraphics[width=0.47\textwidth,viewport=10 5 515 355, clip]
  {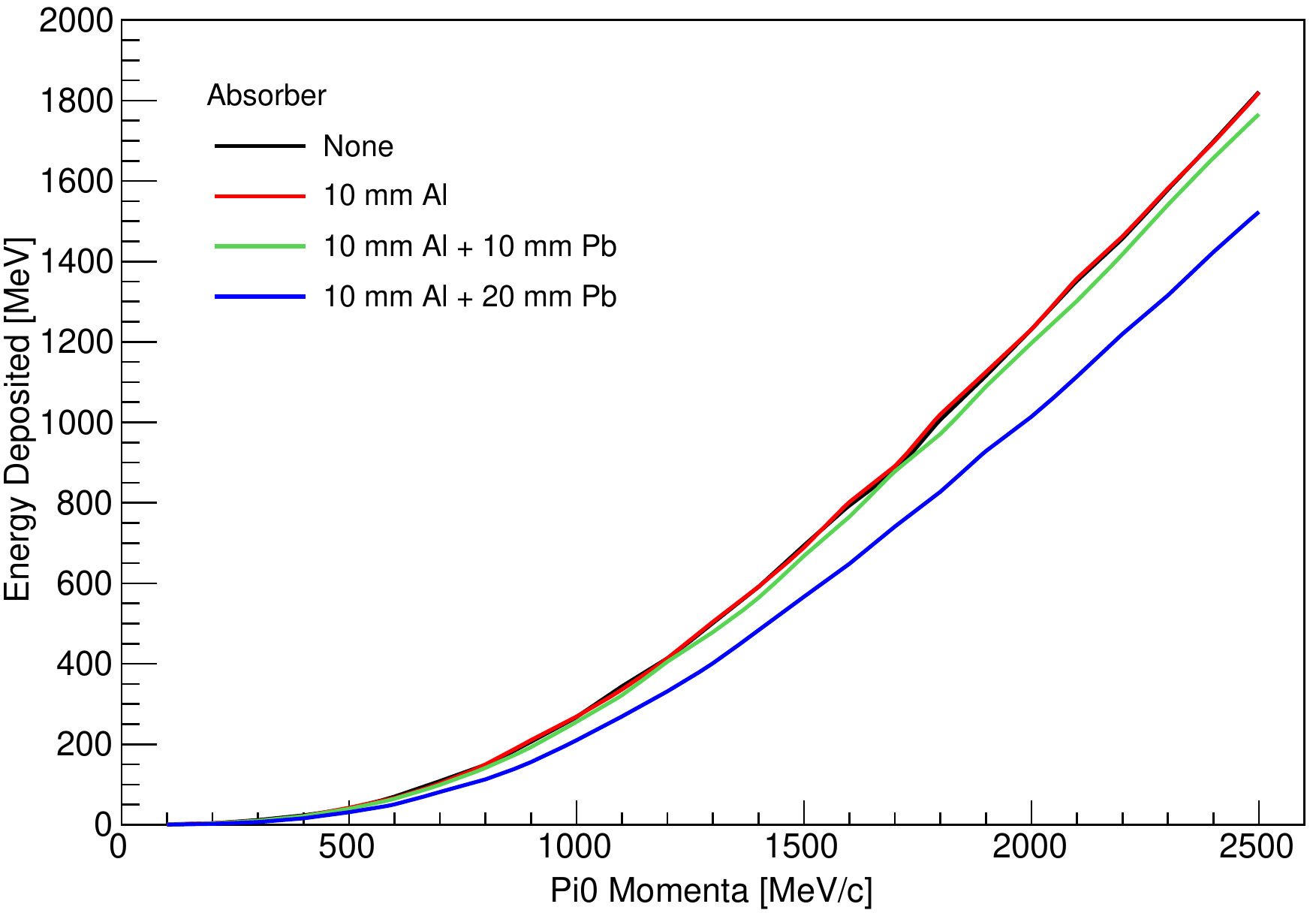}}
\end{subfloat}
\hfill
\begin{subfloat}[][]{
  \includegraphics[width=0.47\textwidth,viewport=10 5 515 355, clip]
  {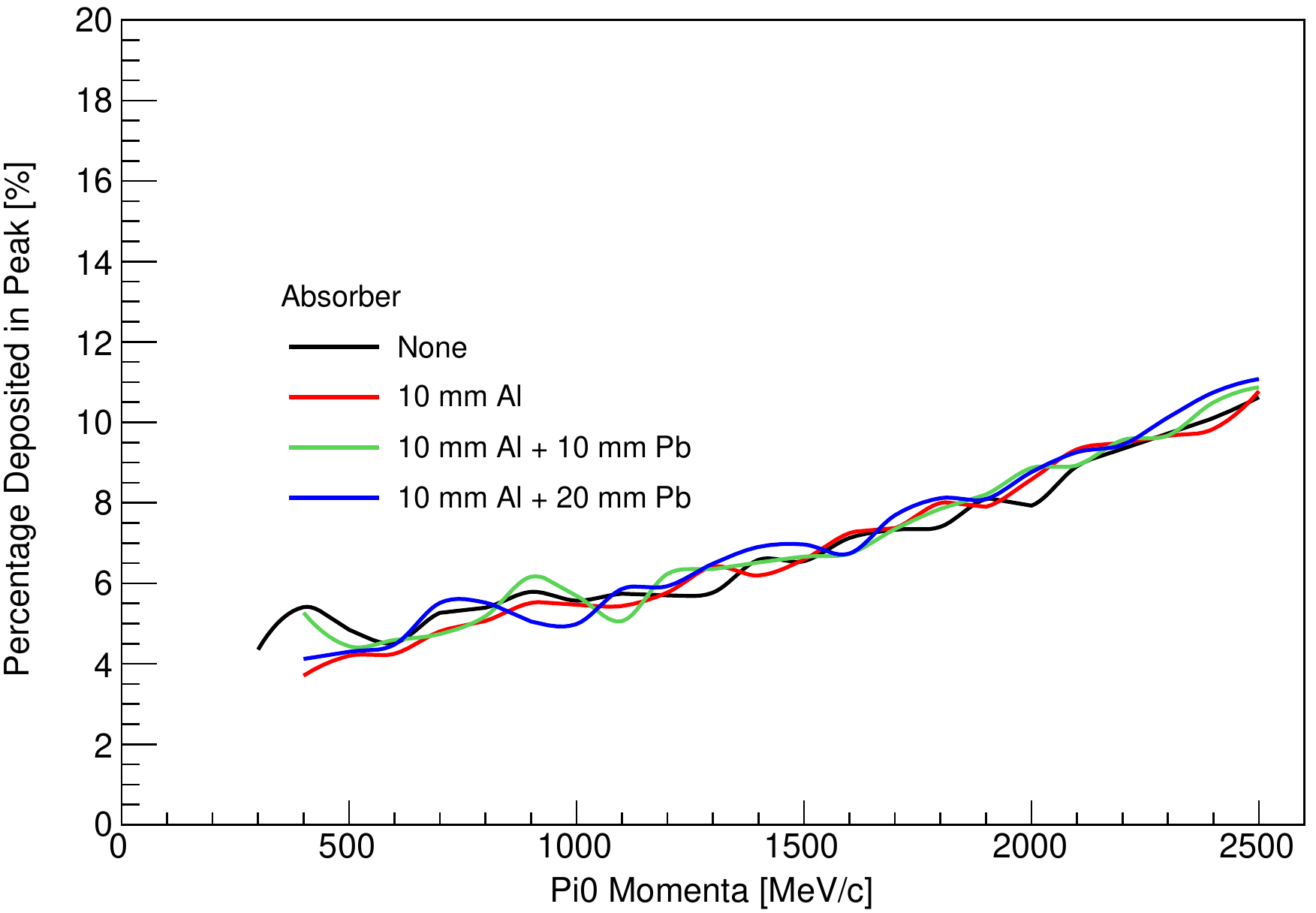}}
\end{subfloat}
\vskip 0.3cm
\begin{subfloat}[][]{
  \includegraphics[width=0.47\textwidth,viewport=10 5 515 355, clip]
  {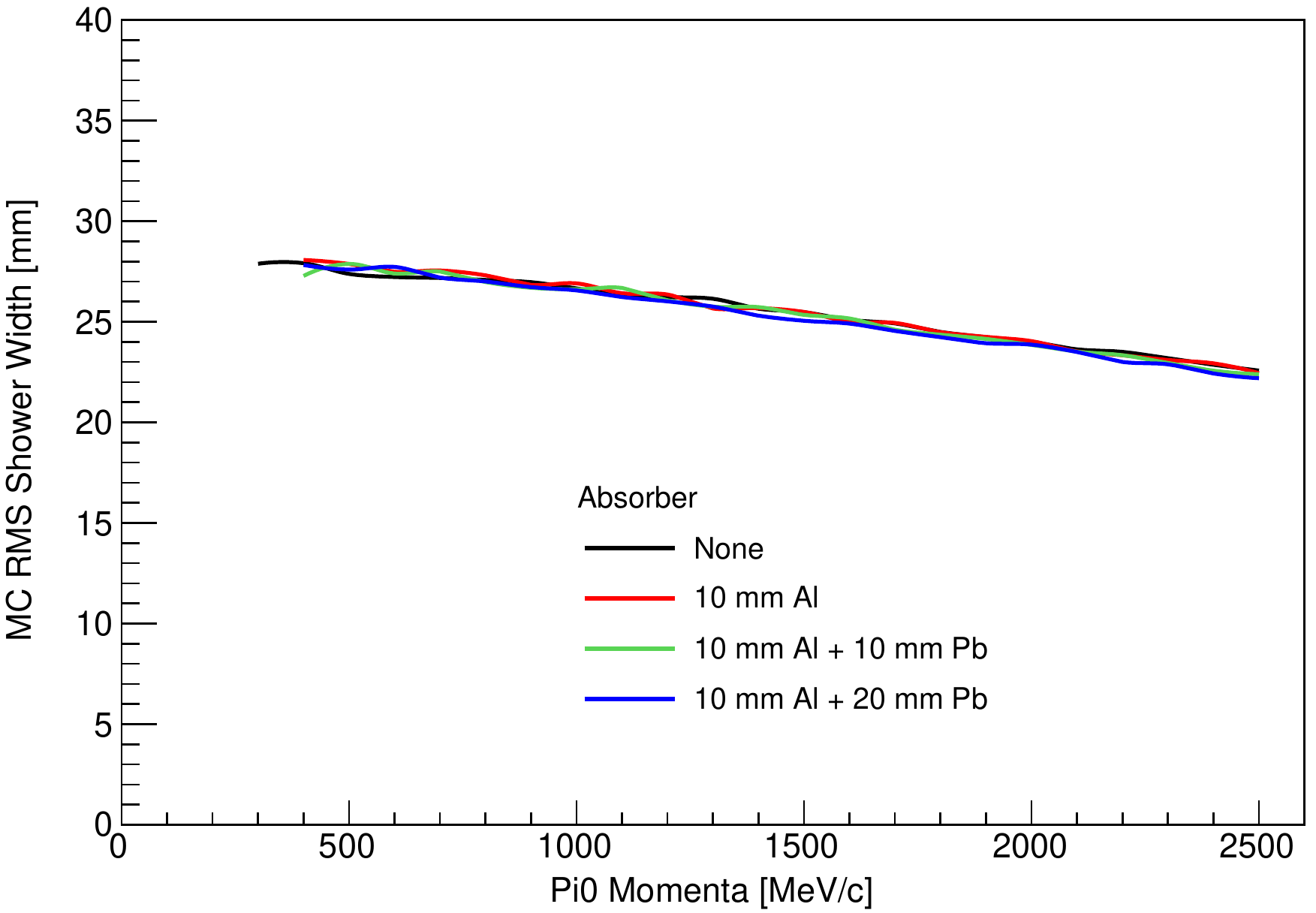}}
\end{subfloat}
\hfill
\parbox[b][0.35\textwidth]{0.47\textwidth} {\caption{$\pi^0$ showering in a $5\times5$ PbWO$_4$ calorimeter
    array as a function of incident pion momentum with different
    absorbers. (a) Sum of energies deposited in all 25 crystals, (b)
    Percentage of energy in the central crystal, and (c) RMS width of
    transverse shower development.
\label{fig:Pi0}}}
\end{figure}

\Cref{fig:Pi0} shows the calorimeter response to $\pi^0$ mesons
originating at the target 1~m away.  The $\pi^0$s primarily decay
isotropically to two photons that may or may not strike the
calorimeter.  For low energies the probability is small and very
little energy is deposited in the calorimeter.  At higher energies the
two photons are boosted in the direction of the calorimeter and
deposit a more significant fraction of their original energy.  The
energy deposited is, in any case, spread over a large area and not
restricted to the central crystal as can be seen from the plots of the
percentage in the central crystal and the RMS width of the shower
development.

\clearpage


\section{Test Beam at DESY}
\label{sec:testbeam}

The Monte Carlo studies discussed in the previous section are
encouraging.  The proposed experiment can make a significant and
direct measurement of the two-photon contribution in a region of $Q^2$
and $\epsilon$ where the discrepancy is clear.  However, the Monte
Carlo studies must be verified.  It is therefore important that the
performance of the calorimeter modules be studied in a test beam.

An initial prototype calorimeter was tested at the DESY test beam
facility in the fall, 2019.  The results are reported here in
\cref{sec:Tb}.  These initial tests with a small $3\times3$ prototype
design are encouraging with good agreement with the Monte Carlo but
further tests are needed.

We propose to perform these measurements at the DESY test beam
facility~\citep{Diener:2018qap} using a $5\times5$ calorimeter array.
The purpose of the test beam activity will be to measure the
performance of the PbWO$_4$ calorimeter array and to verify the Monte
Carlo simulations.  We would use various energies, with and without
the absorber plates, and incident at various positions and angles
across the calorimeter array.  Monte Carlo simulations can
suggest and be used to train reconstruction algorithms but these need
to be verified with actual measurements therefore the proposed test
beam studies are very important.

If the ceramic glass crystals being developed by Tanja Horn (CUA) are
available we would also test these in the prototype calorimeters.
This clearly has links with efforts underway in Europe and the United
States for future detectors for the proposed Electron-Ion Collider.


\section{Conclusion}
\label{sec:concl}

The observed discrepancy in the proton form factor ratio is a
fundamental problem in nuclear physics and possibly in quantum
electrodynamics. Why are the leading order QED radiative corrections
insufficient to resolve the discrepancy?  Are higher order corrections
necessary or are more detailed models for the intermediate hadronic
state needed? Or is some other process responsible?

An extracted positron and electron beam facility at DESY would provide
a unique opportunity to measure the two-photon exchange contribution
to elastic lepton-proton scattering over a kinematic range where the
observed discrepancy is clearly evident. The above proposal outlines
an initial plan for an experimental configuration that could help
resolve this issue and provide insight to the radiative corrections
needed to understand the proton form factors at higher momentum
transfers.

\clearpage


\appendix


\section{Test Beam Results}
\label{sec:Tb}

Over two weeks in September-October, 2019, a calorimeter consisting of
a $3\times3$ array of PbWO$_4$ crystals was studied at the DESY test
beam facility~\citep{Diener:2018qap}. Tests were made scanning the
electron beam across the face of the calorimeter and with different
thicknesses of absorber plates.  We also compared, in parallel, a
traditional, triggered readout with a streaming readout scheme.  Further
test beam studies were made in the fall of 2021 and spring of 2022 using a more 
realistic $5\times5$~array of lead tungstate crystals.  This calorimeter
was also cooled and measured at 25\degree, 10\degree, -10\degree, and -25\degree 
C.  A small number of high-density ceramic glass crystals were also tested.  
The analysis of the 2022 test run is ongoing.

\subsection{Calorimeter Setup and Tests}
\label{sec:Tb_setup}

\begin{figure}[!ht]
\begin{center}
\includegraphics[width=\textwidth]{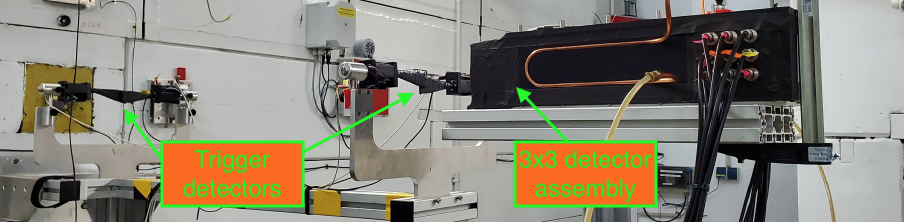}
\caption{Photo of 3x3 lead tungstate calorimeter prototype and trigger detectors used in initial test run at DESY. \label{figIF1}}
\end{center}
\end{figure}

The calorimeter used nine $2\times2\times20$ cm$^3$ lead tungstate
crystals read out using Hamamatsu R1166 PMTs attached to one end of
each crystal.  The crystals were wrapped with one layer of white Tyvek
(0.4~mm thick) and an outer layer of opaque aluminum foil (0.09~mm
thick). The crystal-PMT assemblies were placed inside a black
anodized aluminum housing. See \cref{figIF1}.  Copper tubes for water
cooling were installed on the outside of the aluminum box.  The
calorimeter assembly was mounted on an XY translation table but was
electrically isolated from the table.  A collimator and a set of four
thin scintillators upstream of the calorimeter were using in the
triggered readout.

High voltage for the PMTs was provided by LeCroy 1461N modules.

Signals from the PMTs were divided by a 50~$\Omega$ splitter. One side
of each splitter output was connected through a 100~ns delay cable to
CAEN V792 QDC. The signals from the four thin scintillators were
combined in a coincidence unit requiring a triple coincidence that was
used to trigger the QDC.  The other splitter output was connected to a
CAEN V1725 digitizer.  Since the digitizer had only 8 channels a
decision was made to read out crystals 1 to 7, and use channel 0 to
record the trigger signal in parallel.

The gain from each crystal and PMT was matched using a 5.2~GeV beam
incident on the center of the crystal and the HV adjusted to give a
common value close to the end of the QDC range.

\begin{figure}[!ht]
\begin{center}
\includegraphics[scale=0.4]{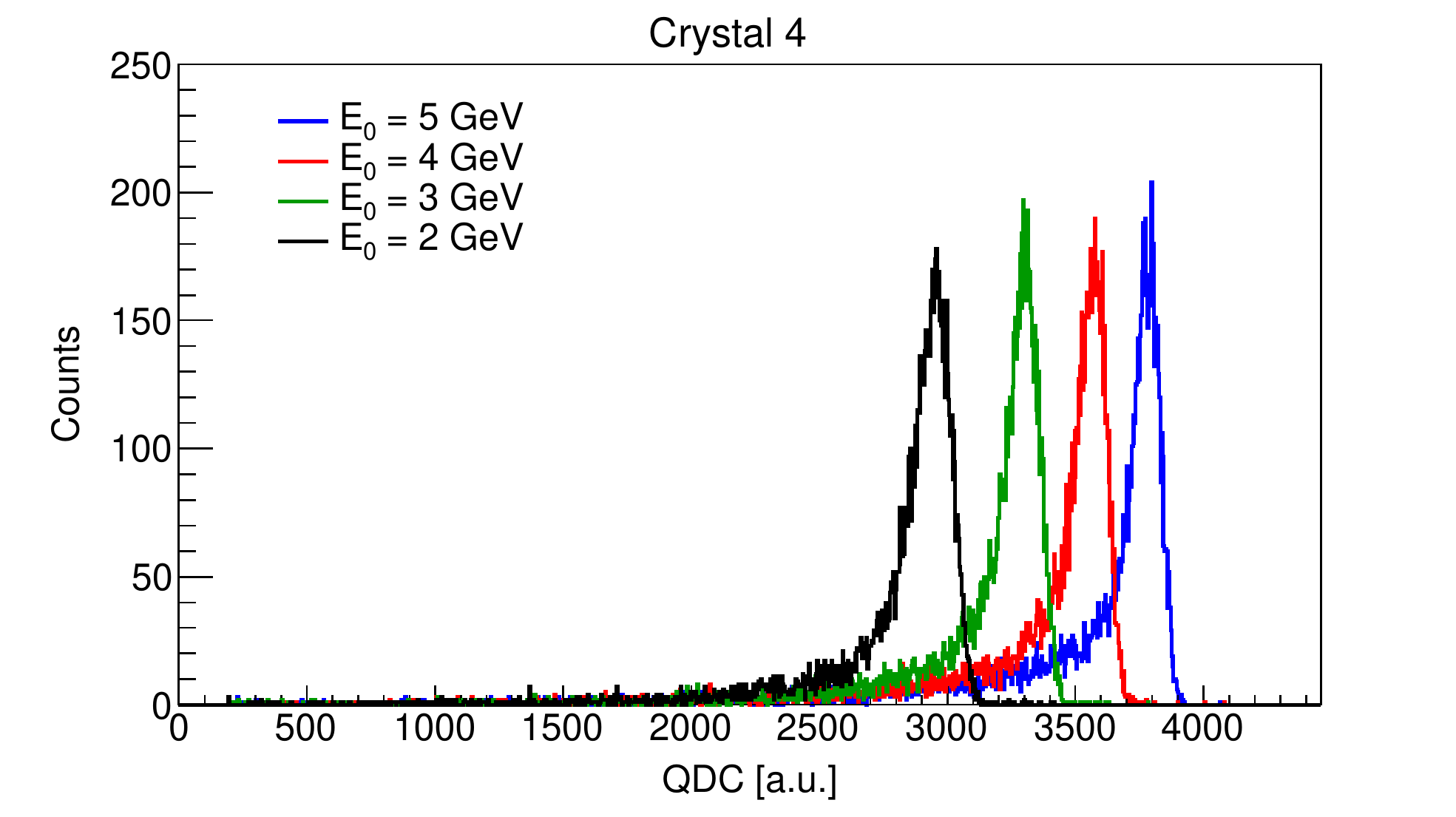}
\includegraphics[scale=0.4]{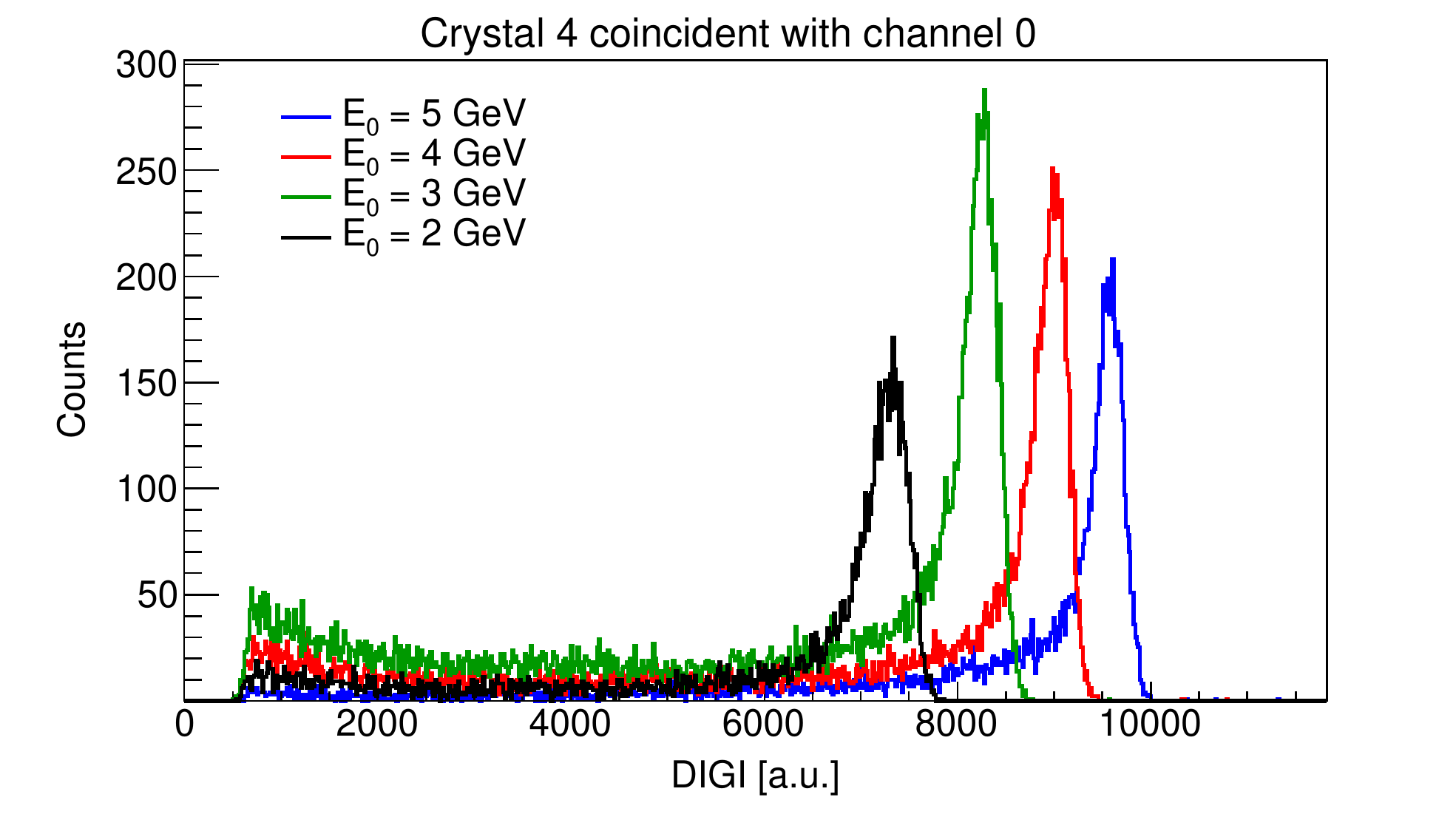}
\caption{Deposited energy in central crystal recorded by QDC (left)
  and digitizer (right). The digitizer spectra also required a
  coincidence with a trigger signal in digitizer
  channel~0. \label{figIF2:spectra}}
\end{center}
\end{figure}

Data were collected at beam energies of 2, 3, 4, and 5~GeV and with
$2\times2$~mm$^2$ and $8\times8$~mm$^2$ collimators.  For all
conditions scans were made over the face of the calorimeter and using
0, 1, and 2~cm thick lead absorber plates before the calorimeter.
Typical energy spectra for all four beam energies can be seen in
\cref{figIF2:spectra}. The left figure shows the QDC spectra (with
pedestal subtraction) and the right figure the digitizer spectra for
events which are in coincidence with the trigger signal.

\begin{figure}[!ht]
\begin{center}
\includegraphics[scale=0.22]{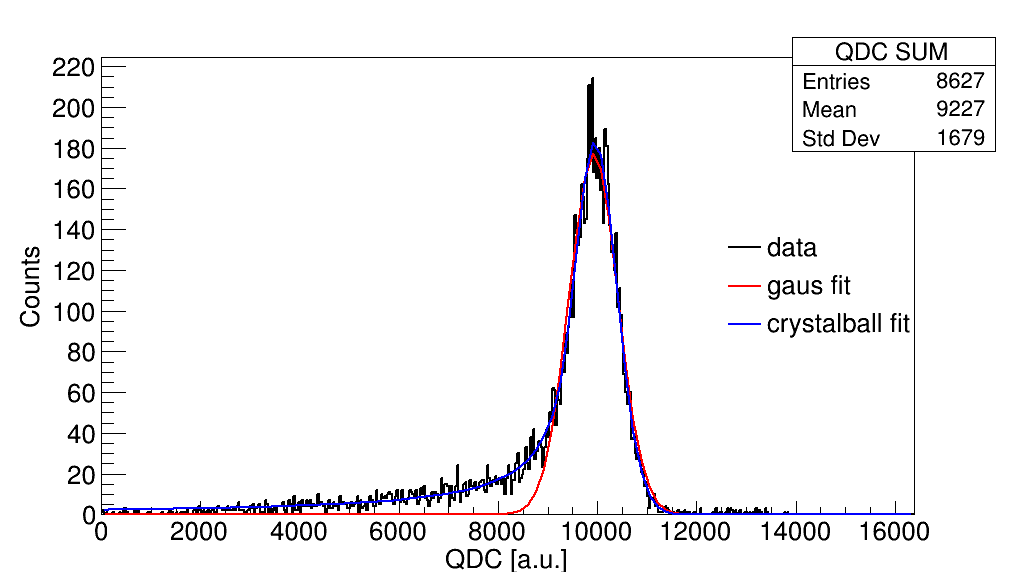}
\includegraphics[scale=0.22]{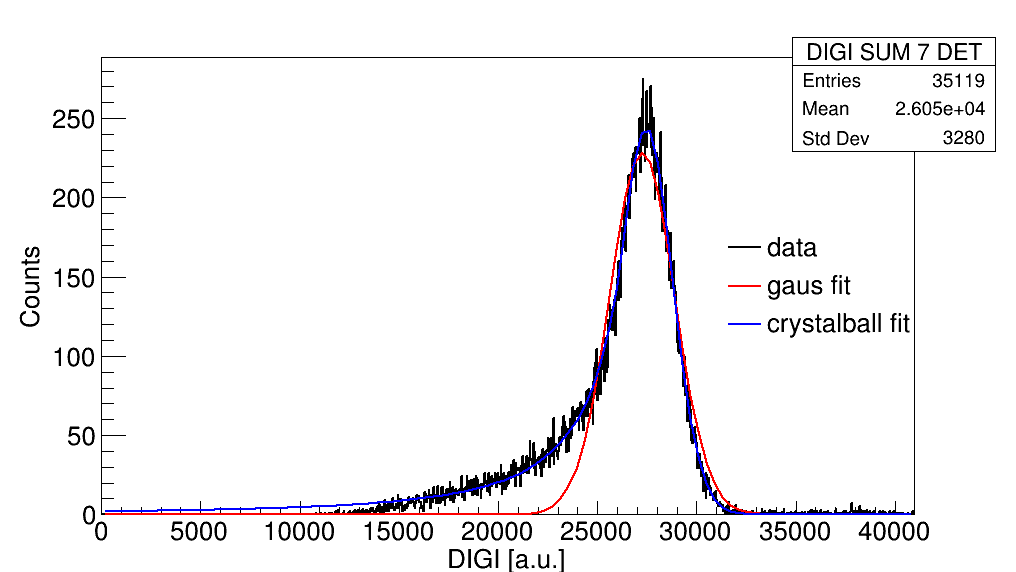}
\caption{Sum of energies deposited in all crystals recorded by QDC
  (left) and digitizer.  The digitizer spectrum also required a
  coincidence with the trigger signal in
  channel~0. \label{figIF3:spectrasum}}
\end{center}
\end{figure}

\begin{figure}[!ht]
\begin{center}
\includegraphics[scale=0.3]{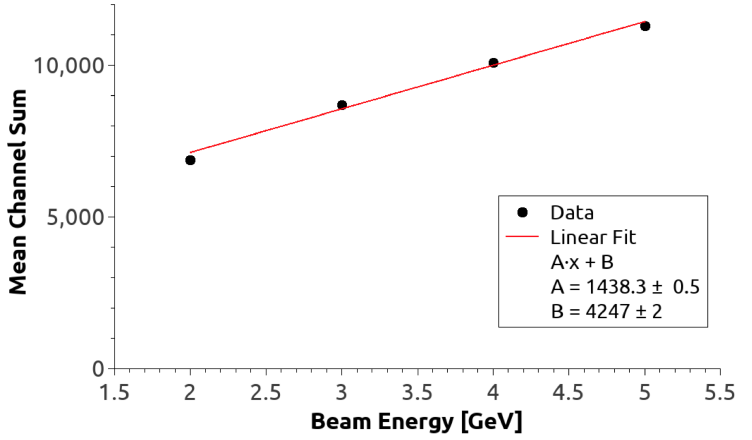}
\includegraphics[scale=0.3]{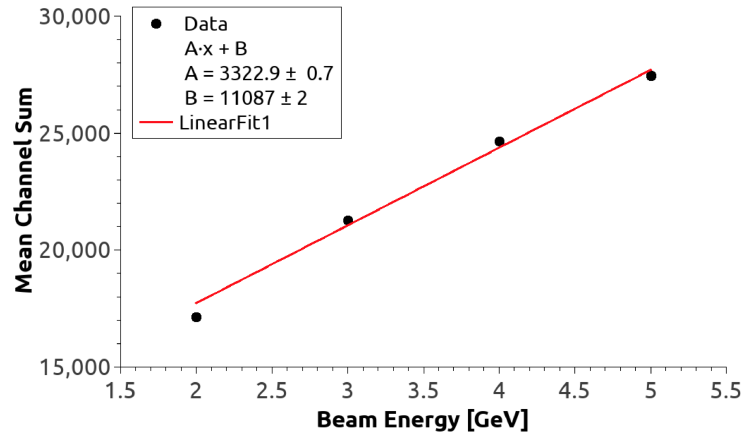}
\caption{Energy dependence of the peak position in QDC (left) and
  digitizer (right). \label{figIF4:mean}}
\end{center}
\end{figure}

\Cref{figIF3:spectrasum} shows the sum of all 25 signals with a 5~GeV
beam centered on the central crystal. The root analysis tool functions
``gaus'' and ``crystalball'' were used to fit the spectra to determine
peak position.
Linearity of the peak position with incident energy is shown in
\cref{figIF4:mean}.

\subsection{Streaming and triggered readout}

\begin{figure}[!ht]
\begin{center}
\includegraphics[scale=0.5]{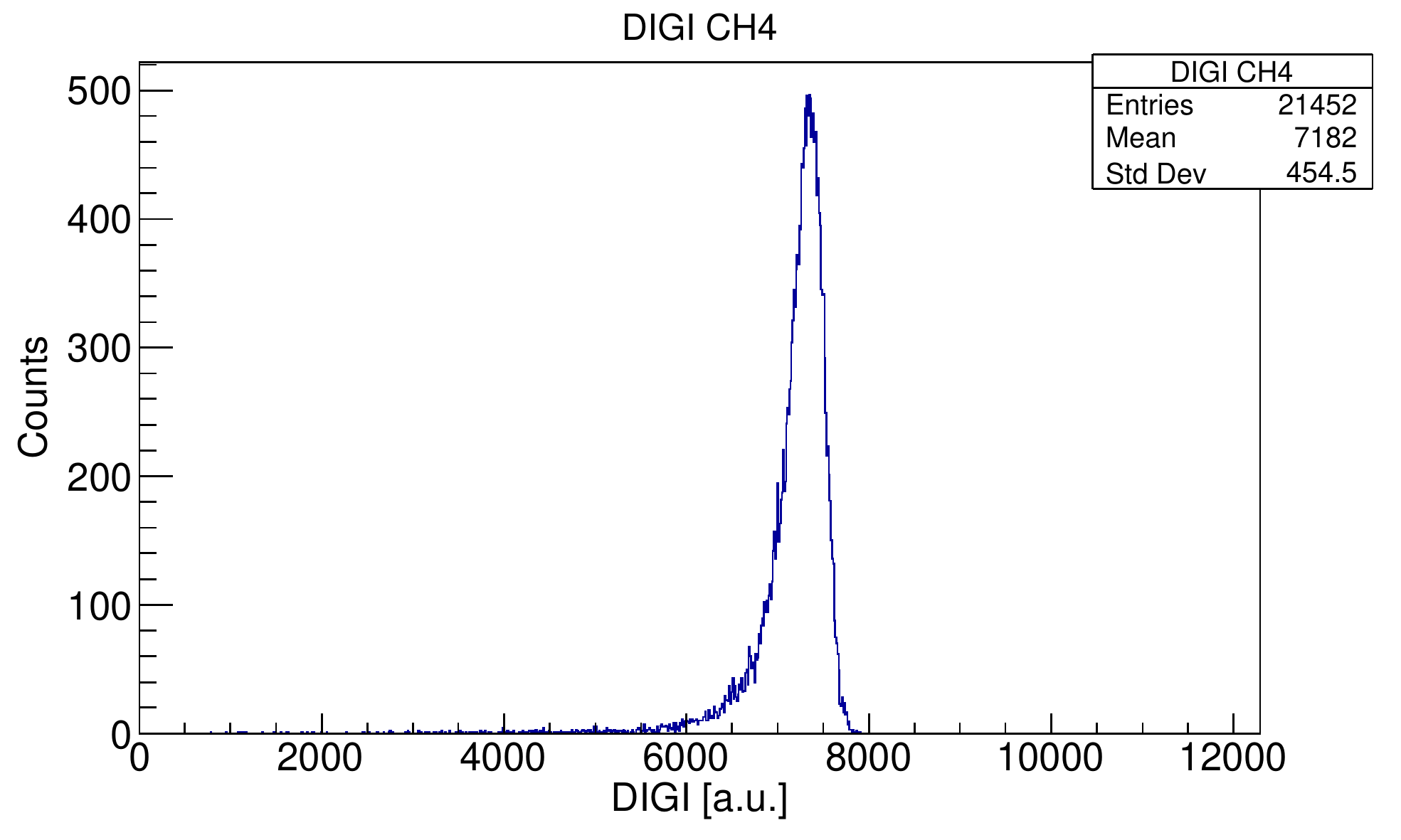}
\caption{Deposited energy in central detector (channel 4) with coincidence signals in at least 6 neighboring crystals. \label{figIF7:digioff}}
\end{center}
\end{figure}

In the triggered readout scheme all channels of the QDC are read out
together after receiving a trigger signal.  This takes some time
during which the QDC is unable to record new events (deadtime).  On
the other hand the streaming readout scheme using the digitizer
continuously records events in all channels.  Thus, the streaming
system records more events in individual channels though many signals
may be uncorrelated from cosmic rays, noise, or background events.  To
make sense of the large amount of data collected by the digitizer it
is necessary to determine the relative timing of all channels.  Then
signals at a common time can be reasonably assumed to arise from the
same event, like corresponding to showering in the calorimeter (see
\cref{figIF7:digioff}).
Similarly, the relative timing to the trigger signal used for the QDC
(connected to channel 0 of the digitizer) can be determined and used
to compare the same event collected by the QDC with that recorded by
the digitizer.
\begin{figure}[!ht]
\begin{center}
\includegraphics[scale=0.5]{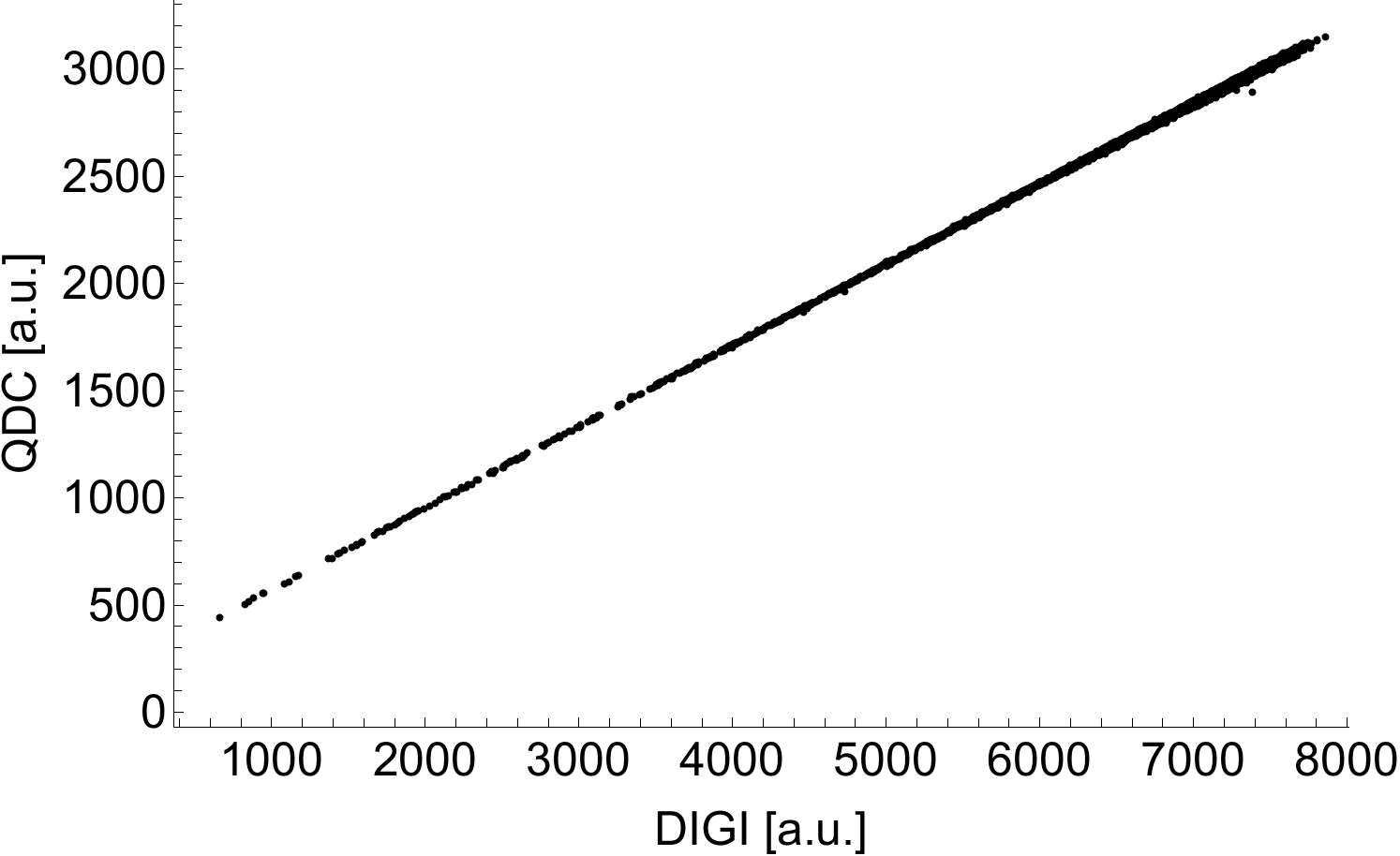}
\includegraphics[scale=0.5]{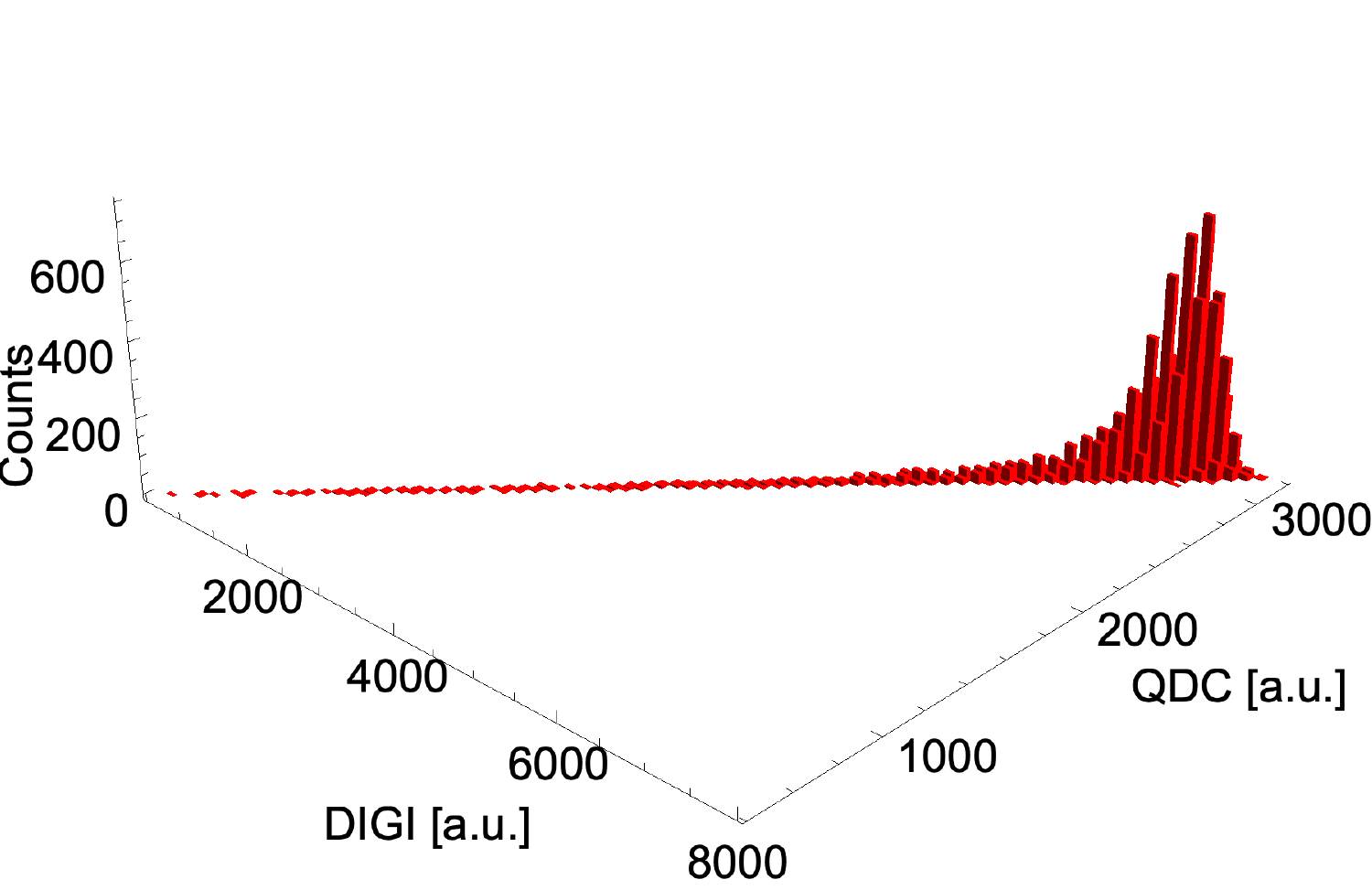}
\caption{Left figure shows the energy deposited in the digitizer
  versus that for the QDC. The right figure is the 3D histogram plot
  of the same data. \label{figIF9:Energy}}
\end{center}
\end{figure}

\subsection{Monte Carlo Simulation of Test Beam}
\label{sec:Tb_sim}
\begin{figure}[!htb]
    \centering
    \includegraphics[width=0.4\linewidth]{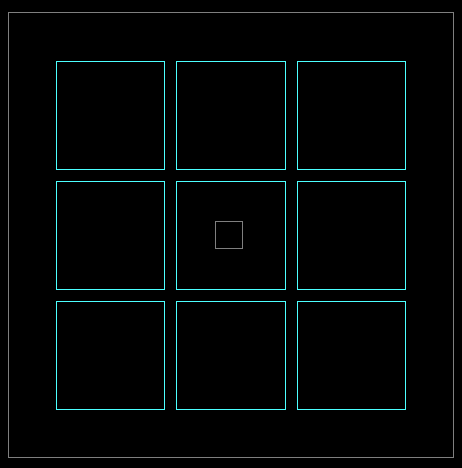} 
    \caption{The Geant4 simulation view of the front face of the
      calorimeter.}
    \label{fig:sim_view}
\end{figure}

A Monte Carlo simulation of the test beam was developed in Geant4
\citep{ALLISON2016186}. We use the FTFP\_BERT physics list provided by
Geant to simulate the showers and energy loss processes in the
crystals. We reproduced the TB24/1 area from the available drawings
and technical details provided by DESY. This included the calorimeter,
absorber plates, trigger scintillators, collimator, and the origin of
the beam source at the DESY~II ring.  This last item was found to be
very important to account for the significant energy straggling
observed in the measured spectra. The Geant4 visualization of the
front face of the $3\times3$ calorimeter array is shown in
Fig.~\ref{fig:sim_view}.

\begin{figure}[!htb]
    \centering
    \includegraphics[width=\linewidth]{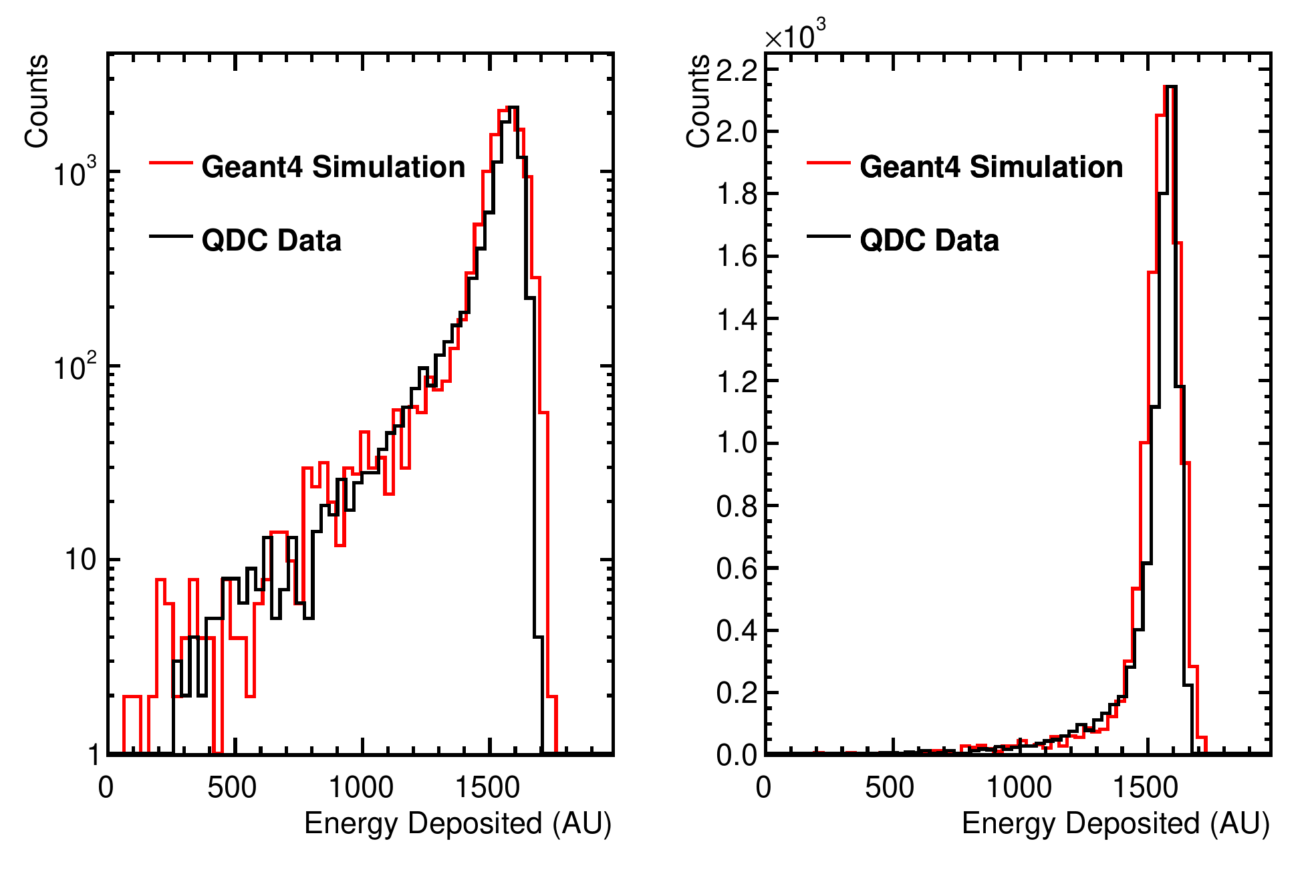}
    \caption{Comparison between Geant4 simulation and data from the
      QDCs using logarithmic (left) and linear (right) scales. The
      simulation is scaled to the height of the data.}
    \label{fig:sim_QDC}
\end{figure}

\Cref{fig:sim_QDC} shows a comparison between the measured QDC data
and the simulation for the central crystal for a 2~GeV incident beam.

\begin{figure}[!htb]
    \centering
    \includegraphics[width=\linewidth]{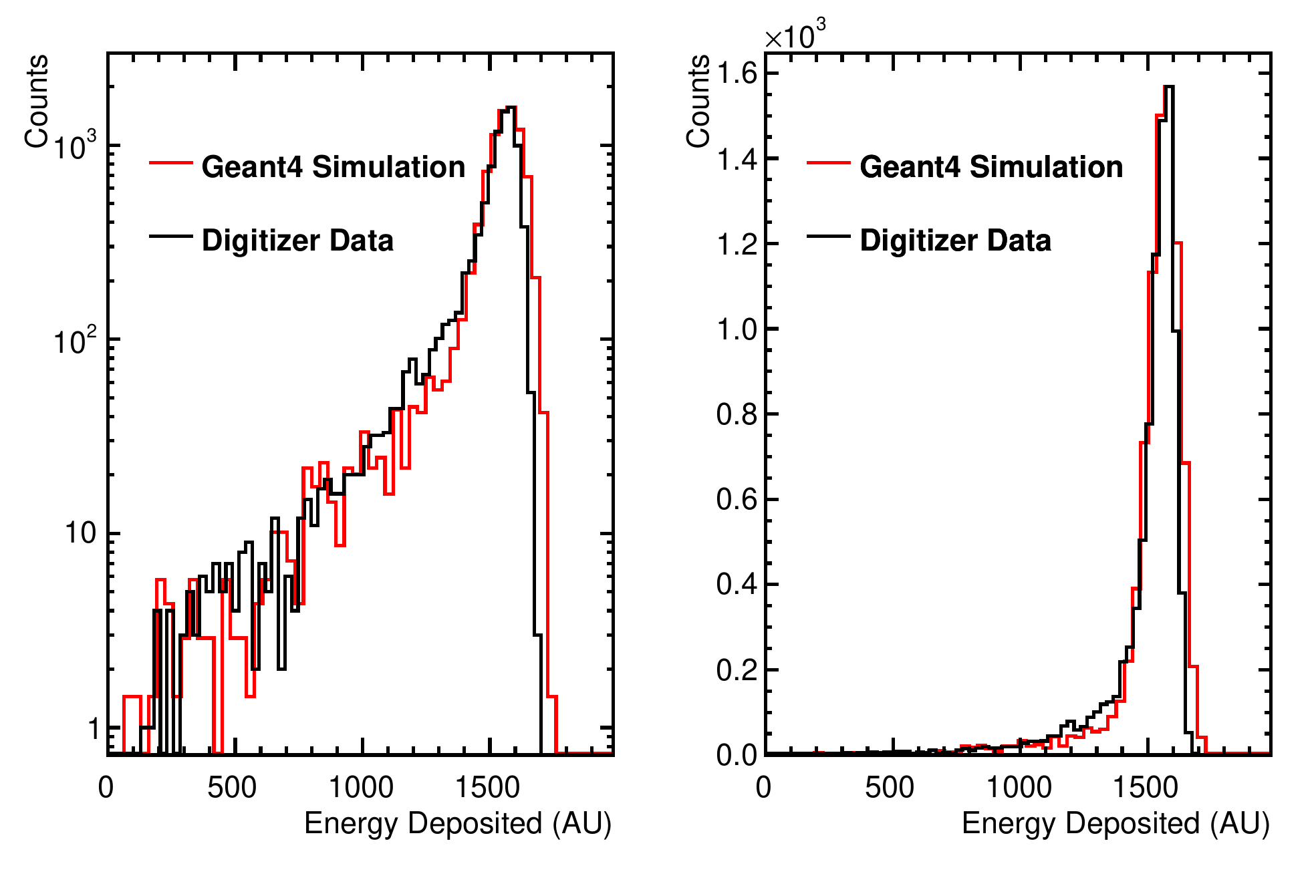}
    \caption{Comparison between Geant4 simulation and data from the
      digitizers. The simulation is scaled to the height of the
      data. }
    \label{fig:sim_digi}
\end{figure}

\Cref{fig:sim_digi} shows a similar comparison between simulation and
the digitizer data. The discrepancies between simulation and data can
be ascribed to an incomplete model of the experimental hall,
specifically any material causing energy loss upstream of the hall. We
believe with improved modeling the agreement could be better.

\clearpage


\section{Monte Carlo Simulation for $e^- + p \rightarrow e^- + p + \pi^0$ at 2~GeV}
\label{sec:Pi0_e_2}

\begin{figure}[!ht]
\begin{subfloat}[][]{
  \includegraphics[width=0.47\textwidth,viewport=10 5 525 390, clip]
  {Plot_Pi0_e_2_30}}
\end{subfloat}
\hfill
\begin{subfloat}[][]{
  \includegraphics[width=0.47\textwidth,viewport=10 5 525 390, clip]
  {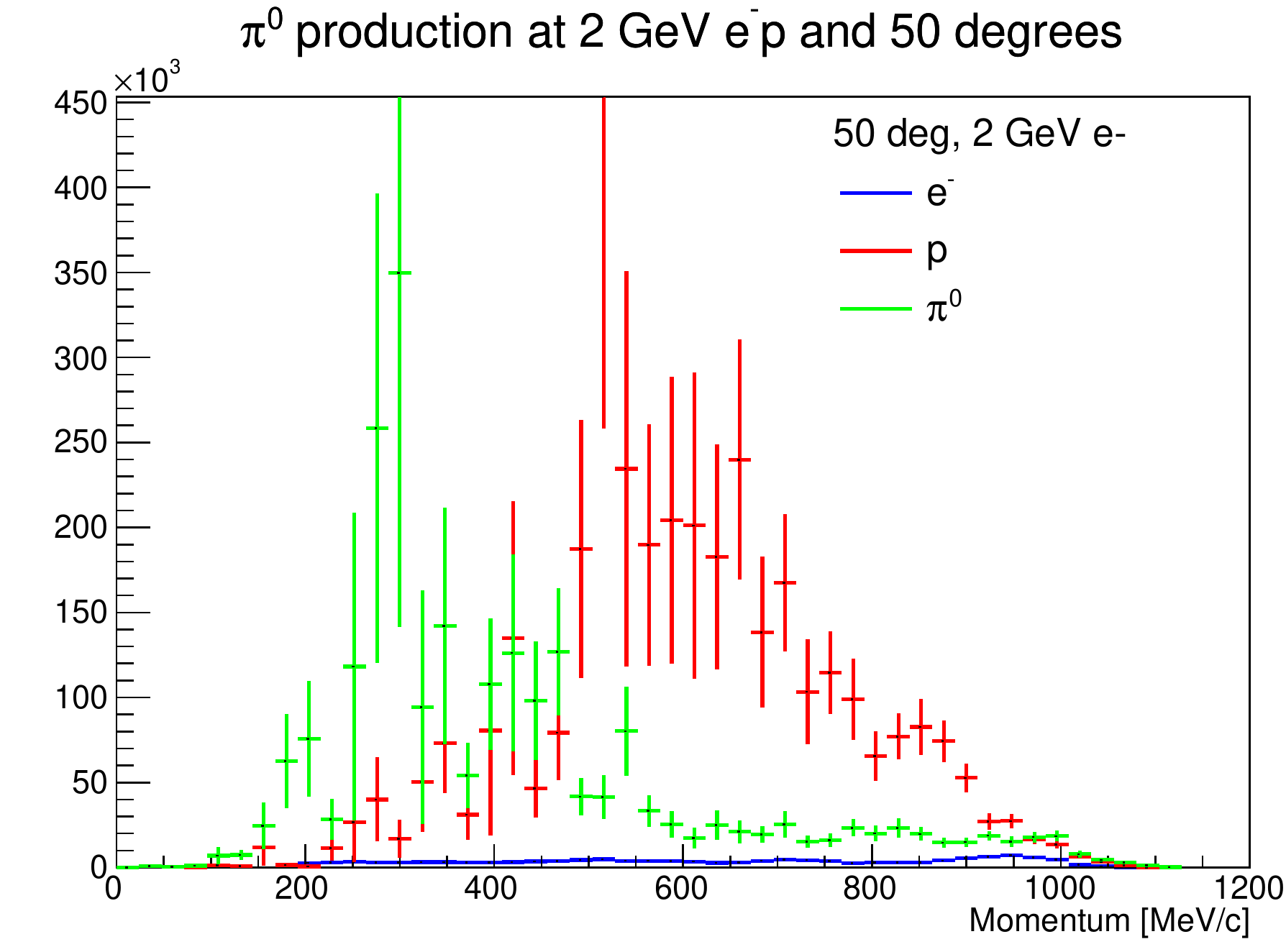}}
\end{subfloat}
\\
\begin{subfloat}[][]{
  \includegraphics[width=0.47\textwidth,viewport=10 5 525 390, clip]
  {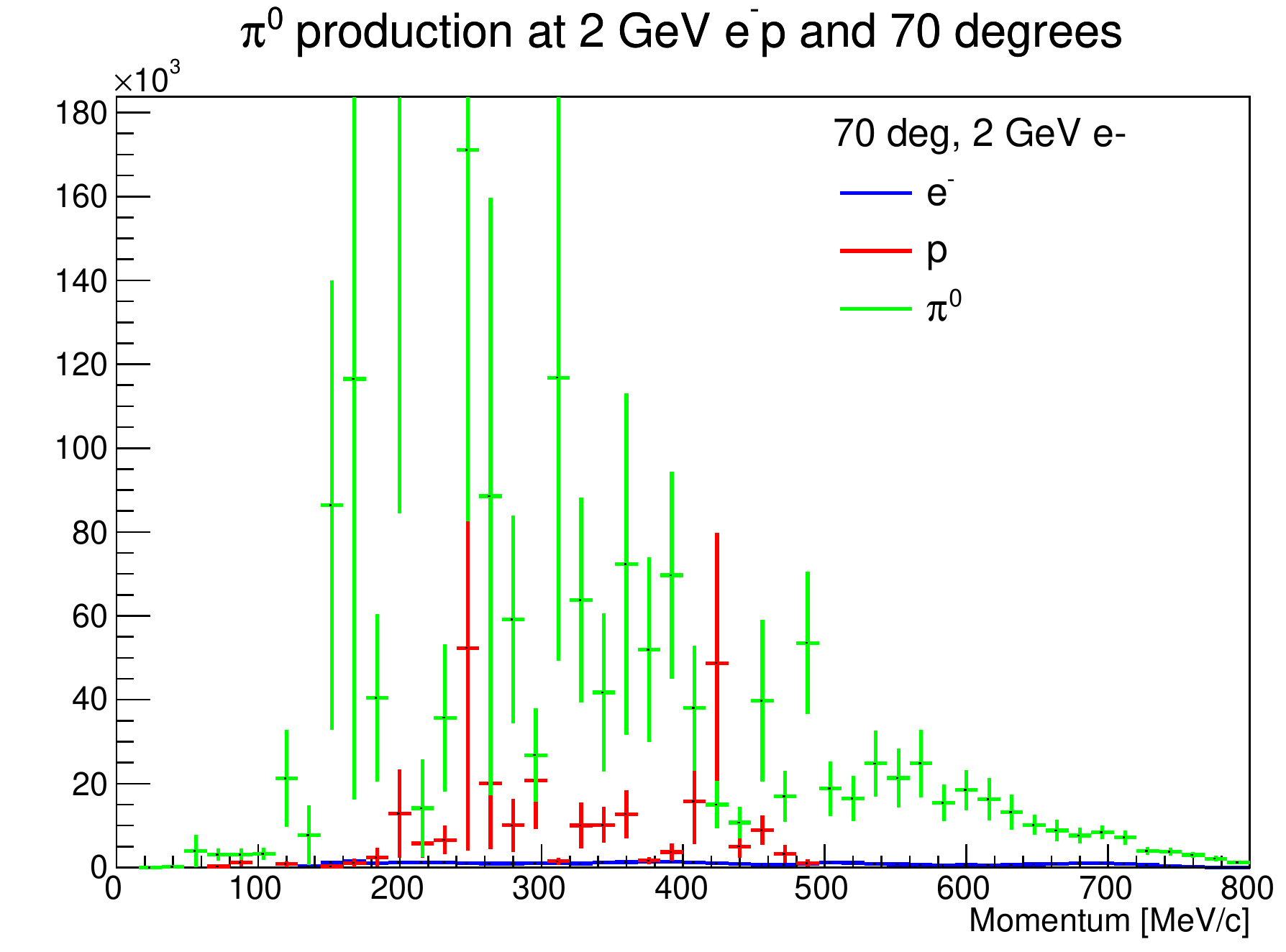}}
\end{subfloat}
\hfill
\begin{subfloat}[][]{
  \includegraphics[width=0.47\textwidth,viewport=10 5 525 390, clip]
  {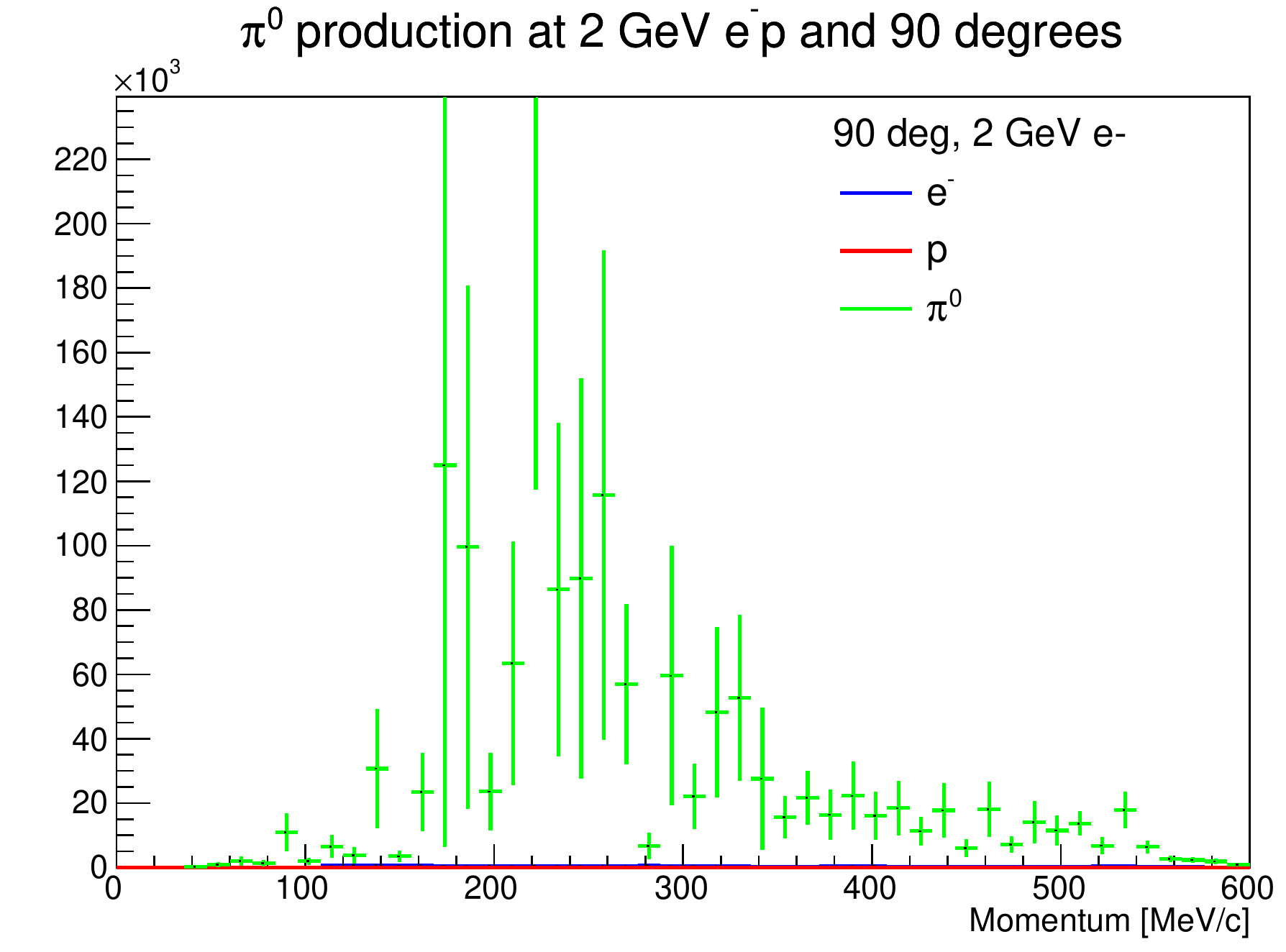}}
\end{subfloat}
\\
\begin{subfloat}[][]{
  \includegraphics[width=0.47\textwidth,viewport=10 5 525 390, clip]
  {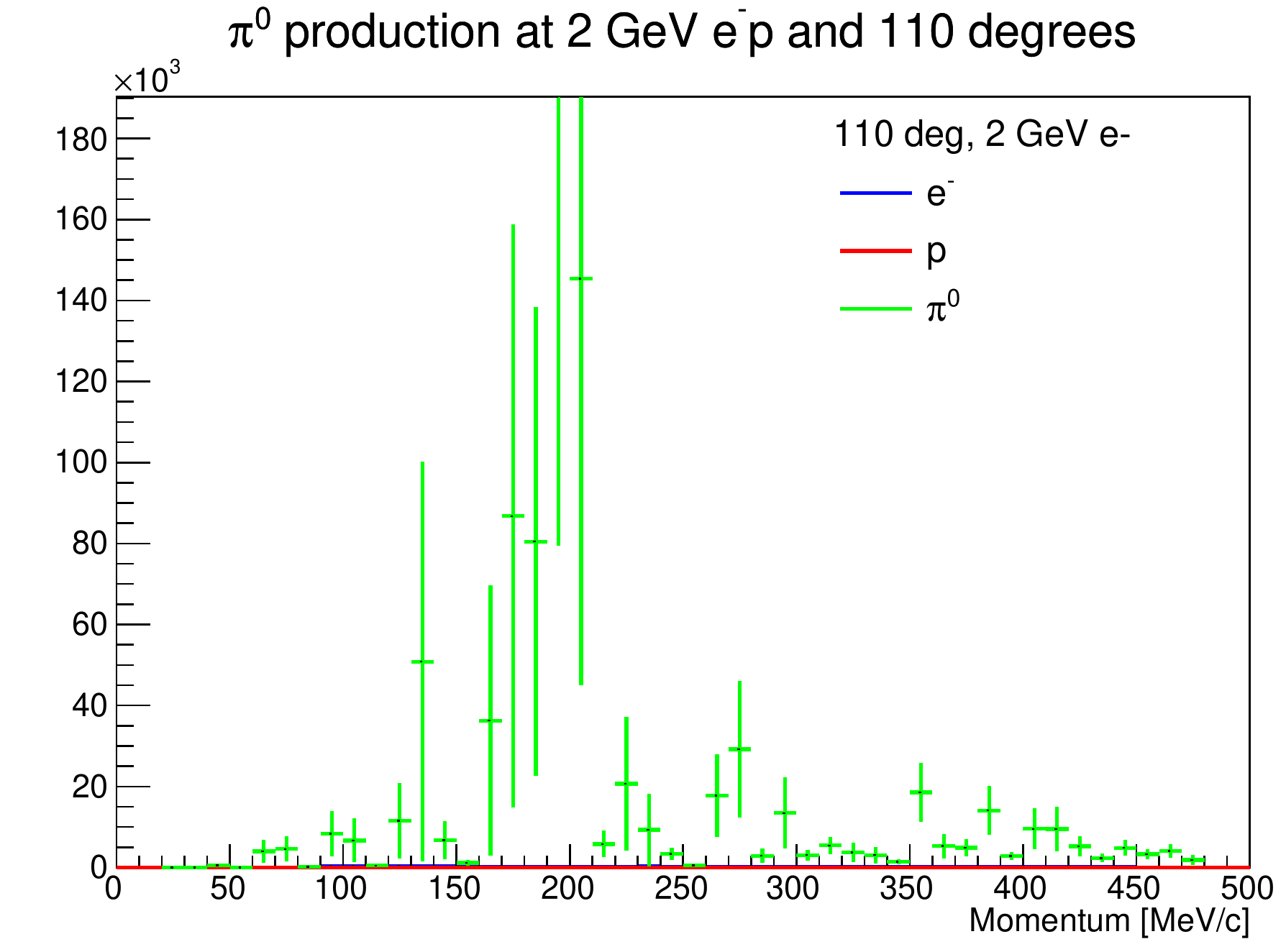}}
\end{subfloat}
\hfill
\parbox[b][0.35\textwidth]{0.47\textwidth} {\caption{Number of
    electrons, protons, and $\pi^0$ directed towards the $5\times5$
    calorimeter arrays at $30\degree$, $50\degree$, $70\degree$,
    $90\degree$, and $110\degree$ during one day of running at the
    nominal luminosity for the reaction $e^- + p \rightarrow e^- + p +
    \pi^0$ at 2~GeV.
\label{fig:Pi0_e_2}}}
\end{figure}


\section{Monte Carlo Simulation for $e^- + p \rightarrow e^- + n + \pi^+$ at 2~GeV}
\label{sec:Pi+_e_2}

\begin{figure}[!ht]
\begin{subfloat}[][]{
  \includegraphics[width=0.47\textwidth,viewport=10 5 525 390, clip]
  {Plot_Pi+_e_2_30}}
\end{subfloat}
\hfill
\begin{subfloat}[][]{
  \includegraphics[width=0.47\textwidth,viewport=10 5 525 390, clip]
  {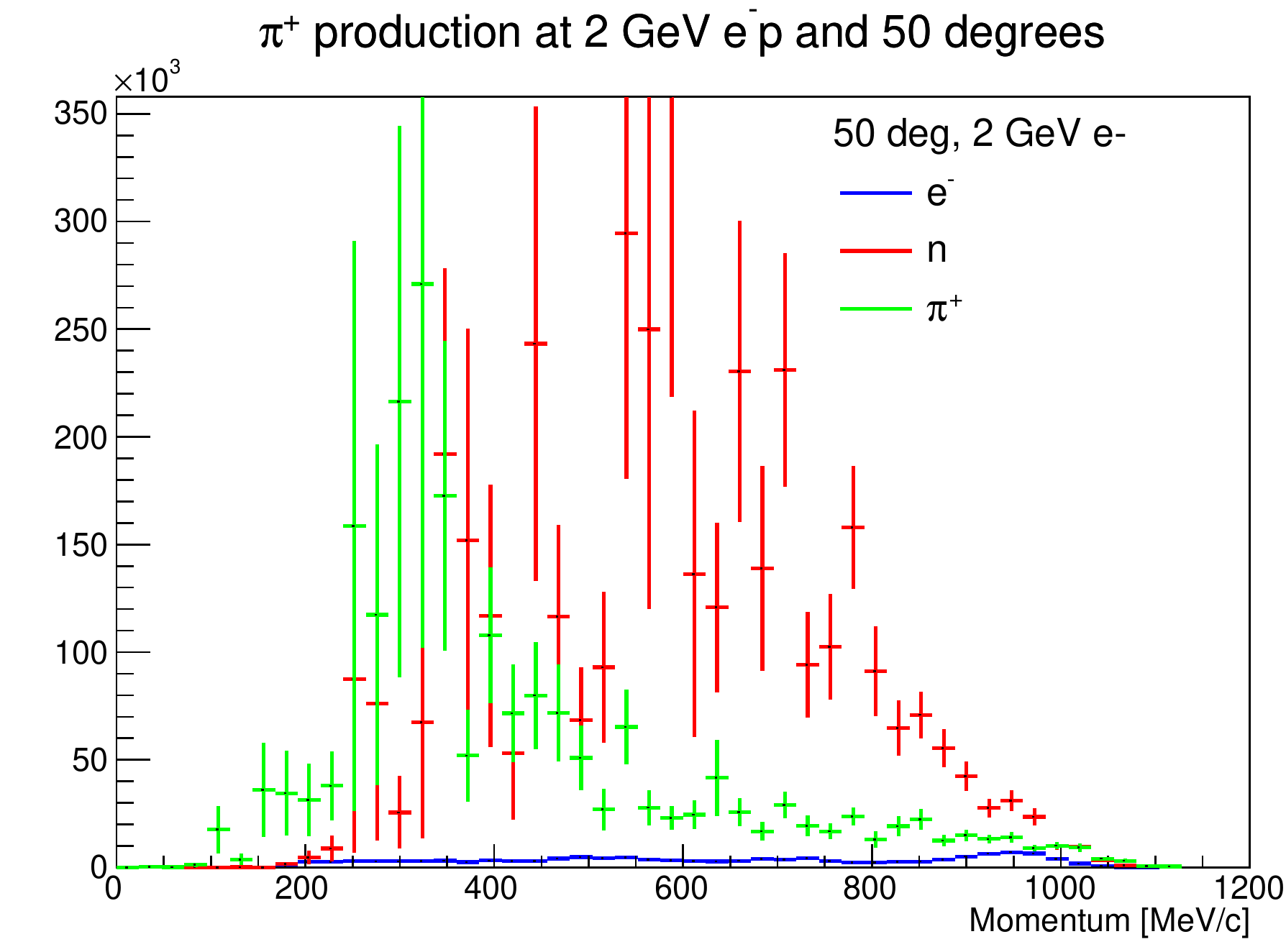}}
\end{subfloat}
\\
\begin{subfloat}[][]{
  \includegraphics[width=0.47\textwidth,viewport=10 5 525 390, clip]
  {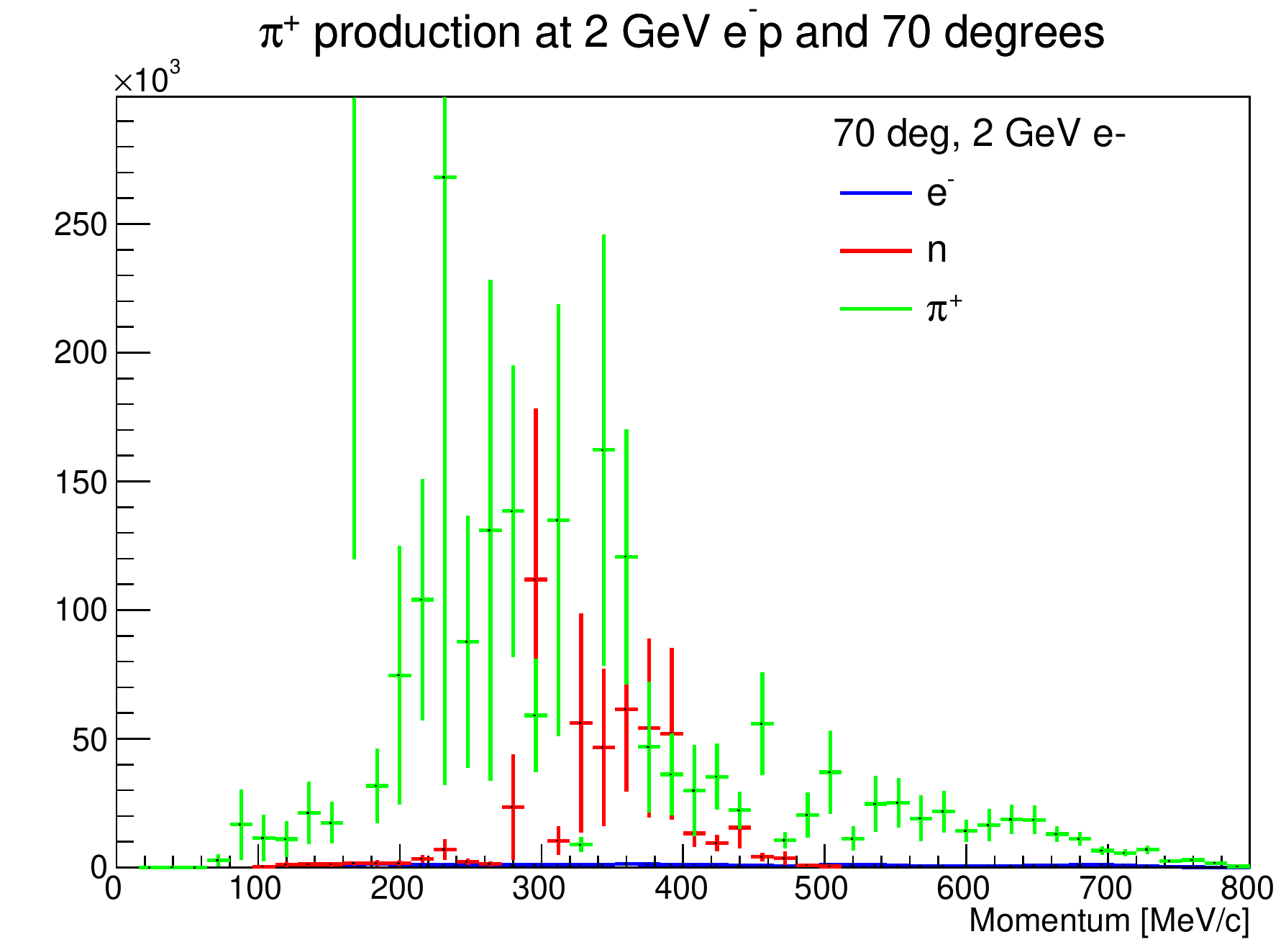}}
\end{subfloat}
\hfill
\begin{subfloat}[][]{
  \includegraphics[width=0.47\textwidth,viewport=10 5 525 390, clip]
  {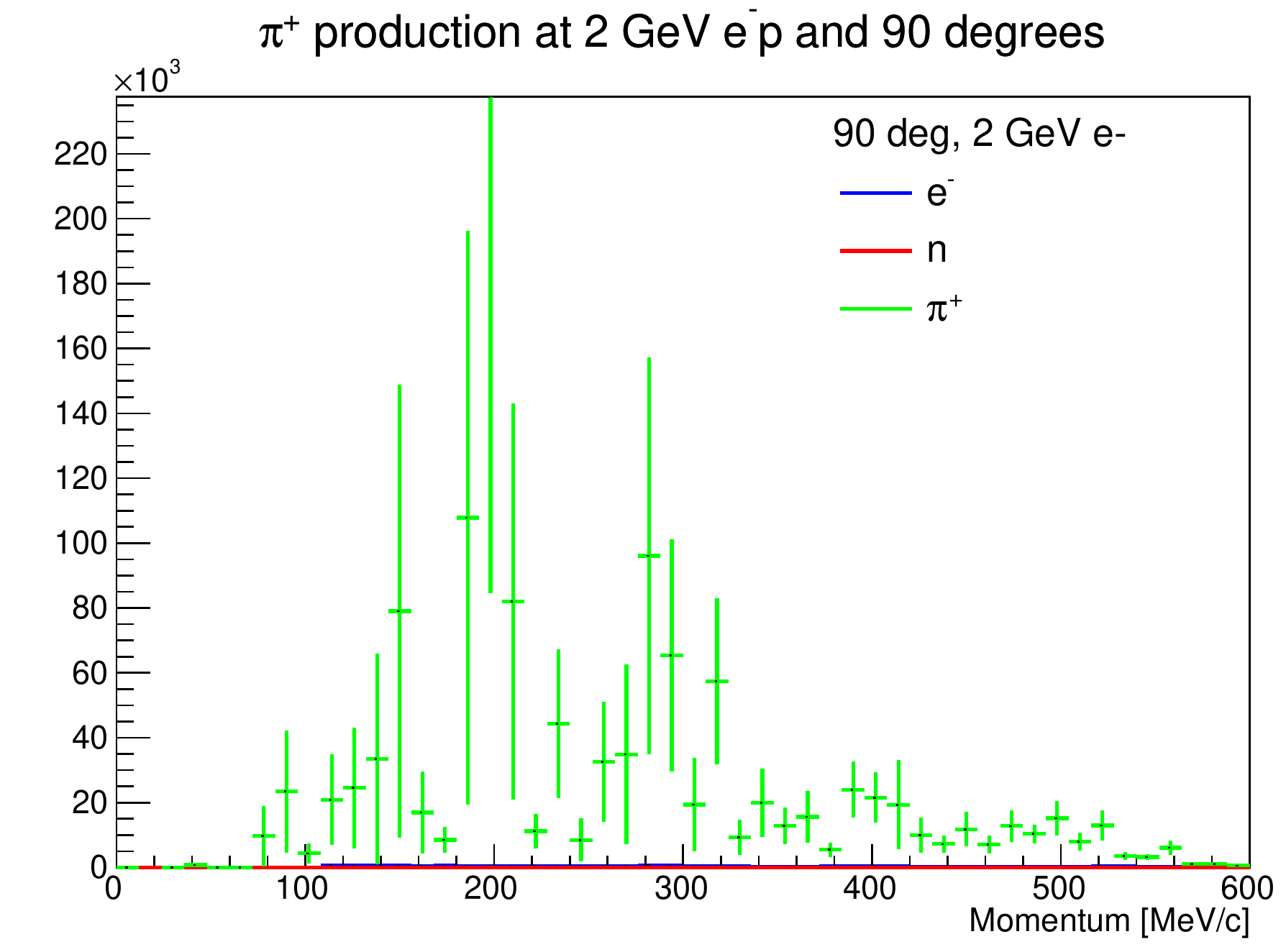}}
\end{subfloat}
\\
\begin{subfloat}[][]{
  \includegraphics[width=0.47\textwidth,viewport=10 5 525 390, clip]
  {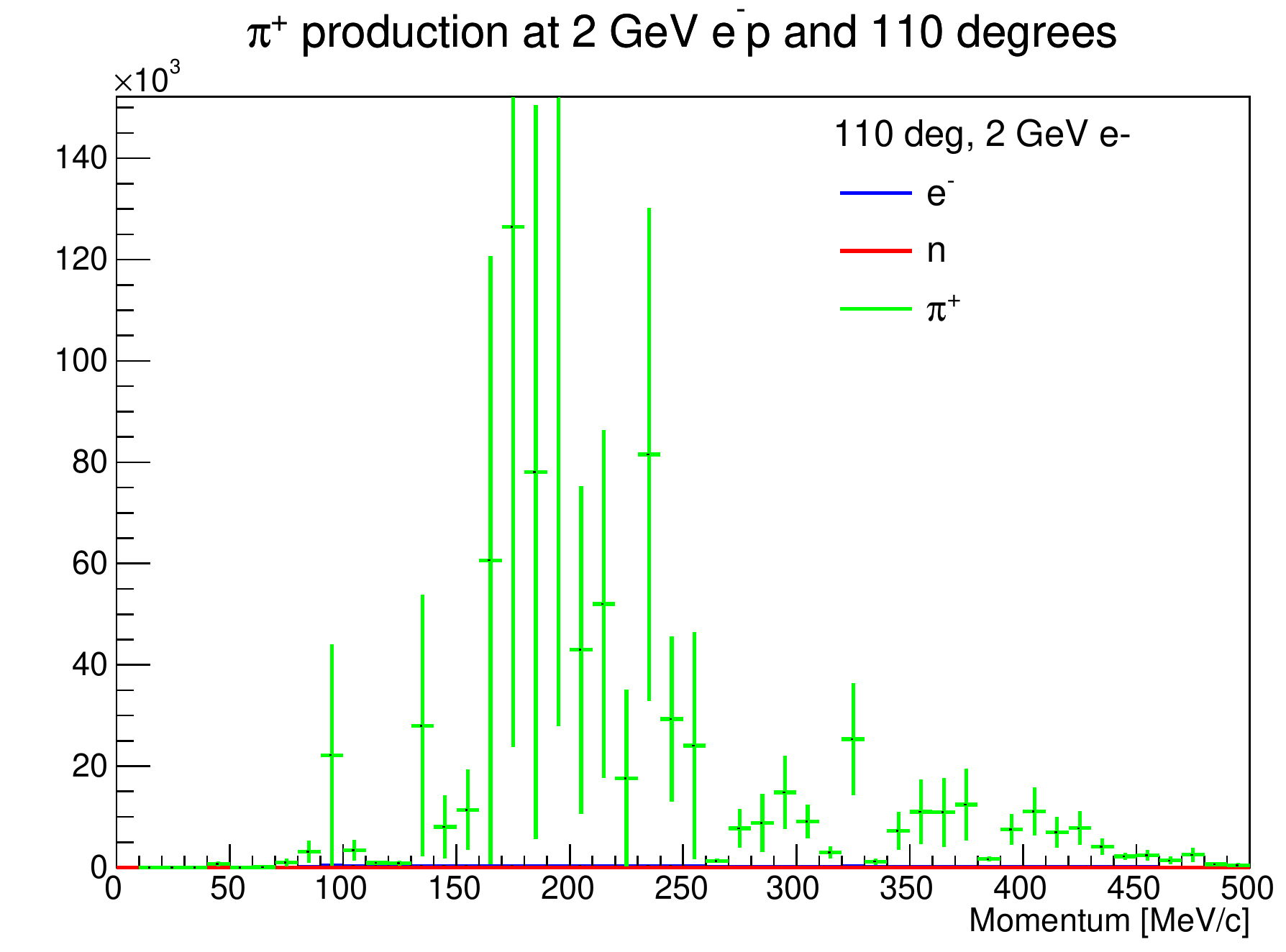}}
\end{subfloat}
\hfill
\parbox[b][0.35\textwidth]{0.47\textwidth} {\caption{Number of
    electrons, neutrons, and $\pi^+$ directed towards the $5\times5$
    calorimeter arrays at $30\degree$, $50\degree$, $70\degree$,
    $90\degree$, and $110\degree$ during one day of running at the
    nominal luminosity for the reaction $e^- + p \rightarrow e^- + n +
    \pi^+$ at 2~GeV.
\label{fig:Pi+_e_2}}}
\end{figure}


\section{Monte Carlo Simulation for $e^+ + p \rightarrow e^+ + p + \pi^0$ at 2~GeV}
\label{sec:Pi0_p_2}

\begin{figure}[!ht]
\begin{subfloat}[][]{
  \includegraphics[width=0.47\textwidth,viewport=10 5 525 390, clip]
  {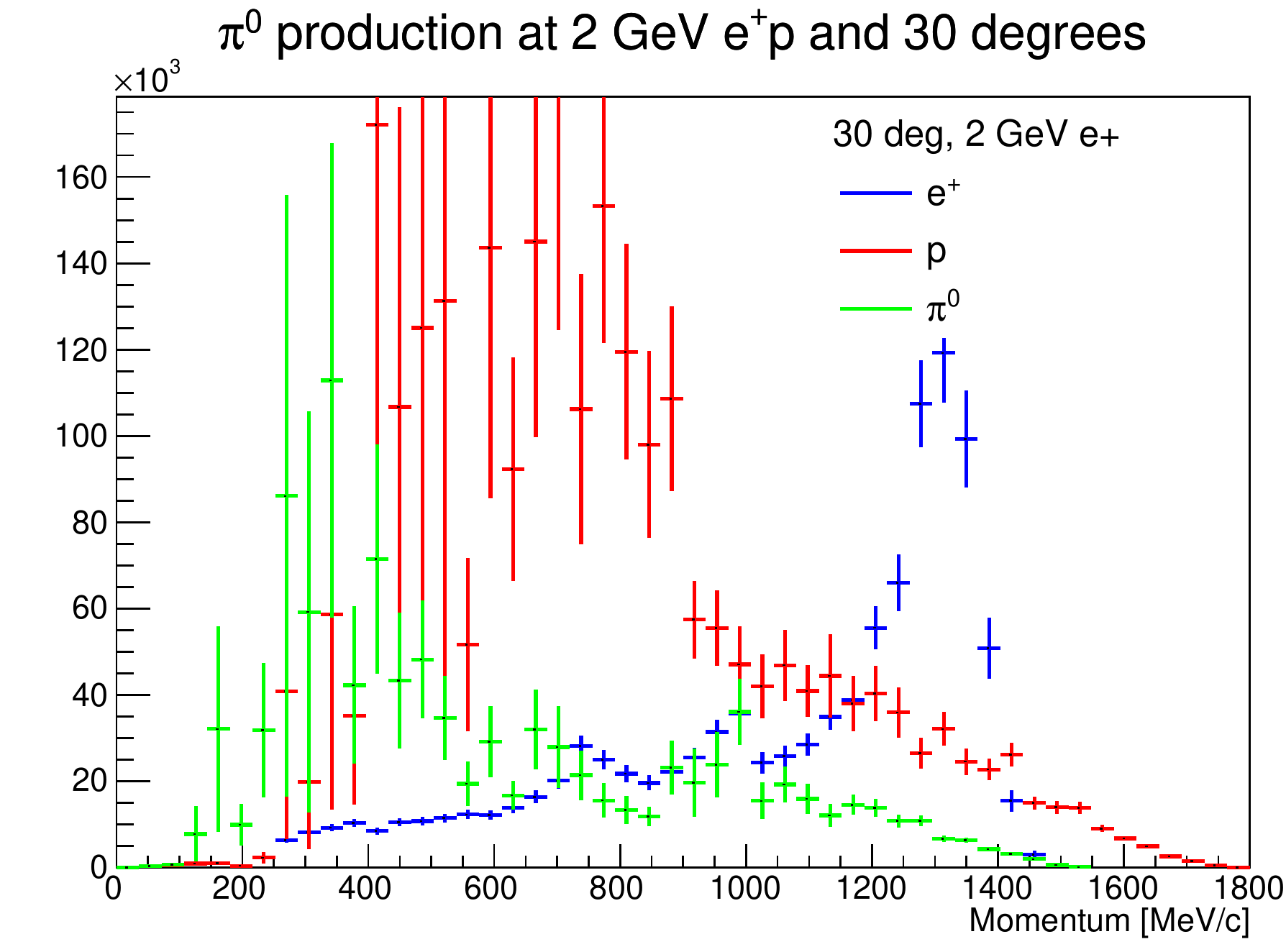}}
\end{subfloat}
\hfill
\begin{subfloat}[][]{
  \includegraphics[width=0.47\textwidth,viewport=10 5 525 390, clip]
  {Plot_Pi0_e_2_50}}
\end{subfloat}
\\
\begin{subfloat}[][]{
  \includegraphics[width=0.47\textwidth,viewport=10 5 525 390, clip]
  {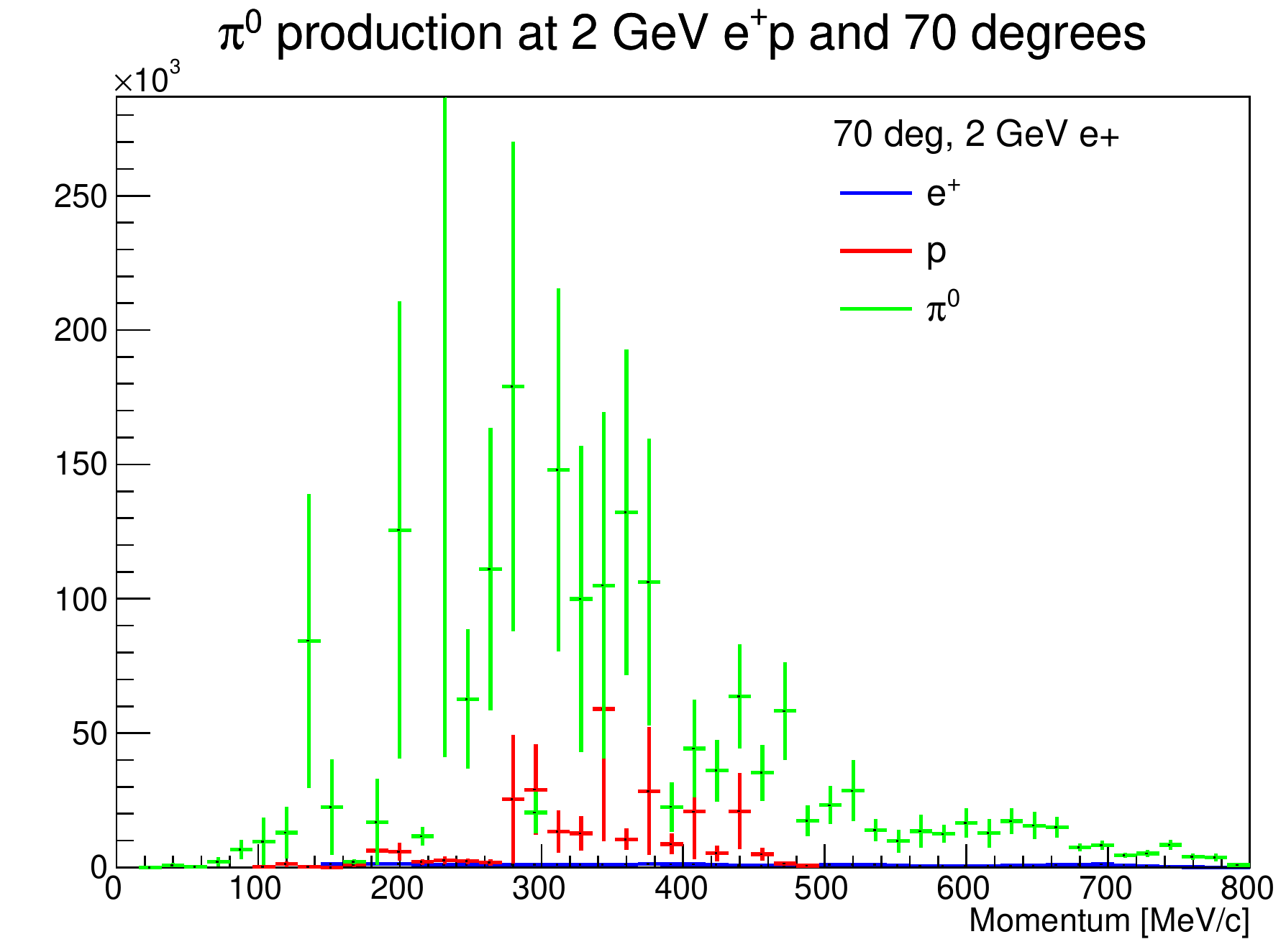}}
\end{subfloat}
\hfill
\begin{subfloat}[][]{
  \includegraphics[width=0.47\textwidth,viewport=10 5 525 390, clip]
  {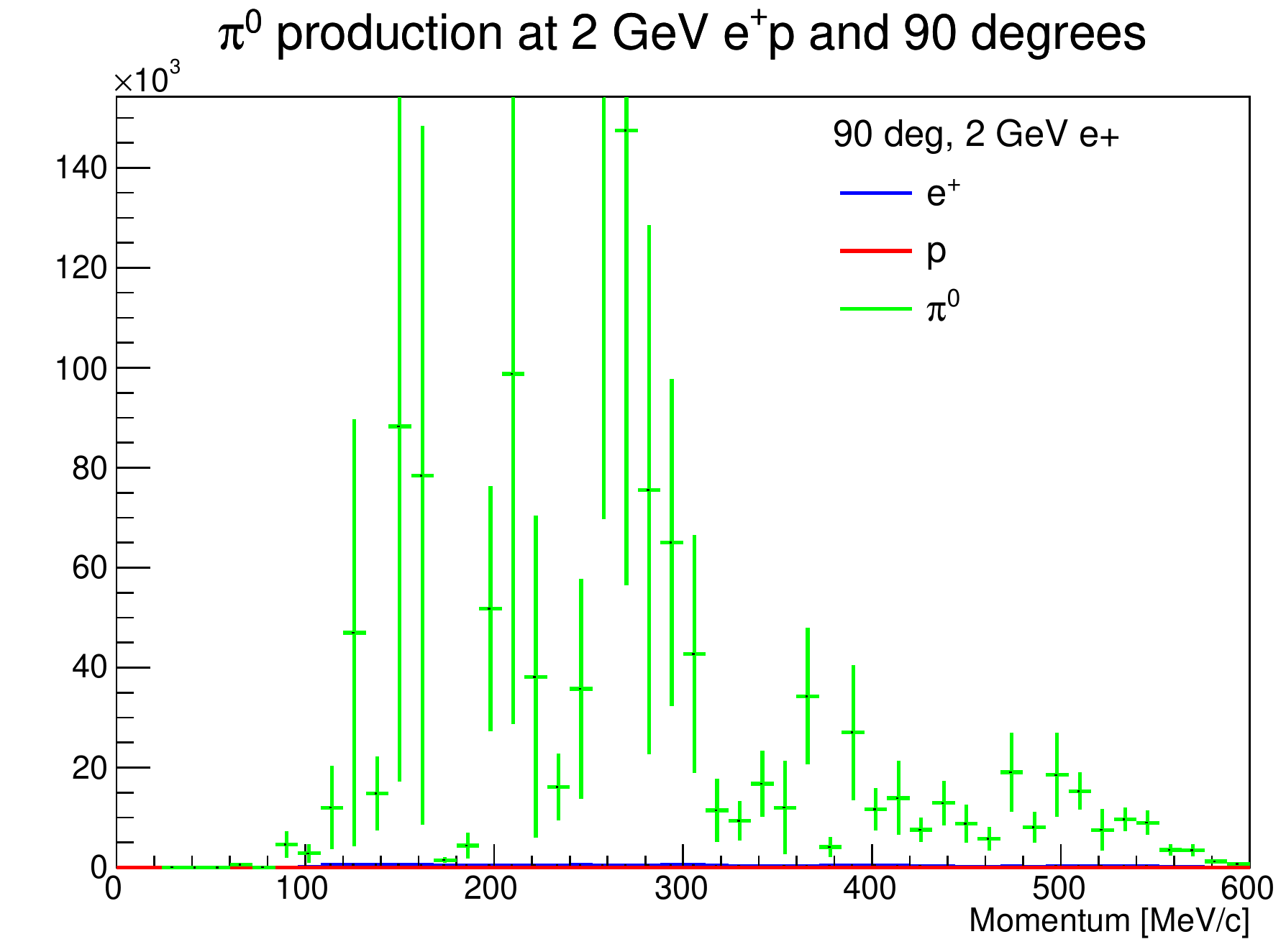}}
\end{subfloat}
\\
\begin{subfloat}[][]{
  \includegraphics[width=0.47\textwidth,viewport=10 5 525 390, clip]
  {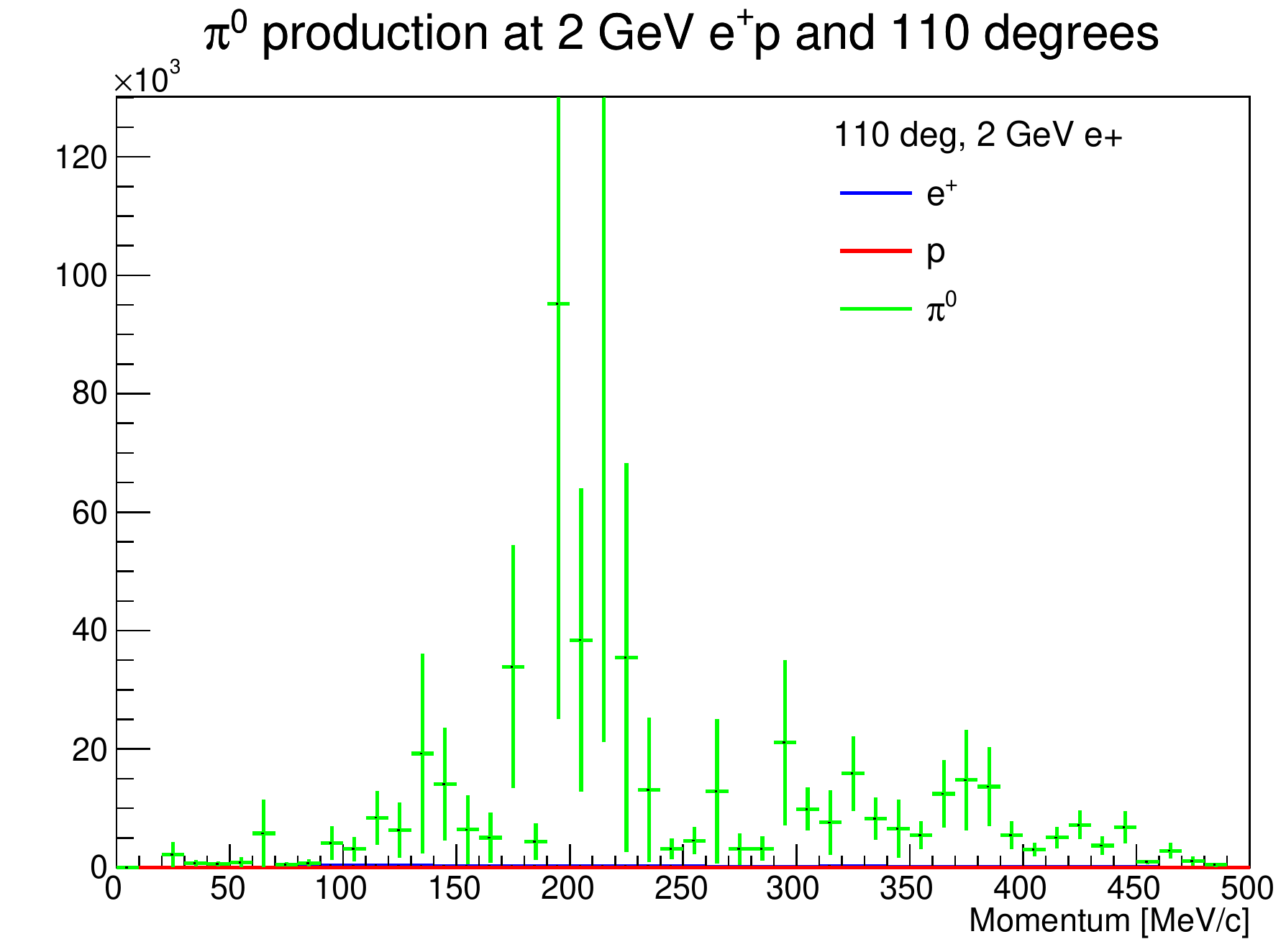}}
\end{subfloat}
\hfill
\parbox[b][0.35\textwidth]{0.47\textwidth} {\caption{Number of
    positrons, protons, and $\pi^0$ directed towards the $5\times5$
    calorimeter arrays at $30\degree$, $50\degree$, $70\degree$,
    $90\degree$, and $110\degree$ during one day of running at the
    nominal luminosity for the reaction $e^+ + p \rightarrow e^+ + p +
    \pi^0$ at 2~GeV.
\label{fig:Pi0_p_2}}}
\end{figure}


\section{Monte Carlo Simulation for $e^+ + p \rightarrow e^+ + n + \pi^+$ at 2~GeV}
\label{sec:Pi+_p_2}

\begin{figure}[!ht]
\begin{subfloat}[][]{
  \includegraphics[width=0.47\textwidth,viewport=10 5 525 390, clip]
  {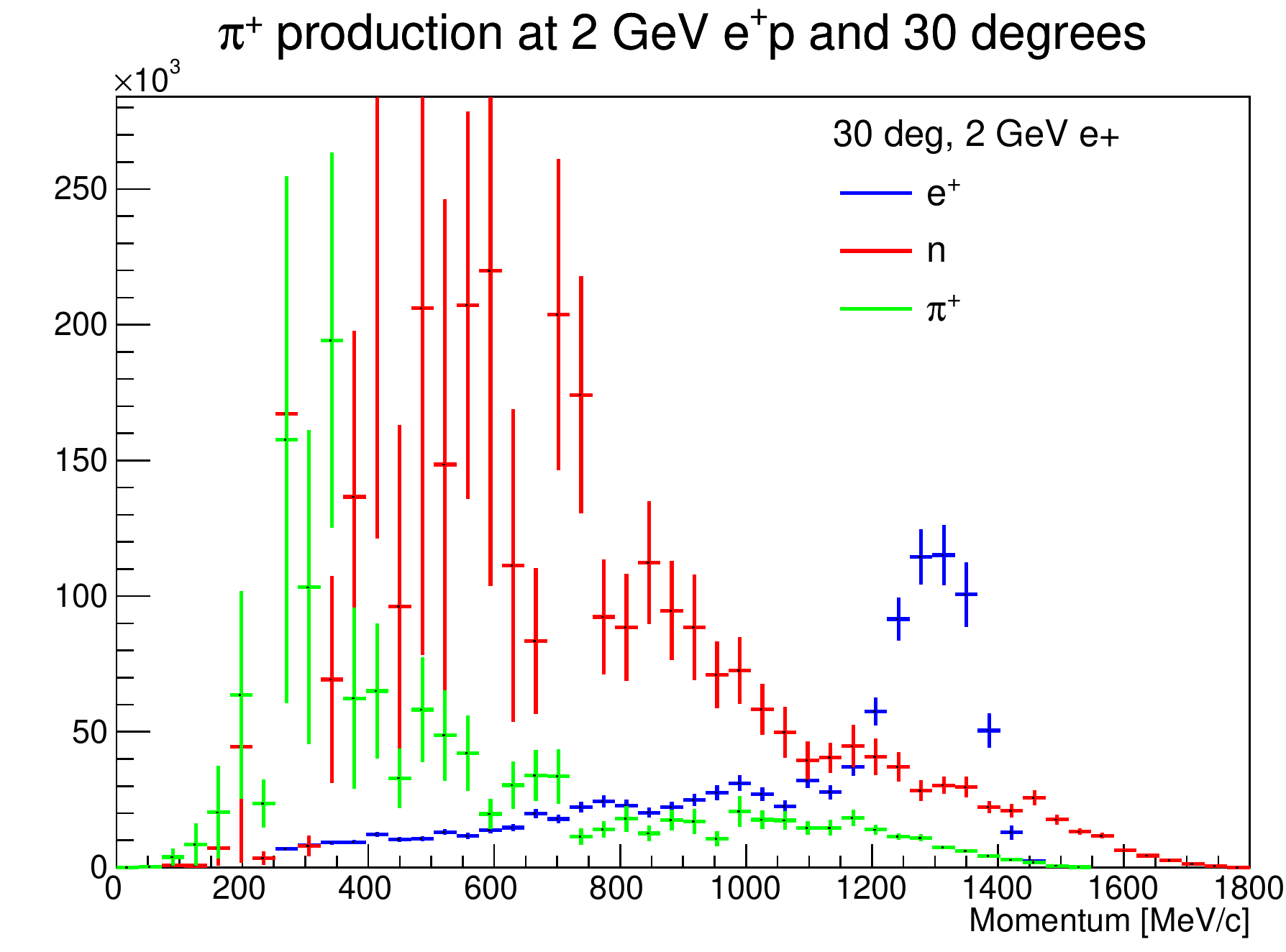}}
\end{subfloat}
\hfill
\begin{subfloat}[][]{
  \includegraphics[width=0.47\textwidth,viewport=10 5 525 390, clip]
  {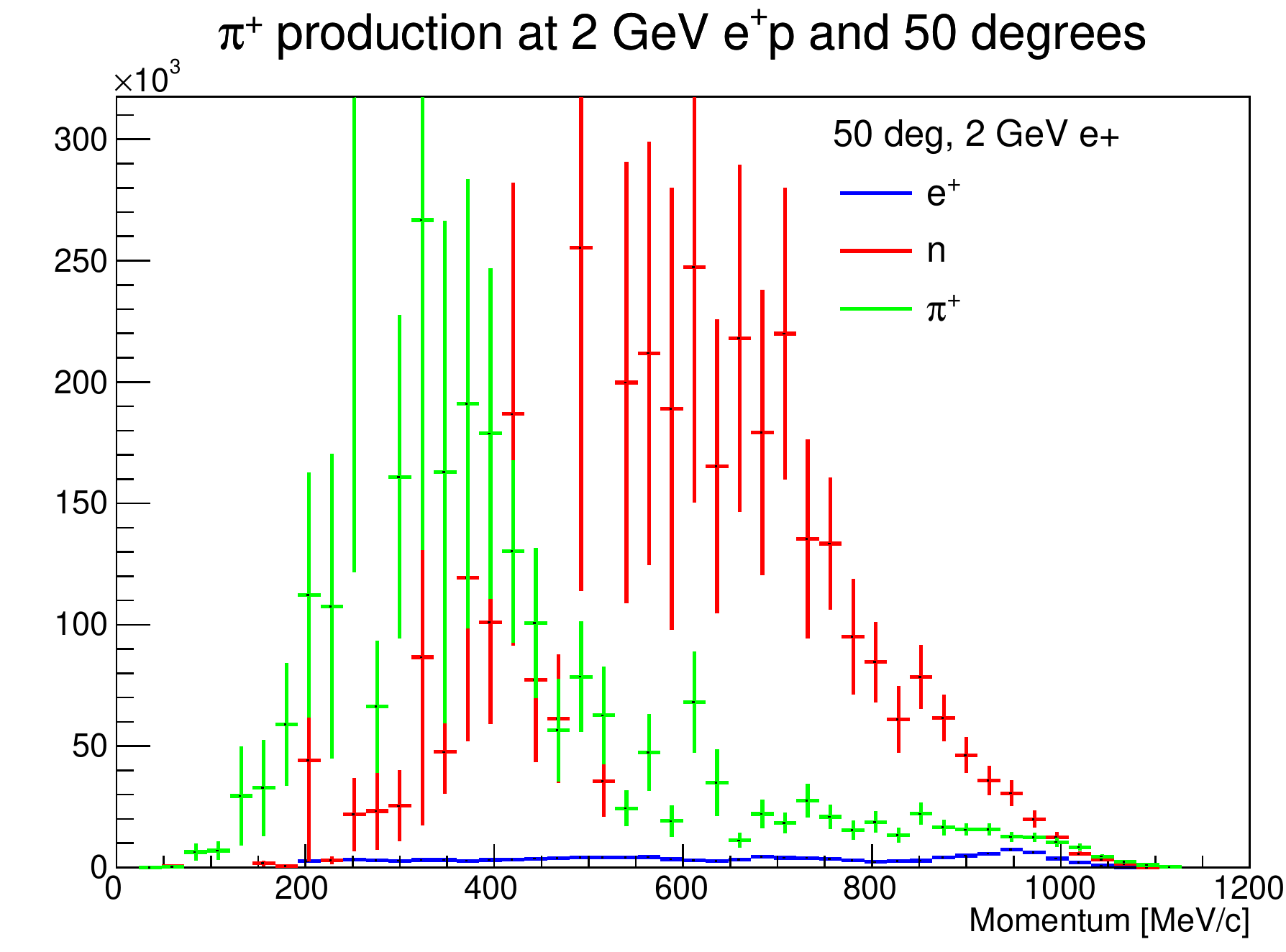}}
\end{subfloat}
\\
\begin{subfloat}[][]{
  \includegraphics[width=0.47\textwidth,viewport=10 5 525 390, clip]
  {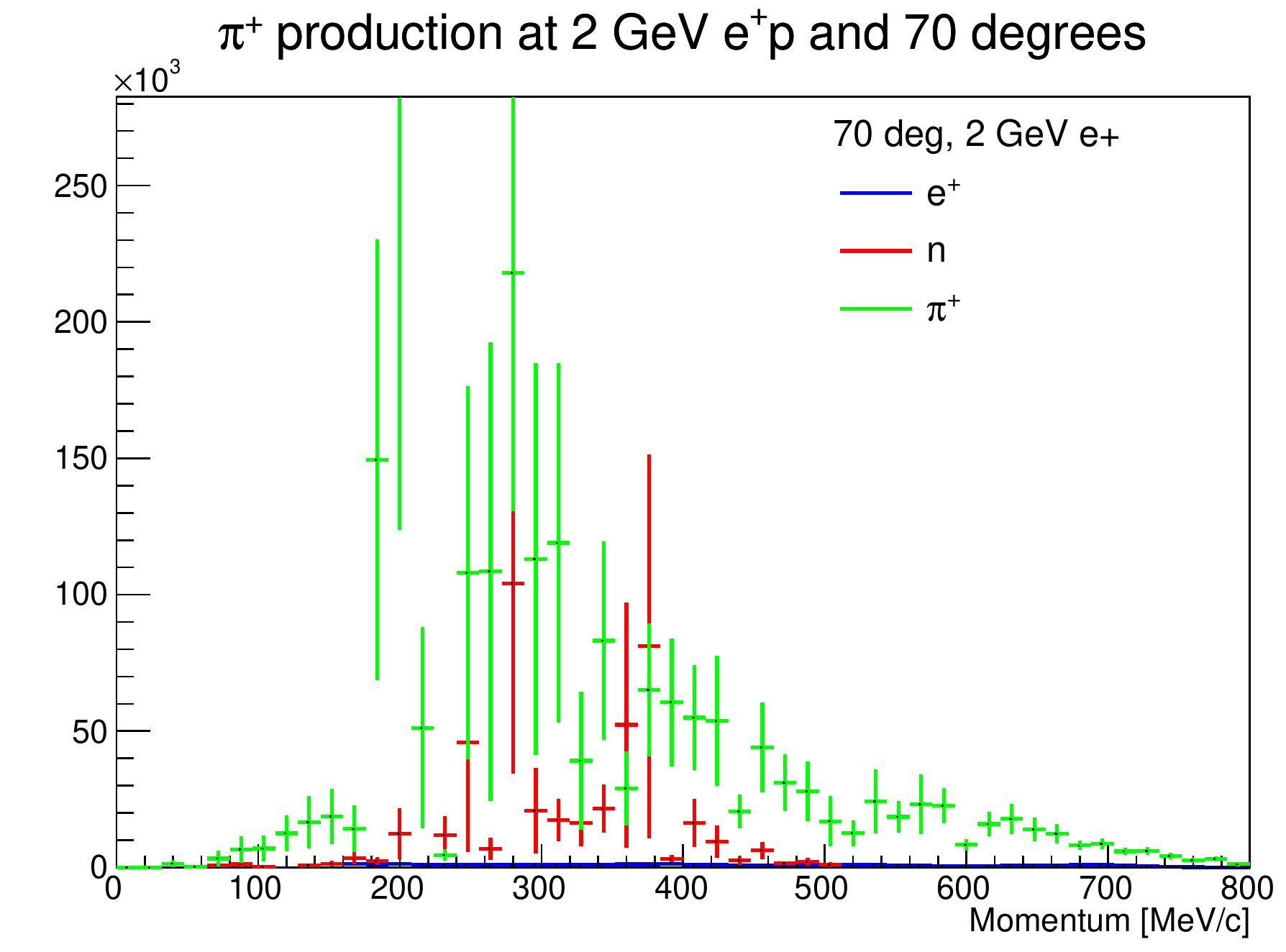}}
\end{subfloat}
\hfill
\begin{subfloat}[][]{
  \includegraphics[width=0.47\textwidth,viewport=10 5 525 390, clip]
  {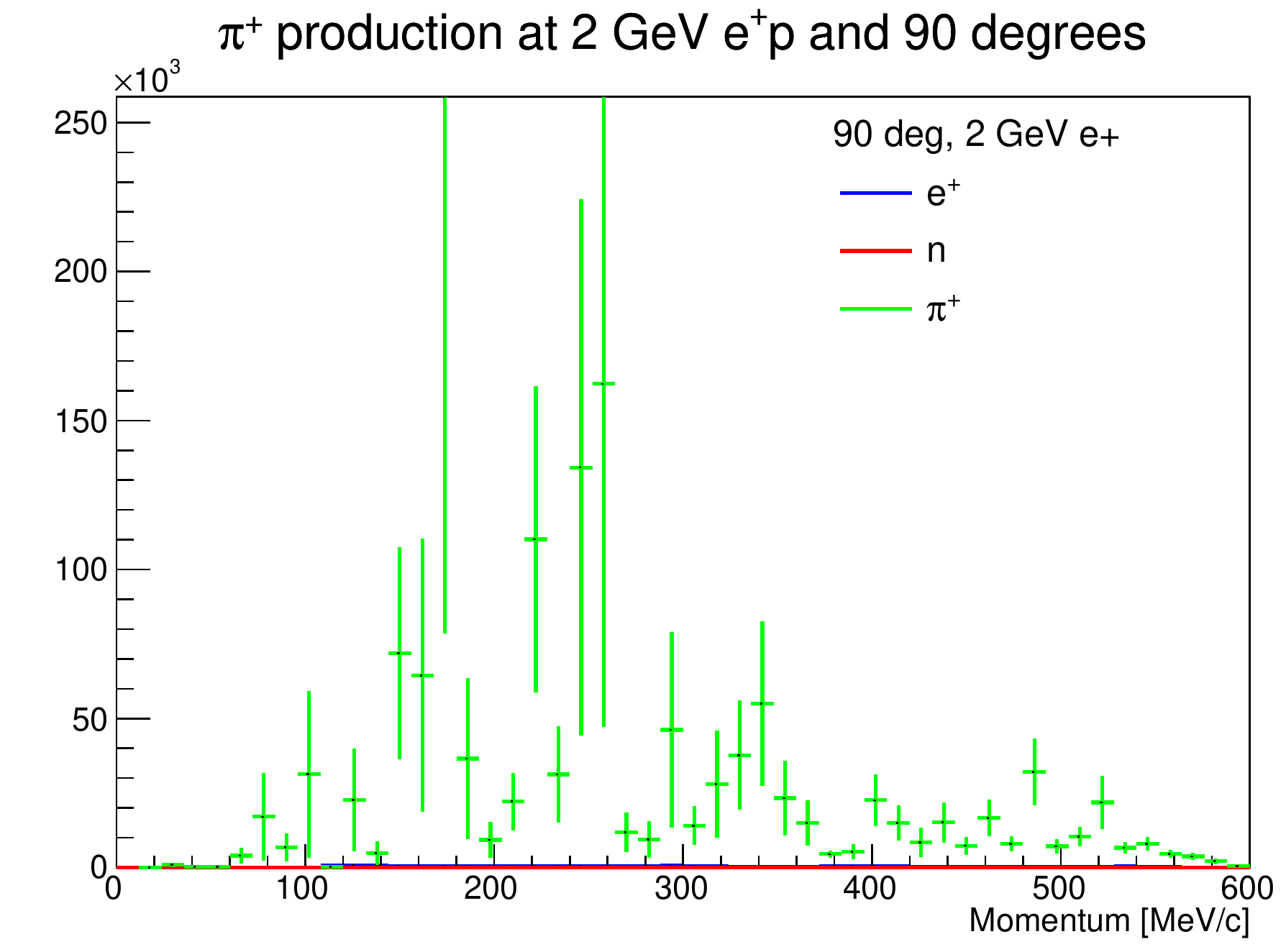}}
\end{subfloat}
\\
\begin{subfloat}[][]{
  \includegraphics[width=0.47\textwidth,viewport=10 5 525 390, clip]
  {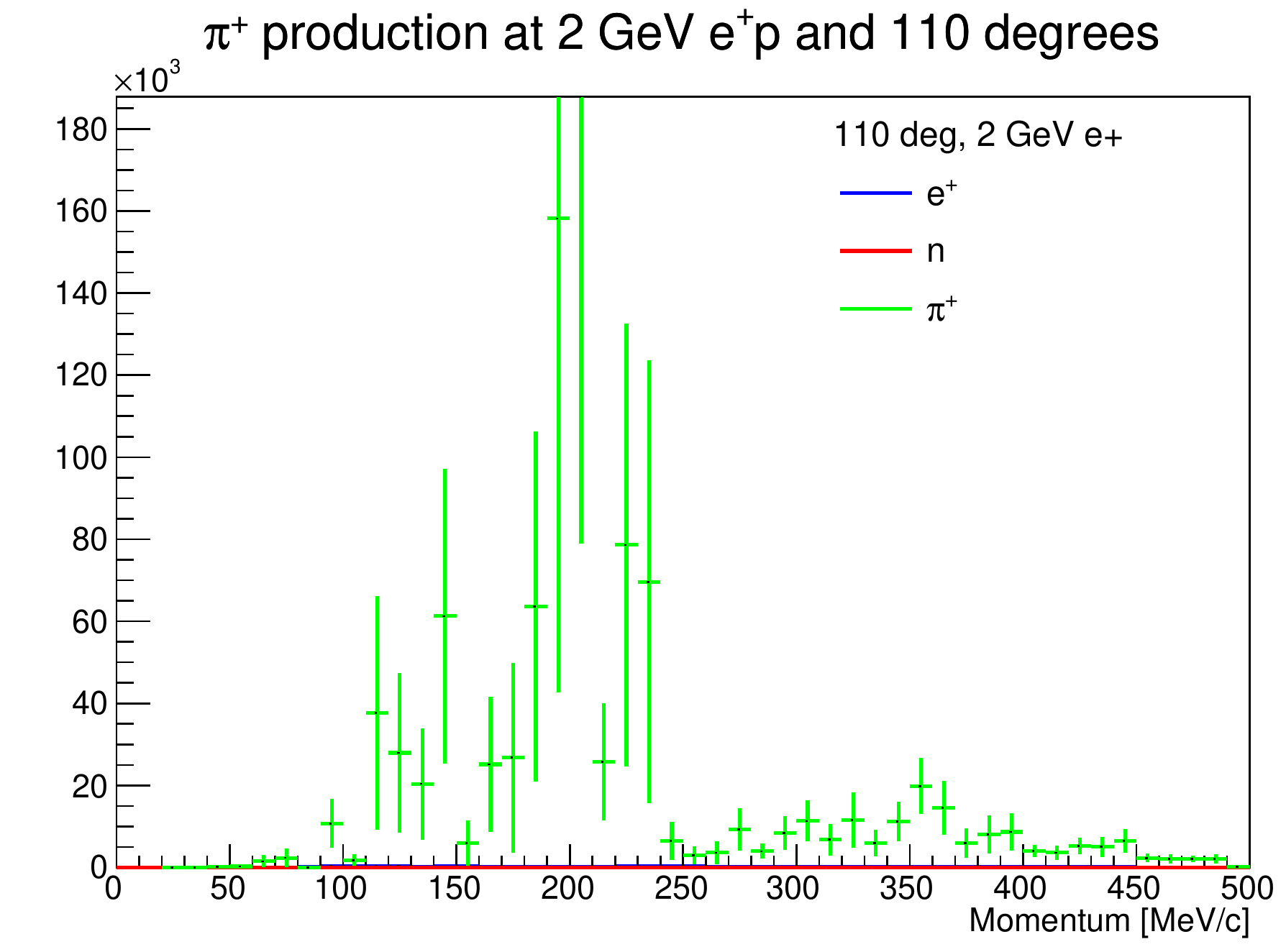}}
\end{subfloat}
\hfill
\parbox[b][0.35\textwidth]{0.47\textwidth} {\caption{Number of
    positrons, neutrons, and $\pi^+$ directed towards the $5\times5$
    calorimeter arrays at $30\degree$, $50\degree$, $70\degree$,
    $90\degree$, and $110\degree$ during one day of running at the
    nominal luminosity for the reaction $e^+ + p \rightarrow e^+ + n +
    \pi^+$ at 2~GeV.
\label{fig:Pi+_p_2}}}
\end{figure}


\section{Monte Carlo Simulation for $e^- + p \rightarrow e^- + p + \pi^0$ at 3~GeV}
\label{sec:Pi0_e_3}

\begin{figure}[!ht]
\begin{subfloat}[][]{
  \includegraphics[width=0.47\textwidth,viewport=10 5 525 390, clip]
  {Plot_Pi0_e_3_30}}
\end{subfloat}
\hfill
\begin{subfloat}[][]{
  \includegraphics[width=0.47\textwidth,viewport=10 5 525 390, clip]
  {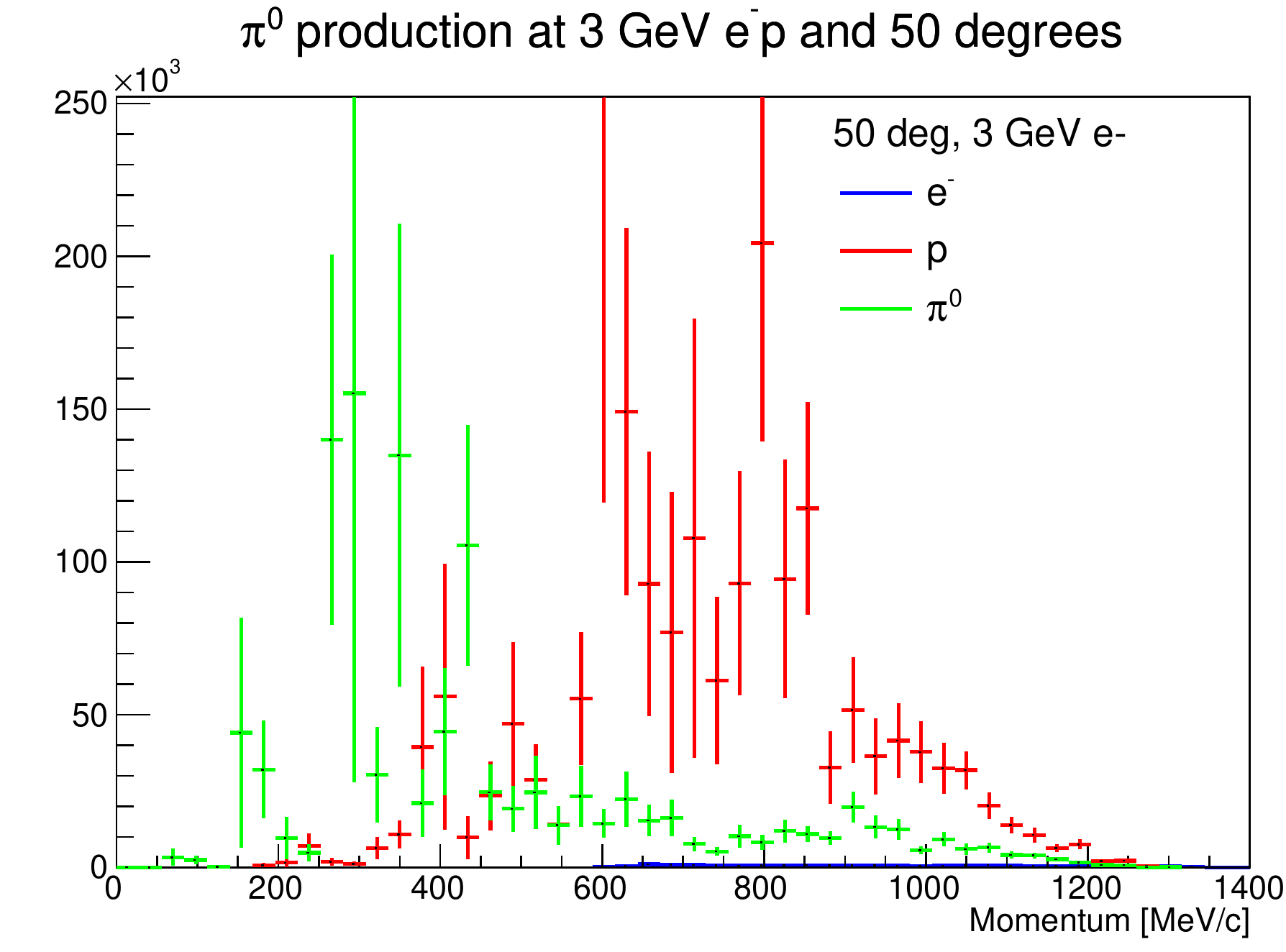}}
\end{subfloat}
\\
\begin{subfloat}[][]{
  \includegraphics[width=0.47\textwidth,viewport=10 5 525 390, clip]
  {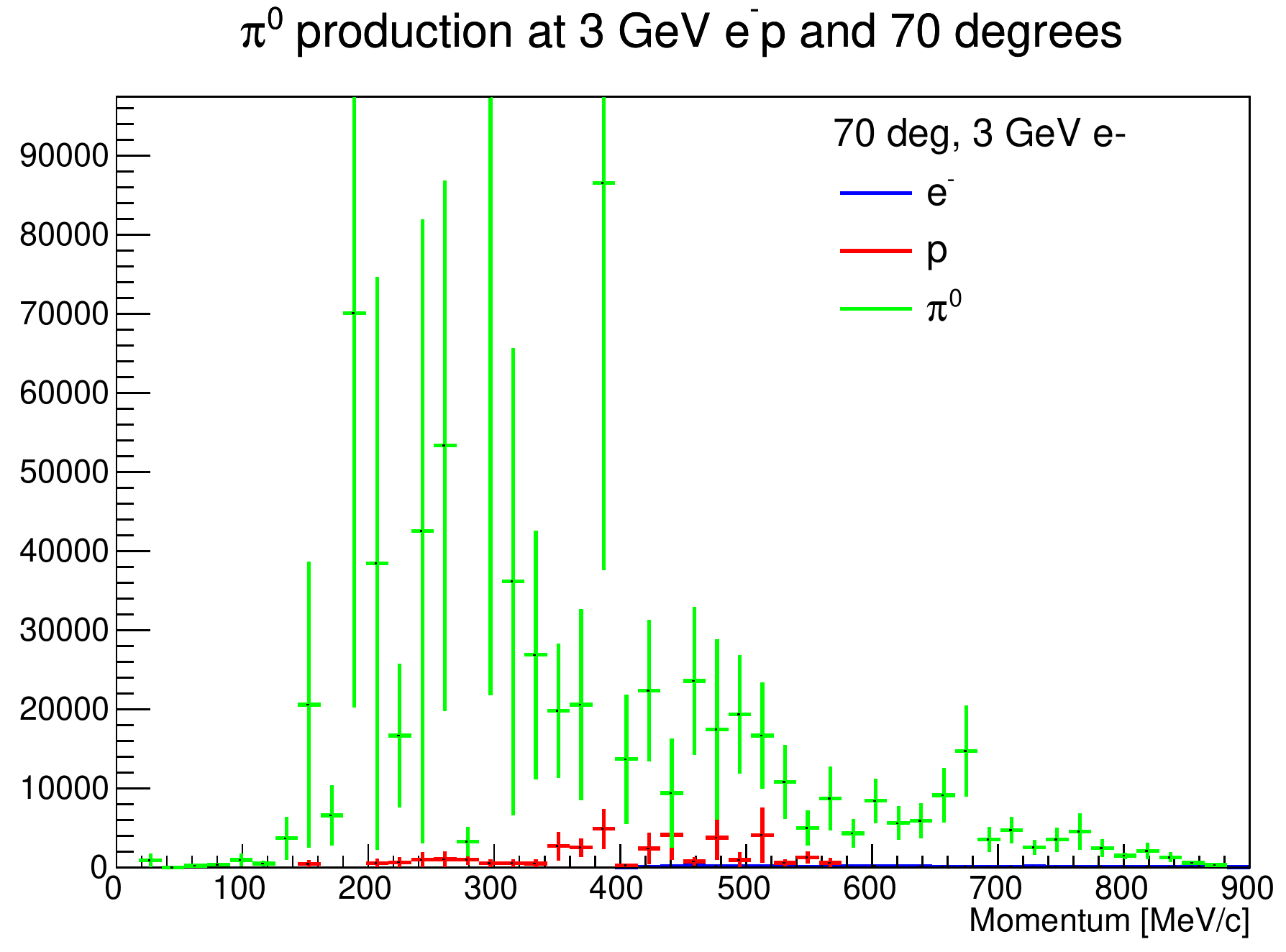}}
\end{subfloat}
\hfill
\begin{subfloat}[][]{
  \includegraphics[width=0.47\textwidth,viewport=10 5 525 390, clip]
  {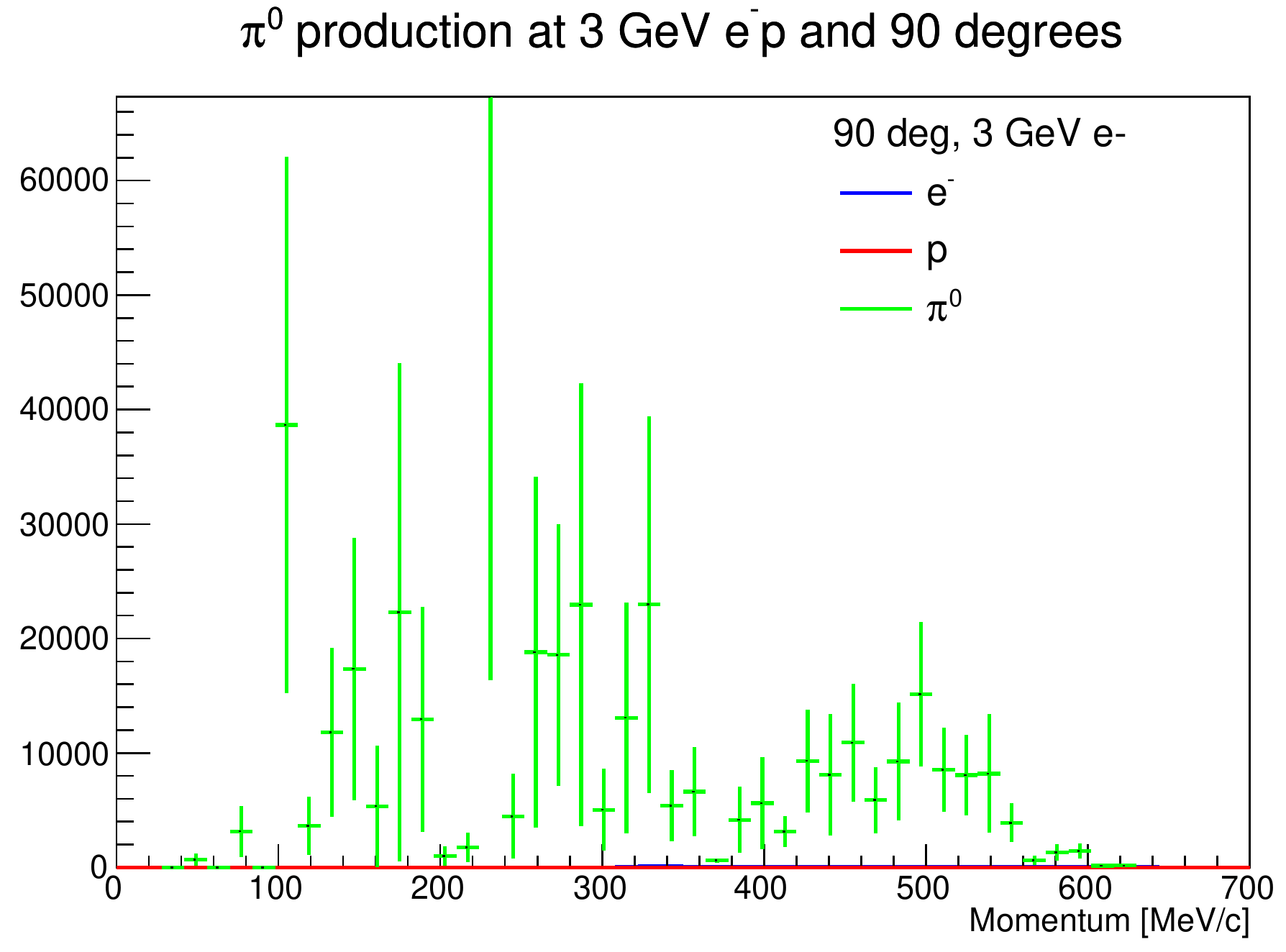}}
\end{subfloat}
\\
\begin{subfloat}[][]{
  \includegraphics[width=0.47\textwidth,viewport=10 5 525 390, clip]
  {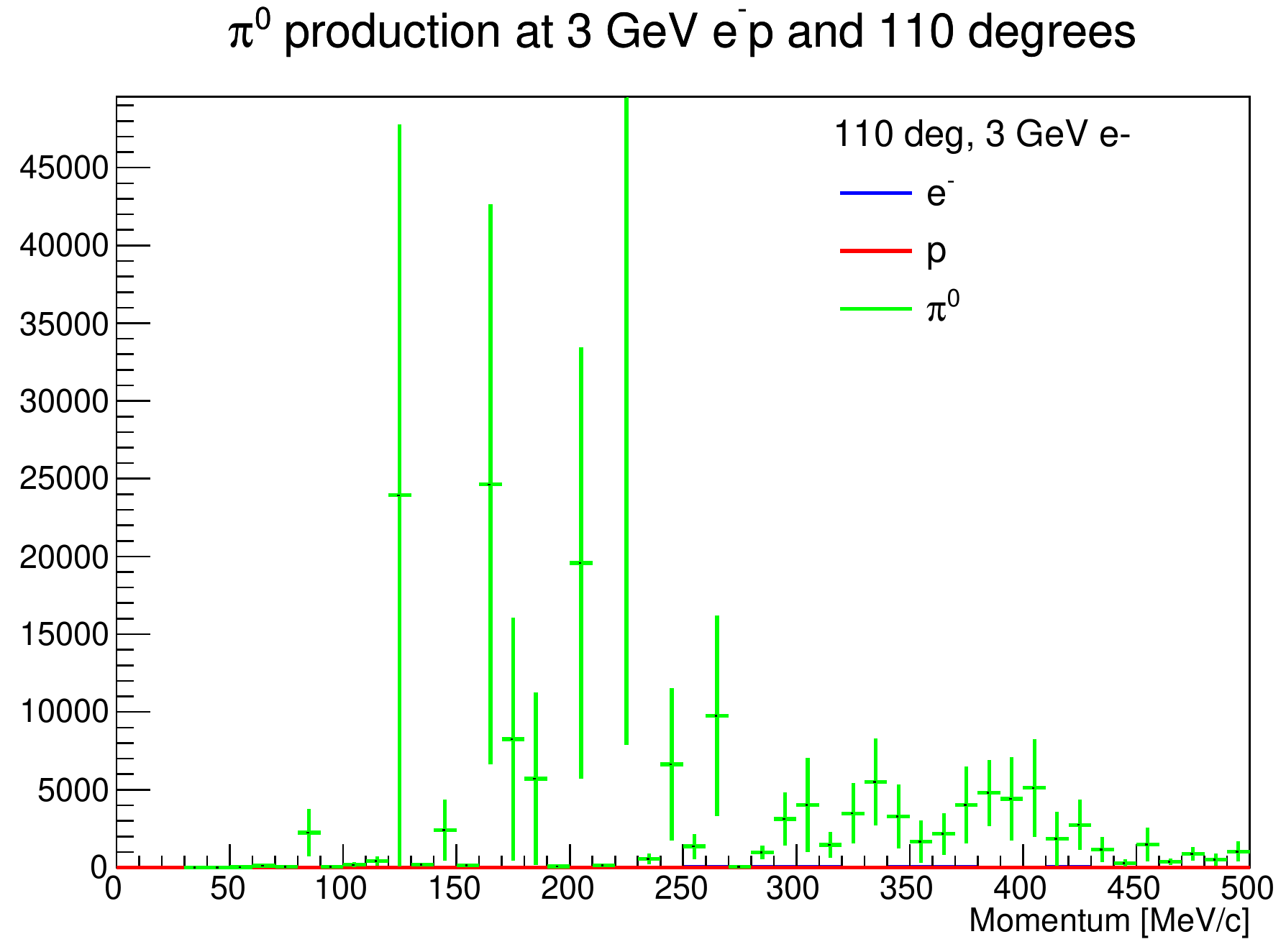}}
\end{subfloat}
\hfill
\parbox[b][0.35\textwidth]{0.47\textwidth} {\caption{Number of
    electrons, protons, and $\pi^0$ directed towards the $5\times5$
    calorimeter arrays at $30\degree$, $50\degree$, $70\degree$,
    $90\degree$, and $110\degree$ during one day of running at the
    nominal luminosity for the reaction $e^- + p \rightarrow e^- + p +
    \pi^0$ at 3~GeV.
\label{fig:Pi0_e_3}}}
\end{figure}


\section{Monte Carlo Simulation for $e^- + p \rightarrow e^- + n + \pi^+$ at 3~GeV}
\label{sec:Pi+_e_3}

\begin{figure}[!ht]
\begin{subfloat}[][]{
  \includegraphics[width=0.47\textwidth,viewport=10 5 525 390, clip]
  {Plot_Pi+_e_3_30}}
\end{subfloat}
\hfill
\begin{subfloat}[][]{
  \includegraphics[width=0.47\textwidth,viewport=10 5 525 390, clip]
  {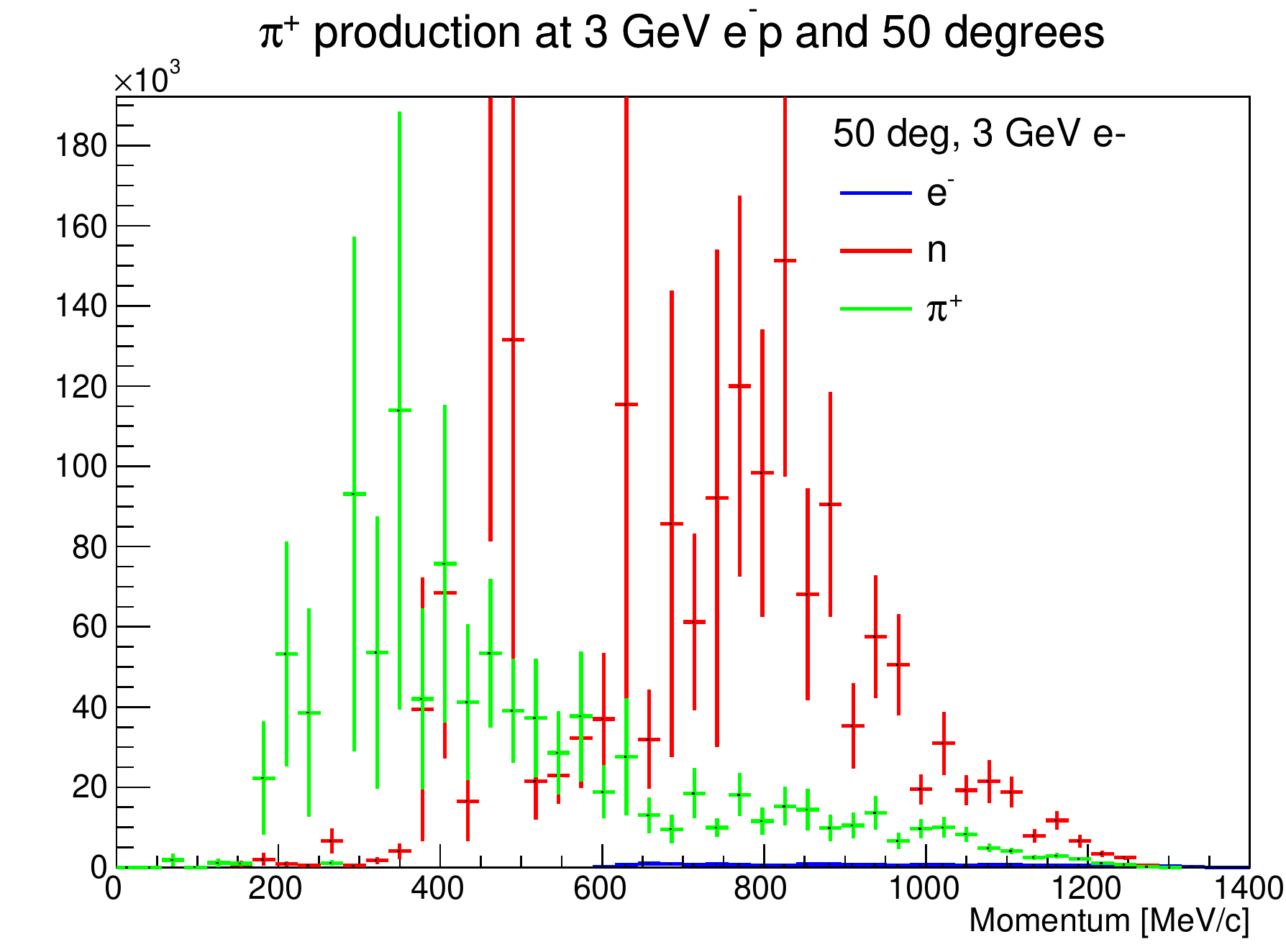}}
\end{subfloat}
\\
\begin{subfloat}[][]{
  \includegraphics[width=0.47\textwidth,viewport=10 5 525 390, clip]
  {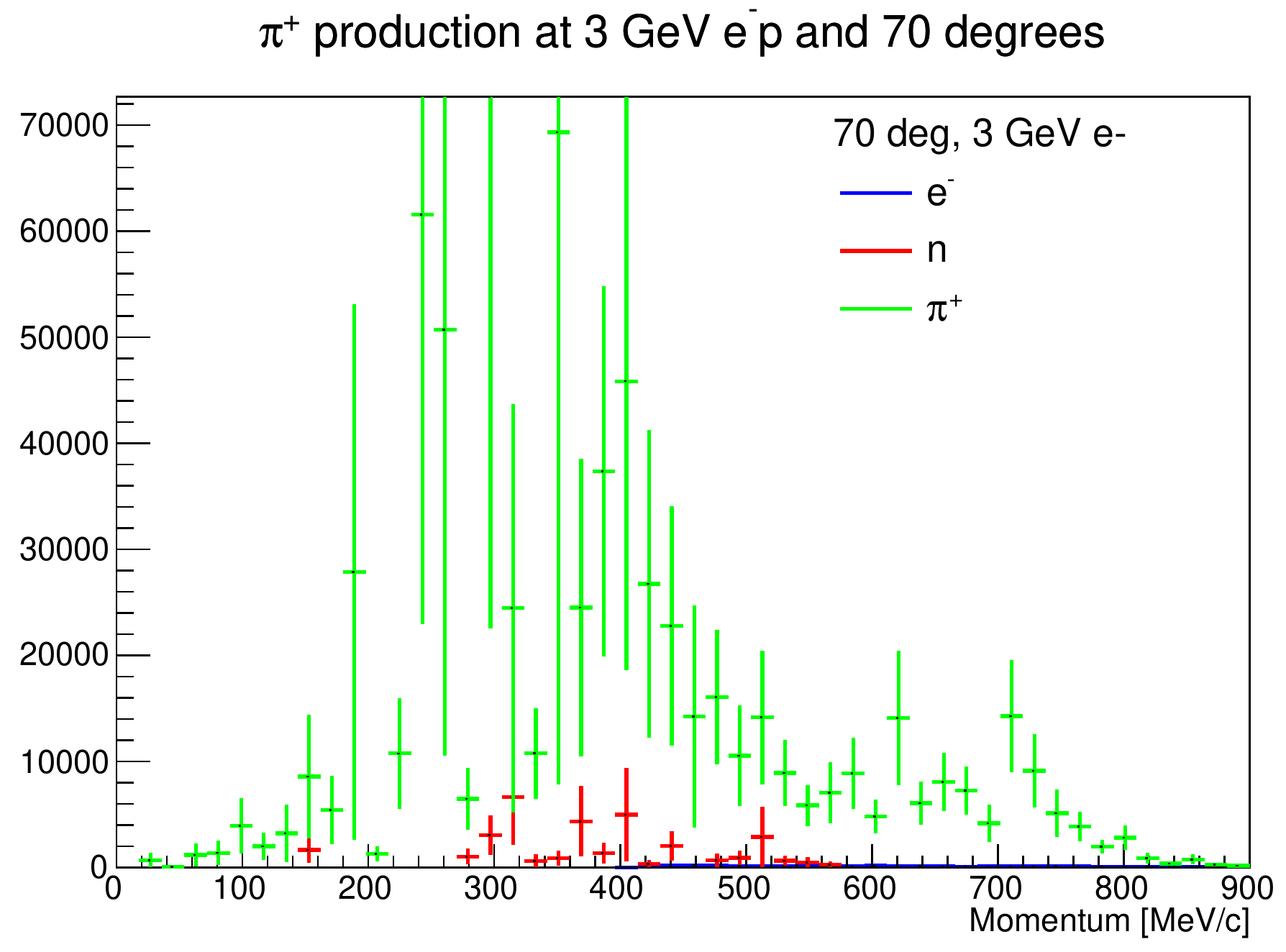}}
\end{subfloat}
\hfill
\begin{subfloat}[][]{
  \includegraphics[width=0.47\textwidth,viewport=10 5 525 390, clip]
  {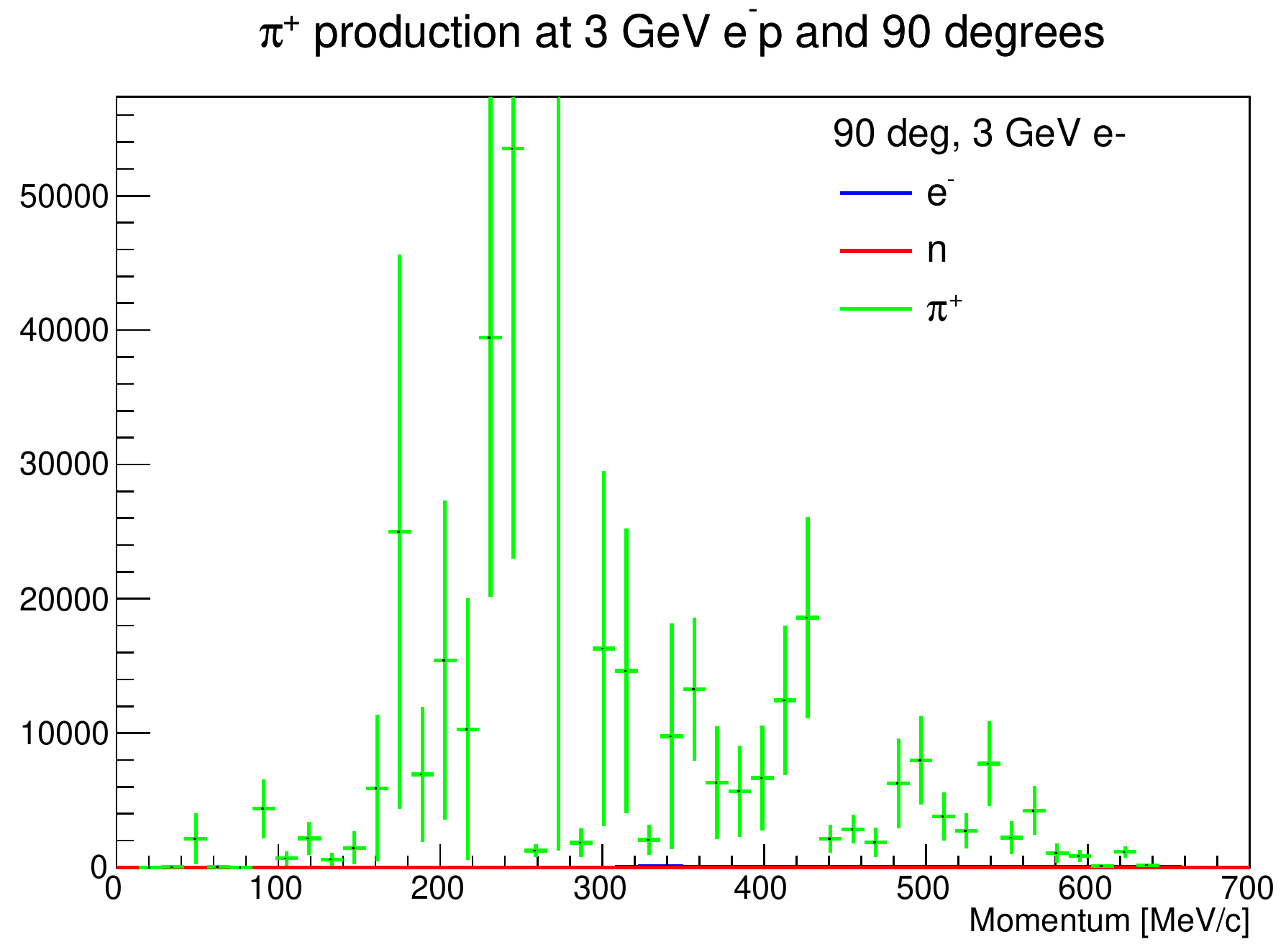}}
\end{subfloat}
\\
\begin{subfloat}[][]{
  \includegraphics[width=0.47\textwidth,viewport=10 5 525 390, clip]
  {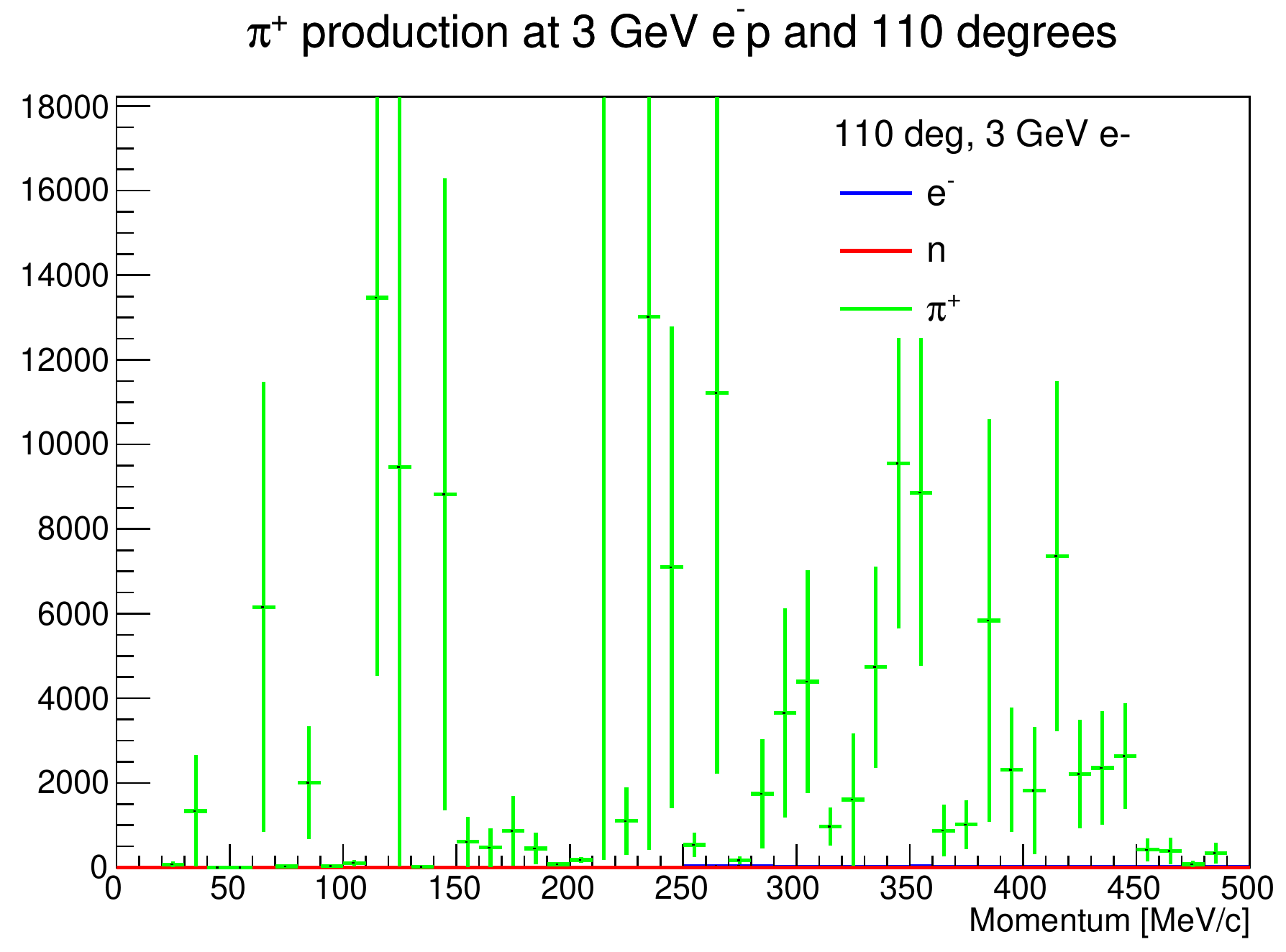}}
\end{subfloat}
\hfill
\parbox[b][0.35\textwidth]{0.47\textwidth} {\caption{Number of
    electrons, neutrons, and $\pi^+$ directed towards the $5\times5$
    calorimeter arrays at $30\degree$, $50\degree$, $70\degree$,
    $90\degree$, and $110\degree$ during one day of running at the
    nominal luminosity for the reaction $e^- + p \rightarrow e^- + n +
    \pi^+$ at 3~GeV.
\label{fig:Pi+_e_3}}}
\end{figure}


\section{Monte Carlo Simulation for $e^+ + p \rightarrow e^+ + p + \pi^0$ at 3~GeV}
\label{sec:Pi0_p_3}

\begin{figure}[!ht]
\begin{subfloat}[][]{
  \includegraphics[width=0.47\textwidth,viewport=10 5 525 390, clip]
  {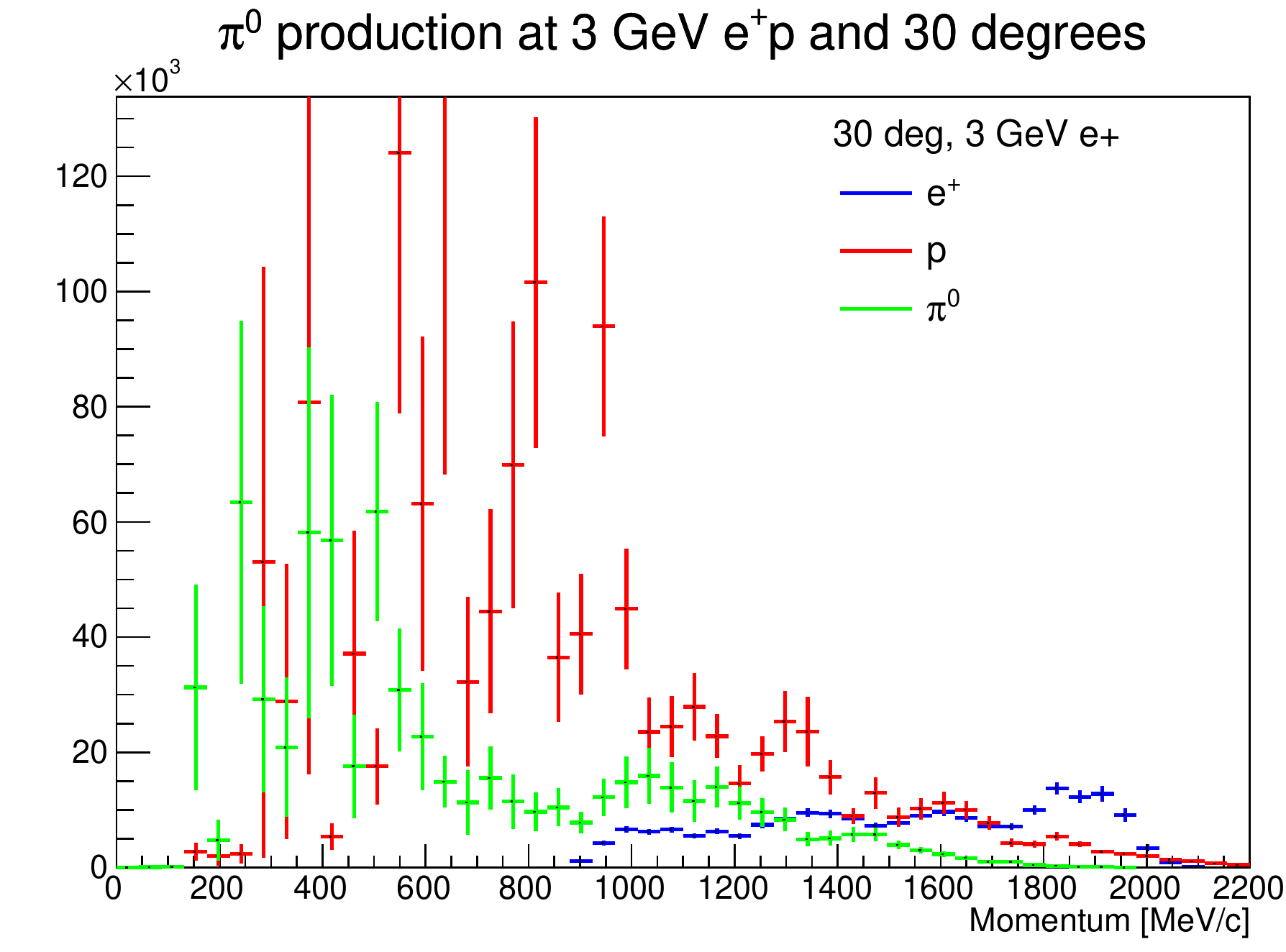}}
\end{subfloat}
\hfill
\begin{subfloat}[][]{
  \includegraphics[width=0.47\textwidth,viewport=10 5 525 390, clip]
  {Plot_Pi0_e_3_50}}
\end{subfloat}
\\
\begin{subfloat}[][]{
  \includegraphics[width=0.47\textwidth,viewport=10 5 525 390, clip]
  {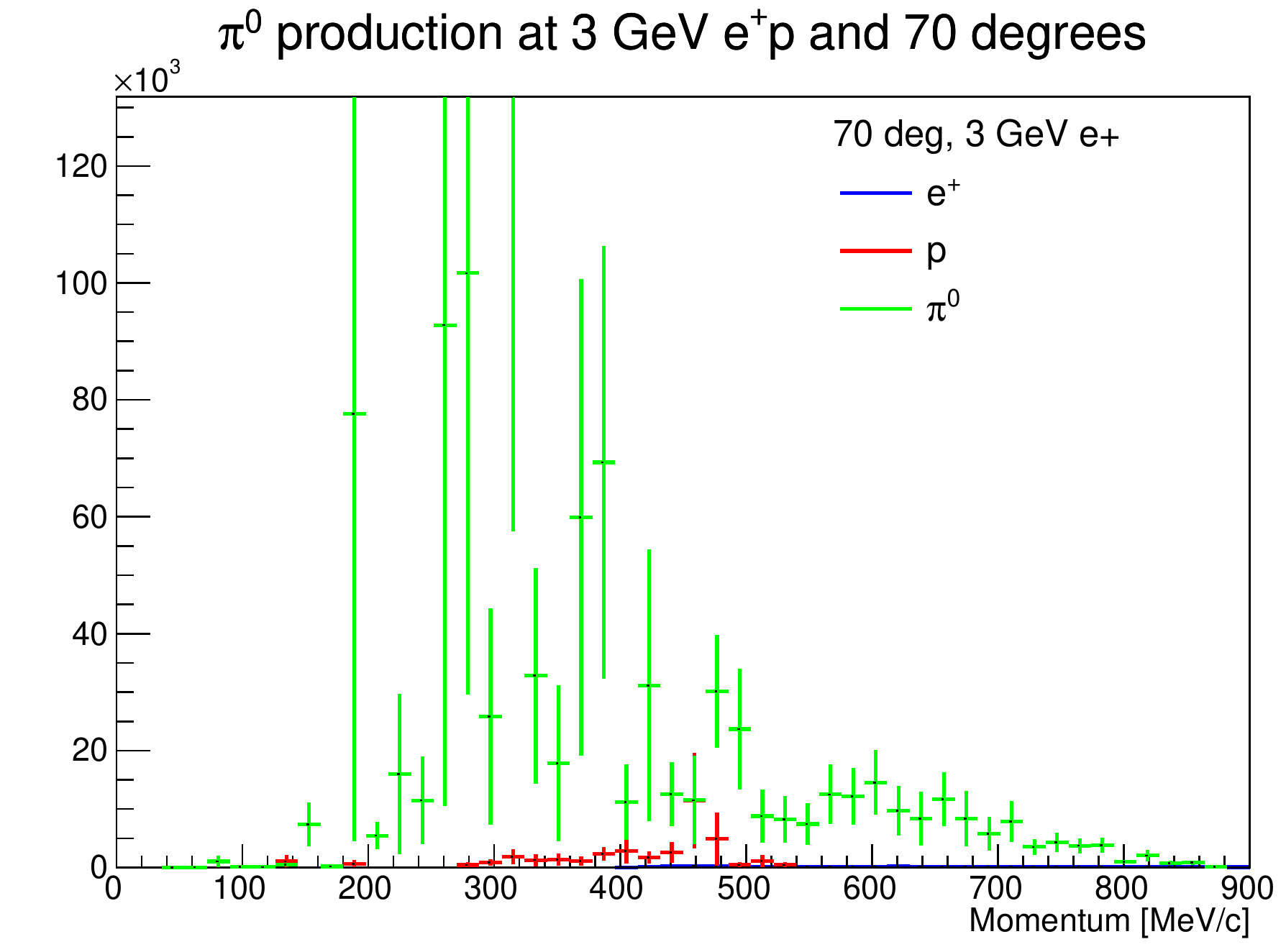}}
\end{subfloat}
\hfill
\begin{subfloat}[][]{
  \includegraphics[width=0.47\textwidth,viewport=10 5 525 390, clip]
  {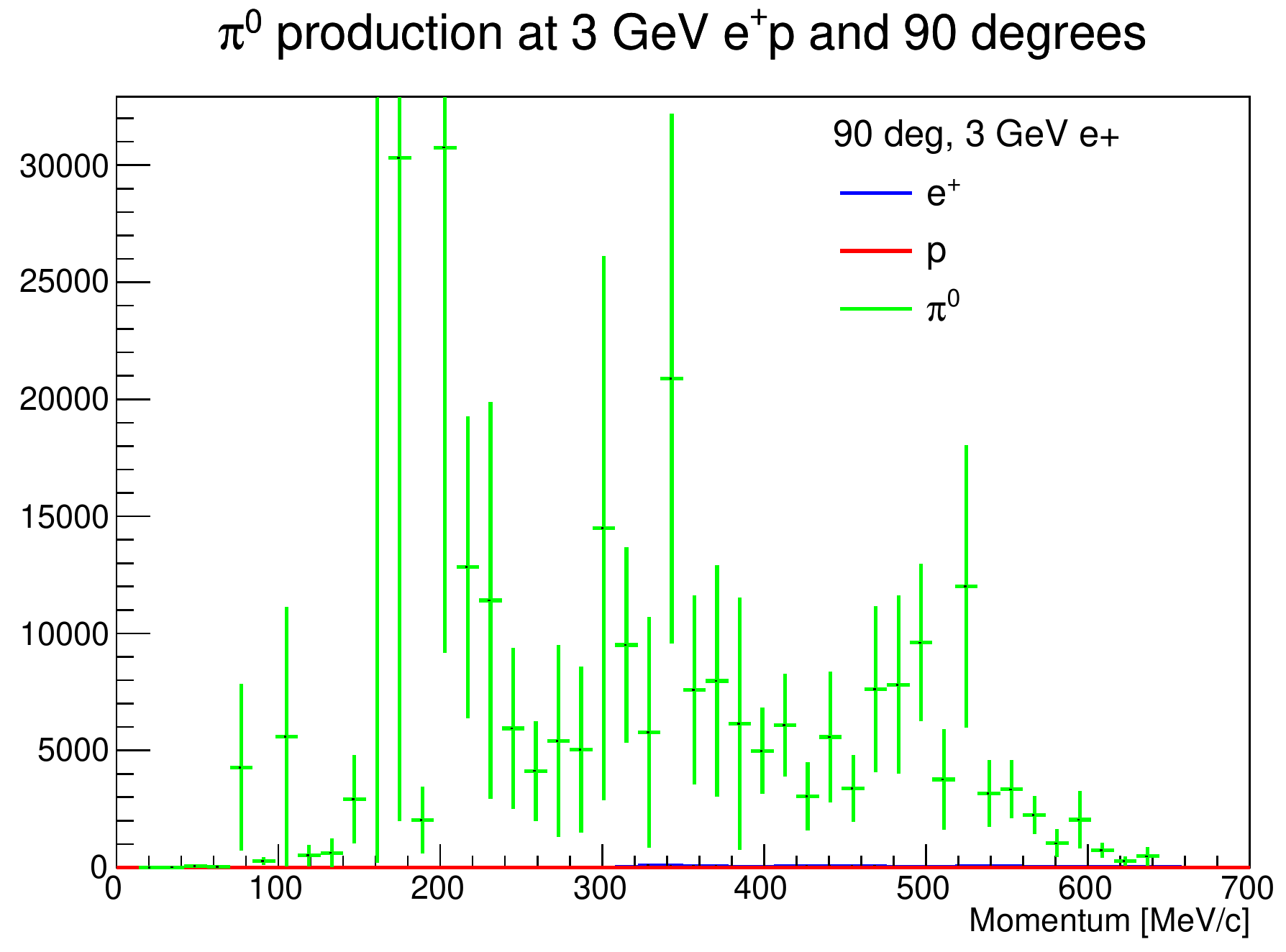}}
\end{subfloat}
\\
\begin{subfloat}[][]{
  \includegraphics[width=0.47\textwidth,viewport=10 5 525 390, clip]
  {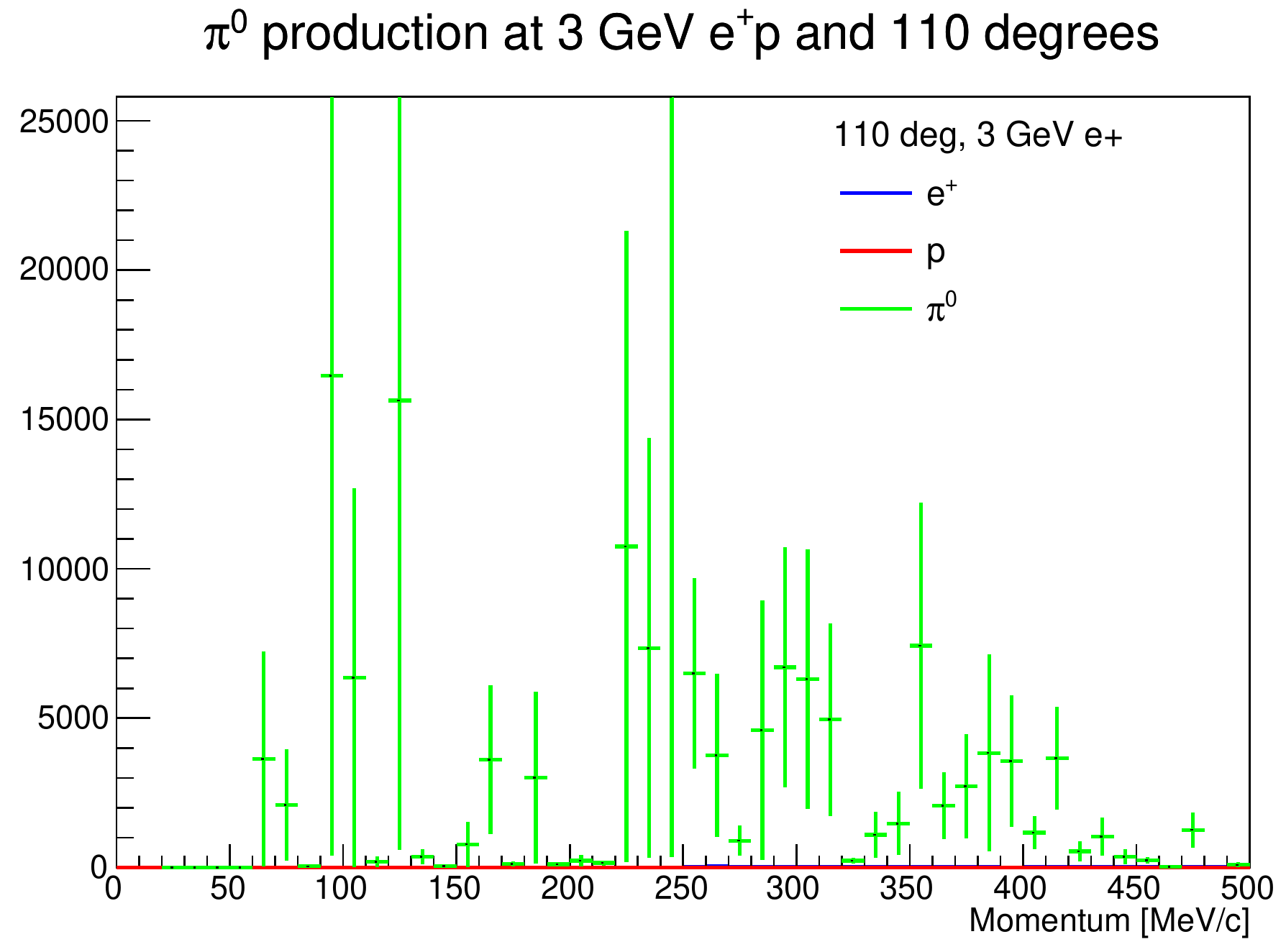}}
\end{subfloat}
\hfill
\parbox[b][0.35\textwidth]{0.47\textwidth} {\caption{Number of
    positrons, protons, and $\pi^0$ directed towards the $5\times5$
    calorimeter arrays at $30\degree$, $50\degree$, $70\degree$,
    $90\degree$, and $110\degree$ during one day of running at the
    nominal luminosity for the reaction $e^+ + p \rightarrow e^+ + p +
    \pi^0$ at 3~GeV.
\label{fig:Pi0_p_3}}}
\end{figure}


\section{Monte Carlo Simulation for $e^+ + p \rightarrow e^+ + n + \pi^+$ at 3~GeV}
\label{sec:Pi+_p_3}

\begin{figure}[!ht]
\begin{subfloat}[][]{
  \includegraphics[width=0.47\textwidth,viewport=10 5 525 390, clip]
  {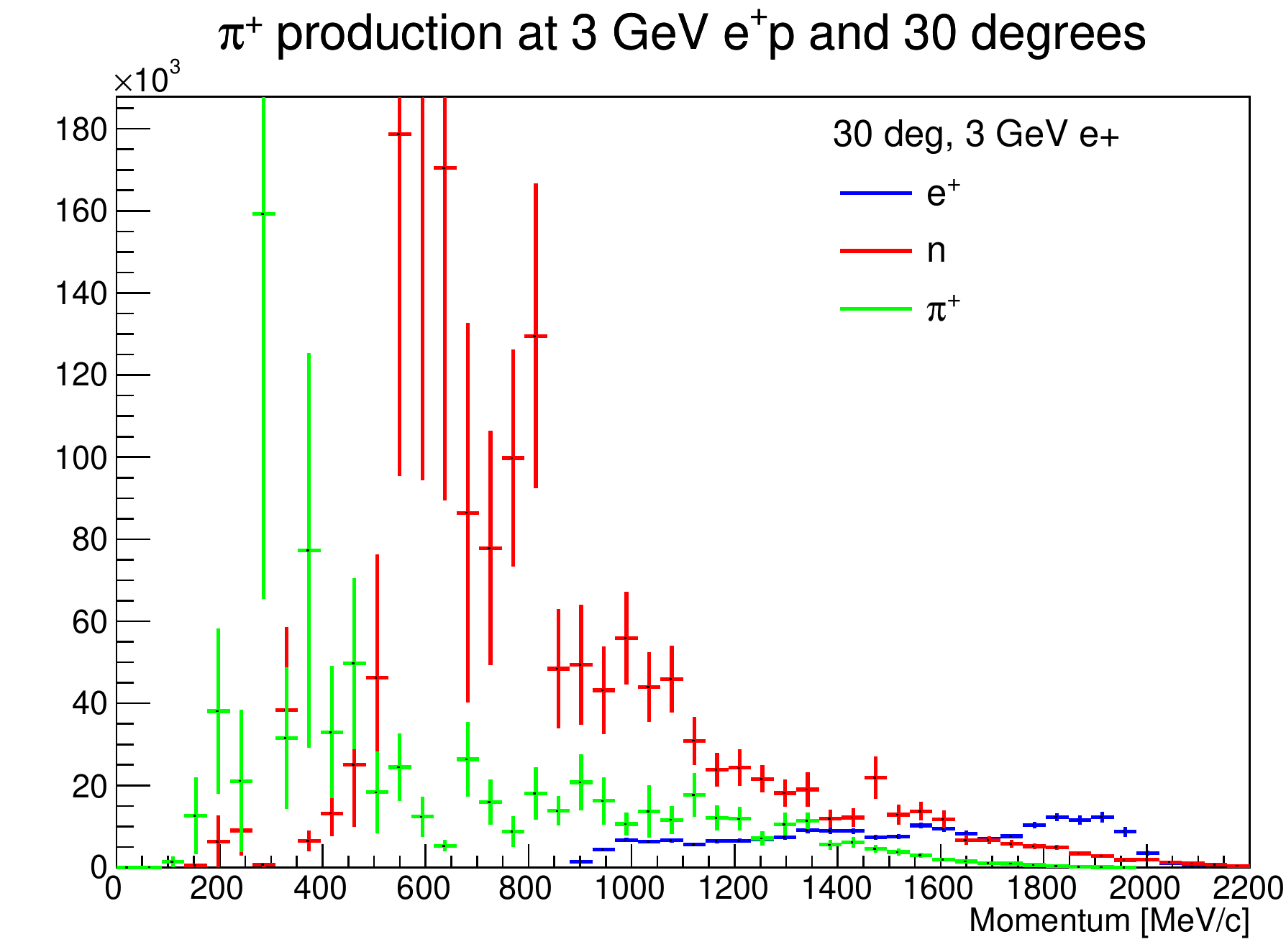}}
\end{subfloat}
\hfill
\begin{subfloat}[][]{
  \includegraphics[width=0.47\textwidth,viewport=10 5 525 390, clip]
  {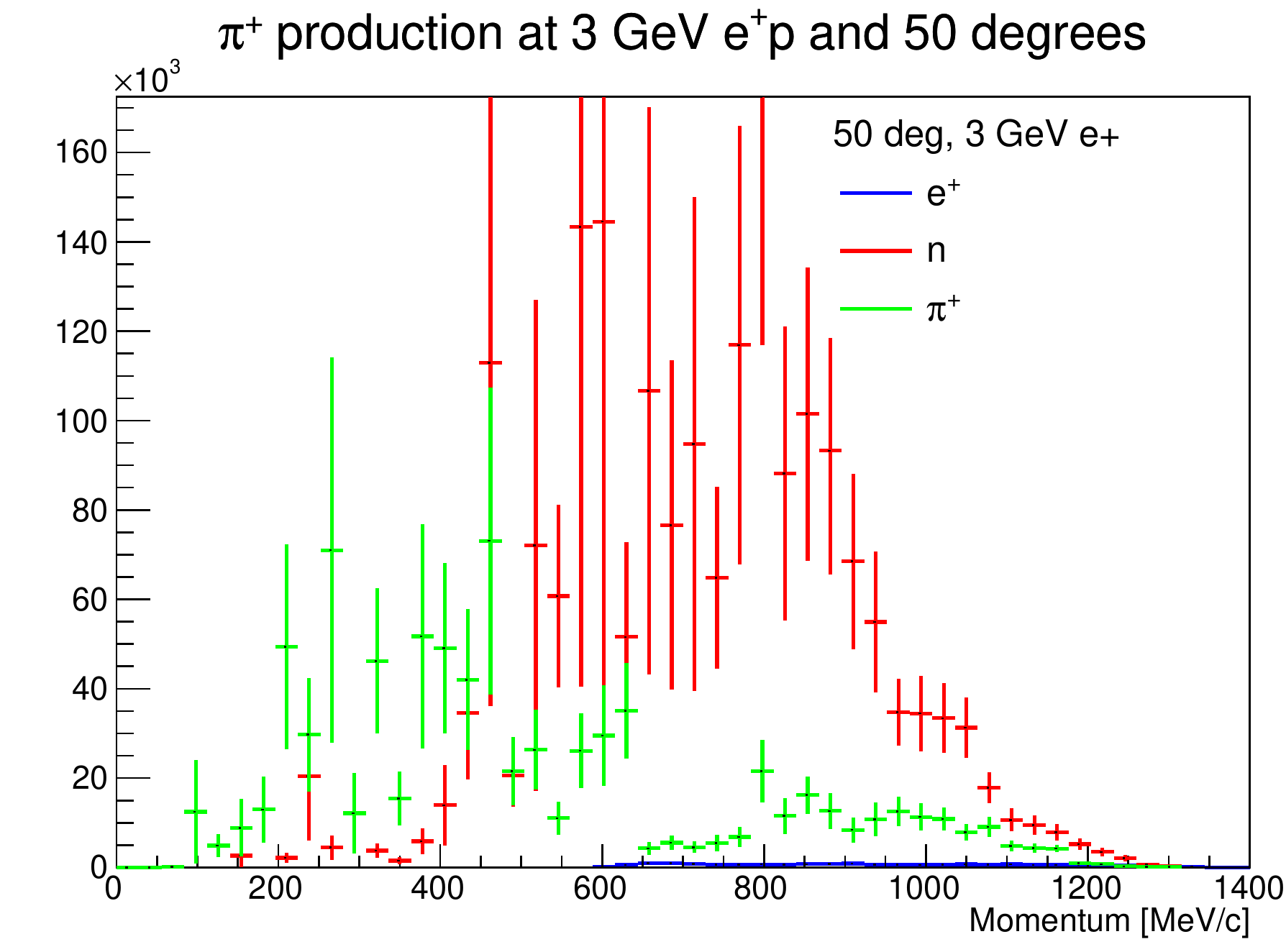}}
\end{subfloat}
\\
\begin{subfloat}[][]{
  \includegraphics[width=0.47\textwidth,viewport=10 5 525 390, clip]
  {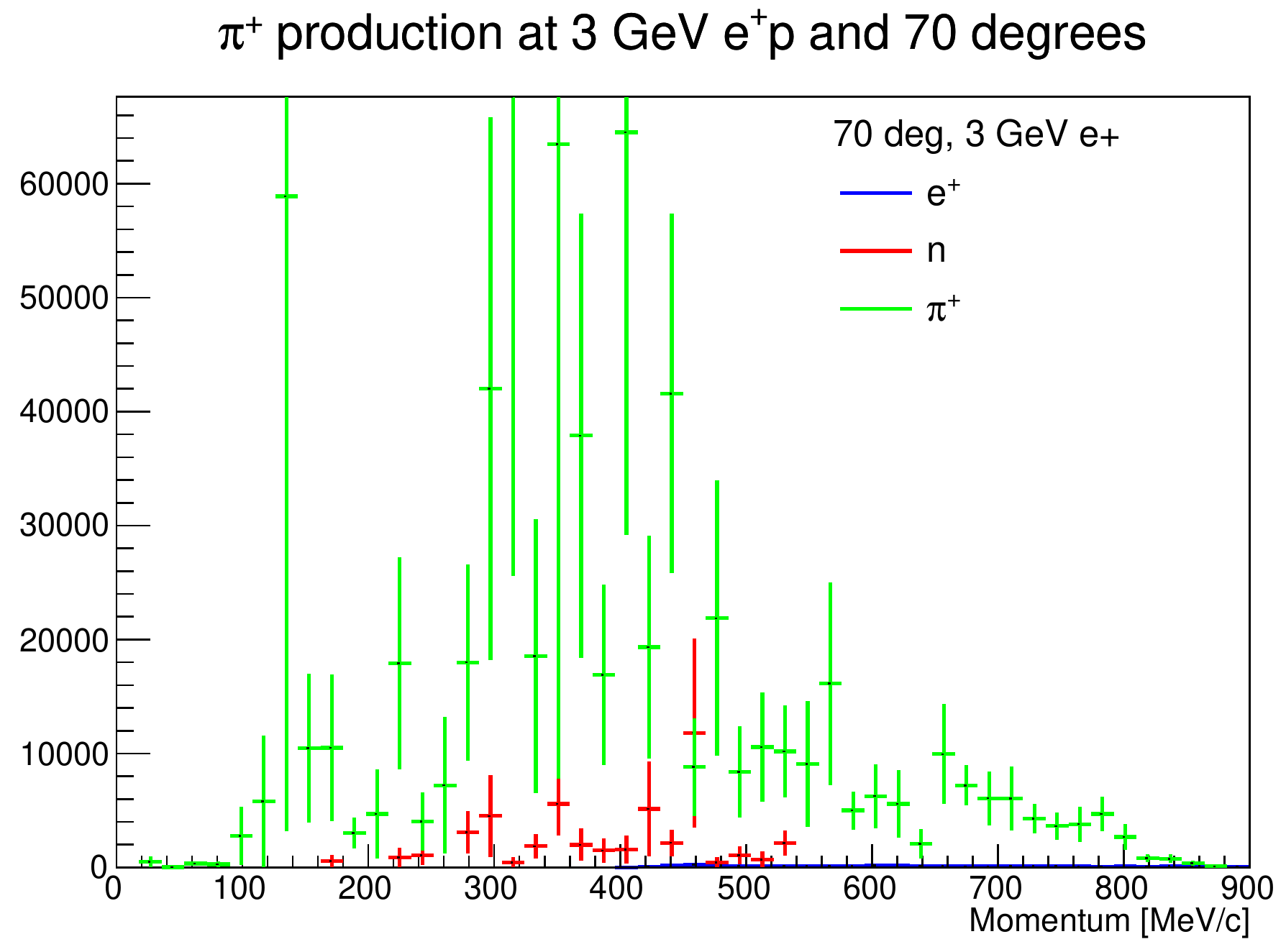}}
\end{subfloat}
\hfill
\begin{subfloat}[][]{
  \includegraphics[width=0.47\textwidth,viewport=10 5 525 390, clip]
  {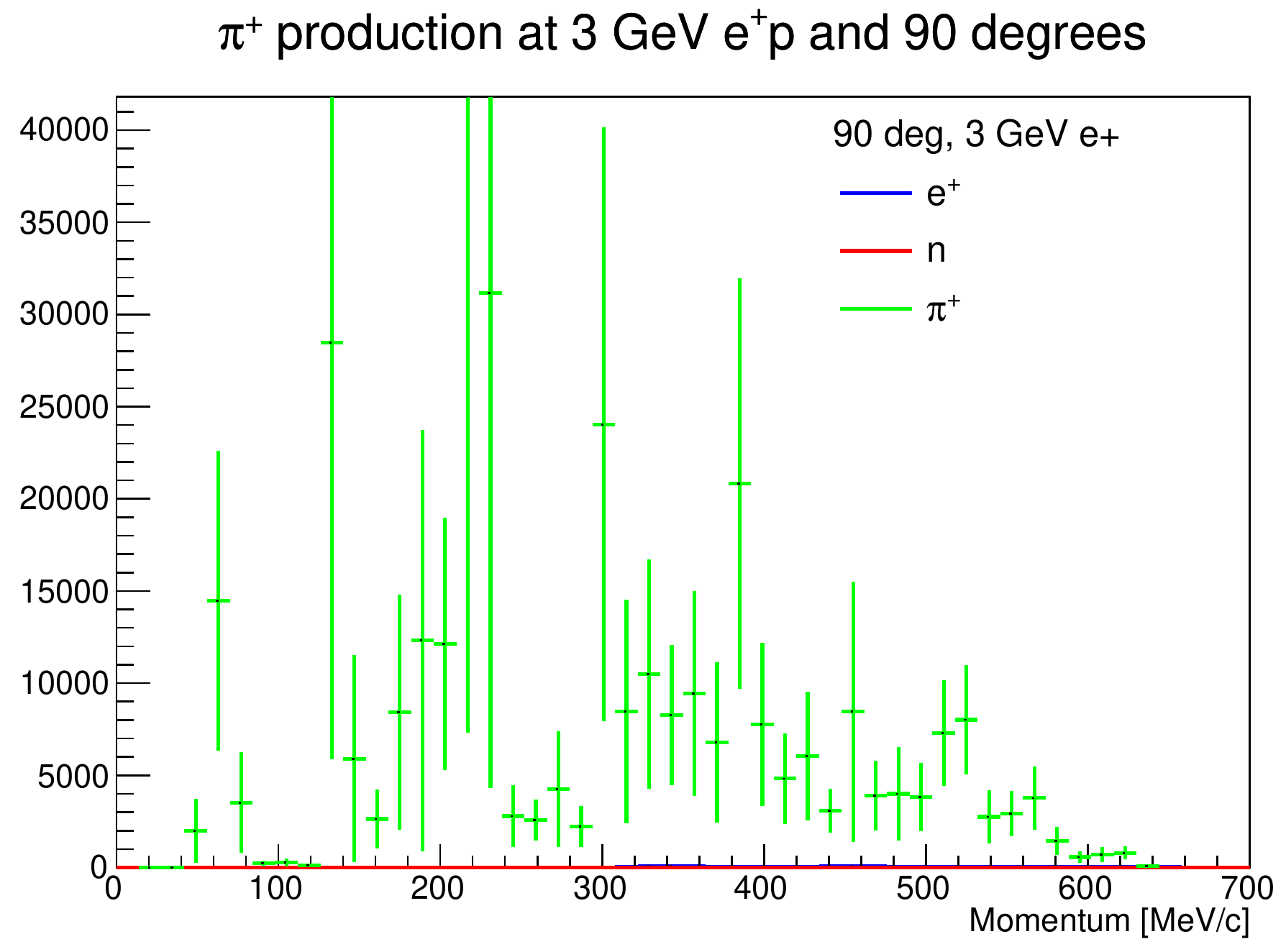}}
\end{subfloat}
\\
\begin{subfloat}[][]{
  \includegraphics[width=0.47\textwidth,viewport=10 5 525 390, clip]
  {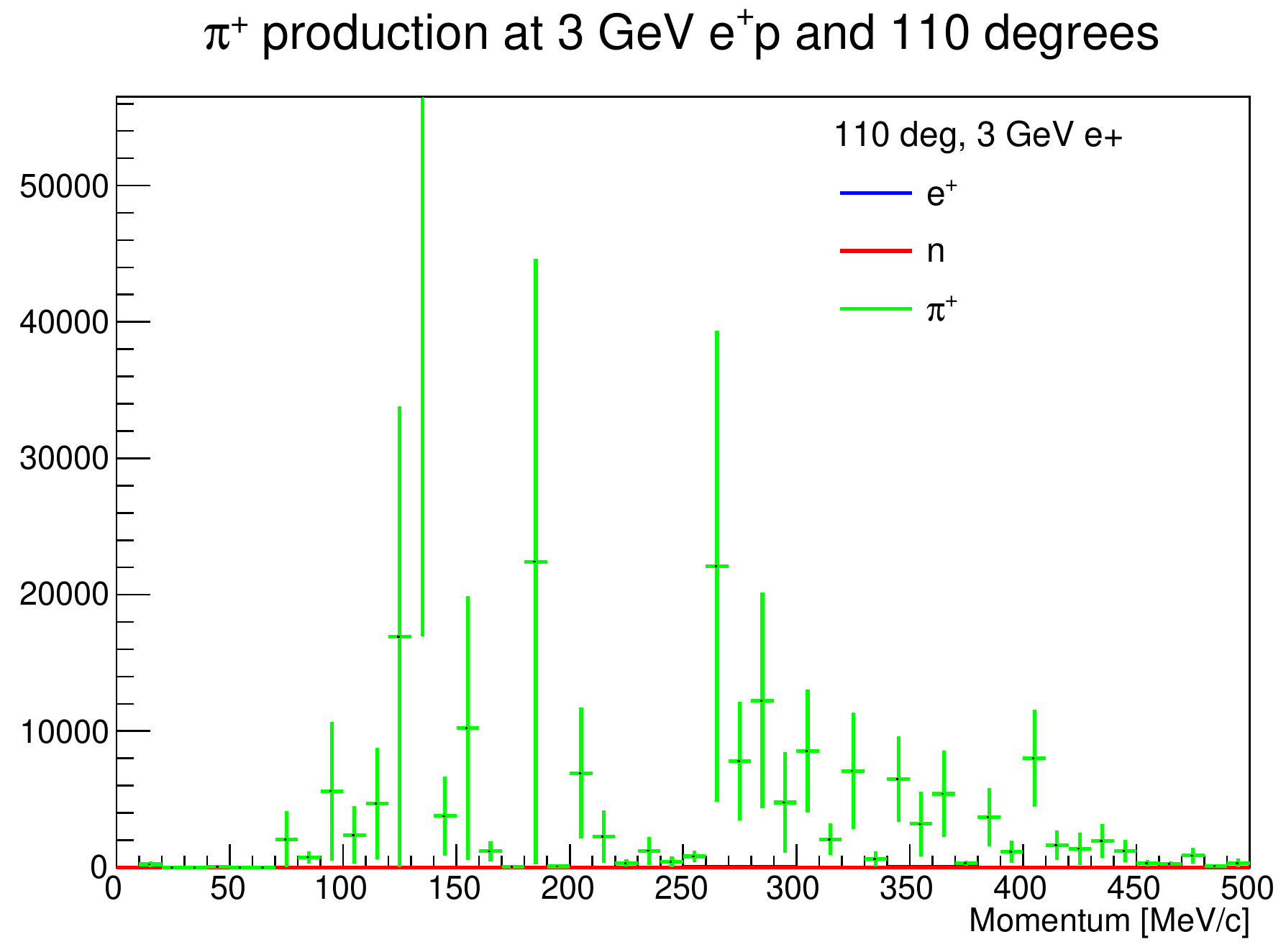}}
\end{subfloat}
\hfill
\parbox[b][0.35\textwidth]{0.47\textwidth} {\caption{Number of
    positrons, neutrons, and $\pi^+$ directed towards the $5\times5$
    calorimeter arrays at $30\degree$, $50\degree$, $70\degree$,
    $90\degree$, and $110\degree$ during one day of running at the
    nominal luminosity for the reaction $e^+ + p \rightarrow e^+ + n +
    \pi^+$ at 3~GeV.
\label{fig:Pi+_p_3}}}
\end{figure}


\section{Hydrogen Properties}
\label{sec:Hprop}

Hydrogen normally exists as a diatomic molecule, H$_2$. The molecule
occurs in two forms or allotropes: orthohydrogen, where the nuclear
spins of the two atoms are parallel ($J=0, 2, 4, \dots$); and
parahydrogen, where the nuclear spins are anti-parallel
($J=1,3,5,\dots$).

The concentrations of the two allotropes vary with temperature.  At
80~K the concentration of each is roughly the same.  At room
temperature and above it is generally 75\% orthohydrogen.  At 19~K it
is 99.75\% parahydrogen.  Various parameters are given in~\cref{tb:Hpara}.
\begin{table}[!hb]
  \centering
  \begin{tabular}{ccc}
&Para-Equilibrium&Normal\\
Critical point\\
\hline
Temperature&32.976~K&33.19~K\rule{0pt}{3ex}\\
Pressure &1.2928~MPa (12.759~atm)&1.315~MPa (12.98~atm)\\
Density&31.43~kg/m$^3$ (15.59~mol/L)&30.12~kg/m$^3$ (14.94~mol/L)\\
Normal boiling point\\
\hline
Temperature&20.268~K&20.39~K\rule{0pt}{3ex}\\
Pressure&0.101325~MPa (1 atm)&0.101325~MPa (1 atm)\\
Density (liquid)&70.78~kg/m$^3$ (35.11~mol/L)&71.0~kg/m$^3$ (35.2~mol/L)\\
Density (vapor)&1.338~kg/m$^3$ (0. 6636~mol/L)&1.331~kg/m$^3$ (0.6604~mol/L)\\
Triple point\\
\hline
Temperature&13.803~K&13.957~K\rule{0pt}{3ex}\\
Pressure&0.00704~MPa (0.0695 atm)&0.00720~MPa (0.0711 atm)\\
Density (solid)&86.50~kg/m$^3$ (42.91~mol/L)&86.71~kg/m$^3$ (43.01~mol/L)\\
Density (liquid)&77.03~kg/m$^3$ (38.21~mol/L)&77.2~kg/m$^3$ (38.3~mol/L)\\
Density (vapor)&0.126~kg/m$^3$ (0.0623~mol/L)&0.130~kg/m$^3$ (0.0644~mol/L)\\
\hline
Molecular Weight&\multicolumn{2}{c}{2.01588}\rule{0pt}{3ex}\\
\hline
  \end{tabular}
  \caption{Parameters for hydrogen}
  \label{tb:Hpara}
\end{table}

\begin{figure}[!htb]
  \centering
  \includegraphics[width=0.75\textwidth] {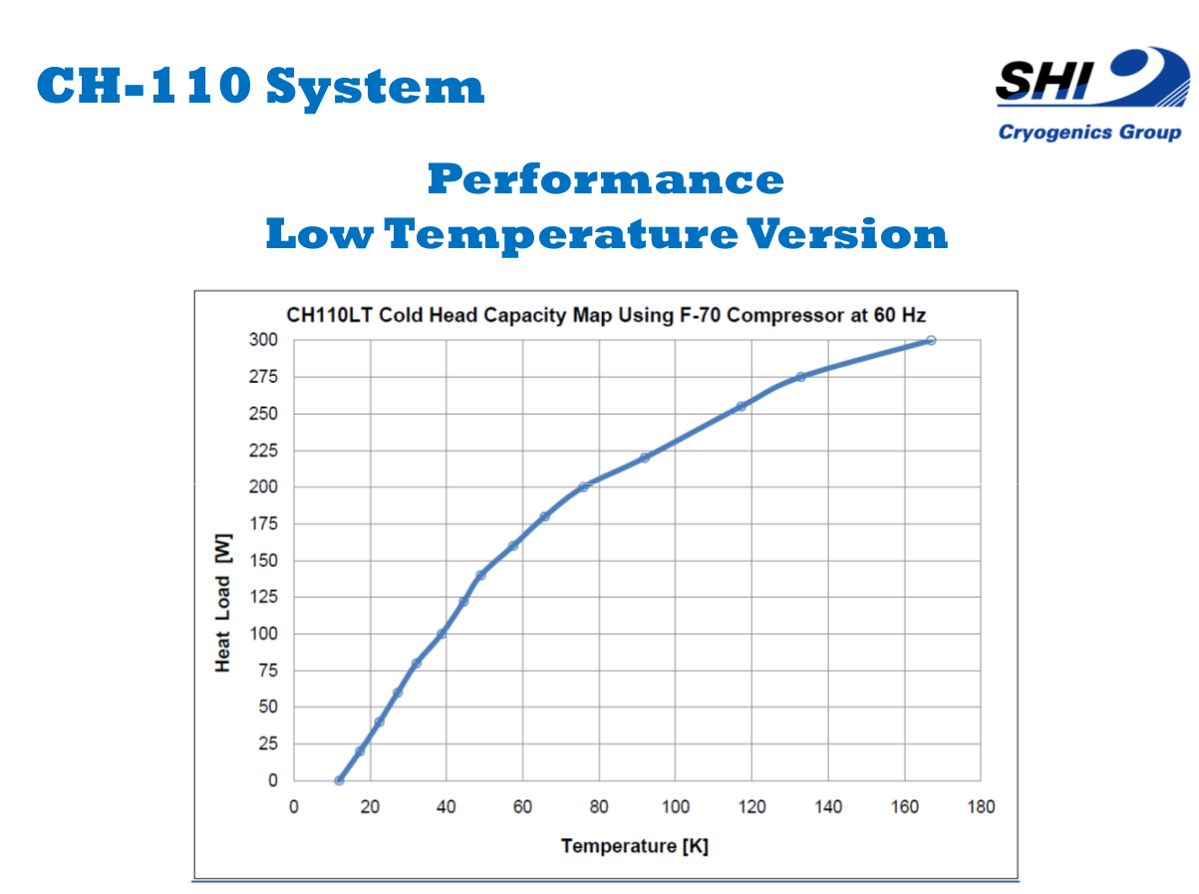}
  \caption{Cooling power of the cold head being considered for the
    TPEX liquid hydrogen target system.}
  \label{fig:CH-110}
\end{figure}

\clearpage


\section{Lead Tungstate, PbWO$_4$, Properties}
\label{sec:PbWOprop}

From the 2020 Particle Data Group Atomic and Nuclear Properties of
Materials:

\begin{tabular}{lcl}
   \\
   Density PbWO$_4$   &=& 8.300\ g$\cdot$cm$^{-3}$\\
   \\
   $2\times2\times20$~cm$^3$ crystal &=& 664.0\ g\\
   \\
   Moli\`{e}re radius &=&1.959~cm\\
   \\
   Nuclear Interaction Length $\lambda_{\rm I}$ &=& 168.3\ g$\cdot$cm$^{-2}$\\
                        &=& 20.28\ cm\\
   \\
   Radiation Length $X_0$ &=& 7.39\ g$\cdot$cm$^{-2}$\\
                &=& 0.8904\ cm\\
   \\
   Energy Loss $dE/dx$ &=& 1.229\ MeV$\cdot$g$^{-1}\cdot$cm$^{2}$\\
                &=& 10.2\ MeV$\cdot$cm$^{-1}$\\
   \\
\end{tabular}


\section{Numbers Used for Calculations in this Proposal}

\begin{tabular}{lcl}
   Density LH$_2$   &=& 0.07078\ g$\cdot$cm$^{-3}$\\
                    &=& 4.2289$\times10^{22}$\ atoms$\cdot$cm$^{-3}$\\
                    \\
20~cm LH$_2$ target &=& 8.4578$\times10^{23}$\ atoms$\cdot$cm$^{-2}$\\
\\
40~nA on 20~cm LH$_2$ target &=& 2.1116$\times10^{35}$\
cm$^{-2}\cdot$s$^{-1}\cdot$sr$^{-1}$\\
                             &=& 2.1116$\times10^{-4}$\
fb$^{-1}\cdot$s$^{-1}\cdot$sr$^{-1}$\\
\\
$1.0\times10^7$~fb &=& 7.6018~events$\cdot$s$^{-1}$~into 3.6~msr\\
\end{tabular}

\vskip1cm


\bibliography{Master.bib}


\end{document}

%% file: cryotarget_WL.tex

The OLYMPUS experiment that previously ran on the DORIS storage ring
at DESY used an internal gas target with typical areal density of
$3\times10^{15}$~atoms$\cdot$cm$^{-2}$. The lepton current averaged
around 60~mA, yielding an instantaneous luminosity about
$1.12\times10^{-6}$~fb$^{-1}\cdot\,$s$^{-1}$.

For this new experiment we propose to build a liquid hydrogen target
that will yield a luminosity about a factor of 200 times higher than
that of the OLYMPUS experiment. This higher luminosity will greatly
shorten the run time needed at 2\,GeV and help to make up for the lower
cross section at higher beam energies.

\begin{table}[ht]
\begin{center}
\caption{Target system requirements.}
 \label{tab:target_requirements}
\begin{tabular}{|l||c|}
\hline
 Parameter & Performance Requirements \\
\hline
\hline
Liquid hydrogen & T$\approx$20 K and P$\geq$1\,atm \\
\hline
Cool down time & $<$ 3 hrs. \\
\hline
Exit windows & scattering into \ang{25} -- \ang{120} and \ang{7} -- \ang{9} allowed \\
\hline
Target Cell &  end cap wall thickness t$_c \le 0.5$ mm, \\
& inner diameter 10\,mm $<$ i.d. $<$ 20\,mm,\\
& wall thickness  t$_{w} \le 1$ mm, \\
& 20 cm in length \\
\hline
\end{tabular}
\end{center}
\end{table}

In order to satisfy the science needs for TPEX, and the safety
requirements that always have to be taken into consideration for
liquid hydrogen targets, we propose to build a liquid hydrogen target
system that is tailored for this new experiment. The experimental
requirements for the target system, detailed in
Table~\ref{tab:target_requirements}, include a single, 20-cm long
liquid hydrogen target with an areal density
of $8.46\times10^{23}$~atoms$\cdot$cm$^{-2}$ that can accommodate
lepton currents up to 60\,nA (30\,nA for positrons and 60\,nA for
electrons); and long, thin scattering chamber windows to allow
particles to be accepted over the large solid angle subtended by the
detectors. Thus, the TPEX cryotarget system requires appropriate
engineering and safety considerations. The Michigan group plans to
work closely with the MIT-Bates engineers and an external company,
Creare, Inc., to design and fabricate the system.

\begin{figure*}[!htp]
\subfloat[]{\label{fig:chamber_1}
    \centering
    \includegraphics[height=6cm]{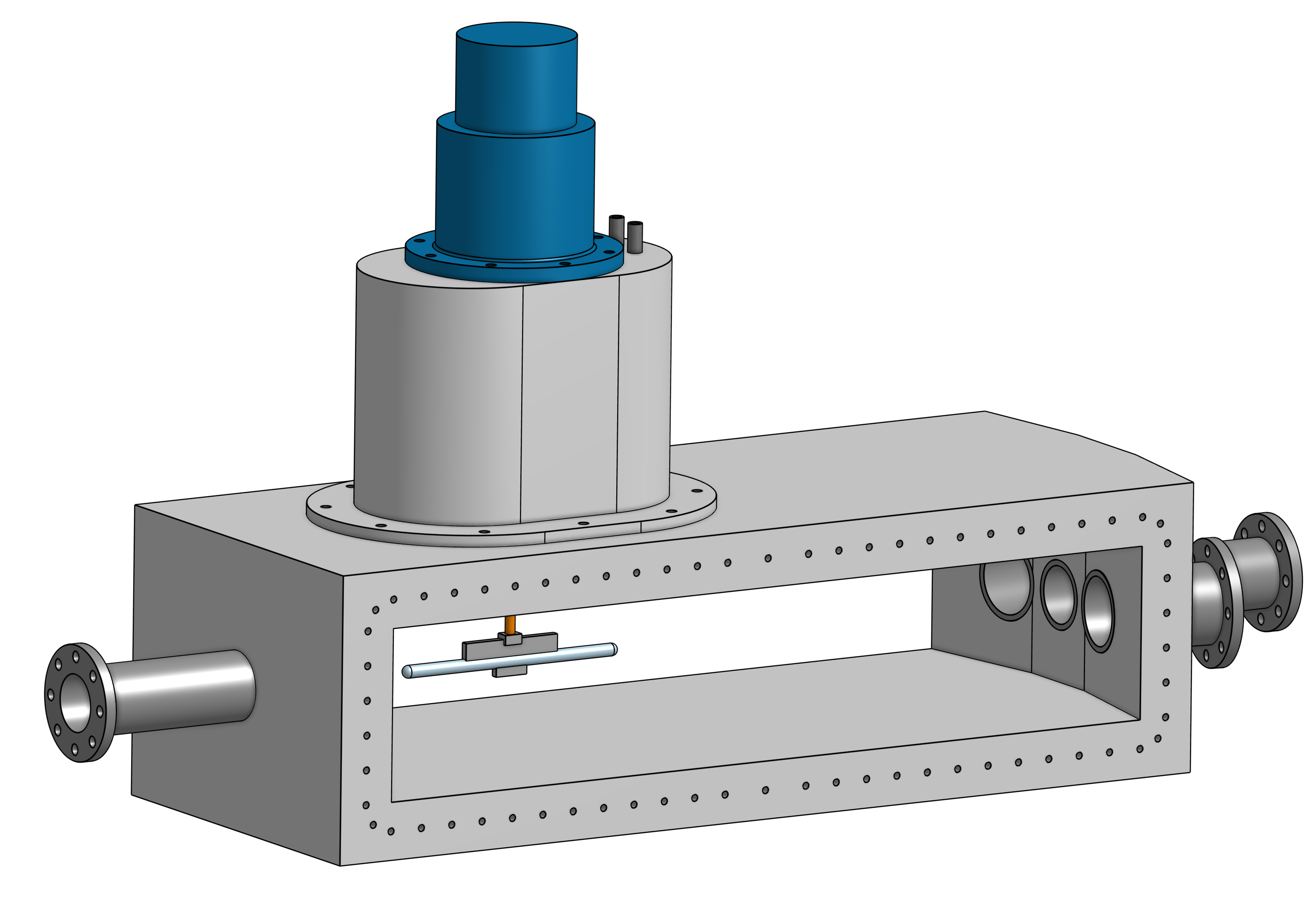}
}\qquad
\subfloat[][]{\label{fig:chamber_2}
    \centering
    \includegraphics[height=6cm]{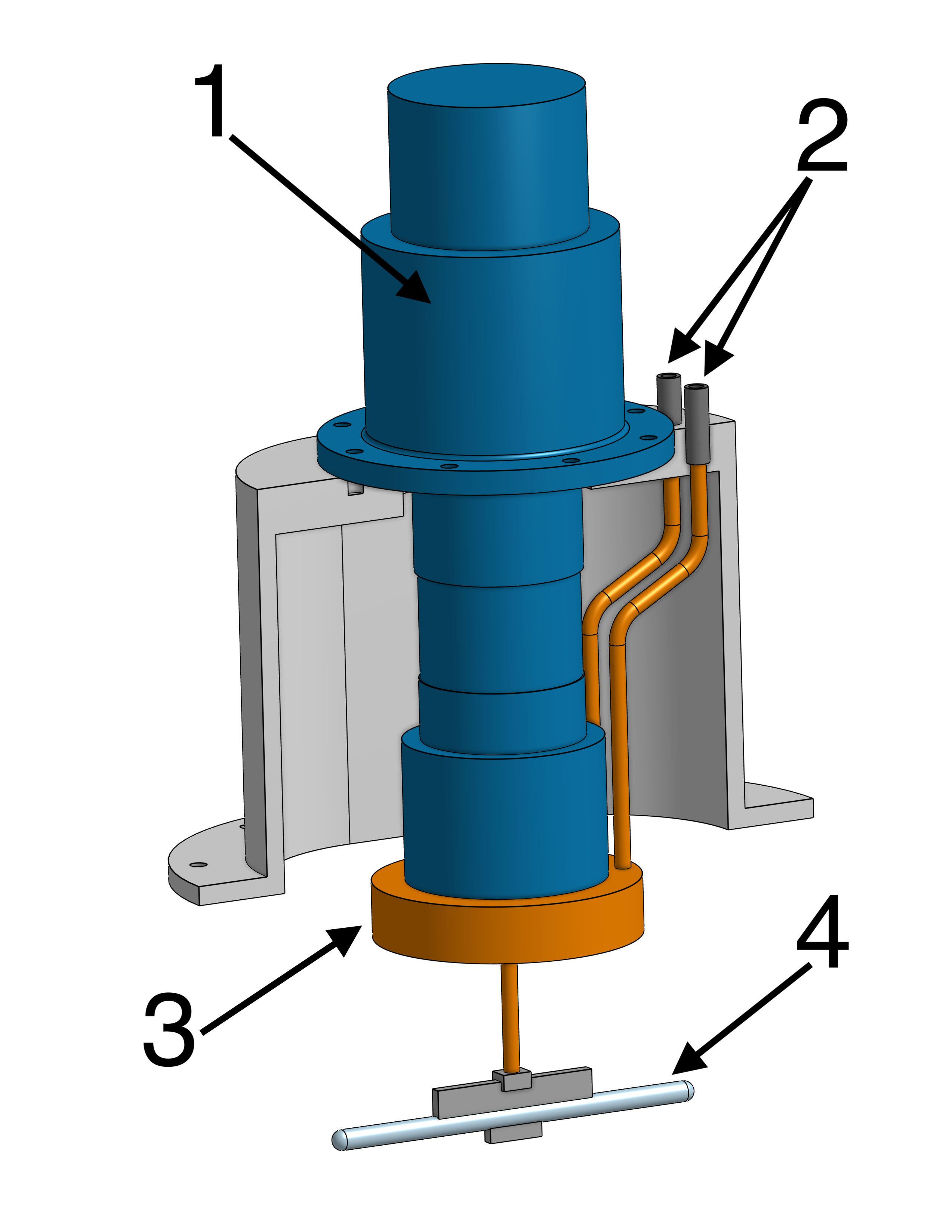}
}
\\
\subfloat[][]{\label{fig:chamber_3}
    \centering
    \includegraphics[width = 0.6\linewidth]{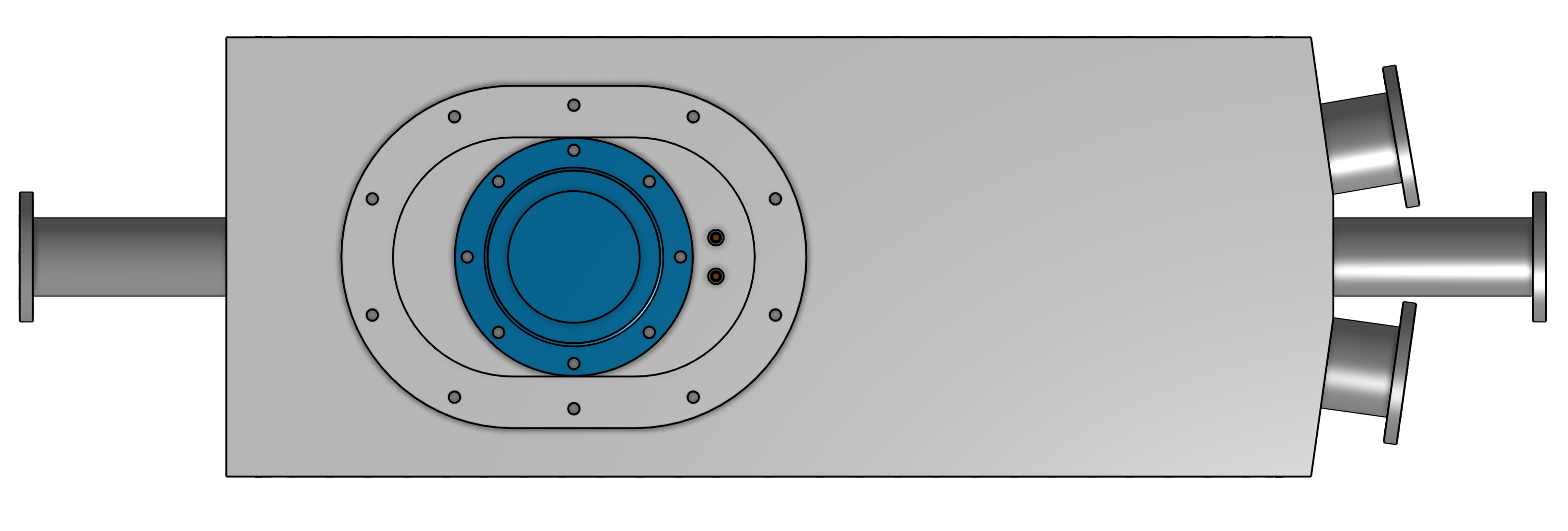}
}
\caption{Conceptual design of the TPEX target chamber:
(a) shows the full chamber view with the lepton beam entering from the
left;
(b) is a sectional drawing of the cryocooler system (1~--~CH110-LT
cryocooler, 2~--~hydrogen supply and exhaust lines, 3~--~condenser
with a cooling loop, 4~--~target cell), and (c) is the top view of the
target chamber.}
\label{fig:target_chamber_design}
\end{figure*}

Figure~\ref{fig:target_chamber_design} presents the conceptual design
of the target system. Panel (a) shows a schematic overview of the
target system, which consists of the scattering chamber, the cryo-cooler
system, and the 20\,cm long, and 2\,cm wide single target cell. 
Details of the target system are shown in panels (b) and (c). In order to maximize rigidity and withstand the enormous force from atmospheric pressure, as well as to avoid welded and bolted joints, we propose to machine the scattering chamber from a single piece of aluminum.

The dimensions of the scattering chamber windows shown in
Fig.~\ref{fig:chamber_1} are determined from the solid angle subtended
by the calorimeters. The two side
exit windows cover the polar angles for the PbWO$_4$ crystal
calorimeters in the range of $\ang{25} < \theta < \ang{120}$. At the end of the
3~m long beampipes leading to the luminosity monitors are two tiny exit windows 
cover a range of $\ang{7} < \theta < \ang{9}$. 
The vertical dimensions of the two side exit windows cover an azimuthal angle of $\phi = \ang{0} \pm \ang{10}$.

\begin{figure*}[!htp]
\centering
\includegraphics[width=.5\linewidth]{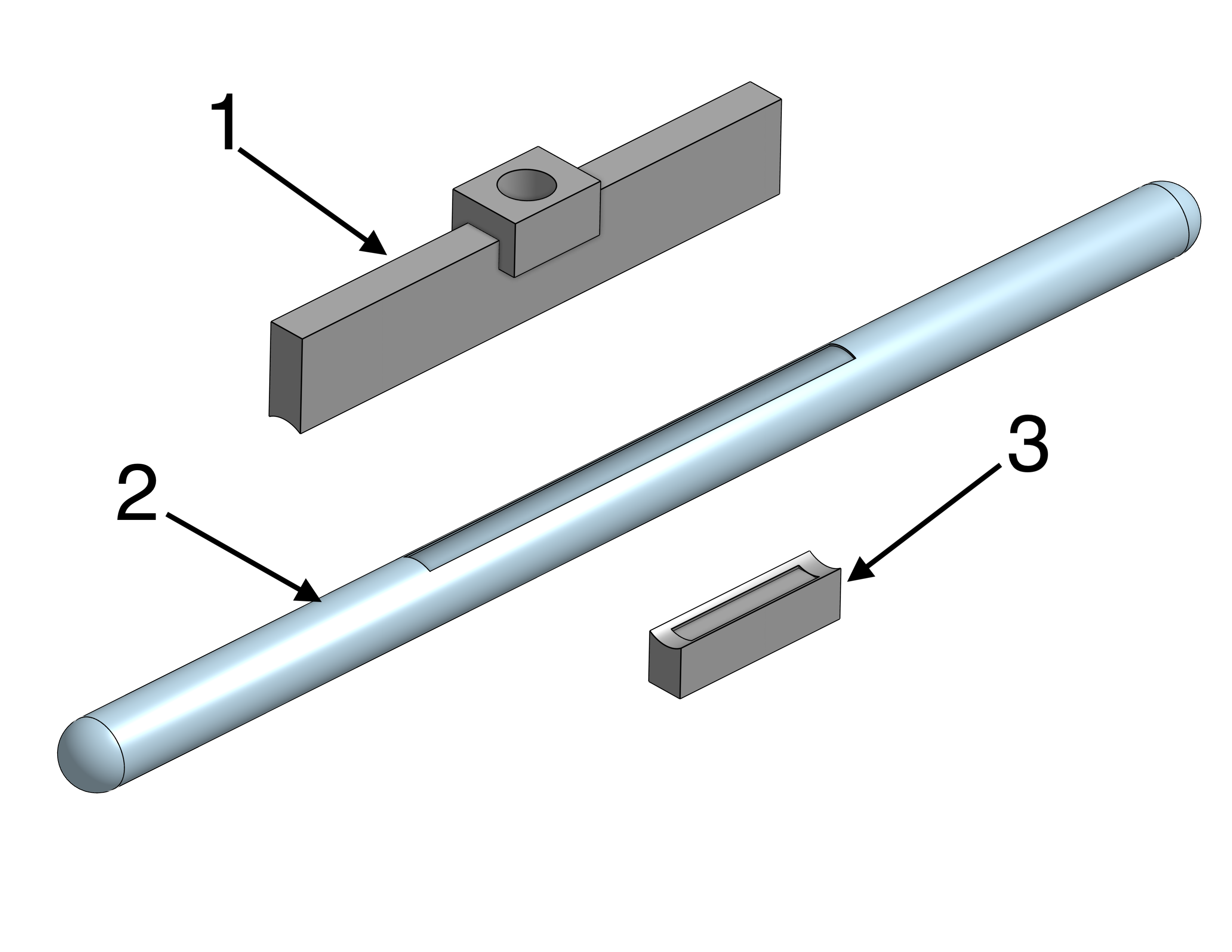}
\caption{Design overview of the 20\,cm long target cell: 
1 -- top block with liquid hydrogen level sensor, 2 -- target cell, 3
-- bottom block with temperature sensor and heater.}
\label{fig:target_cell_design}
\end{figure*}

A schematic drawing of the 20\,cm long target cell is shown in
Fig.~\ref{fig:target_cell_design}. At the time of this proposal, it
had not yet been decided if the target cell 
diameter should be 10\,mm or 20\,mm. This will in part depend on the
lepton beam properties. The general aim is to minimize the target cell
diameter to restrict the amount of hydrogen present in the target
system, while minimizing heat load in the cell walls by the beam
halo. The cell walls are made of 0.25\,mm thick, drawn aluminum tubes,
similar to those used for cigar tube travel cases. 
It is expected that the entire target system contains not more than 60 gas liters (or 75\,ml LH$_2$) of hydrogen gas.

The maximum beam heat load for the 3\,GeV electron/positron beam
impinging on the 20\,cm long cell at 60\,nA is $H= 60\, {\rm nA} \cdot
0.070\,{\rm g/cm}^3 \cdot 20\,{\rm cm} \cdot 30\, {\rm
MeV/(g/cm}^2{\rm )} = 2.5\, {\rm W}$. The thermal, or radiation heat
load on the 20\,cm long and 20\,mm diameter target cell is about
$730\,{\rm mW}/(n+1)$, where $n$ is the number of superinsulation
layers wrapped loosely around the cell. So, for the expected 10 layers
of superinsulation, the thermal heat load will be approximately 70\,mW,
which is much smaller than the beam heat load. We expect some bubble
formation in the liquid H$_2$ due to the heat load from the lepton
beams.  The long, rectangular slot in the target cell, shown in
Fig.~\ref{fig:target_cell_design}, allows the bubbles to escape into
the top aluminum block. This helps to minimize density fluctuations
and target thickness variations. The entrance and exit cups of the
target cell will be thinned by chemical etching 
to reduce the amount of material in the beam, and thus
the background caused by $e^\pm$ scattering off the target cell.

Liquid hydrogen will be filled through a single fill tube that serves
as the return tube for the boiled off hydrogen. The fill/return tube
connects the condenser, which is bolted to a cryo-cooler, with the top
aluminum block also shown in Fig.~\ref{fig:target_cell_design}. This
aluminum block also houses two liquid hydrogen level sensors (with
one serving as a backup). Each sensor is a $100\,\Omega$ Allen Bradley
carbon resistor driven at 20\,mA. One Lakeshore
Cernox$^{\text \textregistered}$ thin film resistance cryogenic
temperature sensor and one ($50\,\Omega$, 50\,W) cartridge heater are
inside the bottom copper block. The temperature sensor, the level
sensors, and the heater are all monitored and controlled by a slow
control system similar to that used in the MUSE
experiment~\cite{MUSE_Target}.

The cryo-cooler/condenser combination will closely follow the successful MUSE design~\cite{MUSE_Target}. We will therefore use the CH110-LT single-stage cryo-cooler from Sumitomo Heavy Industries Ltd~\cite{CH-100_Manual} for refrigeration. This cryo-cooler, in combination with the Sumitomo F-70 compressor~\cite{F-70_Manual}, was chosen for MUSE~\cite{MUSE_TDR} over Cryomech partly because Sumitomo has a service center in Darmstadt, Germany, while Cryomech does not have a service center in Europe. As shown in Fig.~\ref{fig:CH-110}, the cryo-cooler has a cooling power of $25$\,W at $20$\,K, which is more than sufficient to cool down and fill the 70 ml LH$_2$ target cell in approximately
2\,hours~\cite{MUSE_Target}.


Geant4 Monte Carlo simulations will be performed for this conceptual
design to verify that the experimental requirements can be met. These
simulations should tell us whether the basic cell design is
acceptable, or whether modifications to the scattering chamber exit
windows are needed to reduce background.

\subsection{Towards a functional LH$_2$ Target for TPEX}
\label{subsec:LH2-plans}

The Michigan group plans to work closely with the MIT-Bates engineers
and an external company, Creare, Inc., to design and fabricate the
system. The U-M group will start with the current conceptual design,
and improvements informed by Geant4 simulations as well as the many
lessons learned from building the cryogenic targets for the MUSE and
SeaQuest experiments, to complete the engineering design. This will be
done in close collaboration with the MIT-Bates engineers, and the
external company, Creare\footnote{Creare is a relatively large Small
Business of approximately 150 people, including 60 engineers, 50
technicians, machinists and technical specialists, and an in-house
machine shop that is accustomed high-demanding high tolerance work.},
who has performed the engineering design and the construction of the
MUSE target system. It is anticipated that this process will take
about 6 months to allow sufficient time to include periodic reviews by
the DESY for compliance and safety issues.

Safety precautions and the lack of a fully developed slow control
system at Creare will not allow a full-blown cool-down test with LH$_2$
to be performed before shipment to DESY. Instead a cool-down test with
neon, which has a similar boiling point as hydrogen but is not
explosive, has to be performed. Once general cool-down performance and
target operation in vacuum, near 20\,K, has been demonstrated, the
target system will be shipped to DESY where the neon test will be
repeated in a staging area to ensure that all components are still
functioning properly. A cool-down test with about 5\,cm$^3$ of
hydrogen, an amount small enough to be safe even if it exploded in the
cryostat vessel, will then be performed to start testing the slow
control system and the safety procedures. Finally, a complete
integration test in the Hall~2 testbeam area to fully test all
components, including slow controls and safety procedures will be
performed before starting the production run in the 2022.